\documentclass[pdftex,twocolumn,epjc3]{svjour3}          
\pdfoutput=1

\RequirePackage[T1]{fontenc}
\usepackage[utf8]{inputenc}
\usepackage[english]{babel}
\usepackage{ulem}

\RequirePackage{graphicx}
\RequirePackage{mathptmx}
\RequirePackage{mathtools}
\RequirePackage{amsmath,amssymb}
\RequirePackage{multirow}
\RequirePackage{xspace}
\RequirePackage{placeins}
\RequirePackage{appendix}

\usepackage{hhline}
\usepackage{gensymb}
\usepackage[caption=false]{subfig}

\usepackage{cite}

\RequirePackage{flushend}
\RequirePackage[colorlinks,citecolor=blue,urlcolor=blue,linkcolor=blue]{hyperref}

\journalname{Eur. Phys. J. C}

\providecommand{\nubar}{\ensuremath{\bar{\nu}}\xspace}
\providecommand{\numu}{\ensuremath{\nu_{\mu}}\xspace}
\providecommand{\numub}{\ensuremath{\nubar_{\mu}}\xspace}

\providecommand{\bracketbar}{\hbox{\kern-7pt\raise2pt%
    \hbox{{\tiny(}{\lower1.5pt\hbox{\bf--}}{\tiny)}}}}

\providecommand{\eavail}{\ensuremath{E_{\mathrm{avail}}}\xspace}
\providecommand{\ehadreco}{\ensuremath{E_{\mathrm{had}}^{\mathrm{reco}}}\xspace}
\providecommand{\ehadtrue}{\ensuremath{E_{\mathrm{had}}^{\mathrm{true}}}\xspace}
\providecommand{\qz}{\ensuremath{q_{0}}\xspace}
\providecommand{\lownu}{low-$\nu$\xspace}
\providecommand{\emu}{\ensuremath{E_{\mu}}\xspace}

\providecommand{\enutrue}{\ensuremath{E_{\nu}^{\mathrm{true}}}\xspace}
\providecommand{\enupeak}{\ensuremath{E_{\nu}^{\mathrm{peak}}}\xspace}
\providecommand{\enureco}{\ensuremath{E_{\nu}^{\mathrm{reco}}}\xspace}
\providecommand{\argon}{$^{40}$Ar\xspace}
\providecommand{\ch}{C$_n$H$_n$\xspace}
\providecommand{\chtwo}{C$_n$H$_{2n}$\xspace}

\providecommand{\varpm}{\mathbin{\vcenter{\hbox{%
  \oalign{\hfil$\scriptstyle+$\hfil\cr
          \noalign{\kern-.3ex}
          $\scriptscriptstyle({-})$\cr}%
}}}}

\begin{document}

\renewcommand{\tableautorefname}{Tab.}
\renewcommand{\figureautorefname}{Fig.}
\renewcommand{\equationautorefname}{Eq.}
\renewcommand{\sectionautorefname}{Sec.}
\renewcommand{\subsectionautorefname}{Sec.}
\renewcommand{\subsubsectionautorefname}{Sec.}
\renewcommand{\appendixautorefname}{}
\providecommand{\Appendixautorefname}{App.}

\hyphenation{gen-erator}
\hyphenation{gen-erators}
\hyphenation{det-ector}
\hyphenation{det-ectors}
\hyphenation{eff-iciency}
\hyphenation{osc-illation}
\hyphenation{nuc-lear}
\hyphenation{MIN-ERvA}
\hyphenation{GiB-UU}

\title{A substandard candle: the low-$\nu$ method at few-GeV neutrino energies}

\author{C. Wilkinson\thanksref{e1,addr1}
        \and
        S. Dolan\thanksref{addr2}
        \and
        L. Pickering\thanksref{addr3}
        \and
        C. Wret\thanksref{addr4}
}

\thankstext{e1}{e-mail: cwilkinson@lbl.gov}

\institute{Lawrence Berkeley National Laboratory, Berkeley, CA 94720, USA\label{addr1}
          \and
          CERN, 1211 Geneva 23, Switzerland\label{addr2}
          \and
          Royal Holloway, University of London, Egham, TW20 0EX, UK\label{addr3}
          \and
          University of Rochester, Rochester, NY 14627, USA\label{addr4}
}

\date{\today}
\maketitle

\begin{abstract}
As accelerator-based neutrino oscillation experiments improve oscillation parameter constraints with more data, control over systematic uncertainties on the incoming neutrino flux and interaction models is increasingly important. The intense beams offered by modern experiments permit a variety of options to constrain the flux using in situ ``standard candle'' measurements. These standard candles must use very well understood interaction processes to avoid introducing additional interaction model dependence. One option often discussed in this context is the ``\lownu'' method, which is designed to isolate neutrino interactions where there is low energy-transfer to the nucleus, such that the interaction cross section is expected to be approximately constant as a function of neutrino energy. The shape of the low-energy transfer event sample can then be used to extract the flux shape. Applications of the method at high neutrino energies (many tens of GeV) are well understood. However, the applicability of the method at the lower energies of current and future few-GeV accelerator neutrino experiments remains unclear due to the presence of nuclear and form-factor effects inherent in the interaction models.

In this analysis we examine the prospects for improving constraints on the accelerator neutrino fluxes in situ with the \lownu method in an experiment-independent way, using (anti)neutrino interactions on argon and hydrocarbon targets from the GENIE, NEUT, NuWro and GiBUU event generators. We begin by investigating the extent to which deviations from the constant cross-section assumption are dependent on poorly understood aspects of the neutrino interaction model. We then assess whether a low energy-transfer event sample can be confidently identified using experimentally accessible observables. We finally consider how the practicalities of reconstructing the energy spectrum of interacting neutrinos in realistic detectors might further limit the utility of \lownu flux constraints. The results show that flux constraints from the \lownu method would be severely dependent on the interaction model assumptions used in an analysis of neutrinos with energies below 5 GeV, and anti-neutrinos below at least 15 GeV. The spread of model predictions show that a \lownu analysis is unlikely to offer much improvement on typical neutrino flux uncertainties, even with a perfect detector. Notably---running counter to the assumption inherent to the \lownu method---the model-dependence increases with decreasing energy transfer for experiments in the few-GeV region.

\end{abstract}

\section{Introduction}
\label{sec:introduction}
There are major experimental efforts underway aimed at measuring neutrino oscillation parameters with few-GeV accelerator neutrino sources, including searches for associated CP (charge-parity) symmetry violation and additional sterile neutrino states~\cite{Esteban:2020cvm, ParticleDataGroup:2020ssz}. These include the currently operating T2K~\cite{T2K:2011qtm} and NOvA~\cite{NOvA:2007rmc} long-baseline oscillation experiments, the currently operating short-baseline neutrino program~\cite{MicroBooNE:2015bmn} (SBN) and the planned Hyper-K~\cite{Hyper-Kamiokande:2018ofw} and DUNE~\cite{Abi:2020wmh} experiments. All of these experiments measure the neutrino interaction rate at a near detector located close to the neutrino production point---where the probability for the PMNS or sterile oscillations of interest are negligible---and at a far detector located at some distance designed to optimize the sensitivity to the relevant oscillation parameters. The measured rate at each detector is the convolution of the neutrino flux, the interaction cross section, the detector efficiency, and the oscillation probability. Since neutrino oscillation probabilities evolve characteristically as a function of neutrino energy, the measured event rate is usually projected into the observable quantity that best approximates the incoming neutrino energy. To precisely infer the parameters governing neutrino oscillation from far detector data, the flux, neutrino cross-section and detector models all need to be well understood and controlled. The near detector offers an invaluable constraint on the product of these models, but due to neutrino oscillations causing a dramatic change in the neutrino energy and flavor distributions between the two detectors, the rate constraint from the near detector must be extrapolated to the far detector indirectly, using models.

Neutrino fluxes from accelerator sources are generally better predicted, but their precision is limited by hadron production uncertainties in the primary proton beam target, secondary hadron beam focusing uncertainties, and finite tolerances in the engineering of the beamline components. The uncertainty on the absolute neutrino flux is usually dominated by hadron production uncertainties, which are often constrained with dedicated hadron production measurements using thin~\cite{Abgrall:2016jif} and replica~\cite{Abgrall:2019tap} targets. Such measurements have helped reduce the absolute (shape-only) neutrino flux uncertainties to the 5--10\% (2--5\%) level in the peaks for running experiments~\cite{minervaFlux,Vladisavljevic:2018prd}, with plans for similar measurements to support the next-generation experiments, DUNE~\cite{Abi:2020wmh} and Hyper-K~\cite{Hyper-Kamiokande:2018ofw}. While a constraint on the absolute neutrino flux is important, precise modeling of the ratio of the near and unoscillated far fluxes is critical when extrapolating a rate constraint from a near detector. Uncertainties on this ratio are usually dominated by finite engineering tolerances and uncertainties in the secondary hadron beam focusing system and can only be further constrained by in situ measurements~\cite{Abi:2020qib}. 

Neutrino-nucleus interaction cross sections in the few-GeV region are challenging to measure due to a variety of complex nuclear physics effects and a number of distinct interaction channels that each contribute significantly to the total cross section. Uncertainties on the neutrino cross section vary as a function of neutrino energy and affect both the interaction rate and the relationship between the true neutrino energy, \enutrue, and the reconstructed neutrino energy, \enureco. A dedicated program to measure neutrino cross sections is underway~\cite{ParticleDataGroup:2020ssz}, but these experiments face many of the same challenges as described above: they measure the convolution of a broad and uncertain neutrino flux with the cross section, from which it is challenging to extract cross sections with the required precision~\cite{Mahn:2018mai, Betancourt:2018bpu, Avanzini:2021qlx}. Directly measuring the neutrino flux within an experiment in a way which is independent of the cross section is therefore very valuable.

Modern experiments have very intense neutrino sources which offer the possibility to use ``weak standard candles'' to constrain the neutrino flux selecting interactions that have a very well understood cross section. For example, the $\nu e^{-} \rightarrow \nu e^{-}$ neutrino-electron scattering process can be calculated with precision~\cite{Tomalak:2019}, and although the cross section is orders of magnitude lower than that of neutrino-nucleus scattering, intense modern accelerator neutrino beams will produce a sizeable event rate, as has been shown by MINERvA~\cite{Park:2015eqa, Valencia:2019mkf}, and studied in detail for DUNE~\cite{Marshall:2019vdy}. The divergence of the neutrino beam makes it challenging to constrain the neutrino flux shape well with neutrino-electron scattering data, but it can offer a precise measurement of the flux normalization for DUNE~\cite{Marshall:2019vdy}. Similarly, the inverse muon decay (IMD) process $\nu_{\mu} e^{-} \rightarrow \nu_{e}\mu^{-}$ can provide a constraint of the neutrino flux. With a threshold of $E_\nu\approx10.6~\text{GeV}$, IMD constrains mostly the higher energy neutrinos, which is of marginal use for current and planned accelerator-based oscillation experiments that observe oscillations in the $< 4$~GeV region. This has been discussed by MINERvA~\cite{MINERvA:2021dhf} and in the context of DUNE~\cite{Marshall:2019vdy}. The ``\lownu'' method---which is the focus of this work---offers an alternative approach with the same basic principle of trying to isolate neutrino interactions with well-known properties so that a neutrino flux constraint can be extracted from a measured event rate. Additional techniques for constraining the neutrino flux spectra by isolating interactions on hydrogen through kinematic techniques have also recently been proposed~\cite{Lu:2015hea, Duyang:2019prb, Munteanu:2019llq}. These either rely on a good understanding of the neutrino-hydrogen cross section, or also rely implicitly or explicitly on the ``\lownu'' method.

The $\nu$ in the ``\lownu'' method refers to the energy transfer to the nucleus, which for charged-current interactions is,  $\nu \equiv \qz=E_{\nu}-E_{l}$\footnote{In a potentially doomed effort to avoid confusion, we will consistently use ``\qz'' to denote the energy transfer and will use ``\lownu'' to denote the method, as ``$\nu$'' has multiple uses in neutrino physics.}, where $E_{\nu}$ is the incoming neutrino energy, and $E_{l}$ is the energy of the outgoing charged lepton. The \lownu method is motivated by the expression of the inclusive charged-current scattering cross section commonly used in deep inelastic scattering theory, written in terms of nucleon structure functions. The differential cross section in \qz is found by integrating $d^2\sigma/d\qz dx$ over $x = Q^2/(2M \qz)$,
\begin{align}
  \frac{\mathrm{d}\sigma}{\mathrm{d}\qz} &= \frac{G^2_{\mathrm{F}} M}{\pi} \int_0^1 \Big(F_2 - \frac{\qz}{E_\nu}  \left[F_2 \varpm xF_3\right]
      + \frac{\qz}{2E_\nu^2}  \left[\frac{Mx(1-R_{\mathrm{L}})}{1+R_{\mathrm{L}}}F_2\right]  \nonumber \\
      &+ \frac{\qz^2}{2E_\nu^2}  \left[\frac{F_2}{1+R_{\mathrm{L}}} \varpm xF_3\right]  \Big)\,\mathrm{d}x,
\label{eq:low-nu}
\end{align}
\noindent where $M$ is the struck nucleon mass, $F_2$ and $xF_3$ are structure functions and $R_{\mathrm{L}}$ is the structure function ratio $F_2/(2xF_1)$, with $G_{\mathrm{F}}$ being Fermi's constant, and the $+(-)$ is used for (anti-)neutrinos~\cite{Mishra:1990ax}. The central idea behind the \lownu method is that for low values of \qz, the $\qz/E_{\nu}$ terms in \autoref{eq:low-nu} are small, so the cross section is approximately constant in $E_{\nu}$.
An event sample with low \qz---with the signature being a single forward-going muon and no other reconstructed particles and few other observable energy deposits---can then be used to measure the flux shape as a function of the neutrino energy. In practice, the calculated flux shape from the \lownu method is combined with an accurately measured neutrino cross section at high energy to provide a constraint of the neutrino flux spectra.

The method relies on three key requirements:
\begin{enumerate}
\item In the low \qz region of interest, the cross section is either constant as a function of $E_{\nu}$ or the non-constant behaviour is well understood.
\item A low-\qz sample can be isolated experimentally, without introducing significant model-dependent corrections.
\item The reconstructed neutrino energy for events in the selected sample, \enureco, can be related to the true neutrino energy, \enutrue, in a way that does not introduce significant model dependence.
\end{enumerate}
Additionally, as a practical matter, the low-\qz sample should be small compared to the total sample in the analysis to avoid double-counting events in subsequent analyses that use the \lownu flux constraint.

Intense modern neutrino beams and capable detectors provide high statistics datasets which potentially allow the use of the stringent cuts on \qz, but the demanding precision sets strong requirements on the model-independence of any method to measure the flux distribution. There has not been a comprehensive study of multiple generators and models in the context of the \lownu method.
In this paper, we use a variety of initial nuclear state, neutrino--nucleon interaction, and final state interaction models that are included in modern neutrino interaction generators to investigate how well the above requirements can be fulfilled by current and future accelerator neutrino oscillation experiments, and whether the \lownu method is able to provide them with a reliable in situ flux constraint. We focus on the behaviour in the $\enutrue \leq 15$ GeV region, which covers the vast majority of the neutrino energy ranges of interest for current and future accelerator neutrino oscillation experiments: SBN, which uses the booster neutrino beam~\cite{MiniBooNE:2008hfu} with an unoscillated peak neutrino energy, \enupeak, of $\enupeak \approx 0.8$ GeV; NOvA which uses the off-axis NuMI beam~\cite{numi, minervaFlux} with $\enupeak \approx 2.0$ GeV; T2K/Hyper-K which both use or will use the off-axis J-PARC beam~\cite{T2K:2012bge} with $\enupeak \approx 0.6$ GeV; and DUNE which will use the LBNF beam~\cite{DUNE:2020ypp, Abi:2020qib} with $\enupeak \approx 2.5$ GeV. We investigate both \numu and \numub interaction on \argon, relevant for SBN and DUNE~\cite{DUNE:2021tad} (which will also have a \chtwo near detector component), and provide corresponding results on \ch and \chtwo in~\autoref{app:chtarget}, which are also relevant for NOvA and T2K/Hyper-K. We neglect detailed discussion of the statistical uncertainty on potential \lownu samples from each experiment to keep discussion general. Additionally, aside from a few comments which aim to contextualize studies of the \lownu method on hydrogen, we focus on the application of the \lownu method in nuclear targets.

In \autoref{sec:history} we provide a brief history of the \lownu method and motivate this work. In \autoref{sec:models} we discuss the different neutrino interaction simulations used. In \autoref{sec:low-qz-cross-section}, we investigate differences between the low-\qz cross sections from the models introduced in \autoref{sec:models}. In \autoref{sec:variables}, we discuss experimentally accessible observables and their relationship with the true kinematic variables \qz and \enutrue. Finally, we present our conclusions in \autoref{sec:conclusions}.

\section{History and motivation}
\label{sec:history}
The \lownu method was developed in the context of the CCFR experiment~\cite{Auchincloss:1990tu,Seligman:1997fe} and is generally attributed to Ref.~\cite{Mishra:1990ax}, but is very closely related to earlier work by Rein and Belusevic~\cite{Belusevic:1987rn,Belusevic:1988ab} on the ``low-$y$'' and ``$y$-intercept'' method, where $y=\qz/E_\nu$. Various incarnations of these methods have been used by the CCFR~\cite{Seligman:1997fe}, NuTeV~\cite{NuTeV:2005wsg}, NOMAD~\cite{Petti:2006tu}, MINOS~\cite{MINOS:2009ugl} and MINERvA~\cite{MINERvA:2016ing,MINERvA:2021mpk} collaborations. The \lownu method has recently been discussed by the MicroBooNE~\cite{MicroBooNE:2021cue} and DUNE~\cite{DUNE:2021tad} collaborations for use in liquid argon detectors.

A key concern when using the formalism of \autoref{eq:low-nu}---which is motivated by deep-inelastic scattering theory---is in what region it breaks down, and whether corrections incurred at very low values of \qz are model dependent. This is discussed by Rein and Belusevic in the context of the low-$y$ method~\cite{Belusevic:1987rn, Belusevic:1988ab}, where they consider the size of the corrections required if the neutrinos interact on free nucleons via quasielastic, resonant, and coherent channels, and the implications for constraining the neutrino flux. They isolate a region $\qz\lesssim1.67$ GeV, and find the ``low-$y$'' cross section becomes energy independent around $E_\nu\gtrsim20$ GeV; roughly when $E_\nu \gg M_{\mathrm{N}}$ for quasi-elastic events, and $E_\nu \gg M_{\mathrm{R}}$ for resonant events, where $M_{\mathrm{N}}$ and $M_{\mathrm{R}}$ are the nucleon and resonance masses respectively. Their calculation of the cross section for $\qz\lesssim1.67$ GeV has an 11.1\% uncertainty, and is dominated by uncertainties in the data for neutrino induced single pion production on the nucleon.
Importantly, there has been no new neutrino-nucleon single pion production data since these studies.
Much of the uncertainty comes from not understanding the contributions that heavy resonances have on the cross section, which vary between pion production channels. This uncertainty is not captured by current simulations in any meaningful way, and affects attempts to use the low-$\nu$ method with all target materials, including hydrogen.
Both CCFR and NuTeV were investigating high energy neutrinos in the $30 \leq E_\nu \leq 360~\text{GeV}$ range. In their \lownu analyses, both experiments used a sample of events with $\qz \leq 20$ GeV. The largest uncertainty for both experiments comes from an estimate of the $\qz/E_{\nu}$-dependent term in \autoref{eq:low-nu}, which they obtain using data-driven estimates. Interestingly, for these  $\qz/E_{\nu}$ estimates, CCFR (NuTeV) used a restricted $4 \leq \qz \leq 20$ GeV ($5 \leq \qz \leq 20$ GeV) sample due to concerns that quasielastic and resonant interactions are poorly understood, and the formalism of \autoref{eq:low-nu} breaks down.

More recent applications of the \lownu method have extended its use to lower neutrino energies, with significantly lower \qz cuts. NOMAD ($3 \leq E_{\nu} \leq 100$ GeV) isolated a $\qz \leq 3$ GeV event sample for use with the \lownu method~\cite{Petti:2006tu}, stating that correction factors for deviations from a constant cross section of $\mathcal{O}(10\%)$ were needed. However, this flux constraint does not appear to have been used for any published cross-section measurements from NOMAD.
MINOS~\cite{Adamson:2009ju, Bhattacharya:2009zza} was at a significantly lower neutrino energy, $2 \leq E_\nu \leq 10~\text{GeV}$, with the spectrum peaking at $E_\nu\approx4~\text{GeV}$. They imposed cuts on \qz that were dependent on $E_\nu$ to ensure adequate statistics in each sample.
MINOS noted that for the $\qz\leq1~\text{GeV}$ sample, the correction for non-flatness was $+4\%(-22\%)$ for (anti-)neutrinos at $E_\nu = 5~\text{GeV}$.
MINERvA's approach is similar to MINOS', splitting the CC-inclusive sample into ranges of $E_\nu$ with different \qz cuts to reduce the statistical uncertainty of the sample. They found compatible correction factors to MINOS, with the largest correction being $+8.1$\% ($-16.7$\%) for (anti-)neutrinos and the $\qz \leq 2.0$ (0.3) GeV sample~\cite{Devan:2015uak}\footnote{MINERvA's publication~\cite{MINERvA:2016ing} quotes different flux correction factors $\eta$ to MINOS by roughly $1/\eta$. The published MINERvA correction factors are also different to MINERvA analyser J. Devan's PhD thesis~\cite{Devan:2015uak} by exactly $1/\eta$. We conclude there appears to be a transcription error from J. Devan's PhD thesis into the publication by MINERvA. We have contacted the MINERvA collaboration about this apparent discrepancy.}. 
More recently, MINERvA presented an analysis of \lownu events in the ``medium energy'' configuration~\cite{MINERvA:2021mpk}, employing a $\qz<0.8~\text{GeV}$ sample for a wide-band neutrino beam with $\left<E_\nu\right>\sim6~\text{GeV}$. The \lownu data was poorly described by MINERvA's simulations, and the difference was attributed to a 1.8 standard deviation shift in the muon energy scale in MINOS (which provides the muon measurement for MINERvA).

Following the discussion of the breakdown in formalism of \autoref{eq:low-nu} in Ref.~\cite{Belusevic:1987rn,Belusevic:1988ab}, and the treatment of the low-\qz region by CCFR and NuTeV, it is natural to ask how application of the \lownu method for lower energy neutrino experiments can be justified. Bodek et al.~\cite{Bodek:2012uu, Bodek:2012cm} discuss these concerns and argue that the \lownu method can be applied using low-\qz cuts, intentionally isolating the quasielastic and $\Delta$ region, exploring cuts on \qz as low as $\qz \leq 0.25$ GeV. Bodek et al. focus on the applicability of the method to the MINERvA and MiniBooNE experiments in the context of the GENIE model~\cite{Andreopoulos:2009rq, Andreopoulos:2015wxa}\footnote{The GENIE version used in Ref.~\cite{Bodek:2012uu} was GENIE 2.6.6, which is nearly a decade old compared to the GENIE models which are investigated in this work.}, with additional theoretical arguments. They conclude that the \lownu method can be applied, with systematic uncertainties as low as $\sim$2--3\%, asserting that the model uncertainties are known and under control. In the decade since the publication of Refs.~\cite{Bodek:2012uu, Bodek:2012cm} there has been widespread development in the modeling of quasielastic and resonant region neutrino interactions, and the various nuclear effects that may play an important role~\cite{hayato_review_2014, Mosel:2016cwa, Katori:2016yel, NuSTEC:2017hzk}. Of particular relevance for the \lownu technique, there has been a lot of discussion about the neutrino energy dependence of some of these effects. The MINERvA \lownu analysis includes some of these additional effects, motivating uncertainties of a few percent~\cite{MINERvA:2016ing}.

\section{Models and simulations}
\label{sec:models}
To meaningfully explore the utility of the \lownu method for few-GeV accelerator neutrino experiments, it is crucial to assess the applicability of each of the three requirements outlined in \autoref{sec:introduction}. Doing this requires an analysis of a range of models which broadly cover plausible differences in: the neutrino energy dependence of the cross section for different ranges of \qz; how true \qz relates to observable quantities; and the extent to which the true neutrino energy can be reconstructed for different ranges of \qz (true and experimentally observable \qz proxies). The modeling of such aspects of neutrino interactions is inextricably linked to the treatment of a variety of different nuclear physics processes. For instance, in the charged-current quasi-elastic (CCQE) process\footnote{Also referred to as one-particle one-hole (1p1h) in the literature.}, which dominates at the low-\qz phase space relevant for the \lownu method at few-GeV energies, the behavior of the cross section for $\qz<100~\text{MeV}$ is strongly influenced by the treatment of deviations from the impulse approximations through collective nuclear-medium effects, often described using approaches based on the ``Random Phase Approximation'' (RPA)~\cite{Nieves:2004wx,Graczyk:2003ru,Martini:2013sha,Jachowicz:2002rr}. To simulate the accuracy of observable proxies for \qz or neutrino energy, it is additionally crucial to model the proportion of the energy in an interaction that can not be reliably observed inside a detector, for example the energy which is lost to neutrons or to unobserved charged pion masses~\cite{Alvarez-Ruso:2017oui}. This depends on the treatment of hadron production mechanisms both ``at the vertex'', i.e. the relative contribution of CCQE over other relevant interaction channels, such as multi-nucleon interactions or pion production, and through final state interaction (FSI) processes. Overall, the relevant physics for assessing the utility of the \lownu method spans a wide range of complex nuclear physics processes that are exceptionally challenging to describe and the modeling of which are under active development within the nuclear theory community~\cite{Katori:2016yel}. For this reason it is not currently feasible to take any single model and define a comprehensive array of uncertainties to cover plausible variations.

Given the difficulty in motivating robust uncertainties that cover all possible model choices, investigating the spread of predictions from multiple models has proven to be a useful, if incomplete, approach. The so-called \textit{model spread} approach can be used to estimate the potential for bias in neutrino oscillation and neutrino-cross section measurements. It reflects the range of possible values that could be obtained were a different model assumed in an analysis, and is useful when there are multiple available models and no clear path to selecting any one correct model. In this work we consider a variety of cross-section models with significantly different approaches to modeling the pertinent processes as a way to explore the spread of current predictions. Whilst there are a wide variety of models which alter relevant nuclear effects for CCQE interactions (which are the most important contribution in this work) the same level of model sophistication and diversity does not exist for other interaction modes. This is particularly important to consider when comparing how the different models predict migration from higher \qz interactions modes into low \qz regions identified by realistic, but inexact, \qz proxy variables---as are discussed in \autoref{sec:variables}. This approach has further important limitations as we can only consider the spread of models that have been written down and subsequently implemented in generators, and there is no guarantee that the range of predictions \textit{cover} nature. As a result, model-spread can only ever provide a lower bound for the uncertainty on deriving a flux constraint from the \lownu method, given the state of neutrino interaction modeling at the present time.

In this work, the GiBUU~\cite{Buss:2011mx}, NEUT~\cite{Hayato:2002sd,Hayato:2009,Hayato:2021heg}, NuWro~\cite{Golan:2012wx,Golan2012nuwro} and GENIE~\cite{Andreopoulos:2009rq, Andreopoulos:2015wxa} neutrino interaction event generators are used with NUISANCE~\cite{Stowell:2016jfr}, which processes generator predictions in a common event framework. Within GENIE, five different model configurations are considered. In general, we select generators and models that have fundamental differences critical to the low \qz region used in the \lownu method which represent state-of-the-art implementations of nuclear theory calculations and/or are widely used by ongoing or upcoming experiments. For instance, GiBUU's sophisticated hadron transport model is significantly different to the commonly used Salcedo-Oset semi-classical cascade~\cite{Salcedo:1987md,Oset:1987re}, which critically impacts the observable energy transfer in the detector after FSI have taken place. We also compare state-of-the-art models, such as SuSAv2 and CRPA, to the historically commonly used GENIEv2, which has simpler 1p1h, 2p2h and nuclear models. For few-GeV experiments like DUNE, MINERvA and NOvA, the transition region between single pion production and deep inelastic scattering (DIS)---sometimes referred to as ``shallow inelastic scattering'' (SIS)---is a significant interaction channel, and we select generators with different treatments of the process. However, all generators interface to PYTHIA~\cite{SJOSTRAND199474,Sjostrand:2006za} for high energy interactions, although the PYTHIA version and implementation details differ. Furthermore, the generators' neutrino-nucleon interaction are generally tuned to similar light-target data for CCQE and CC$1\pi$ interactions, leading the neutrino-nucleon cross-section to be similar in neutrino energy. An outline of each model is given below.

\subsection{GiBUU}
The GiBUU theory framework is described in detail in Ref.~\cite{Buss:2011mx}. GiBUU uses a nuclear ground state based on the local density approximation and incorporates a momentum-dependent mean-field nuclear potential into cross-section calculations to describe all interaction modes in a consistent way, as further detailed in Refs.~\cite{Leitner:2008ue,Gallmeister:2016dnq}. GiBUU does not explicitly include any RPA correction, although its careful treatment of bound nucleons would seem to require a much weaker RPA correction than used in some other models~\cite{Lalakulich:2012hs}. GiBUU uses a phenomenological 2p2h model based on Ref.~\cite{OConnell:1972edu} and driven by electron scattering data, as described in Ref.~\cite{Gallmeister:2016dnq}. For the extension to neutrino interactions, the model involves a normalization scaling that is linear with the difference between the proton and neutron content of the target nucleon. This behaviour has been validated with exclusive \ch cross-section measurements in Ref.~\cite{Dolan:2018sbb} and makes GiBUU's 2p2h prediction significantly larger than that of other models. GiBUU's treatment of FSI also differs substantially from that of other models considered in this work, using quantum-kinetic transport theory to propagate outgoing hadrons through the nucleus. GiBUU's FSI describes the evolution of the phase space density for each hadron under the influence of the same mean field potential that is used for the initial nuclear ground state in the cross-section calculation. 

\subsection{NEUT}

NEUT is the primary neutrino interaction simulation used by the T2K and Super-Kamiokande collaborations, and is detailed in Refs.~\cite{Hayato:2002sd,Hayato:2009,Hayato:2021heg}. For this work, we generate events NEUT 5.5.0. CCQE and 2p2h are simulated using the NEUT implementation of the Valencia group's models~\cite{Nieves:2011yp, Nieves:2011pp}, with a custom approach to the treatment of nuclear removal energy~\cite{Bourguille:2020bvw}. Pion production in the invariant mass region most relevant to this work, $W\leq 2~\text{GeV}$, is described using the Rein–Sehgal resonant model~\cite{Rein:1980wg}, with improvements to the nucleon axial form factors~\cite{Graczyk:2014dpa,Graczyk:2007bc} and the inclusion of final-state lepton mass effects~\cite{Berger:2007rq,Graczyk:2007xk,Kuzmin:2003ji}. SIS/DIS hadron production is simulated using PYTHIA 5.72~\cite{SJOSTRAND199474} or a custom model based on KNO scaling (see Sec.~V.C of Ref.~\cite{Aliaga:2020rqb}) for interactions with a hadronic invariant mass above and below 2 GeV respectively. Pion FSIs are described using the semi-classical intranuclear cascade model by Salcedo and Oset~\cite{Salcedo:1987md,Oset:1987re}, tuned to modern $\pi$--$A$ scattering data~\cite{PinzonGuerra:2018rju}. Nucleon FSIs are described in an analogous cascade model~\cite{Hayato:2009}. Within intranuclear cascade models, outgoing hadrons are individually stepped through the remnant nucleus where they can rescatter through various processes, sometimes producing additional hadrons that then are also stepped through the cascade. A comparison of FSI models in NEUT, NuWro and GENIE is presented in Ref.~\cite{Dytman:2021ohr}.

\subsection{NuWro}

The version of the NuWro event generator~\cite{Golan:2012wx,Golan2012nuwro} (v.~19.02.2) used for this work is configured to simulate CCQE interactions using the standard Llewellyn-Smith approach~\cite{LLEWELLYNSMITH1972261} with custom RPA corrections~\cite{Graczyk:2003ru} and a local-Fermi gas nuclear ground state model combined with an effective nuclear momentum-dependent potential~\cite{Juszczak:2003zw}. The 2p2h model in NuWro implementation from the Valencia group, also used in NEUT. Resonant pion production is calculated using the Adler model~\cite{Adler:1975mt, ADLER1968189}, extending only to $W \leq 1.6~\text{GeV}$ since only the $\Delta$(1232) resonance is considered. DIS is modelled using PYTHIA 6~\cite{Sjostrand:2006za} above $W = 1.6~\text{GeV}$, although a linear transition between resonant pion production and DIS is considered from $W = 1.3~\text{GeV}$. FSIs are modeled using a similar intranuclear cascade model to NEUT, but which has been separately tuned. Further details concerning the FSI and pion production models can be found in Ref.~\cite{Gonzalez-Jimenez:2017fea,Niewczas:2019fro, Sobczyk:2014xza,Golan:2012wx,Dytman:2021ohr}.

\subsection{GENIEv3}

The GENIE neutrino interaction event generator is the primary generator of most Fermilab neutrino-beam experiments. In this work we consider a variety of model configurations in GENIE v3. These configurations alter the CCQE, 2p2h and FSI models, whilst the modeling of other interaction modes remains the same. Single pion production is simulated similarly to NEUT, employing the Rein-Seghal model with lepton mass corrections~\cite{Berger:2007rq} up to $W = 1.7~\text{GeV}$. DIS and SIS are described using the custom ``AGKY'' model~\cite{Yang:2009zx} for $W \leq 2.3~\text{GeV}$, PYTHIA 6~\cite{Sjostrand:2006za} is used for $W > 3.0~\text{GeV}$ and a linear transition is considered between.

\subsubsection{10a and 10b}

GENIE configurations 10a and 10b differ only in their treatment of FSI. For both configurations, the CCQE and 2p2h models of the Valencia group are used~\cite{Nieves:2011yp, Nieves:2011pp}, based on a Local-Fermi gas nuclear ground state. In the 2p2h case, a custom approach to producing nucleon kinematics from the inclusive model predictions is employed, as described in Ref.~\cite{Schwehr2017}. GENIE 10b uses the ``\textit{hN}'' intranuclear cascade model, similar to NEUT and NuWro, whilst 10a uses the ``\textit{hA}'' empirical approach in which the overall ``fate'' of hadrons in the cascade is decided in a single interaction rather than in a stepped process~\cite{Dytman:2021ohr}. For these configurations GENIE v3.0.6 is used. 

\subsubsection{SuSAv2}
The SuSAv2 GENIE configuration uses the same models as 10b but the CCQE and 2p2h are replaced by predictions from the SuSAv2-MEC model~\cite{Gonzalez-Jimenez:2014eqa,Megias:2016fjk,RuizSimo:2016rtu,RuizSimo:2016ikw}, as implemented in GENIE in Ref.~\cite{Dolan:2019bxf}. SuSAv2 CCQE acts as an inclusive parameterization of the sophisticated relativistic mean field model~\cite{Caballero:2006wi,Caballero:2005sj,Caballero:2007tz,Meucci:2009,Gonzalez-Jimenez:2019ejf}, which has been well validated using electron-nucleus scattering data. It also includes a detailed treatment of FSI and its impact on the inclusive scattering cross section, which is neglected in the 10a and 10b configurations. No RPA effects are included. The 2p2h model is fully relativistic and, unlike the Valencia model used in most other generators which is cut off at 1.2 GeV energy transfer, is capable of predicting the full-energy range of interest for this study. It should be noted that inexact methods are used to determine the hadron kinematics during event generation from an input inclusive cross section\footnote{Whilst every generator model implementation resorts to inexact methods at some point, the CRPA and SuSAv2 implementation uses techniques which ensure an exact reproduction of the inclusive cross-section model predictions but which use a ``factorized'' approach to predict outgoing hadron kinematics~\cite{Dolan:2019bxf}.}. For this GENIE configuration a preliminary version of the upcoming GENIE v3.2.0 is used.

\subsubsection{CRPA}

The CRPA GENIE configuration uses the same models as 10b but with the CCQE replaced by predictions from the GENIE implementation of the CRPA model from the Ghent group~\cite{Jachowicz:2002rr,Pandey:2014tza}, as detailed in Ref.~\cite{Dolan:2021rdd}. CRPA has been successful in describing electron-scattering data and, at low-energy transfers, differs substantially from the predictions of other commonly used models due to its detailed modeling of low energy excitations~\cite{Nikolakopoulos:2018sbo}. As in the SuSAv2 case, CRPA includes the impact of the FSI on the inclusive cross-section predictions but, within the GENIE implementation, similar inexact methods are used to determine the hadron kinematics during event generation. The current version of CRPA used in this work has a transition to the SuSAv2 CCQE cross section at high energy transfers and uses an unregularised nucleon-nucleon interaction within the RPA calculation~\cite{Nikolakopoulos:2020alk,Jachowicz:2021ieb}, which may be better suited to low energy and momentum transfer interactions. For this GENIE configuration we use a specially modified version that includes the CRPA model which is otherwise built on the same preliminary version of the upcoming GENIE v3.2.0 used for the SuSAv2 model predictions.

\subsection{GENIEv2}

In addition to using the latest GENIE, a widely used historical version of the generator is also considered. For this GENIE v2.12.10\footnote{Using configuration \texttt{ValenciaQEBergerSehgalCOHRES}.} is used in a similar configuration to GENIEv3 10a. As in 10a the Valencia CCQE and 2p2h models are used, and resonant pion production is simulated with the modified Rein-Seghal mode and the FSI is based on an older version of the hA model. The primary differences stem from the implementation details of the models, particularly the removal energy and nucleon ejection treatment for CCQE interactions~\cite{geniev304TN}, in addition to the tuning and interaction modes considered in the hA FSI~\cite{Dytman:2021ohr}. Whilst GENIEv3 should generally be considered an improvement with respect to GENIEv2, we opt to include this older version in our studies due to its past and current widespread use. For example, a similar configuration of GENIEv2 was used for the MINERvA collaboration's \lownu analyses~\cite{MINERvA:2016ing,MINERvA:2021mpk}.

\section{Low-\qz cross section predictions}
\label{sec:low-qz-cross-section}
In this section, we investigate the first of the three requirements for using the \lownu method as set out in \autoref{sec:introduction}: that the cross section of a low-\qz sample is either constant as a function of $E_{\nu}$ or any non-constant behaviour is well understood.
We produced large event samples using each of the different models described in \autoref{sec:models}, including \numu and \numub scattering on \argon, \ch and \chtwo targets, with a uniform neutrino flux in the range $E_\nu=0$--$20~\text{GeV}$. A minimum of 100 million events\footnote{For GiBUU this corresponds to the number of {\it unweighted} events.} were generated for all configurations to ensure high statistics over the phase space of interest.

\begin{figure*}[htbp]
  \centering
  \captionsetup[subfloat]{captionskip=-1pt}
  \subfloat[\numu--\argon, $\qz \leq 0.1$ GeV]  {\includegraphics[width=0.3\linewidth]{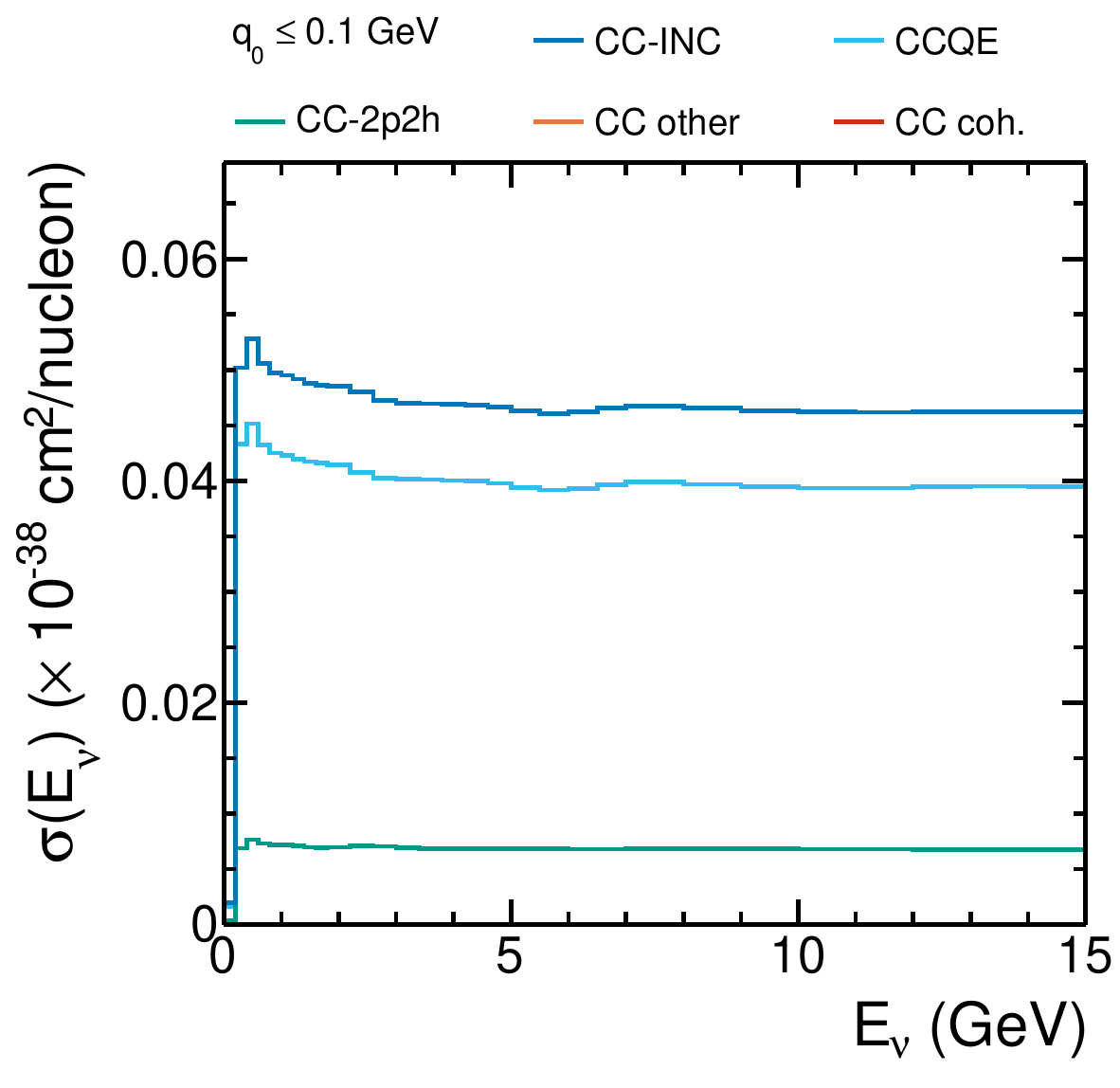}}
  \subfloat[\numu--\argon, $\qz \leq 0.3$ GeV]  {\includegraphics[width=0.3\linewidth]{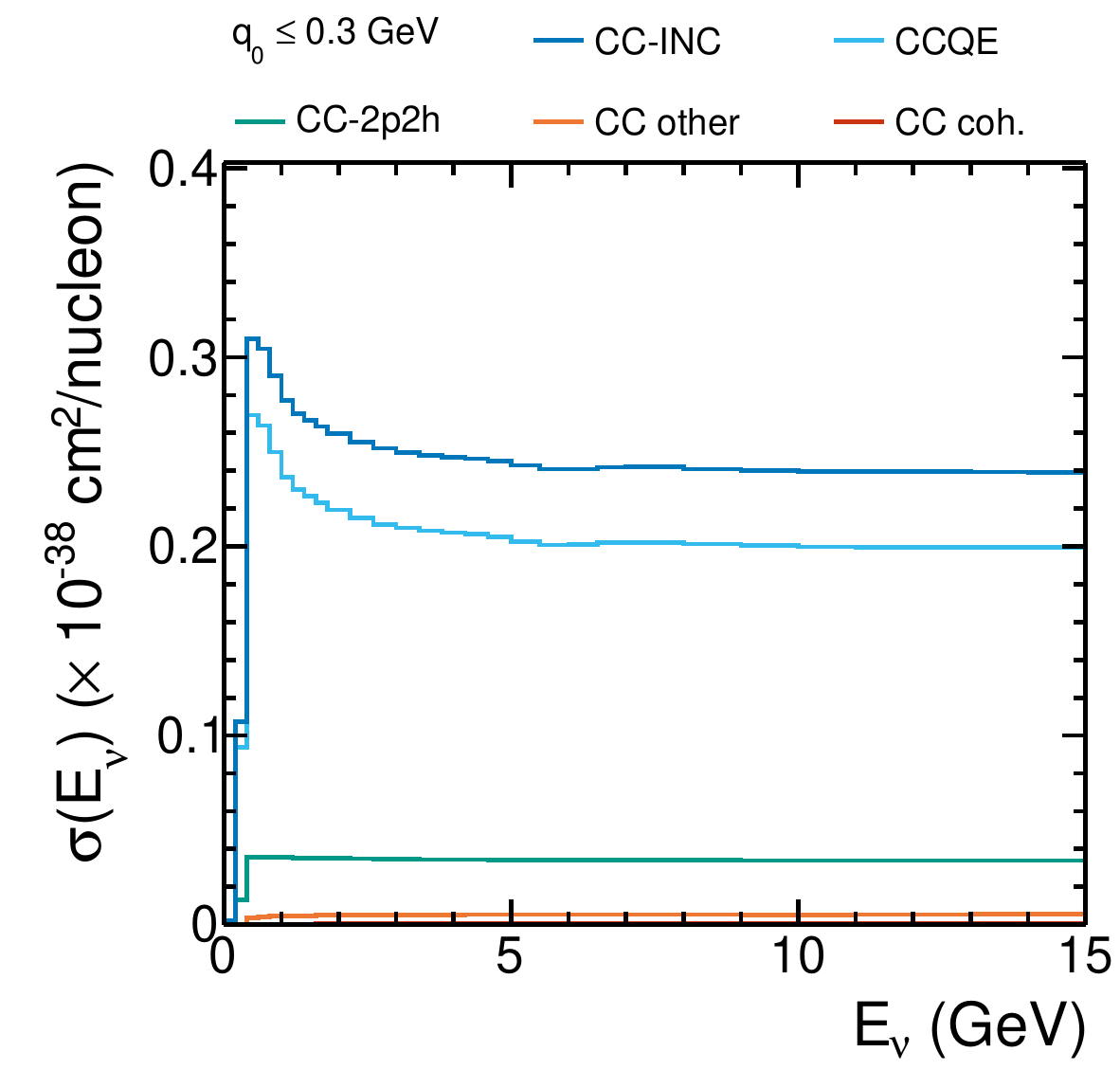}}
  \subfloat[\numu--\argon, $\qz \leq 0.5$ GeV]  {\includegraphics[width=0.3\linewidth]{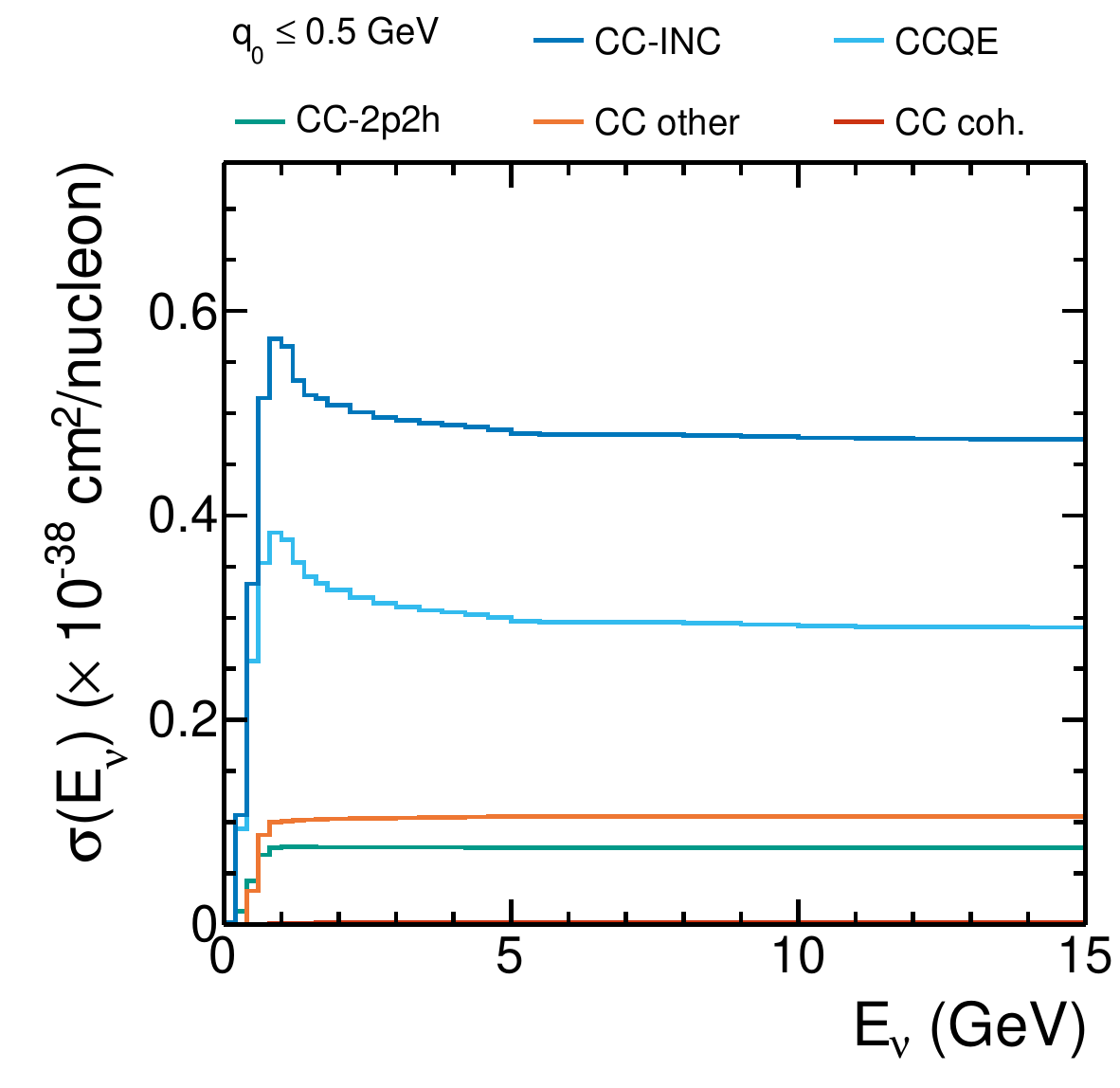}}\\\vspace{-2pt}
  \subfloat[\numub--\argon, $\qz \leq 0.1$ GeV]  {\includegraphics[width=0.3\linewidth]{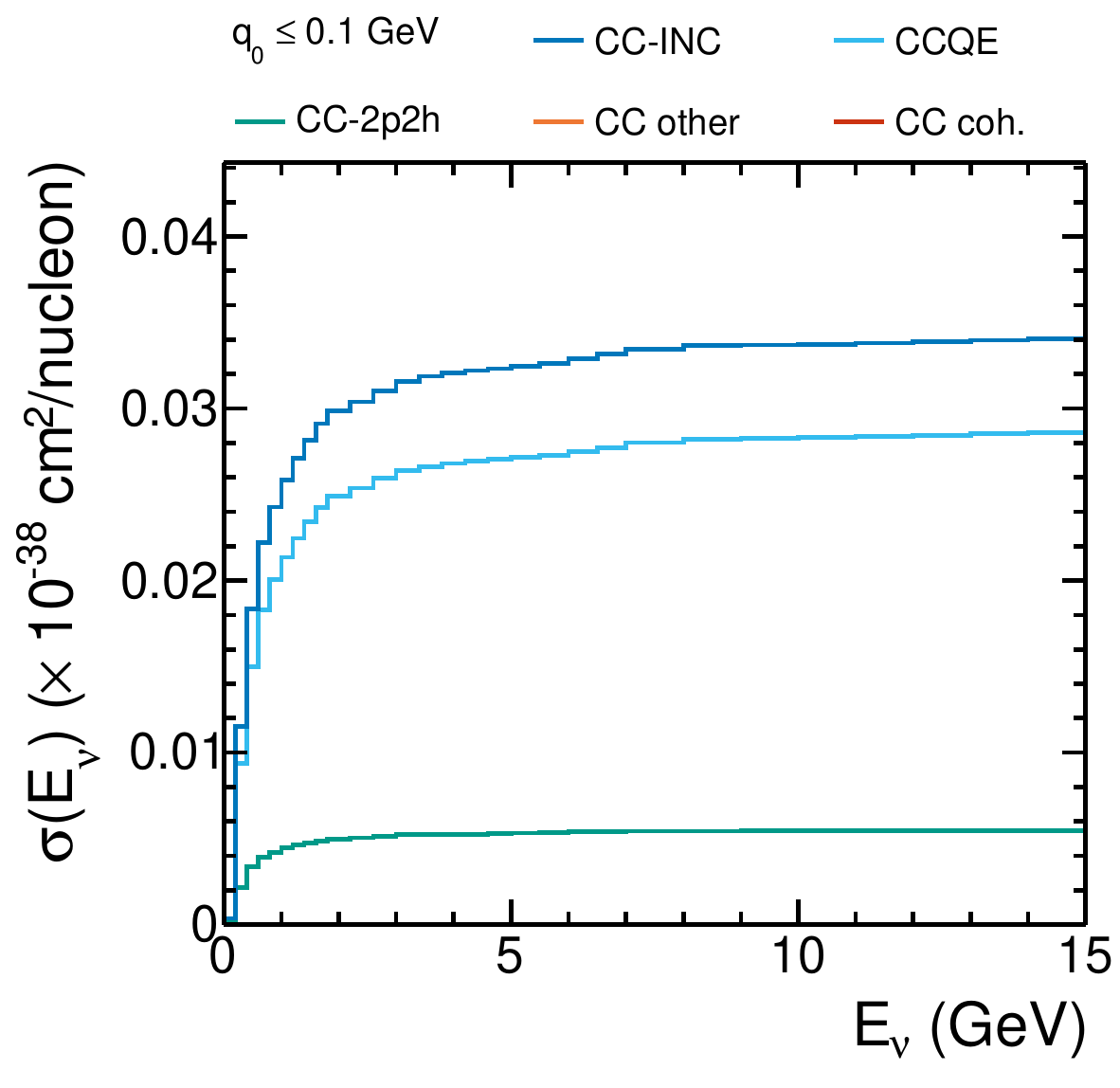}}
  \subfloat[\numub--\argon, $\qz \leq 0.3$ GeV]  {\includegraphics[width=0.3\linewidth]{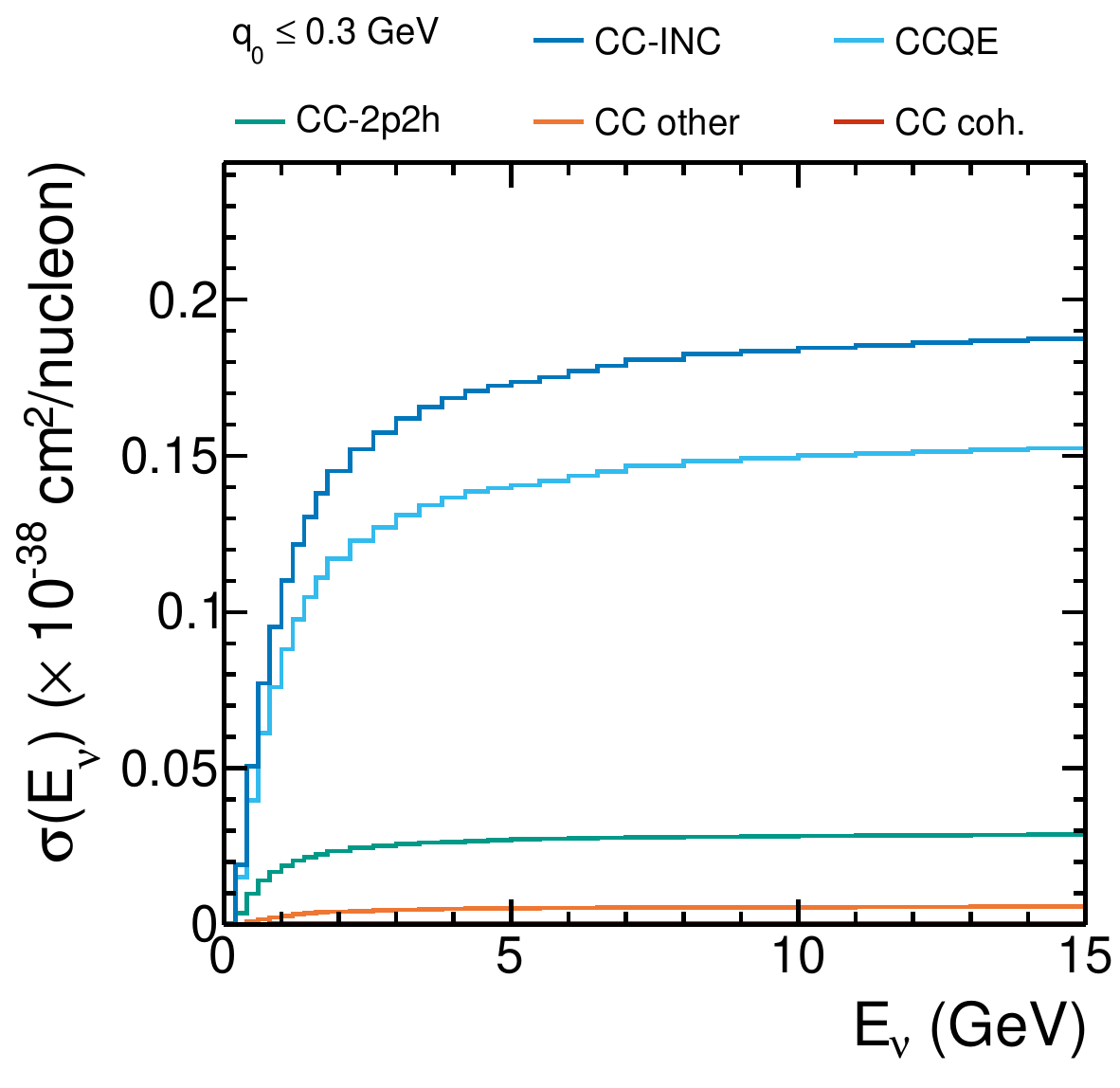}}
  \subfloat[\numub--\argon, $\qz \leq 0.5$ GeV]  {\includegraphics[width=0.3\linewidth]{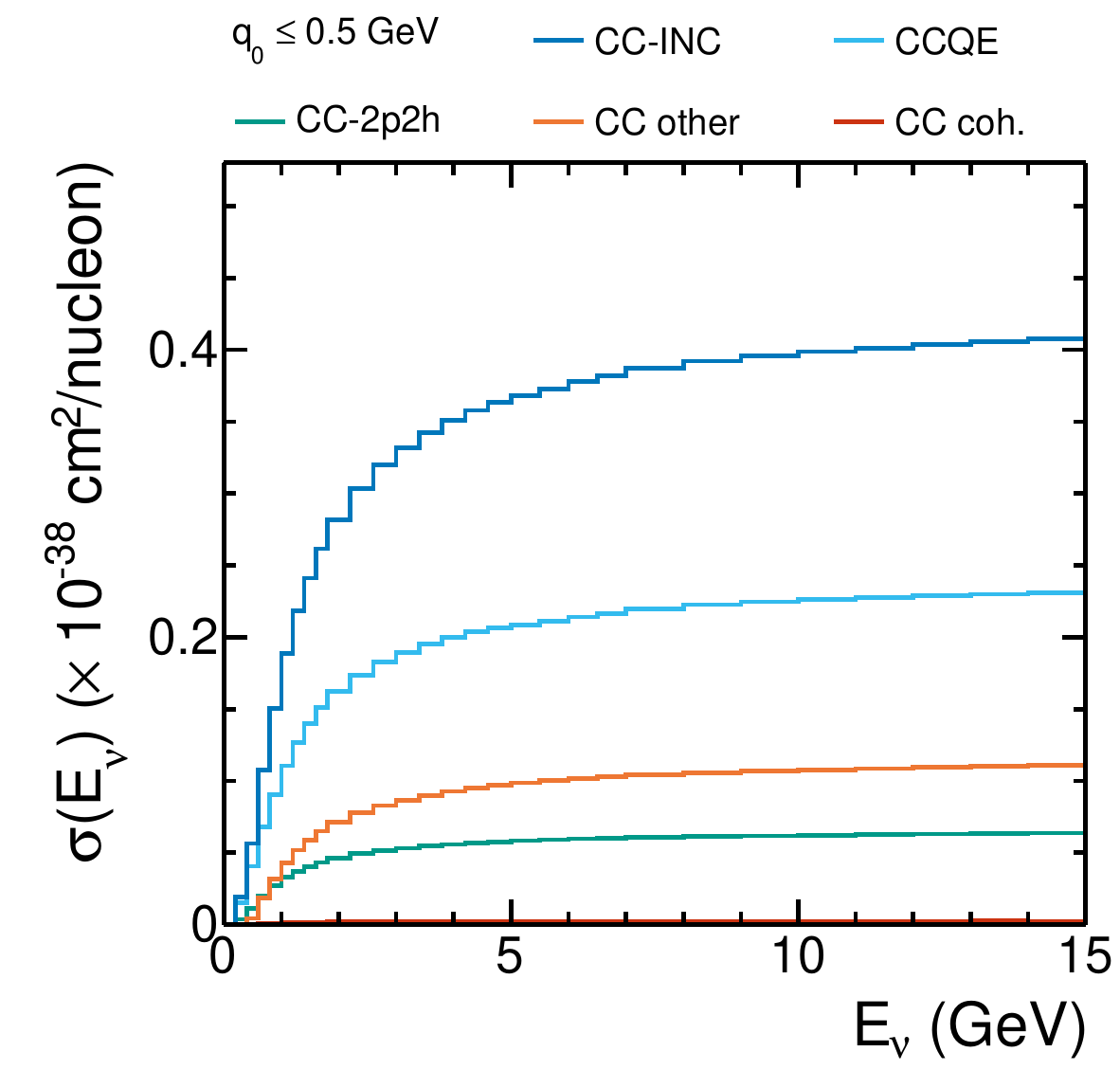}}
  \caption{Contributions to the GENIE 10a charged-current cross section with \qz $\leq$ 0.1, 0.3 and 0.5 GeV as a function of \enutrue, for both \numu--\argon and \numub--\argon, generated with a flat neutrino flux. Contributions from separate interaction channels are also shown.}
  \label{fig:contrib_vary_q0}
\end{figure*}
\begin{figure*}[htbp]
  \centering
  \captionsetup[subfloat]{captionskip=-1pt}
  \subfloat[GENIEv3 10a] {\includegraphics[width=0.33\linewidth]{{figures/flat_numu_Ar40_q0_0.3_GENIEv3_G18_10a_02_11a}.pdf}}
  \subfloat[GENIEv3 10b] {\includegraphics[width=0.33\linewidth]{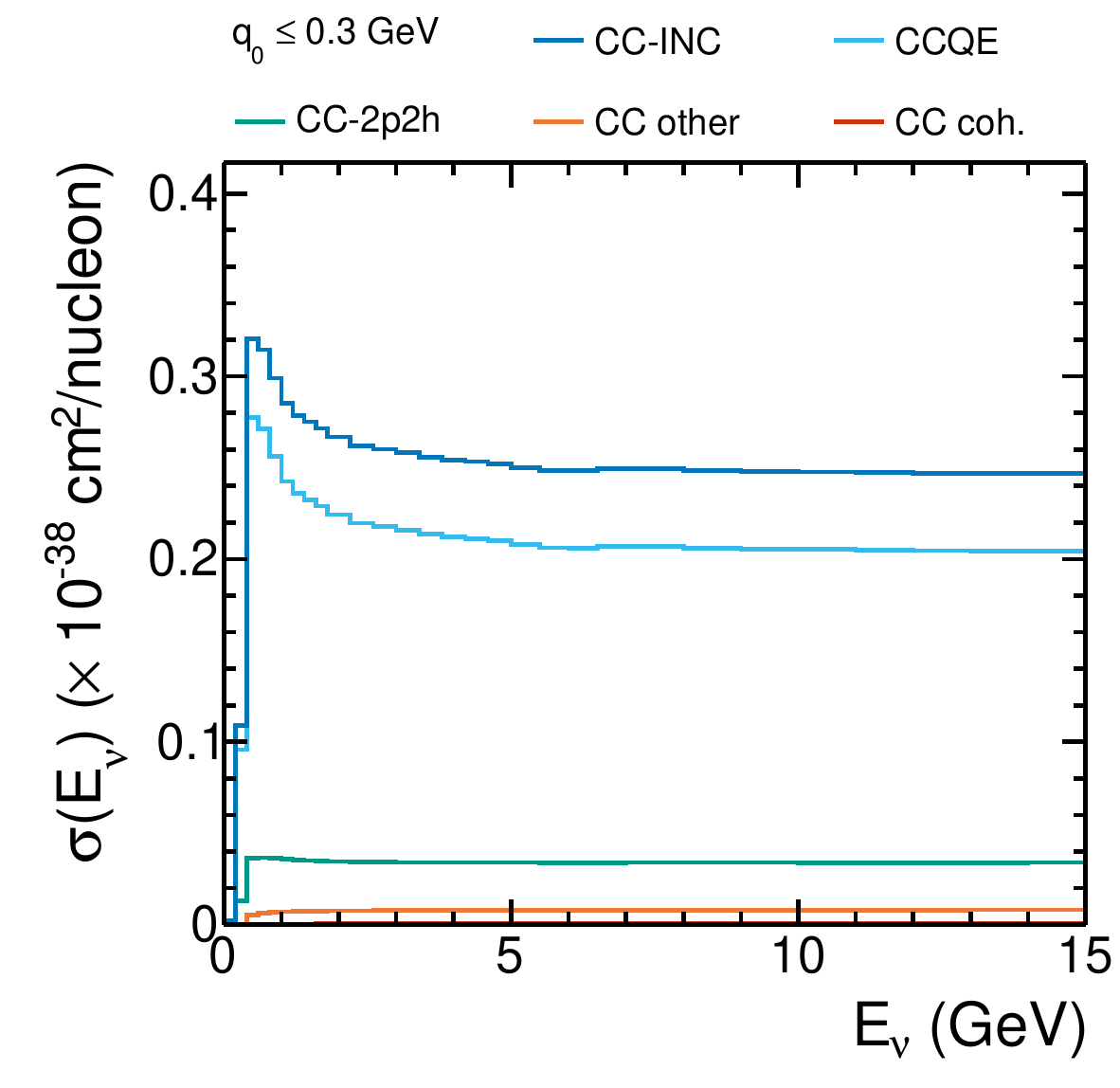}}
  \subfloat[CRPA]        {\includegraphics[width=0.33\linewidth]{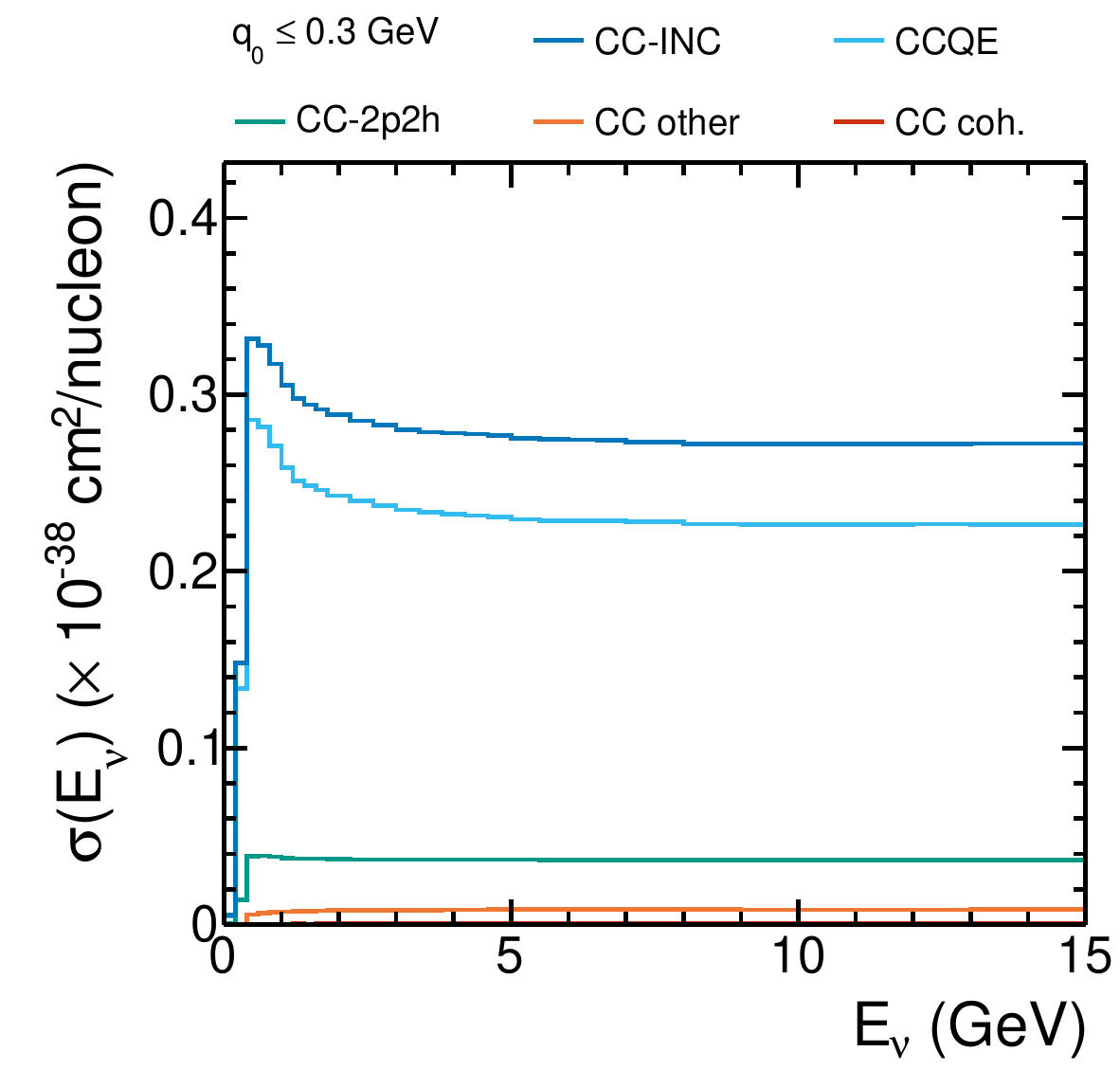}}\\\vspace{-2pt}
  \subfloat[SuSAv2]      {\includegraphics[width=0.33\linewidth]{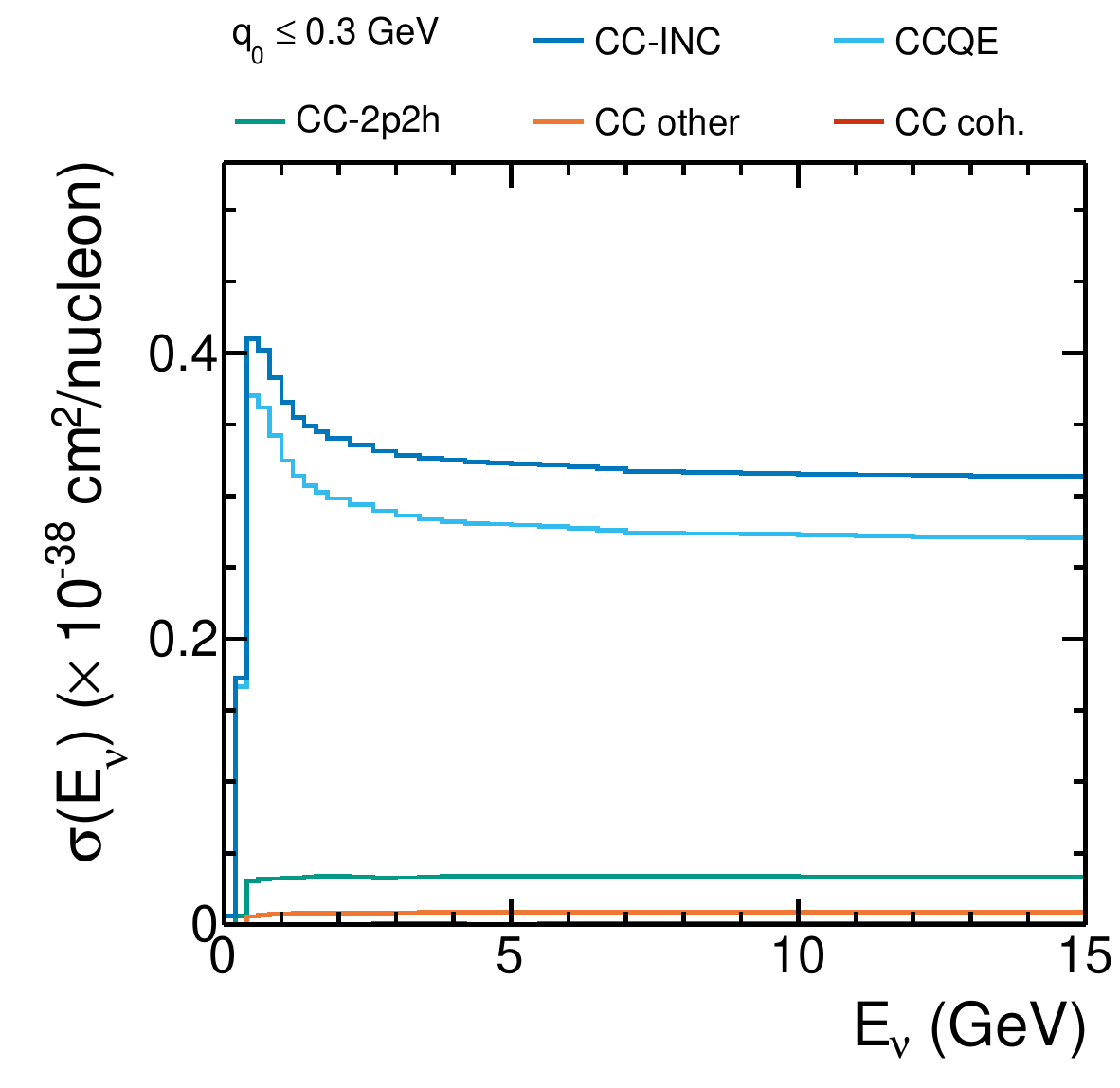}}
  \subfloat[GENIEv2]     {\includegraphics[width=0.33\linewidth]{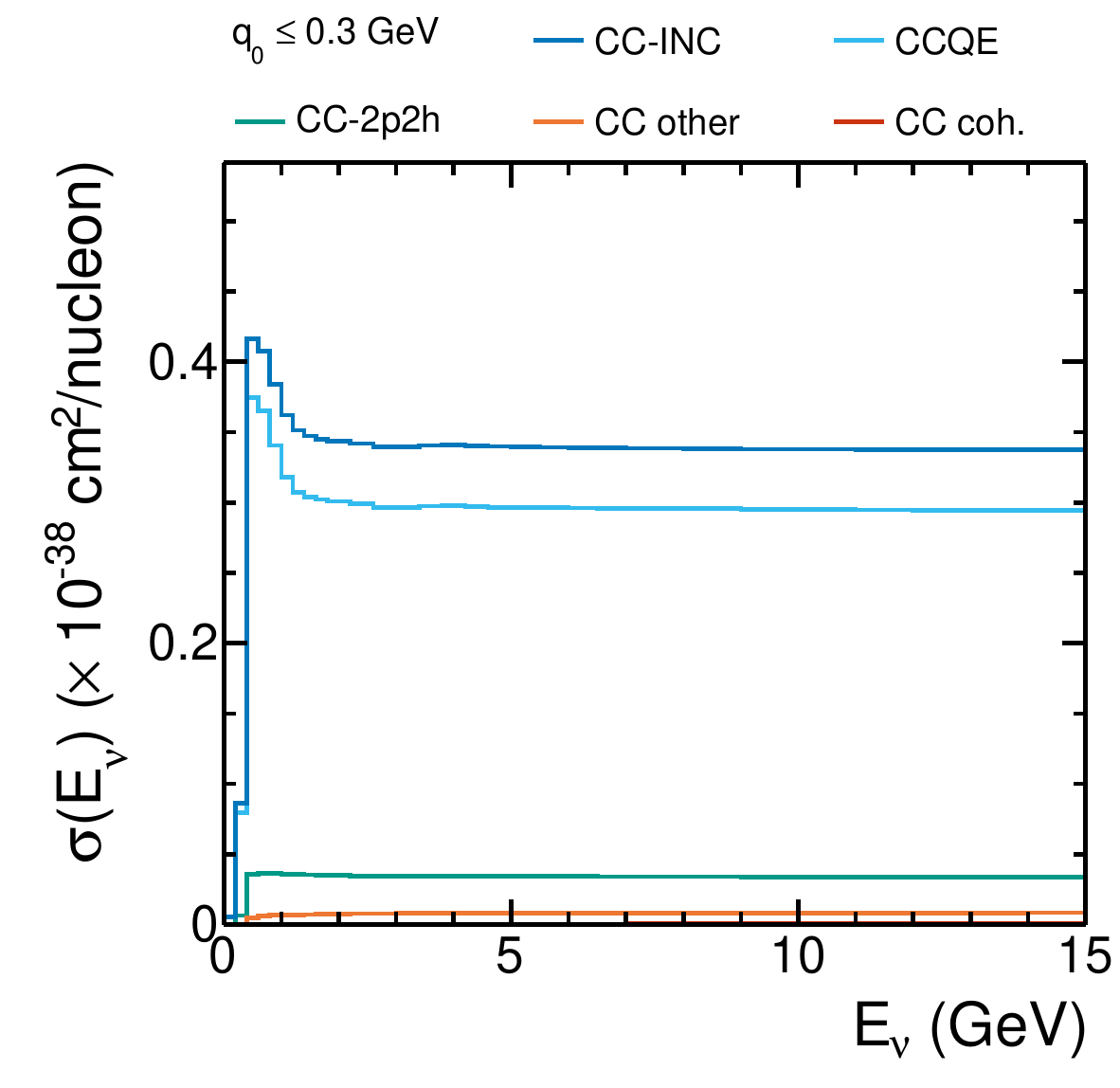}}
  \subfloat[NuWro]       {\includegraphics[width=0.33\linewidth]{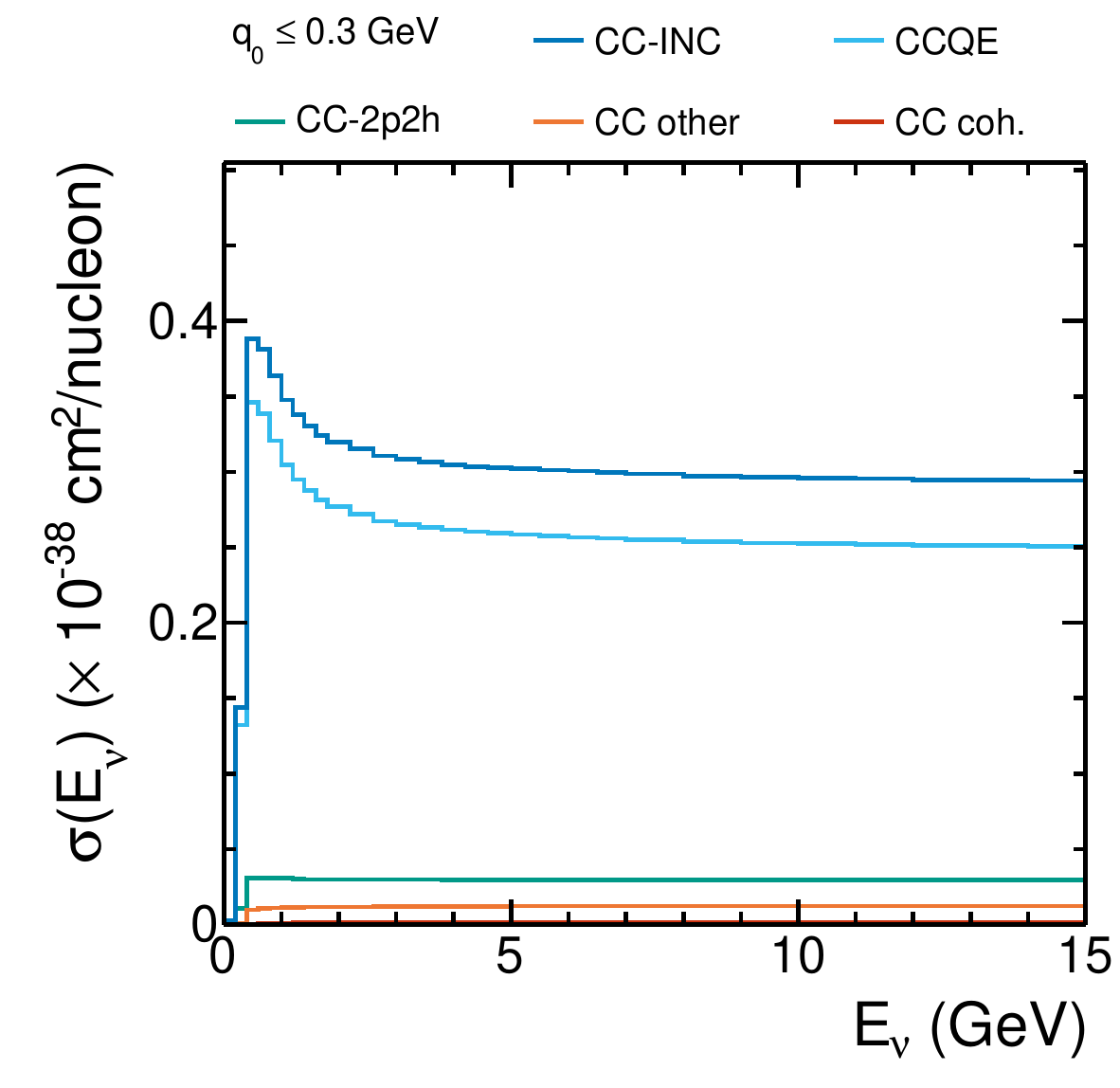}}\\\vspace{-2pt}
  \subfloat[GiBUU]       {\includegraphics[width=0.33\linewidth]{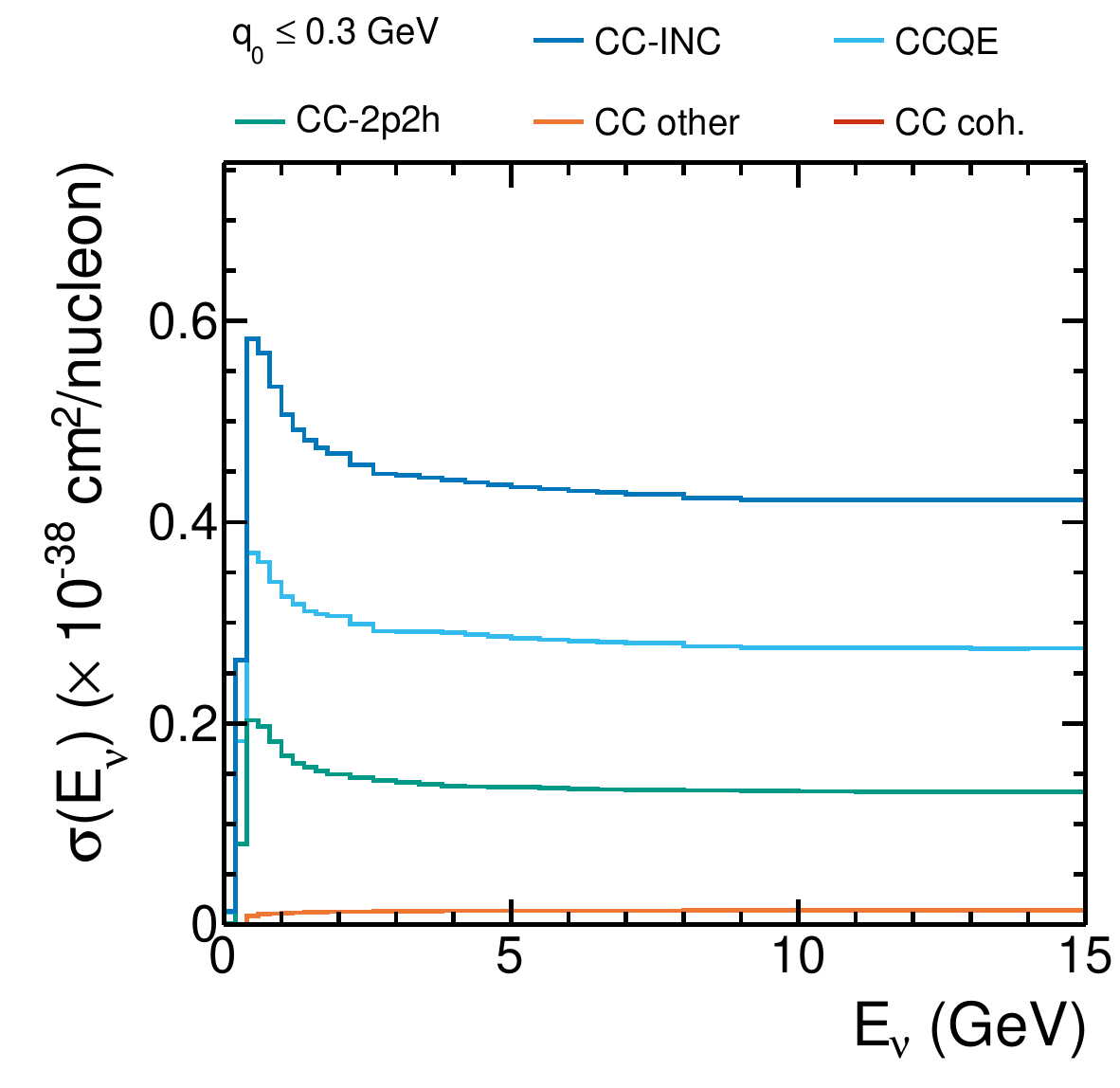}}
  \subfloat[NEUT]        {\includegraphics[width=0.33\linewidth]{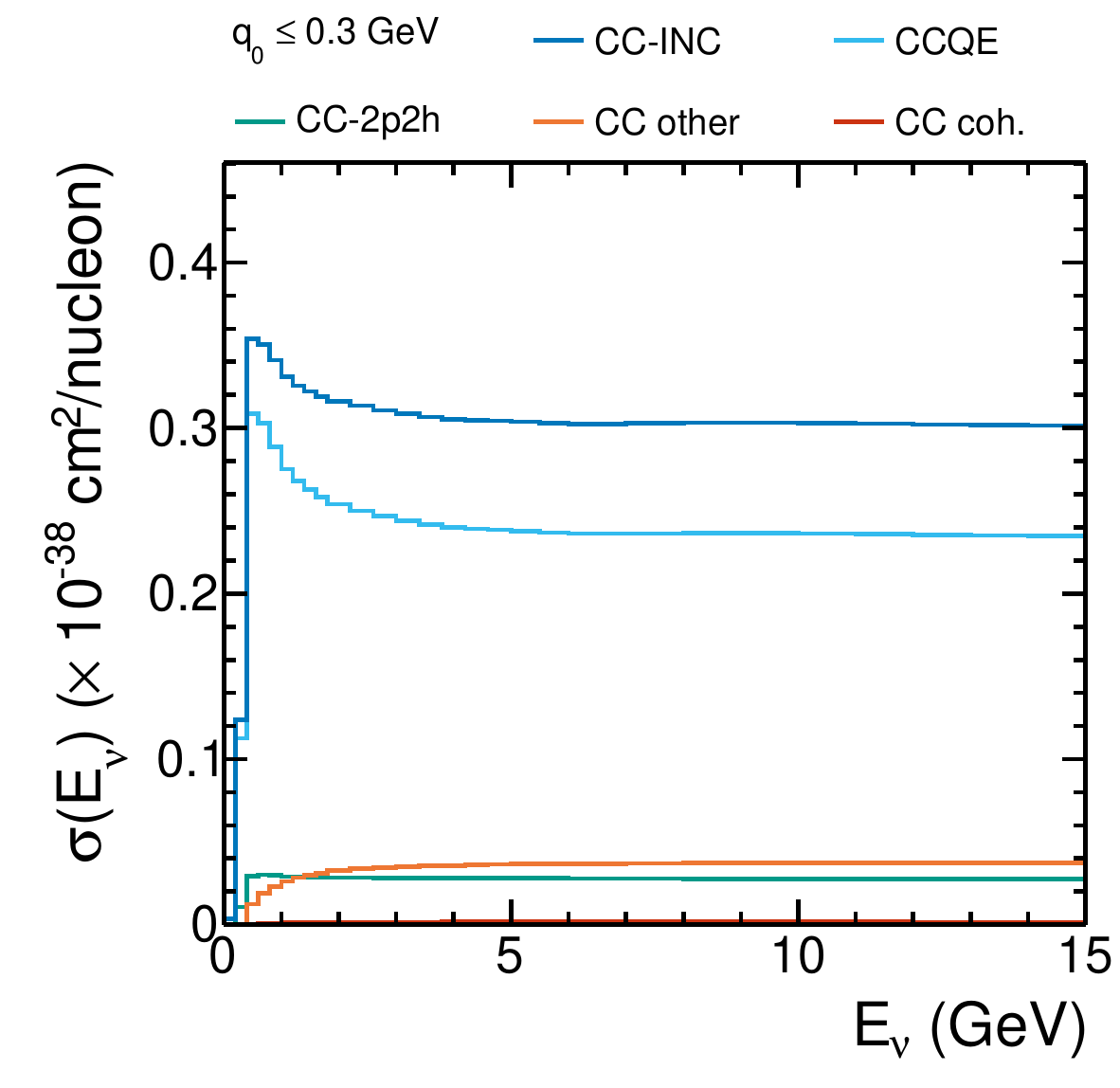}}  
  \caption{Contributions to the charged-current cross section with $\qz \leq 0.3$ GeV as a function of \enutrue for \numu--\argon and a variety of generator models, generated with a flat neutrino flux. Contributions from separate interaction channels are also shown.}
  \label{fig:contrib_vary_gen}
\end{figure*}

\autoref{fig:contrib_vary_q0} shows the contributions to the charged-current cross section for $\qz \leq$ 0.1, 0.3 and 0.5 GeV, for both \numu--\argon and \numub--\argon calculated with the GENIEv3 10a model. For all the \qz cuts, the total \numu--\argon charged-current cross section plateaus to be approximately constant with neutrino energy, after an initial rise and low energy hump. At $\qz \leq 0.1$ GeV, the GENIEv3 10a cross section is mostly CCQE, with some CC-2p2h contribution. The ``CC-other'' category consists of all other CC interactions that are not CCQE, CC-2p2h or CC coherent, and is dominated by pion production via resonances when low-\qz cuts are imposed. This contribution increases with higher \qz cut values and is also approximately constant with increasing neutrino energy.
In the GENIEv3 10a model shown in \autoref{fig:contrib_vary_q0}, these become approximately constant in \enutrue before the CCQE contribution does. The situation is rather different in the \numub--\argon case; there is no hump after the initial rapid rise in the charged-current cross section for any of the \qz cuts shown, and the cross section does not become constant as a function of neutrino energy for $E_{\nu} \leq 15$ GeV.

\autoref{fig:contrib_vary_gen} shows the contribution to the \numu--\argon charged-current cross section with a cut of $\qz \leq 0.3$ GeV, for all the models discussed in \autoref{sec:models}. Broadly, the models exhibit similar behaviour to GENIEv3 10a: in each case the cross section becomes {\it approximately} constant above $\enutrue\sim5~\text{GeV}$. The energy at which the asymptotic behavior is reached varies between models. The normalization of the $\qz \leq 0.3$ GeV cross section is different between models, but that is not an issue for the \lownu method, which only claims to provide shape information about the flux. However, the contributions from different interaction channels are very different between the different generator models, which is likely to cause differences when we move away from true \qz and \enutrue to observable proxy variables. This is because the physical processes they represent distribute four-momentum between final-state particles in very different ways. A particularly extreme example of this can be seen in the GiBUU model, which predicts a much stronger 2p2h cross-section contribution compared to other models due to the aforementioned scaling based on the proton and neutron content difference.

\begin{figure*}[htbp]
  \centering
  \captionsetup[subfloat]{captionskip=-2pt}
  \subfloat[Model comparison]  {\includegraphics[width=0.3\linewidth]{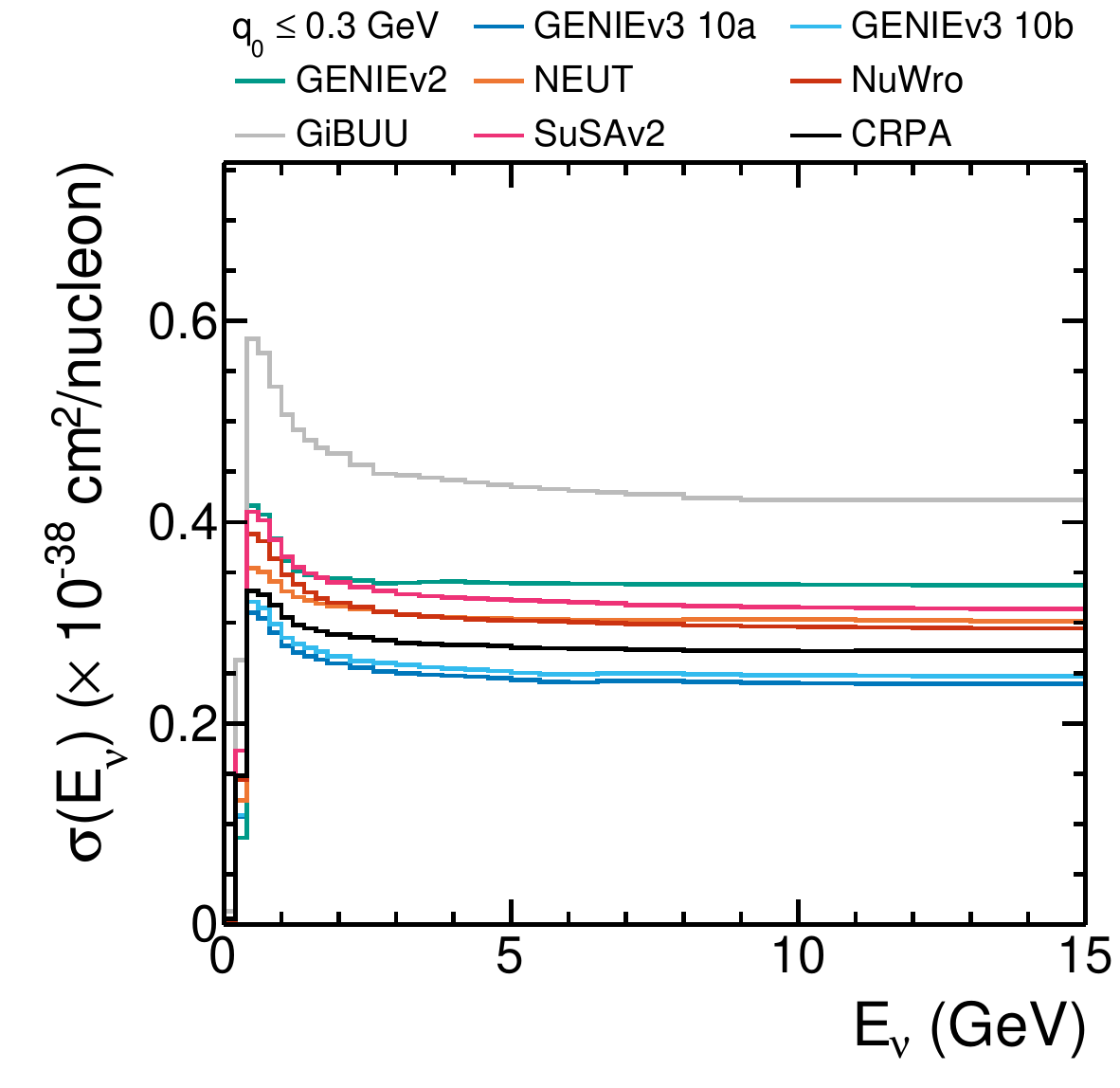}}
  \subfloat[Shape-only]        {\includegraphics[width=0.3\linewidth]{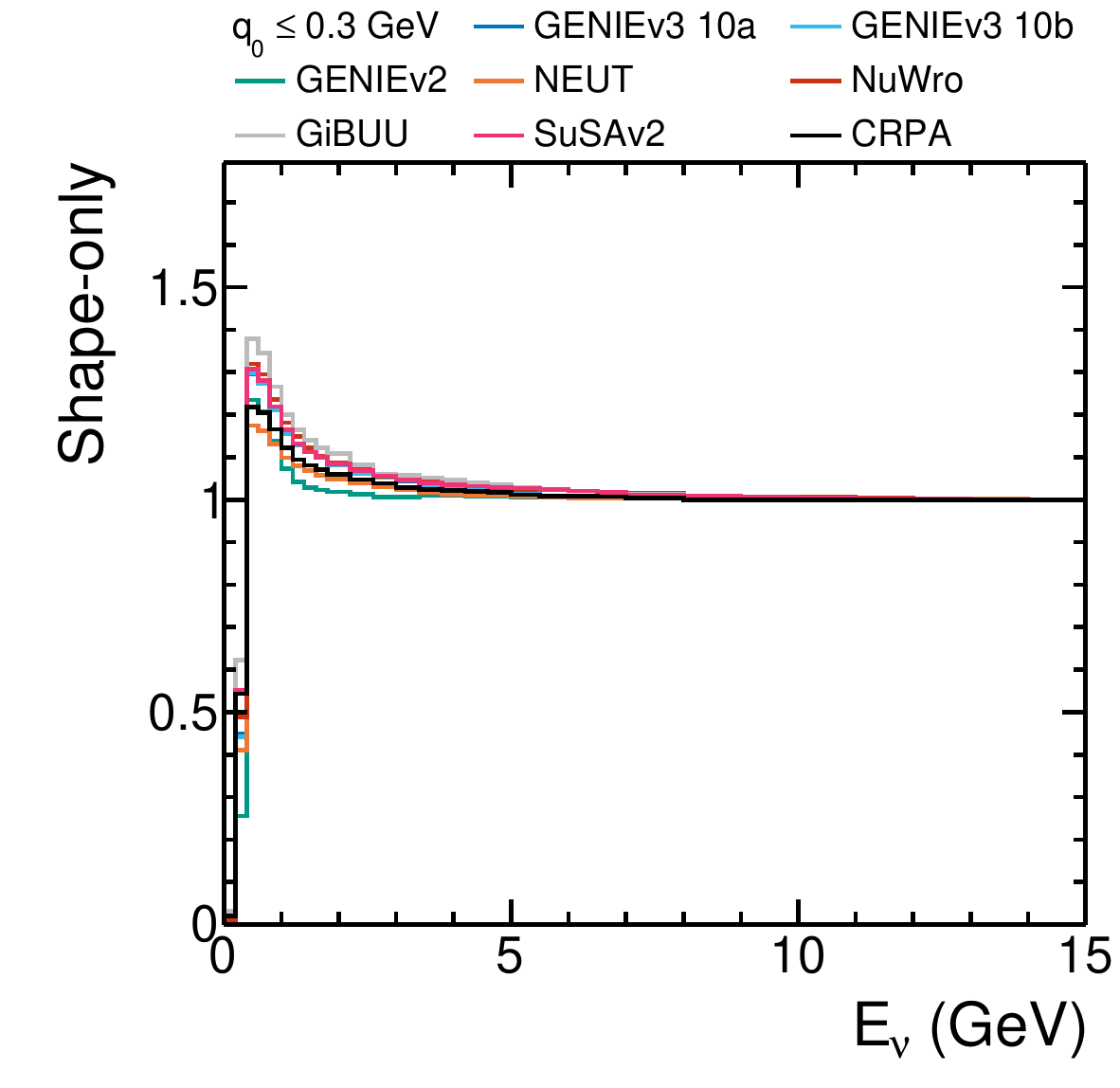}}
  \subfloat[Shape-only ratio]  {\includegraphics[width=0.3\linewidth]{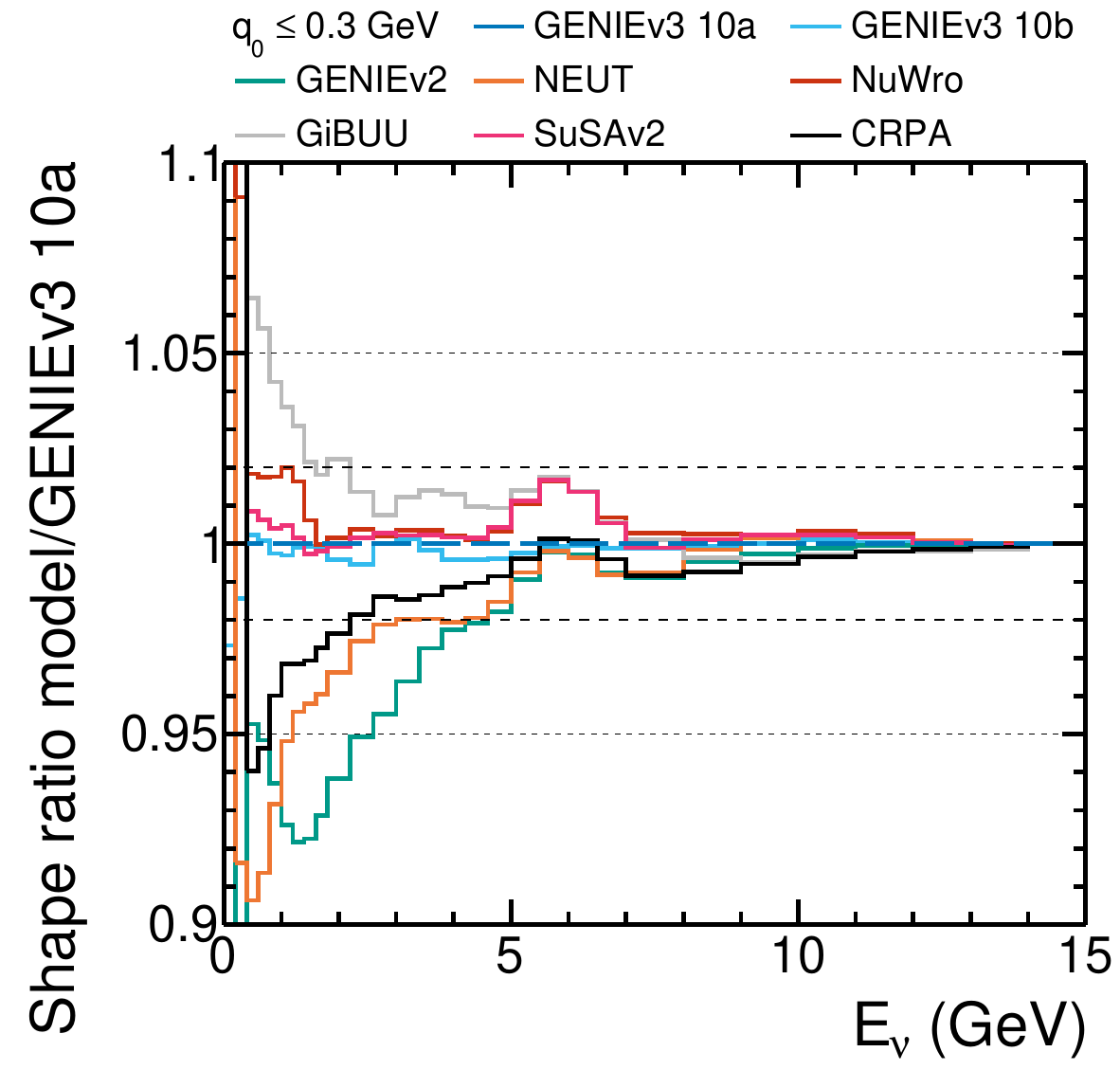}}
  \caption{Comparison of the total \numu--\argon charged-current cross section with true $\qz \leq 0.3$ GeV as a function of \enutrue for all generator models considered (left); comparison of the shape of the same cross section, normalized such that the 14--15 GeV bin is 1 (middle); comparison of the ratio of the shape-only model predictions with respect to GENIEv3 10a as a reference model (right). In the latter plot, horizontal long (short) dashed lines have been added at $\pm$2\% ($\pm$5\%), to guide the eye in assessing the scale of the bias.}
  \label{fig:model_comp_progression}
\end{figure*}

\begin{figure*}[htbp]
  \centering
  \captionsetup[subfloat]{captionskip=-2pt}
  \subfloat[\numu--\argon, $\qz \leq 0.1$ GeV]  {\includegraphics[width=0.3\linewidth]{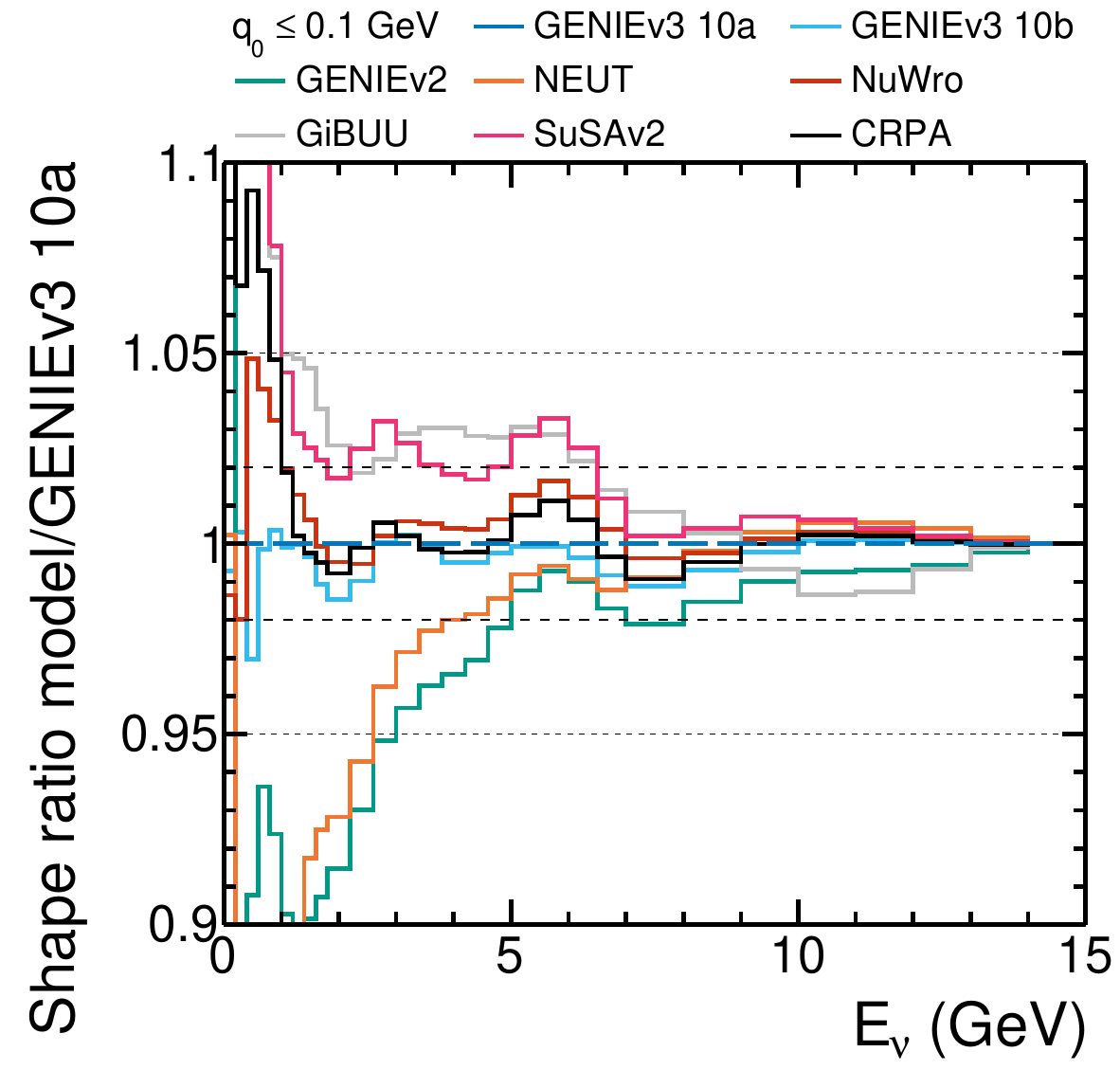}}
  \subfloat[\numu--\argon, $\qz \leq 0.3$ GeV]  {\includegraphics[width=0.3\linewidth]{{figures/flat_numu_Ar40_q0_0.3_GENCOMP_SHAPE_ratio}.pdf}}
  \subfloat[\numu--\argon, $\qz \leq 0.5$ GeV]  {\includegraphics[width=0.3\linewidth]{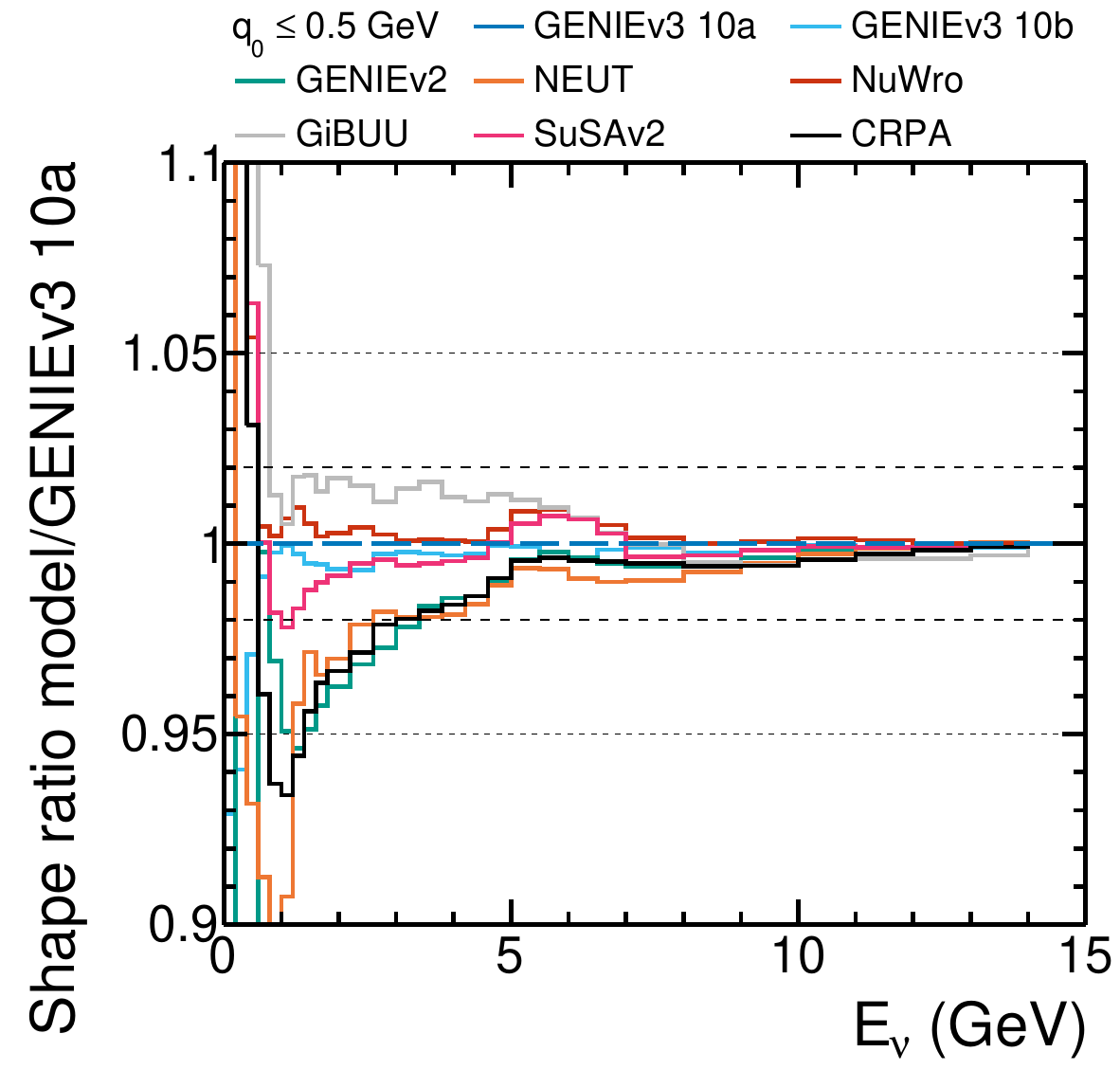}}\\\vspace{-5pt}
  \subfloat[\numub--\argon, $\qz \leq 0.1$ GeV]  {\includegraphics[width=0.3\linewidth]{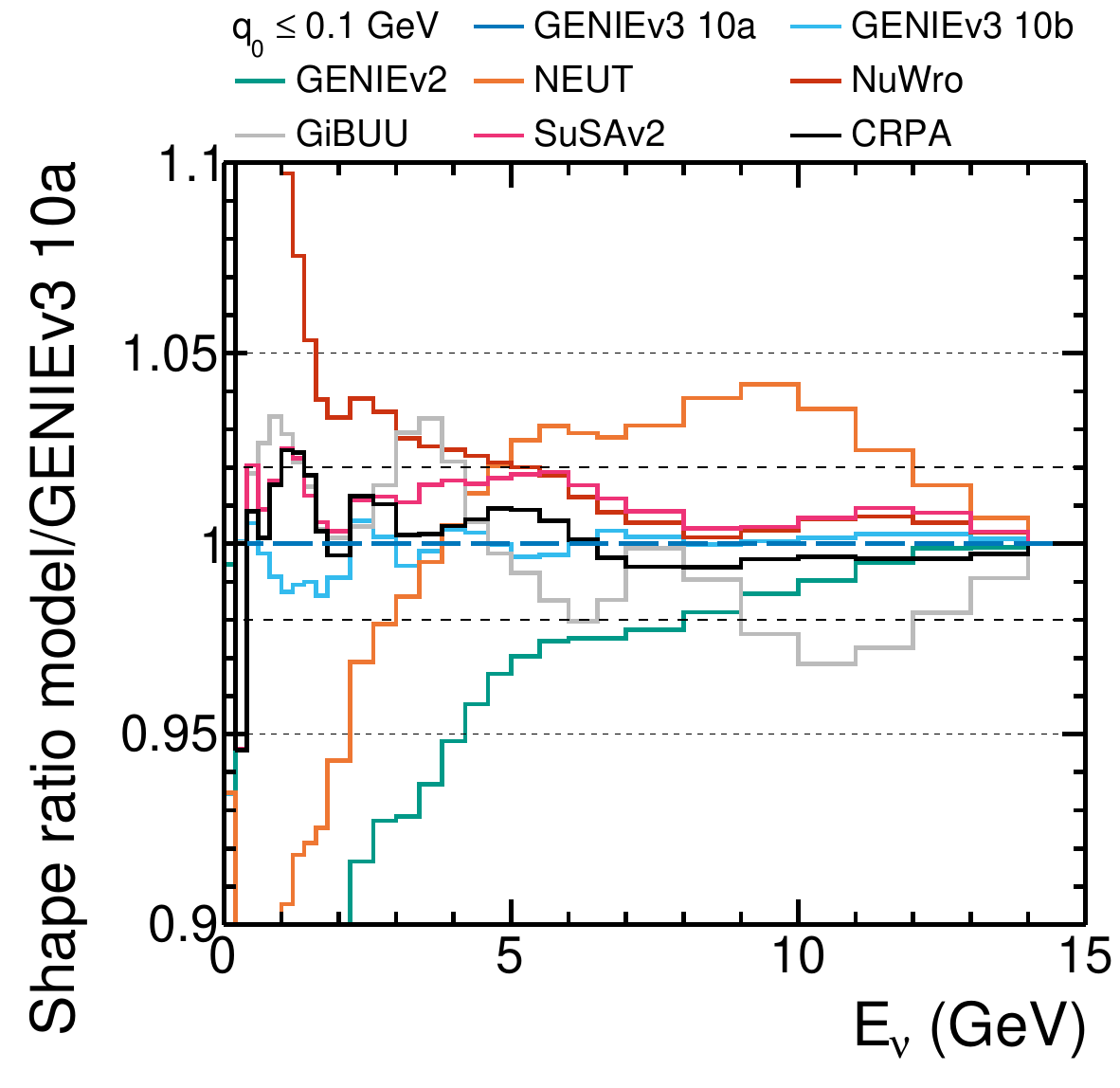}}
  \subfloat[\numub--\argon, $\qz \leq 0.3$ GeV]  {\includegraphics[width=0.3\linewidth]{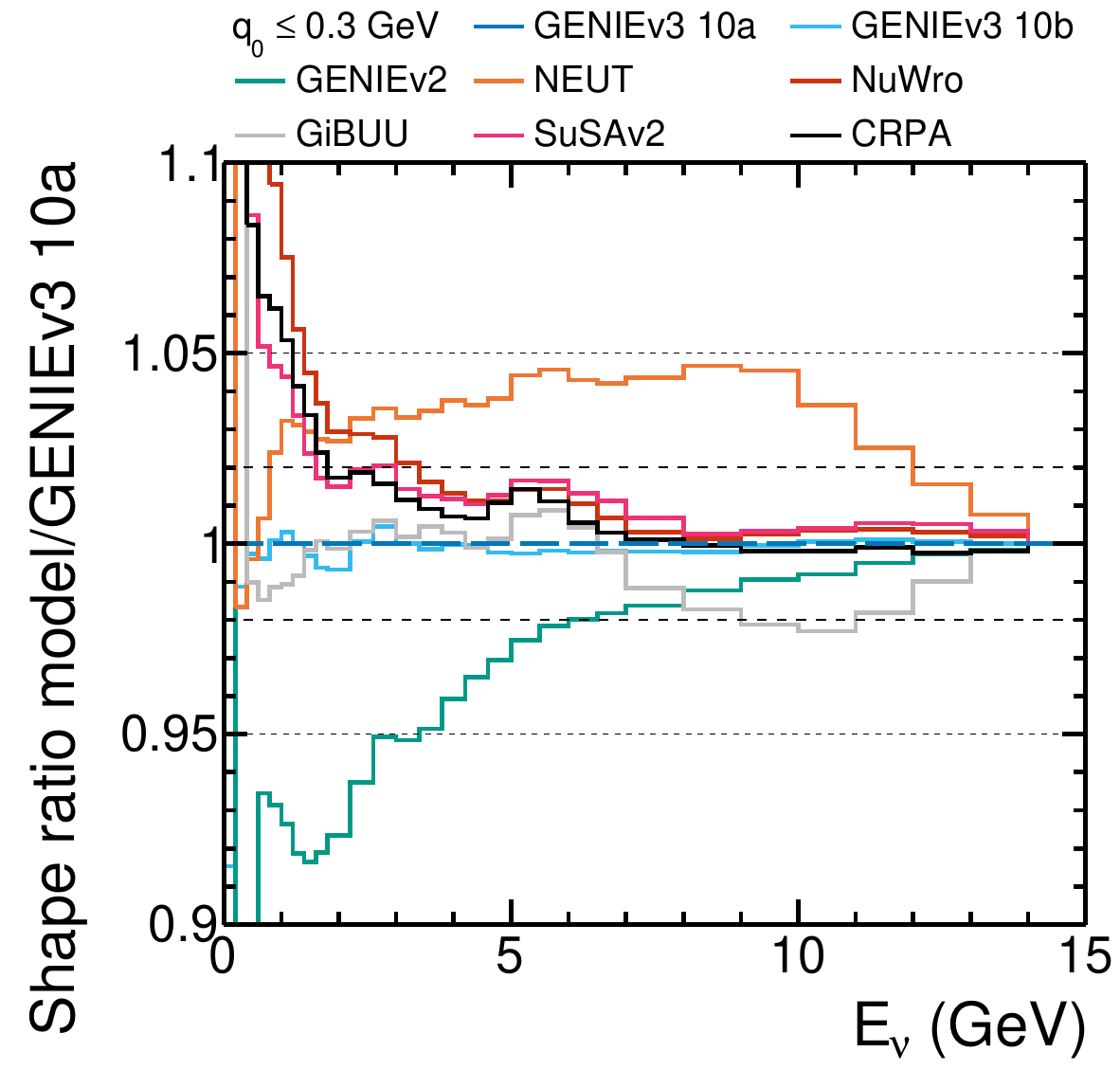}}
  \subfloat[\numub--\argon, $\qz \leq 0.5$ GeV]  {\includegraphics[width=0.3\linewidth]{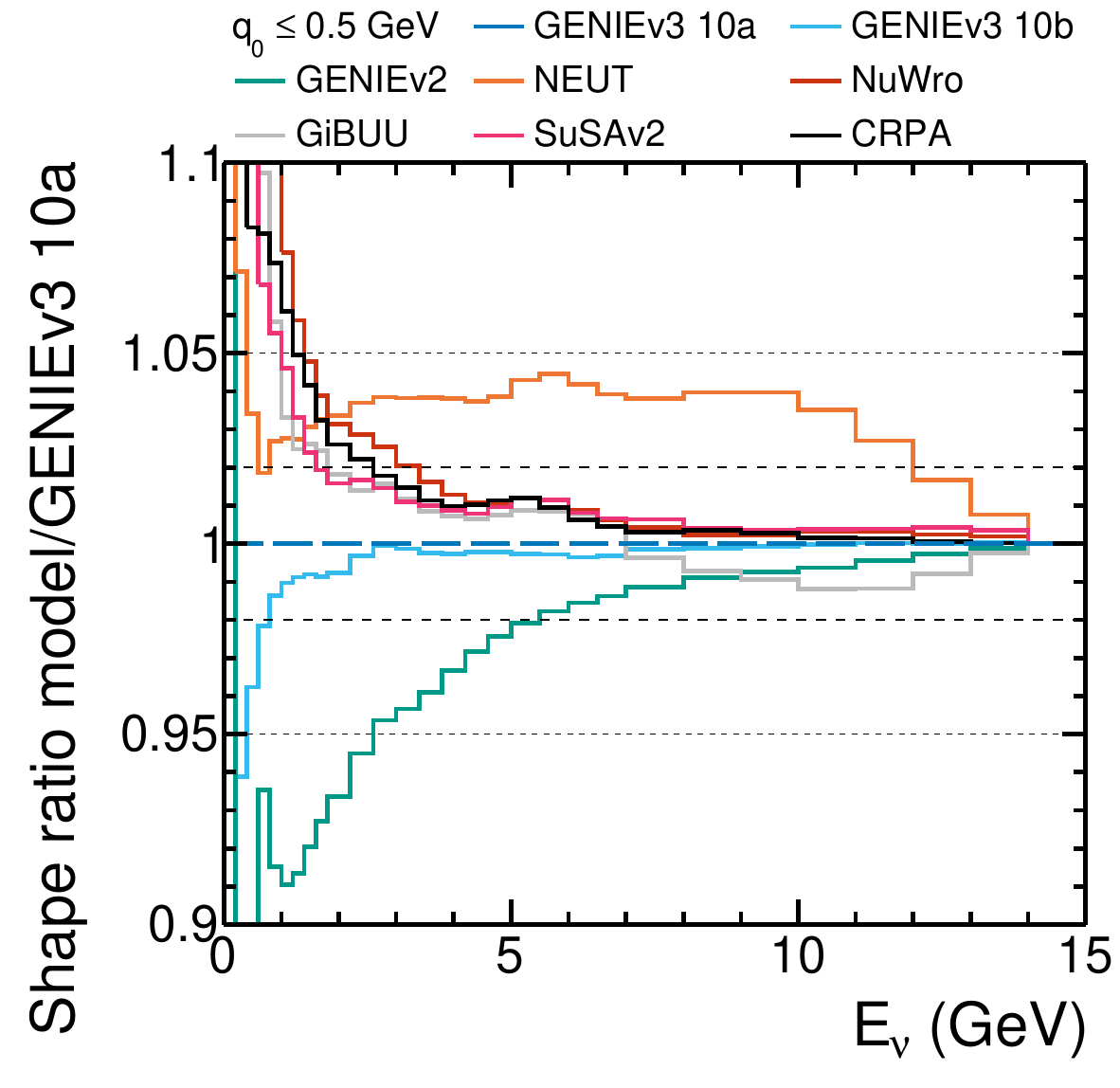}}
  \caption{Comparison of the shape-only ratios of the charged-current cross section with \qz $\leq$ 0.1, 0.3 and 0.5 GeV as a function of \enutrue, with respect to the GENIEv3 10a prediction. Shown for both \numu--\argon and \numub--\argon, and generated with a flat neutrino flux. Horizontal long (short) dashed lines have been added at $\pm$2\% ($\pm$5\%), to guide the eye in assessing the scale of the bias.}
  \label{fig:shape_vary_q0}
\end{figure*}

Although \autoref{fig:contrib_vary_gen} is illustrative, it is difficult to asses subtle differences between models. \autoref{fig:model_comp_progression} introduces the plot style which will be used to compare model differences pertinent to the application of the \lownu method from now on, for the example of $\qz \leq 0.3$ GeV. First, we show the \numu--\argon charged-current cross section predictions for all the models investigated in this work, with a specific \qz cut applied. Second, we show the ``shape-only'' comparisons, where the normalization of the $E_\nu=14-15~\text{GeV}$ bin is set to unity and all other bins are scaled relative to it. Although this shape-only definition seems unusual at first, it is motivated by that the \lownu method has recognized issues at low energies, with increasingly constant behaviour at higher energies. It follows the usage of ``accurately measured'' high-energy neutrino cross section data as a normalization factor when extracting a flux constraint with the method.
Finally, we show ratios of the ``shape-only'' comparisons to a reference model, chosen to be GENIEv3 10a due to its current widespread use. This final plot highlights the fractional difference between models as a function of the neutrino energy variable (here true neutrino energy). The spread of the predictions provides an estimate of the systematic uncertainty from cross-section modeling that would be applicable to a flux measurement extracted using the \lownu method with this sample and \qz cut value. It reflects the range of different values that would be obtained were different cross-section models assumed in the analysis. To guide the eye, long (short) dashed horizontal lines are shown at $\pm$2\% ($\pm$5\%) deviations from the reference GENIEv3 10a model. These indicate the approximate size of the shape-only (total) flux uncertainty around the peak neutrino energy from current and planned accelerator neutrino sources~\cite{minervaFlux, Vladisavljevic:2018prd, Abi:2020qib}, including information from hadron production experiments. In order to provide a useful flux constraint, it would be necessary to improve upon these {\it a priori} flux uncertainties.

\autoref{fig:shape_vary_q0} shows the shape-only ratio to GENIEv3 10a for all the models of interest, for a variety of true \qz cuts, for \numu--\argon and \numub--\argon. There are a number of striking features. In general the model differences appear largest for lower \qz cuts---counter to the arguments made in favor of the \lownu method for lower \enutrue experiments---particularly pronounced in the $E_\nu<5~\text{GeV}$ region. For $E_\nu<2~\text{GeV}$ the differences between generators are larger than 20\%. For \numu--\argon and a cut of $\qz \leq 0.3$ GeV, the models only come within 2\% of the reference GENIEv3 10a model for $E_{\nu} \gtrsim 5$ GeV. The variations are larger in the \numub--\argon case, with significantly different behavior for NEUT, GENIEv2 and GiBUU. \autoref{app:chtarget} presents a similar analysis using \ch and \chtwo targets, which shows the same trends. It is clear that, regardless of the choice of true \qz cut, any flux extracted from the \lownu method would risk introducing a large model-dependent bias at the peak neutrino energies of any current and planned long-baseline oscillation experiment, even with an ideal detector that could be used to observe true \qz and \enutrue.

\begin{figure*}[htbp]
  \centering
  \captionsetup[subfloat]{captionskip=-2pt}
  \subfloat[\numu--\argon, $\qz \leq 0.8$ GeV]  {\includegraphics[width=0.3\linewidth]{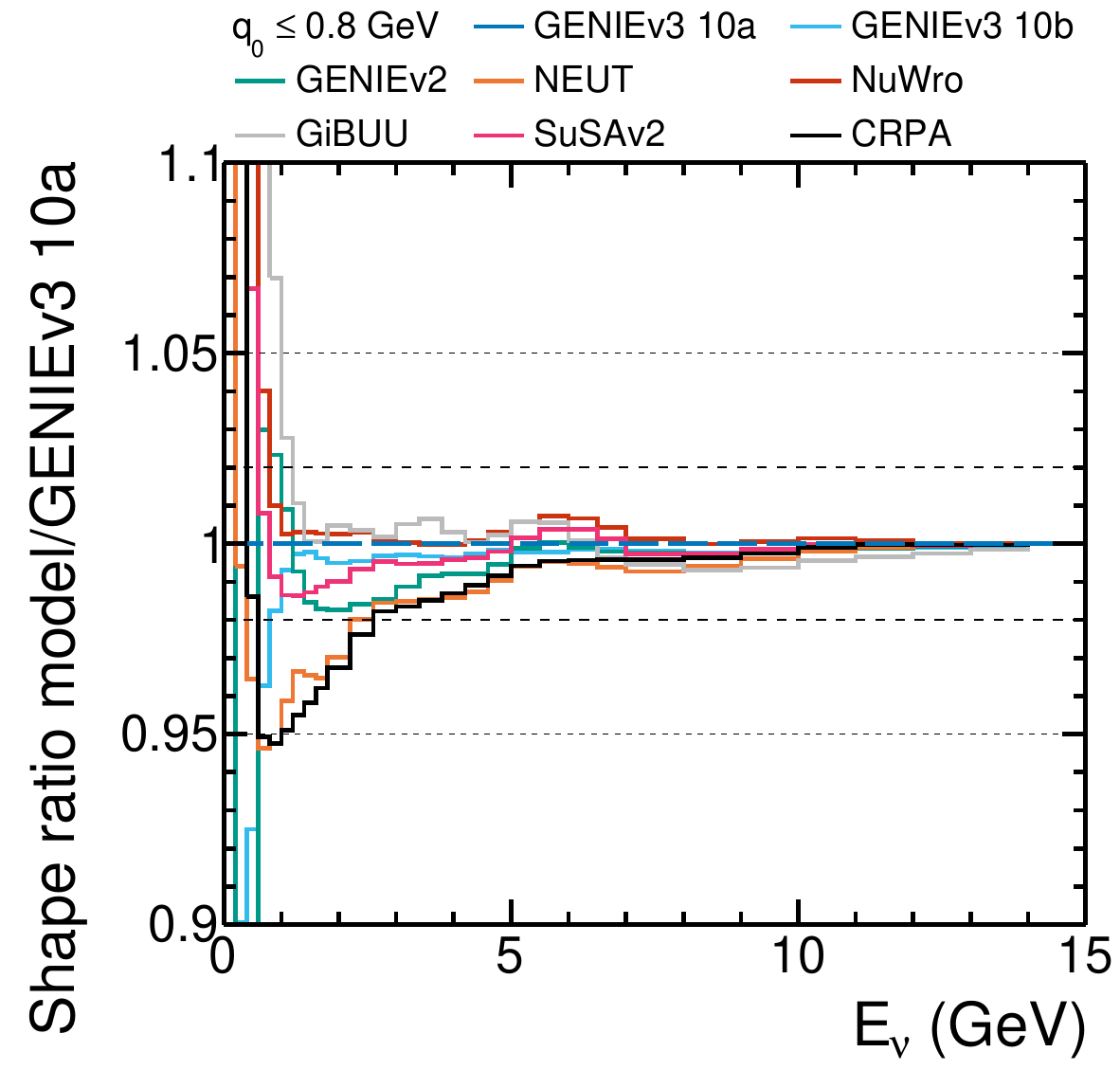}}
  \subfloat[\numu--\argon, $\qz \leq 2.0$ GeV]  {\includegraphics[width=0.3\linewidth]{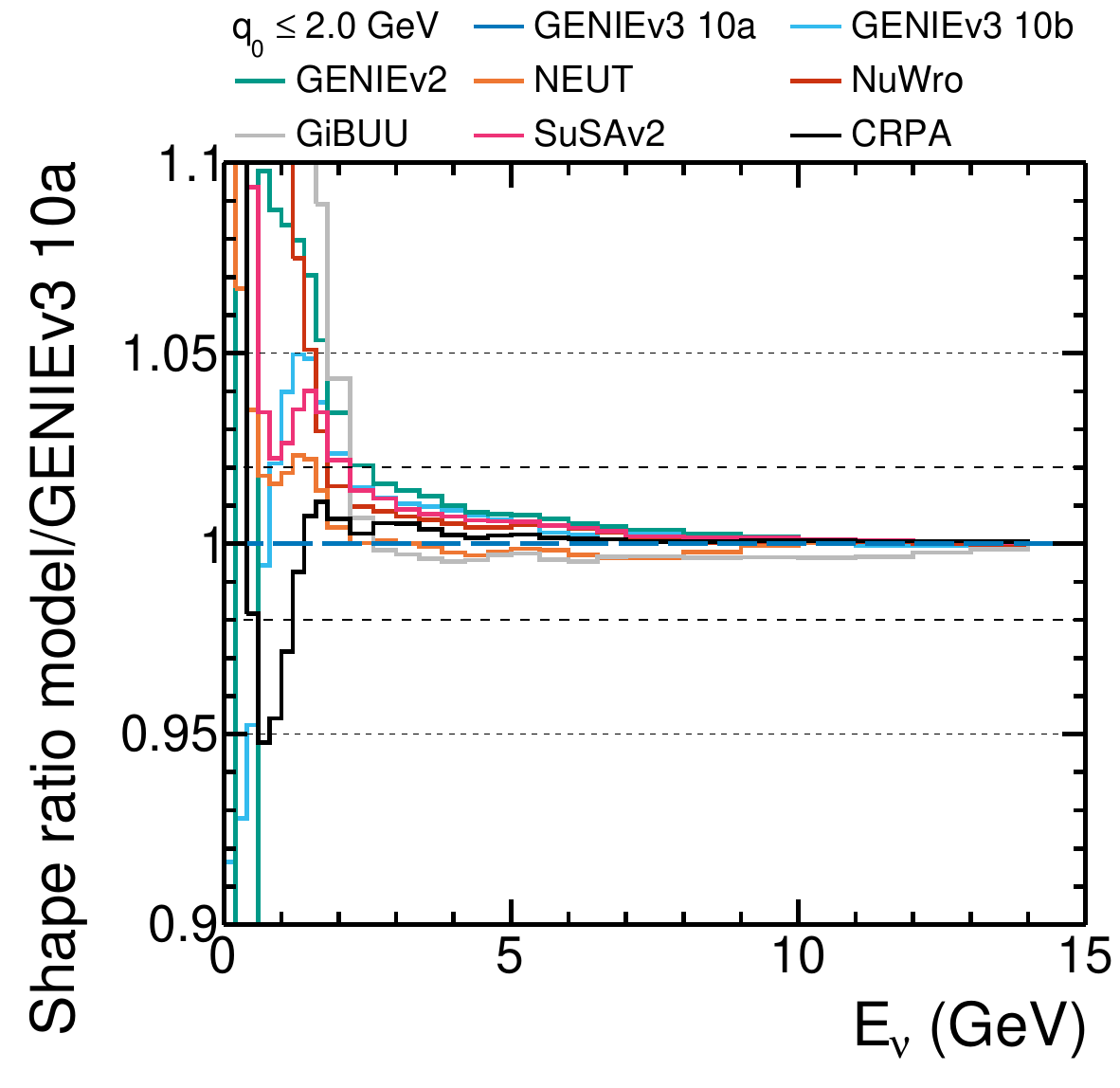}}
  \subfloat[\numu--\argon, all-\qz]             {\includegraphics[width=0.3\linewidth]{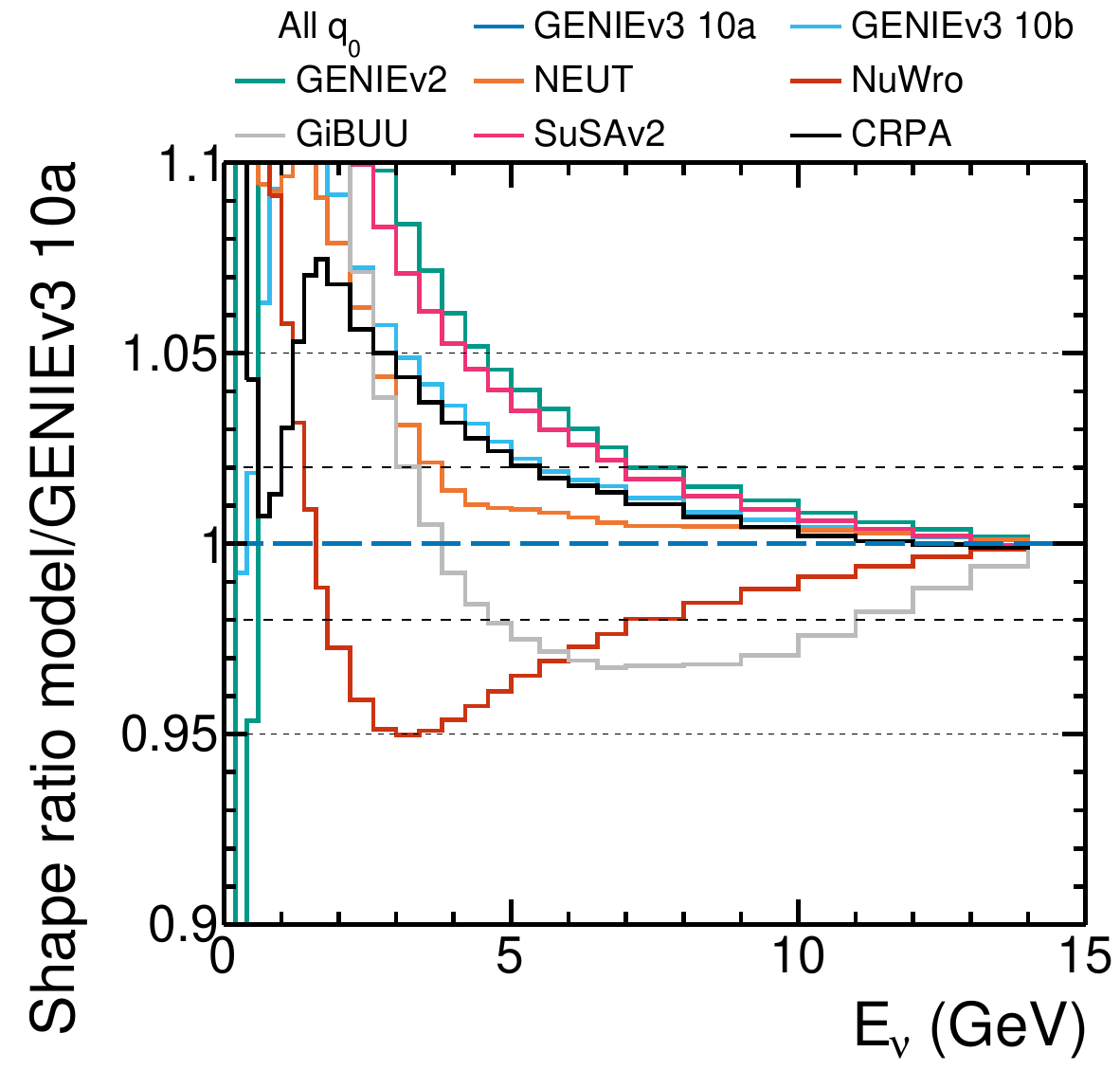}}
  \caption{Comparison of the shape-only ratios of the charged-current \numu--\argon cross section with \qz $\leq$ 0.8, 2.0 GeV and all-\qz as a function of \enutrue, with respect to the GENIEv3 10a prediction. Horizontal long (short) dashed lines have been added at $\pm$2\% ($\pm$5\%), to guide the eye in assessing the scale of the bias.}
  \label{fig:shape_vary_highq0}
\end{figure*}

To further investigate the observation made from \autoref{fig:shape_vary_q0} that the model spread decreases with higher \qz cuts, \autoref{fig:shape_vary_highq0} shows the ratios between models for $\qz \leq$ 0.8, 2 GeV and all-\qz (no cut). It is striking that by far the best agreement between models is for $\qz \leq 2$ GeV, with deviations less than 2\% for $\enutrue \gtrsim 2$ GeV. However, this is more likely to be a consequence of the relative lack of model diversity in the higher \qz regions more than any indication that the cross section is significantly better understood there. It important to note that for all current or planned neutrino oscillation experiments, a cut of $\qz < 2.0~\textrm{GeV}$ cannot be considered ``low \qz'', and any arguments made for a constant cross section from \autoref{eq:low-nu} completely break down. Moreover, such a high \qz cut would also lead to considerable overlap between the flux-constraining sample and any analysis samples, creating statistical problems when double-counting observations. When no \qz cut is applied, significant disagreements at low \enutrue for all-\qz are seen, largely driven by the normalization to the highest \enutrue bin.

Even for an idealized analysis with a perfect detector, in which \enutrue can be observed, and a sample can be selected based on true \qz cuts, we already see significant issues with the \lownu method. There are sizeable differences between the models investigated, which should be a concern of experiments seeking to use the \lownu method at these neutrino energies, and with correspondingly low cuts on \qz. The observation that the model differences increase with cuts at lower \qz highlights that the DIS-motivated formalism in \autoref{eq:low-nu} is not an appropriate approximation of the cross section at these values of \enutrue and \qz. The nucleon structure function treatment that motivates the \lownu method neglects explicit descriptions of the complex nuclear dynamics that become important at $\qz \sim \mathcal{O}(100~\text{ MeV}$). 

The observed differences are perhaps unsurprising. The models implement very different approaches to the simulation of pertinent nuclear physics processes, which become increasingly important for the low \qz region that occupies a larger fraction of the available phase space at lower neutrino energies. Furthermore, the low \qz region is also sensitive to kinematic threshold effects, recognized elsewhere~\cite{Belusevic:1987rn,Belusevic:1988ab, Bodek:2012uu}.  It is also worth remarking that the model differences observed for \numu interactions do not generally bare much resemblance to those observed for the \numub case, therefore suggesting that the model dependence inherent in the \lownu method has a strong potential to introduce biases in the modelling \numu/\numub cross-section asymmetry. Such biases have the potential to mimic or diminish a CP violation signal and so should be of particular concern for long-baseline neutrino oscillation experiments. 

\section{Experimentally accessible variables}
\label{sec:variables}

\begin{figure*}[htbp]
  \centering
  \captionsetup[subfloat]{captionskip=-1pt}
  \subfloat[\enutrue = 2 GeV, \ehadtrue]  {\includegraphics[width=0.3\linewidth]{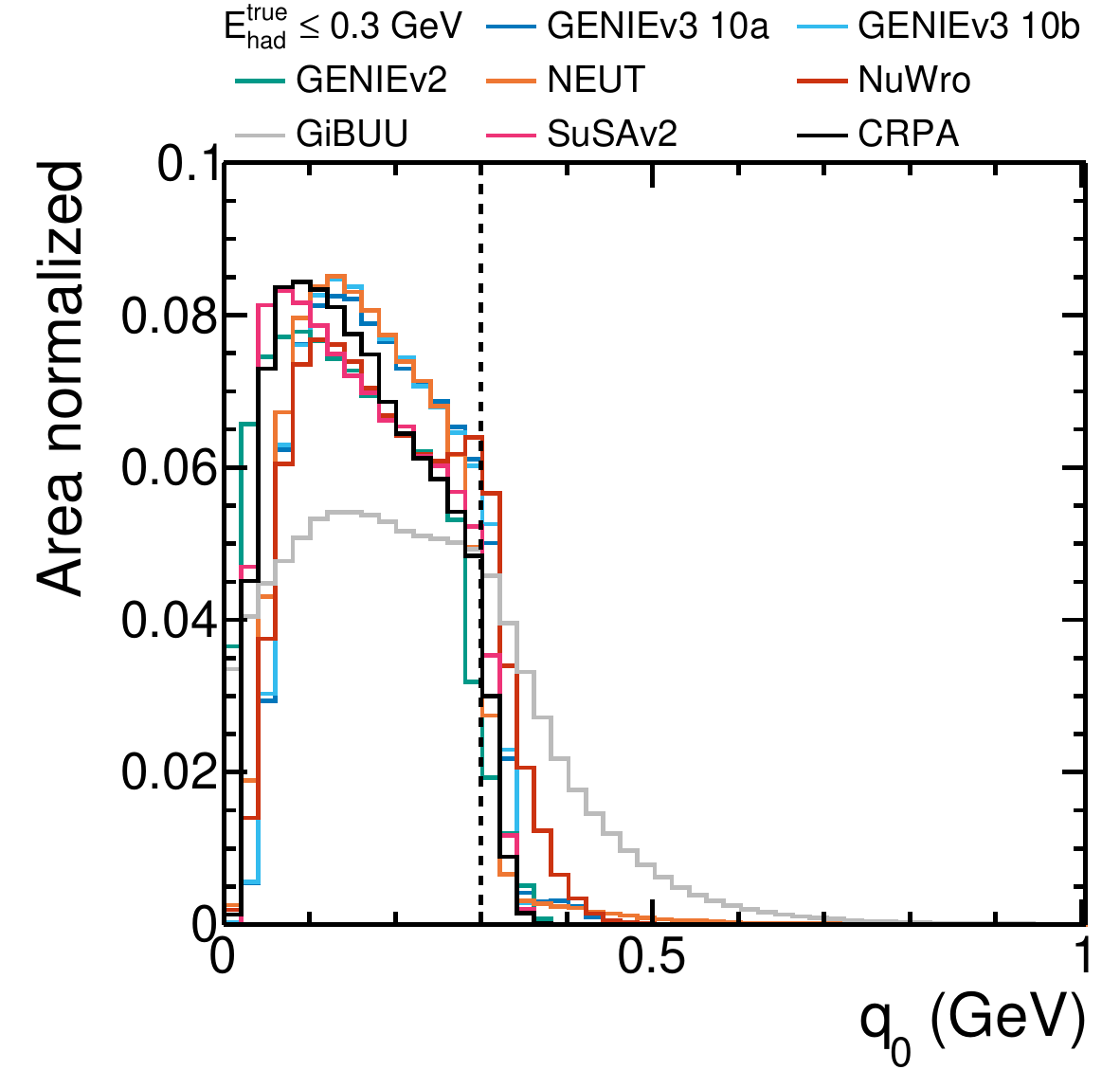}}
  \subfloat[\enutrue = 5 GeV, \ehadtrue]  {\includegraphics[width=0.3\linewidth]{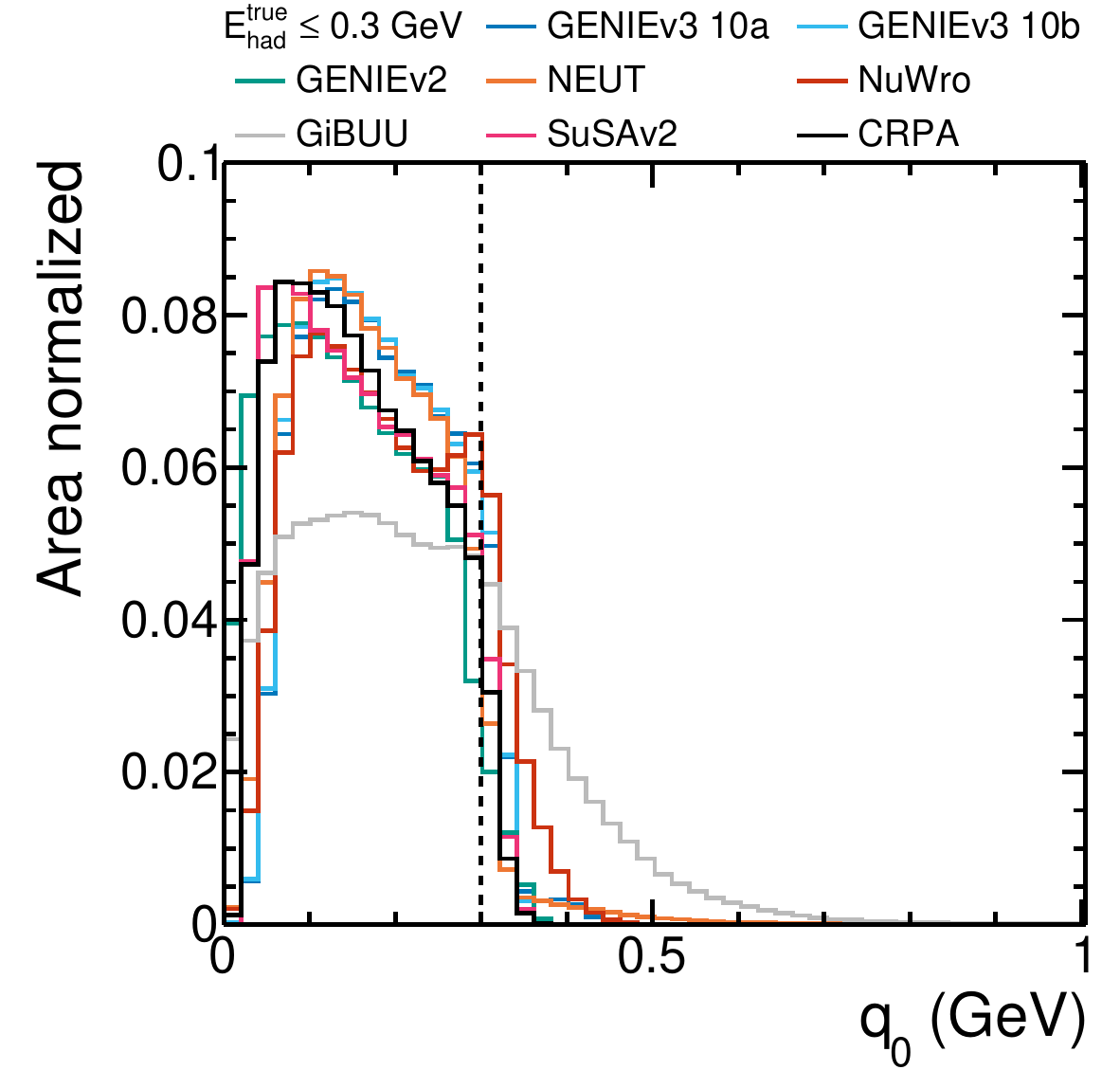}}
  \subfloat[\enutrue = 10 GeV, \ehadtrue]  {\includegraphics[width=0.3\linewidth]{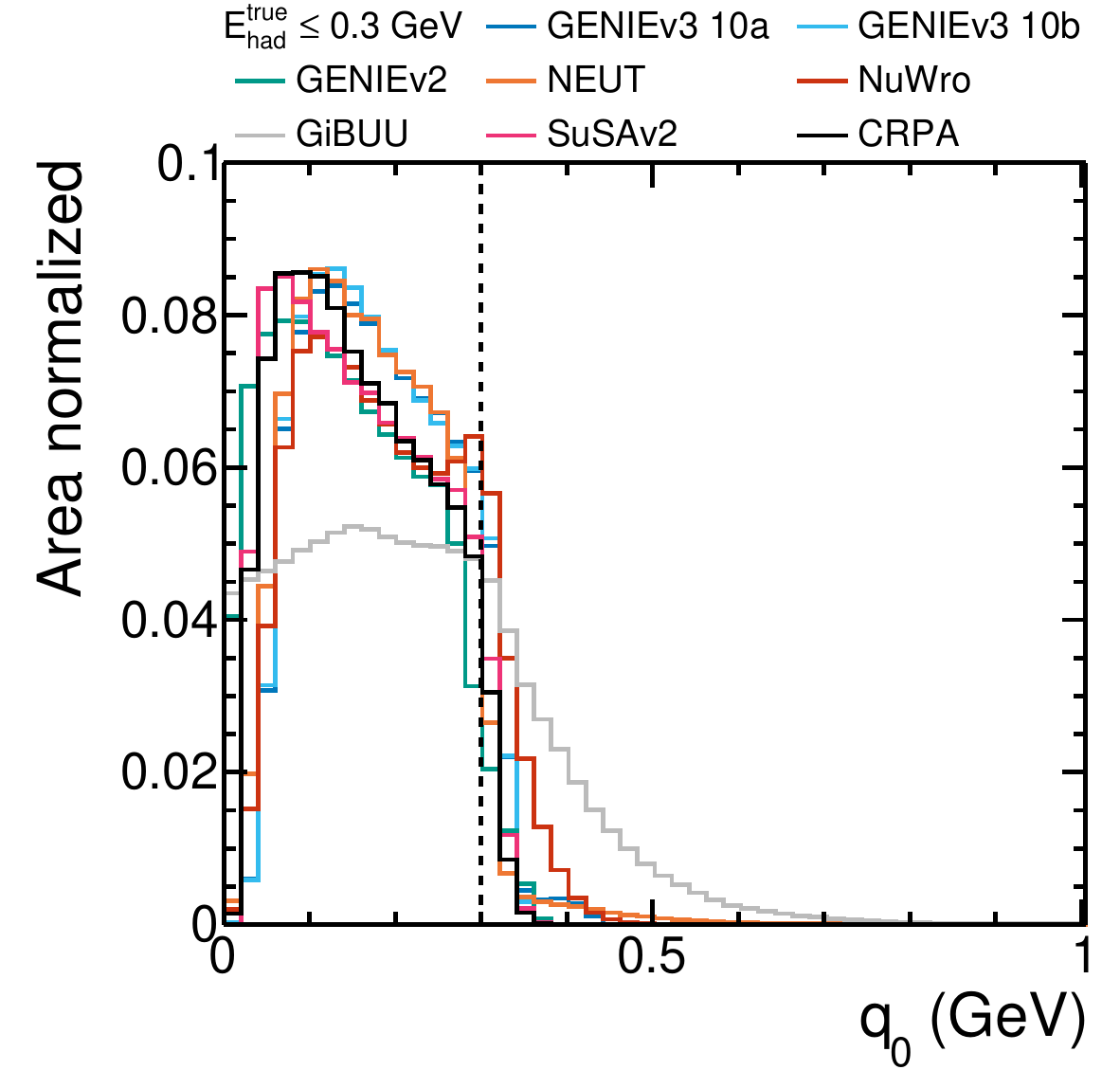}}\\\vspace{-2pt}
  \subfloat[\enutrue = 2 GeV, \ehadreco]  {\includegraphics[width=0.3\linewidth]{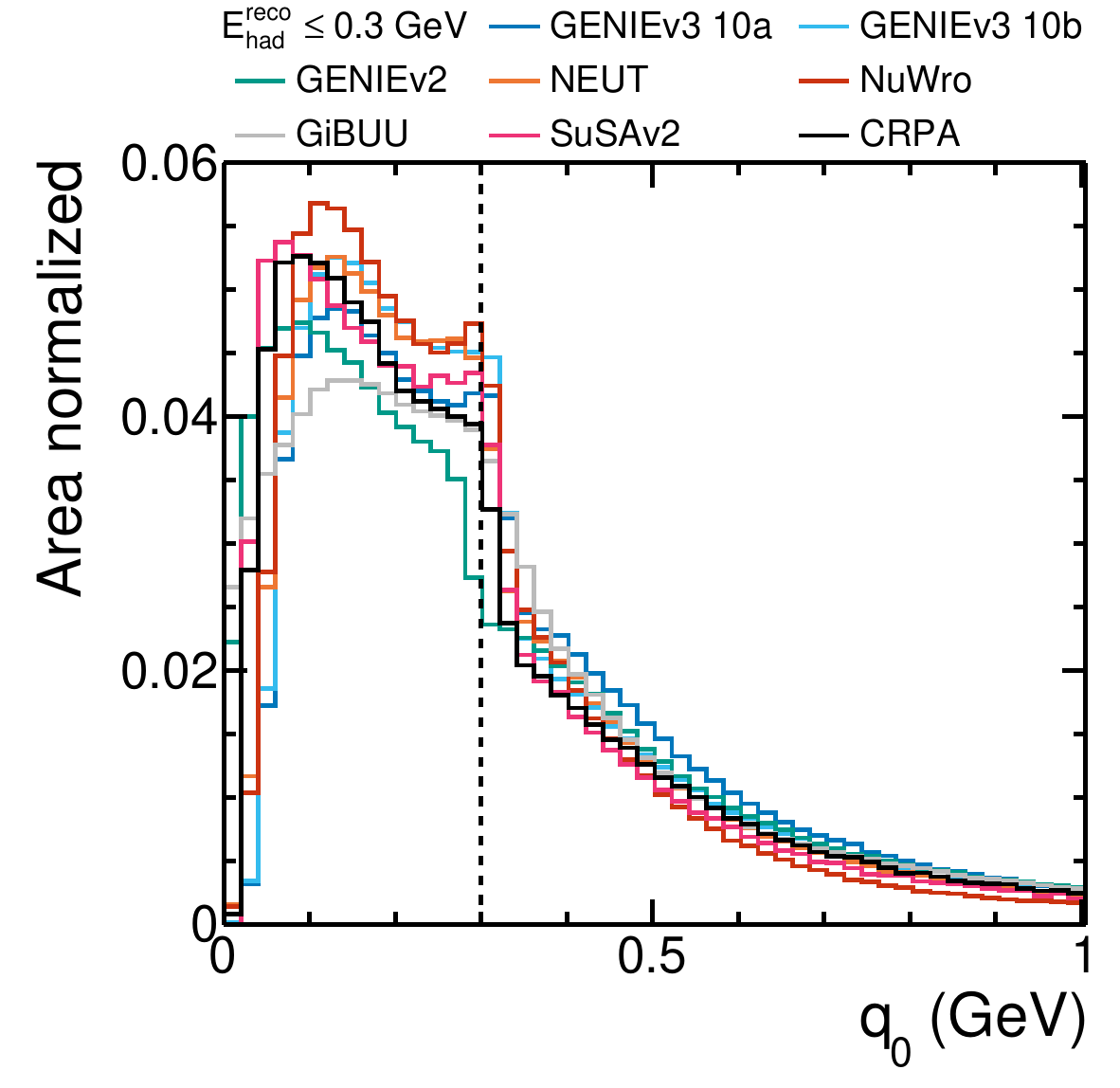}}
  \subfloat[\enutrue = 5 GeV, \ehadreco]  {\includegraphics[width=0.3\linewidth]{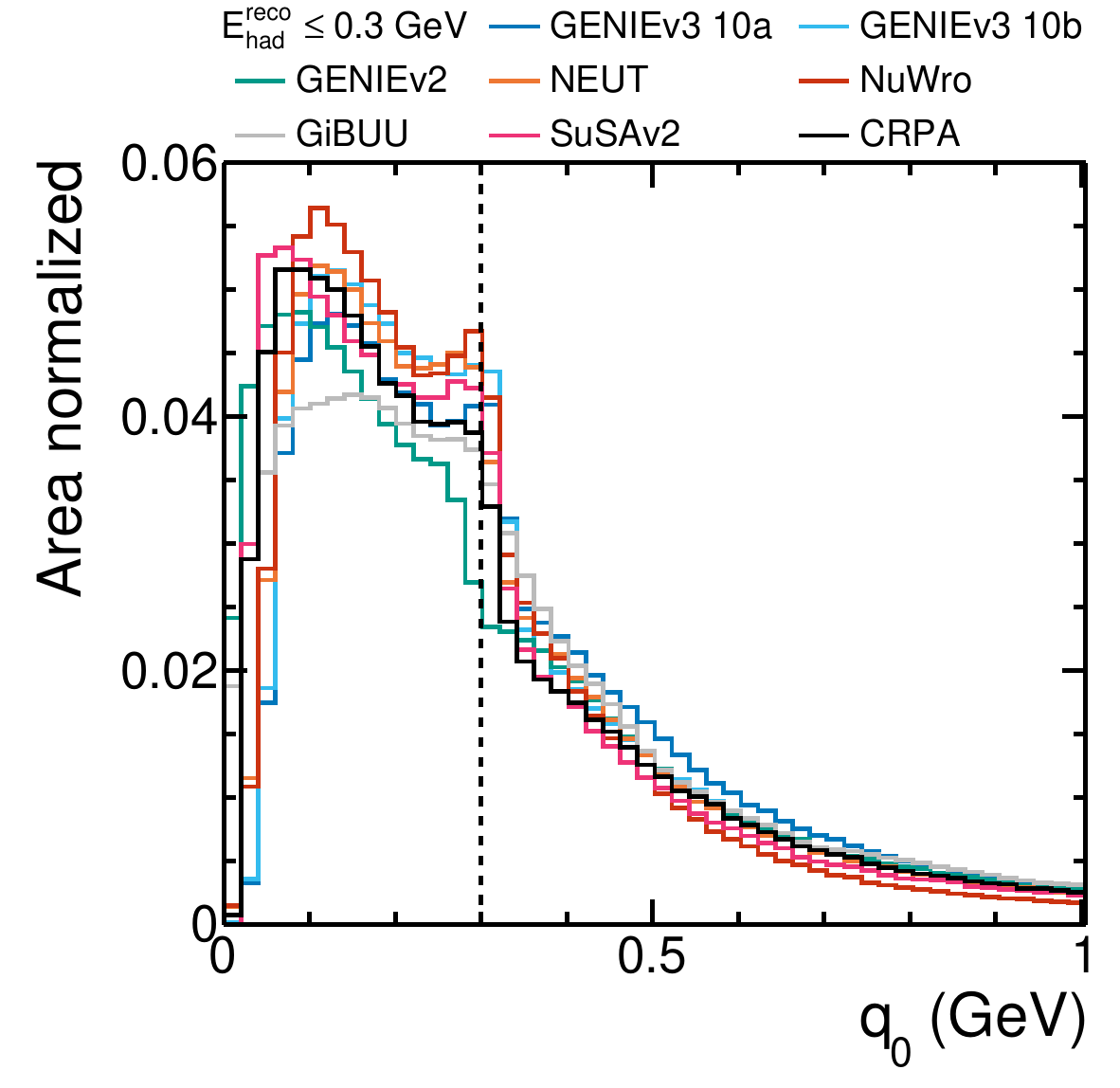}}
  \subfloat[\enutrue = 10 GeV, \ehadreco]  {\includegraphics[width=0.3\linewidth]{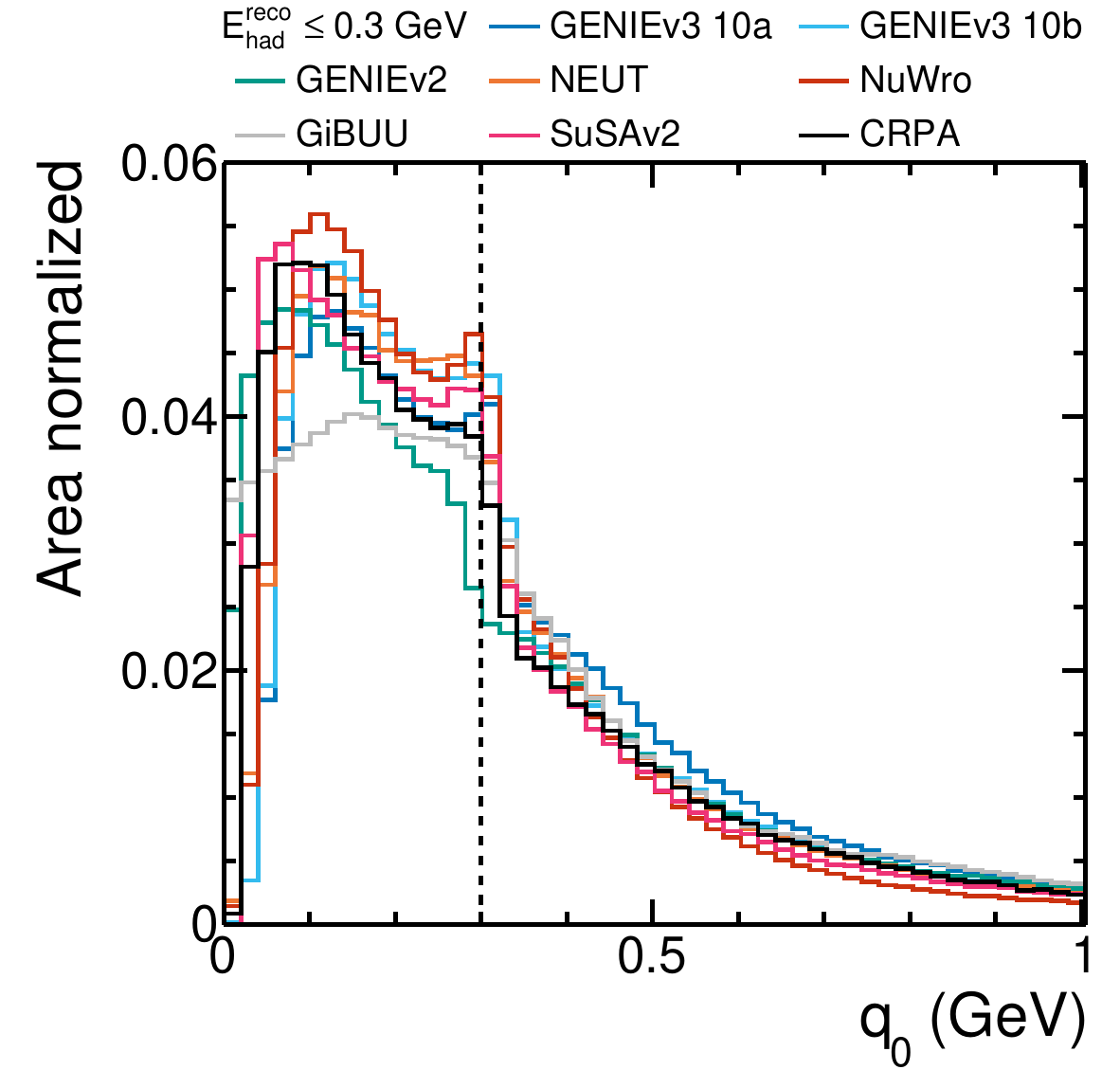}}\\\vspace{-2pt}
  \subfloat[\enutrue = 2 GeV, \eavail]    {\includegraphics[width=0.3\linewidth]{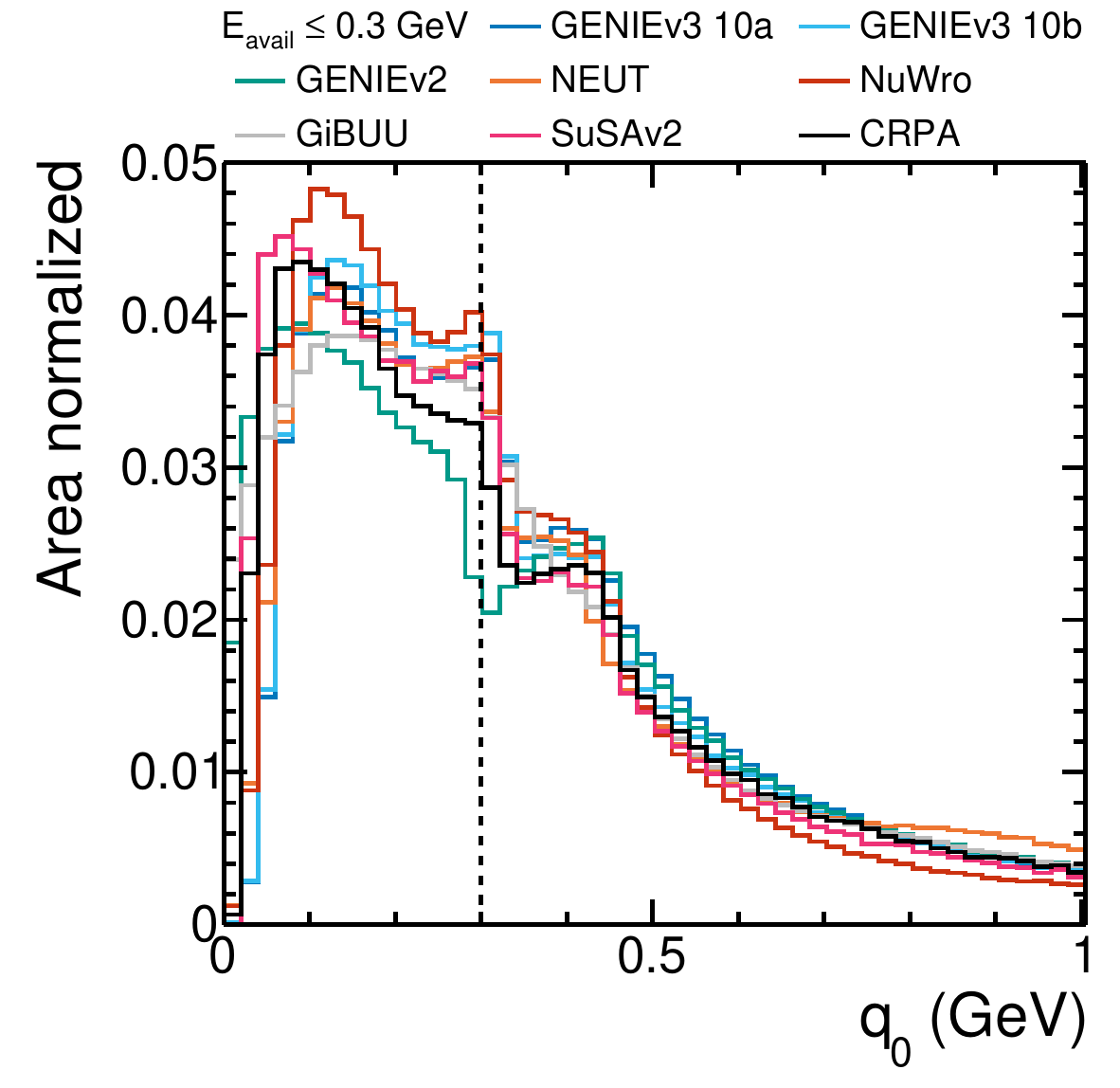}}
  \subfloat[\enutrue = 5 GeV, \eavail]    {\includegraphics[width=0.3\linewidth]{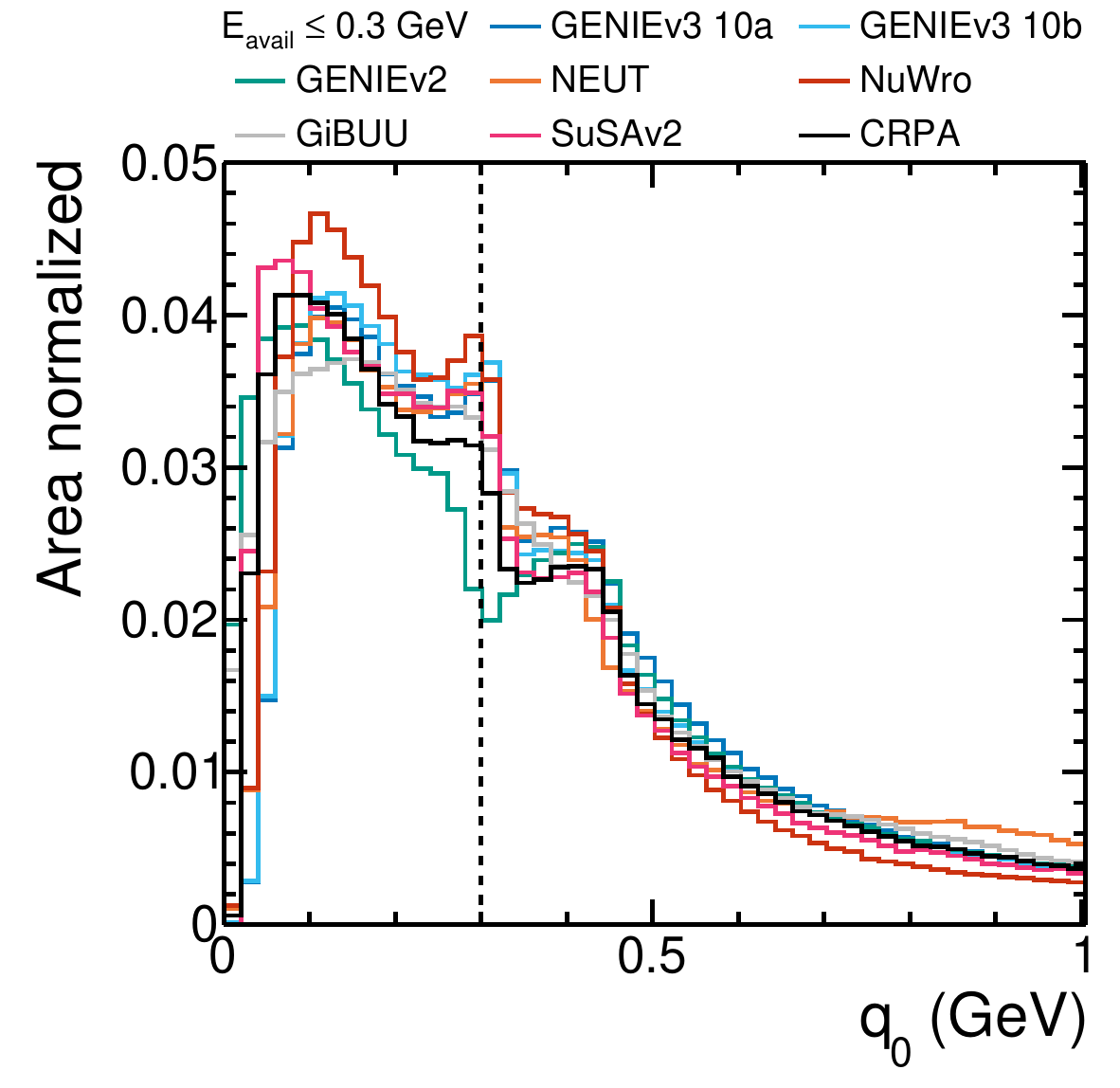}}
  \subfloat[\enutrue = 10 GeV, \eavail]    {\includegraphics[width=0.3\linewidth]{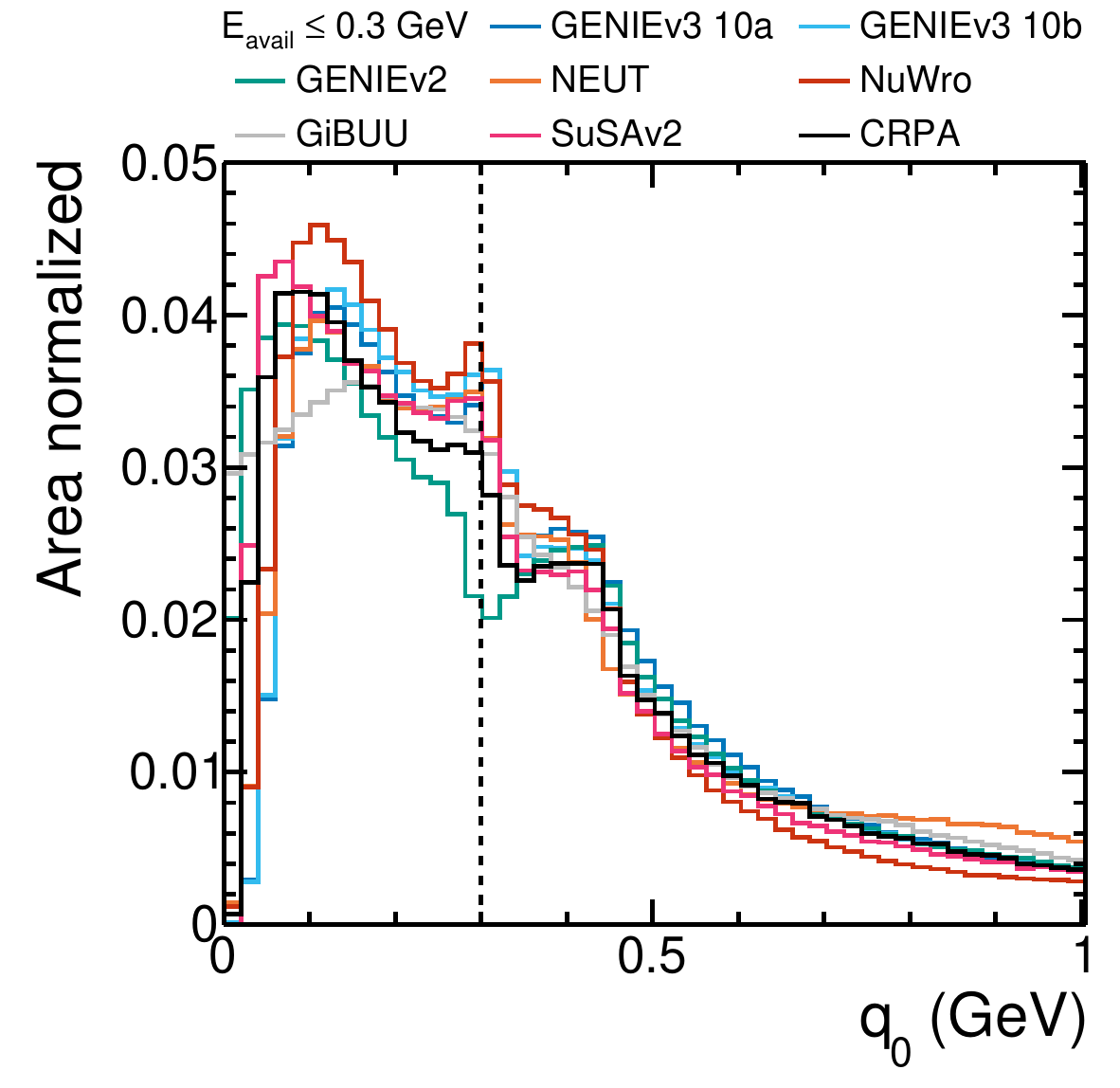}}\\\vspace{-2pt}
  \caption{The \qz distributions for CC-inclusive \numu--\argon events which pass cuts on the three proxy variables which are experimentally observable: \ehadtrue, \ehadreco and \eavail$ \leq$ 0.3 GeV, shown for all the generators models considered, for three different neutrino distributions, which are uniform and centred on 2, 5 and 10 GeV with a $\pm$0.5 GeV width.}
  \label{fig:smear_q0_proxy_numu}
\end{figure*}

\begin{figure*}[htbp]
  \centering
  \captionsetup[subfloat]{captionskip=-1pt}
  \subfloat[\enutrue = 2 GeV, \ehadtrue]  {\includegraphics[width=0.3\linewidth]{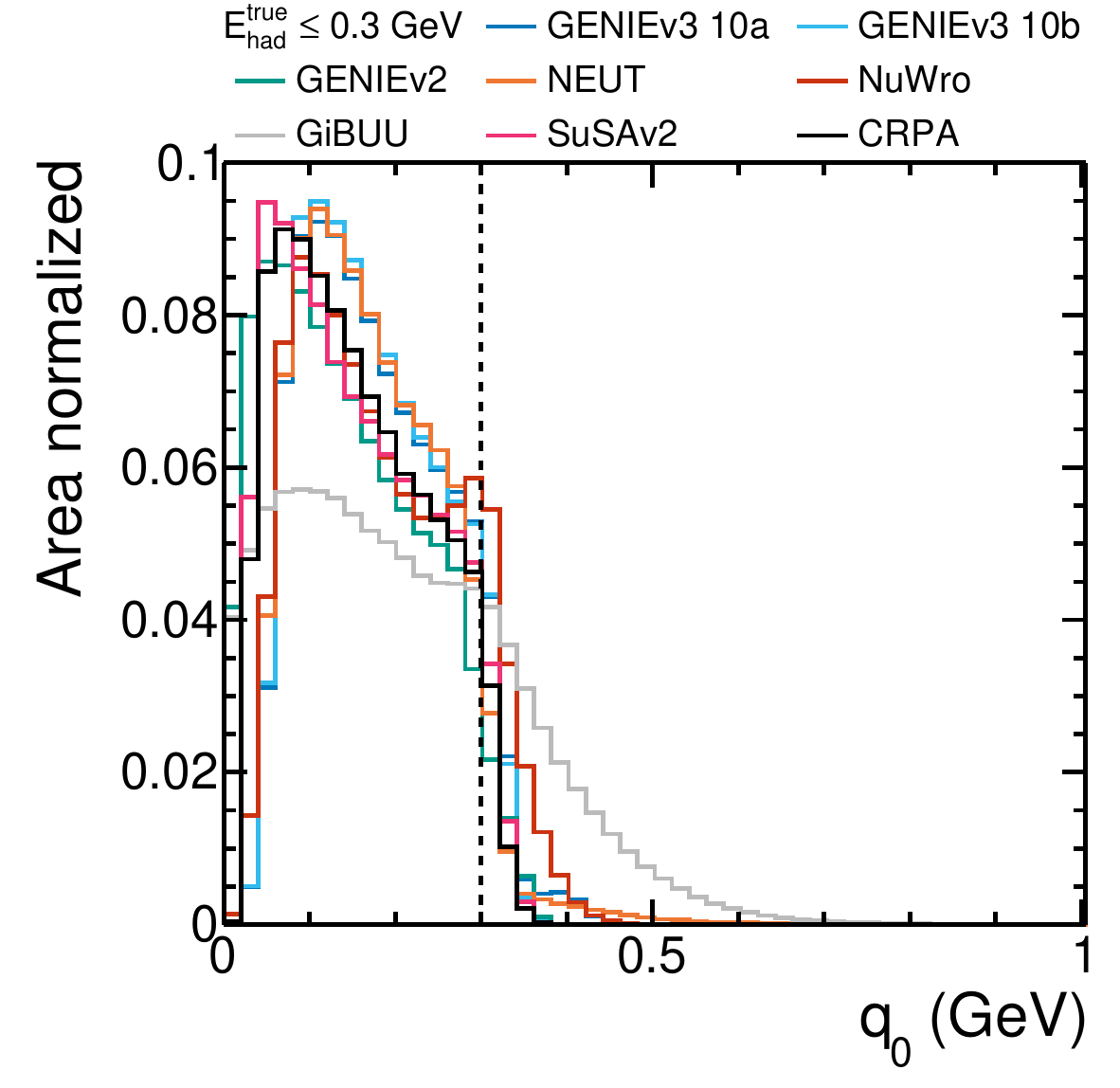}}
  \subfloat[\enutrue = 5 GeV, \ehadtrue]  {\includegraphics[width=0.3\linewidth]{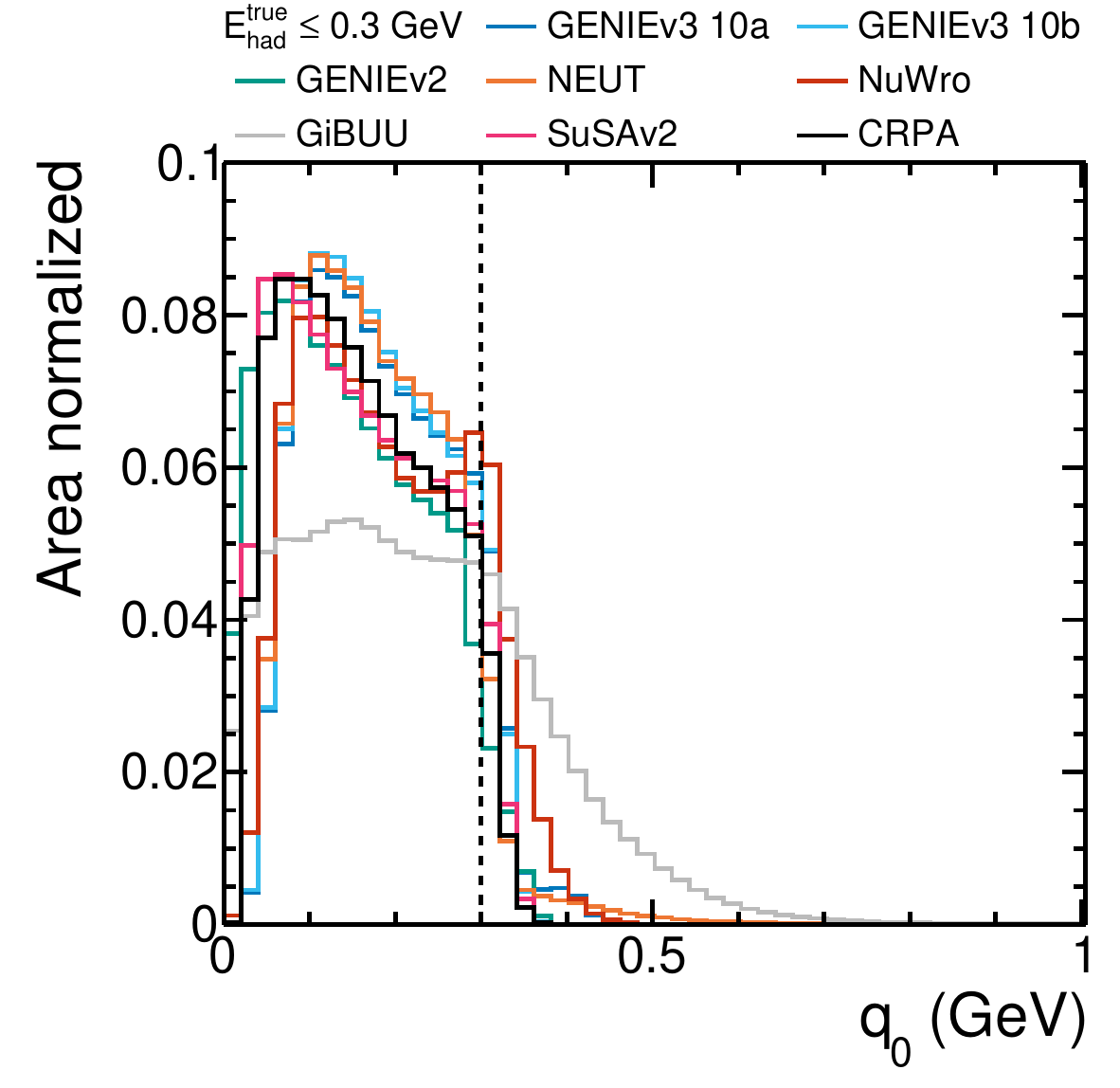}}
  \subfloat[\enutrue = 10 GeV, \ehadtrue]  {\includegraphics[width=0.3\linewidth]{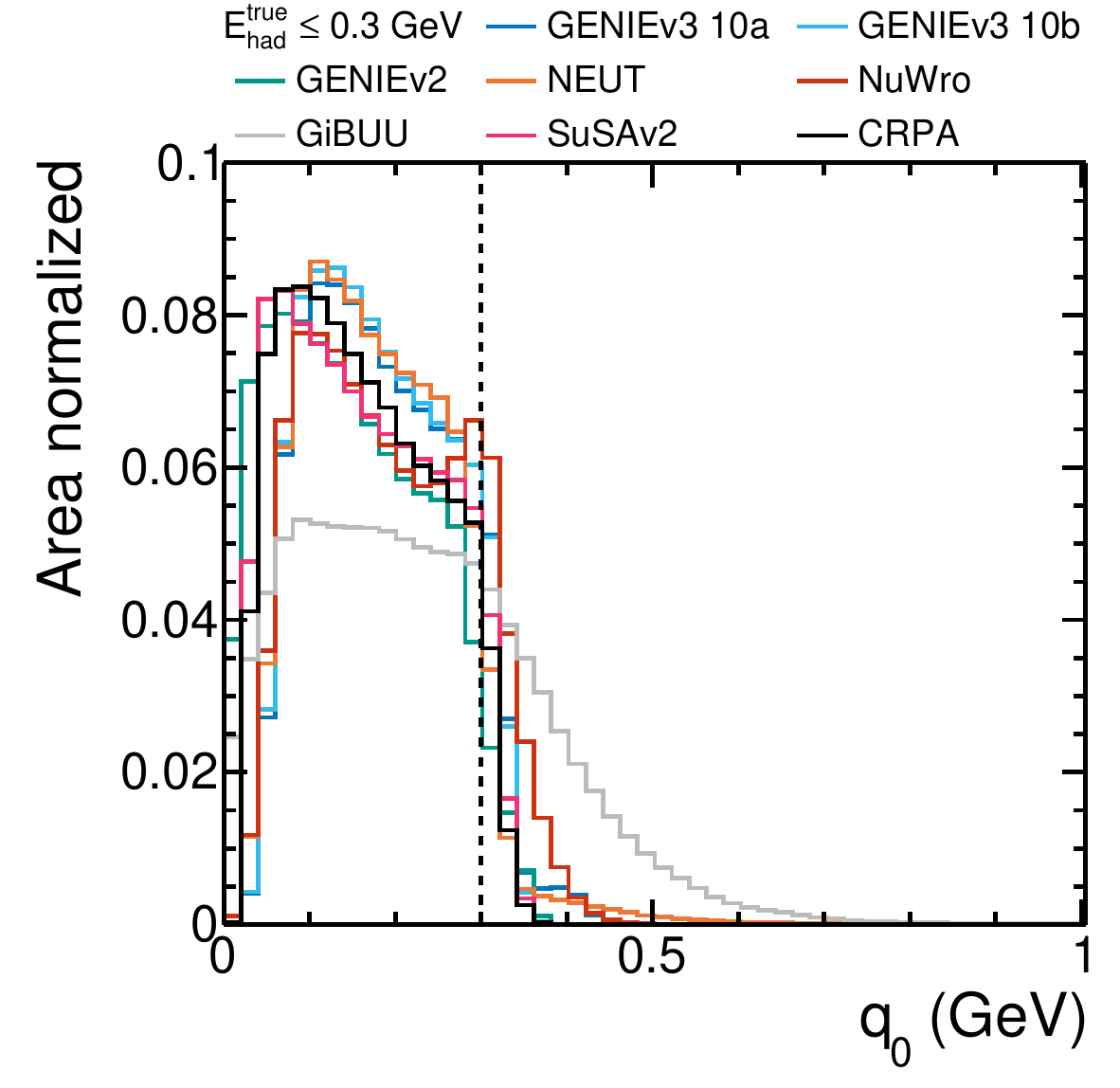}}\\\vspace{-2pt}
  \subfloat[\enutrue = 2 GeV, \ehadreco]  {\includegraphics[width=0.3\linewidth]{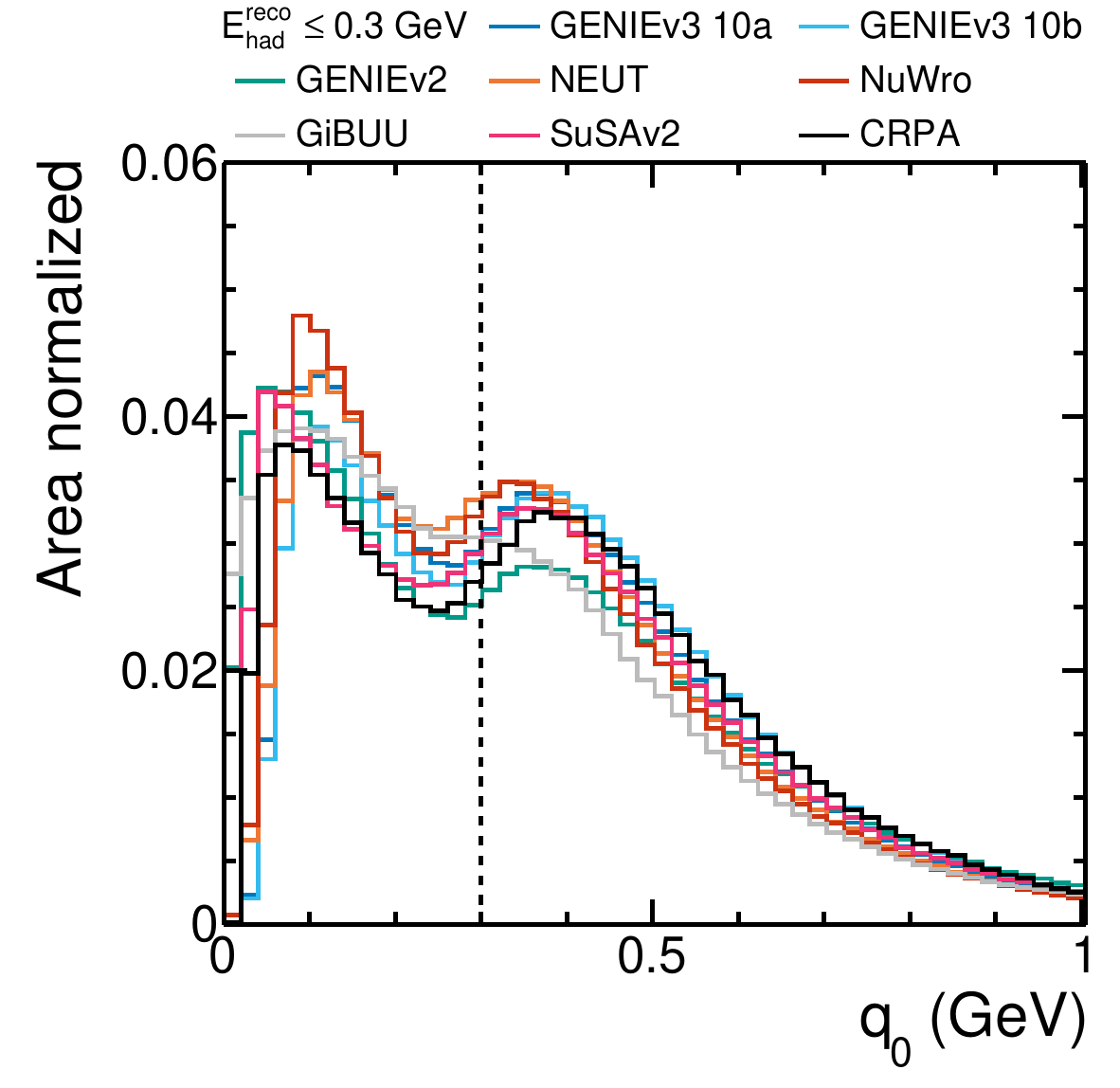}}
  \subfloat[\enutrue = 5 GeV, \ehadreco]  {\includegraphics[width=0.3\linewidth]{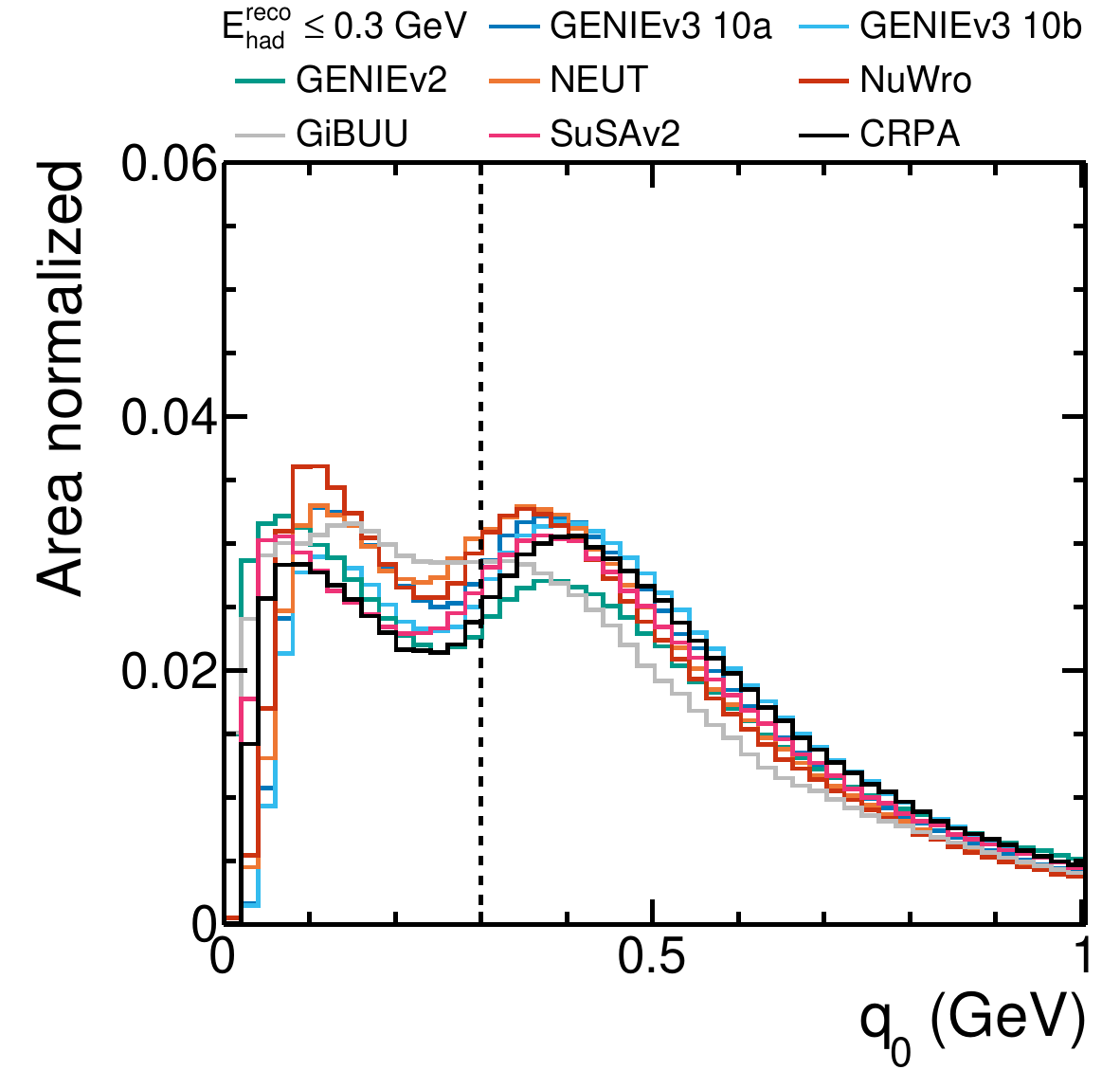}}
  \subfloat[\enutrue = 10 GeV, \ehadreco]  {\includegraphics[width=0.3\linewidth]{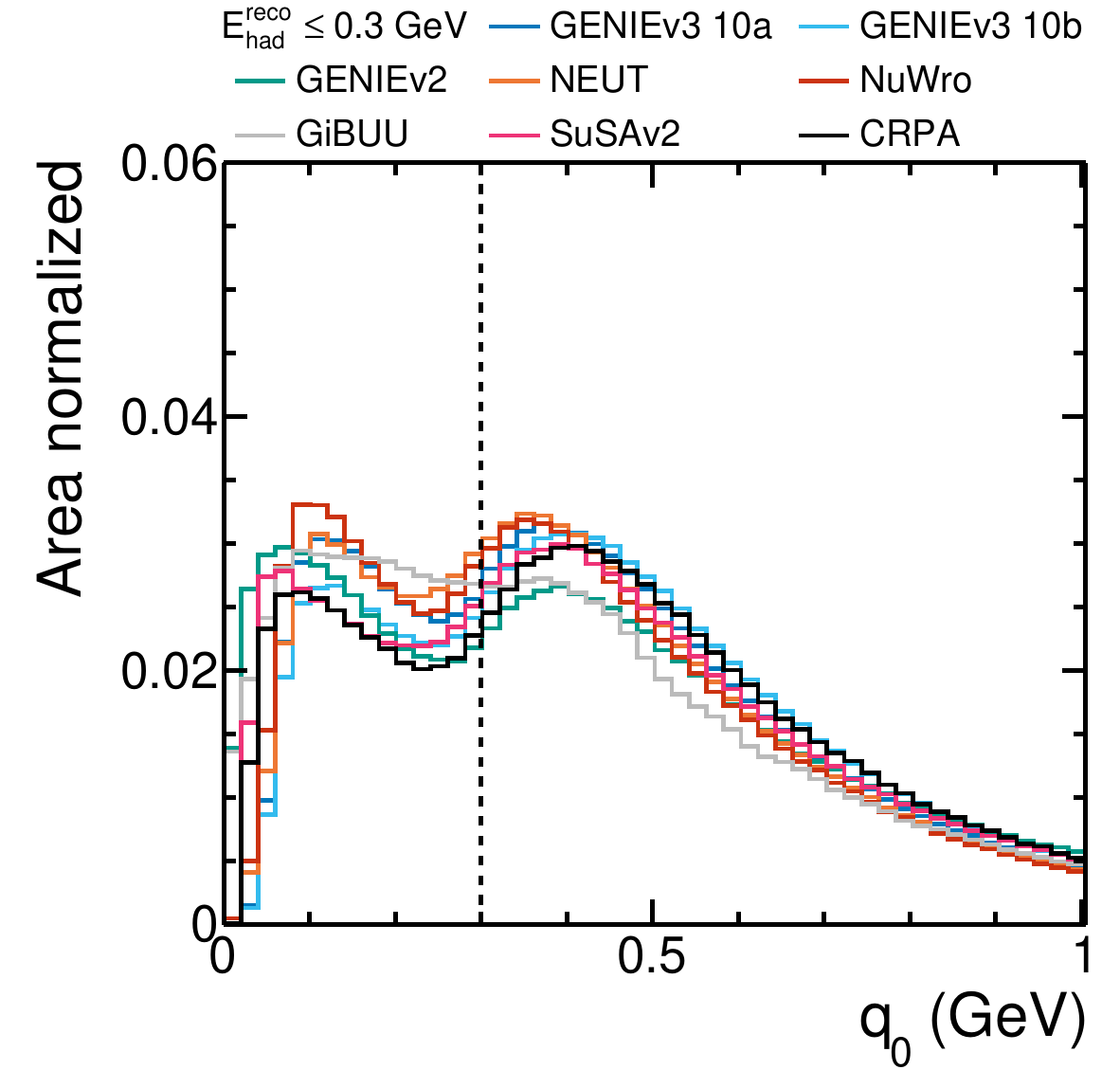}}\\\vspace{-2pt}
  \subfloat[\enutrue = 2 GeV, \eavail]    {\includegraphics[width=0.3\linewidth]{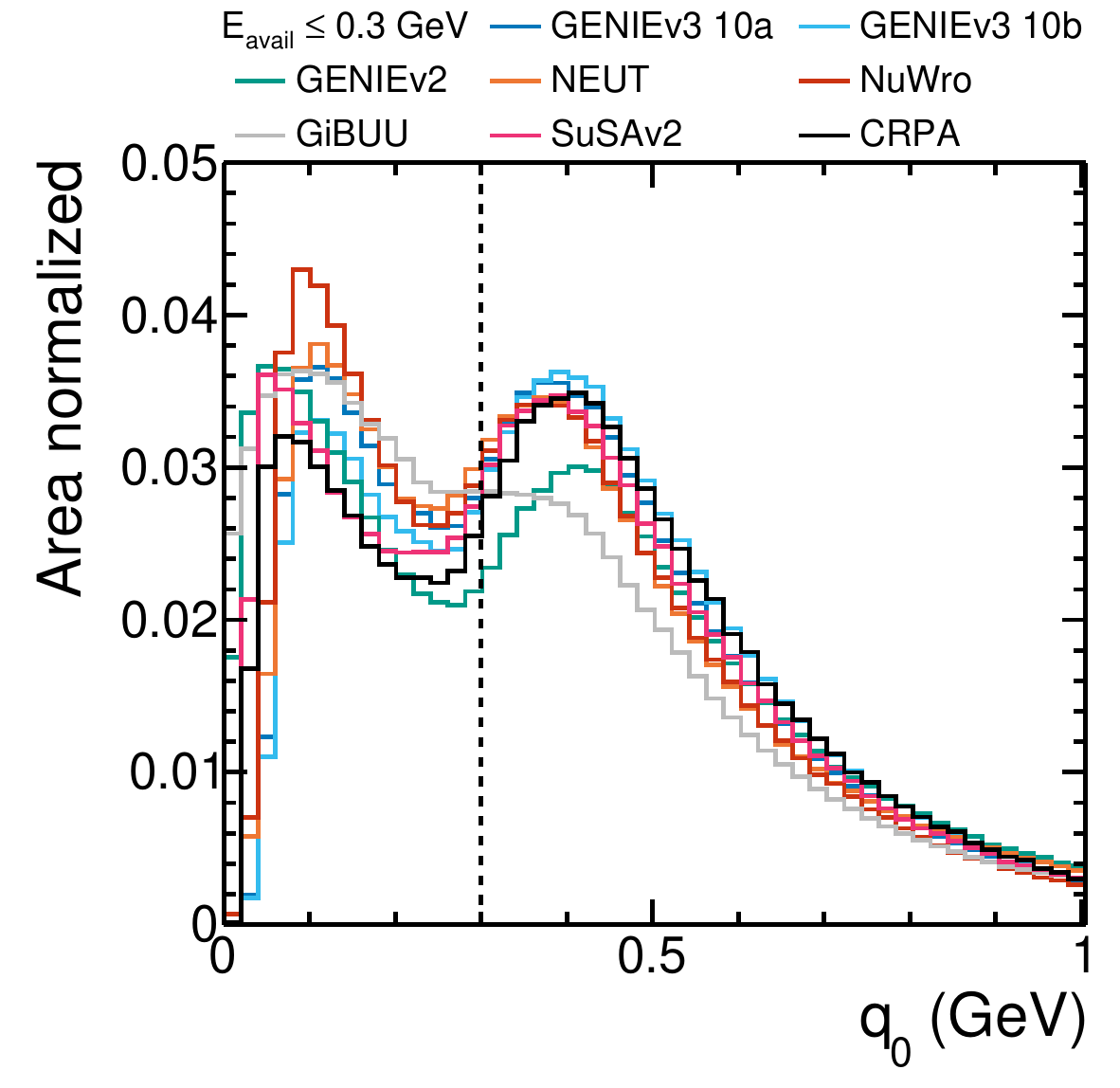}}
  \subfloat[\enutrue = 5 GeV, \eavail]    {\includegraphics[width=0.3\linewidth]{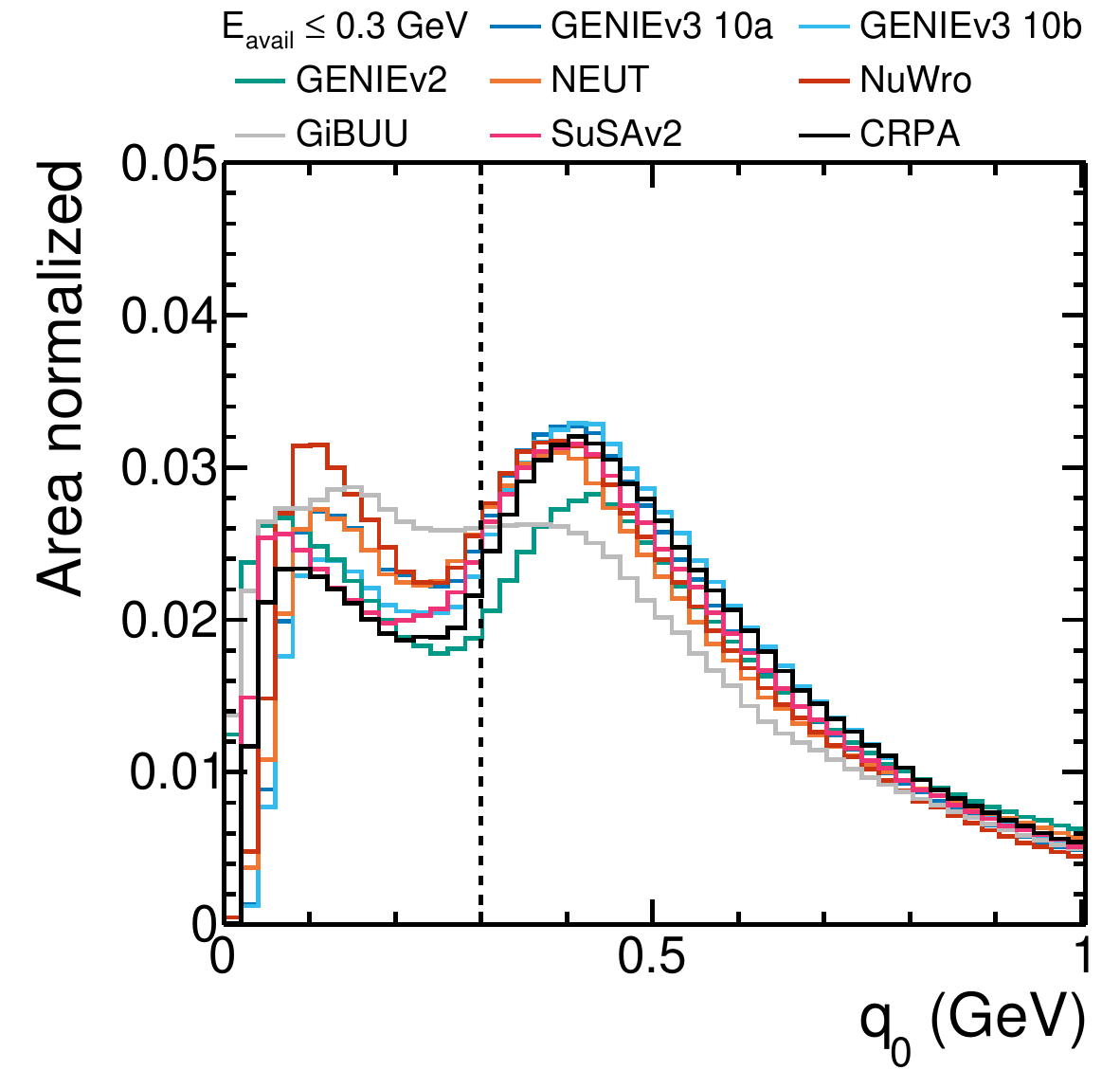}}
  \subfloat[\enutrue = 10 GeV, \eavail]    {\includegraphics[width=0.3\linewidth]{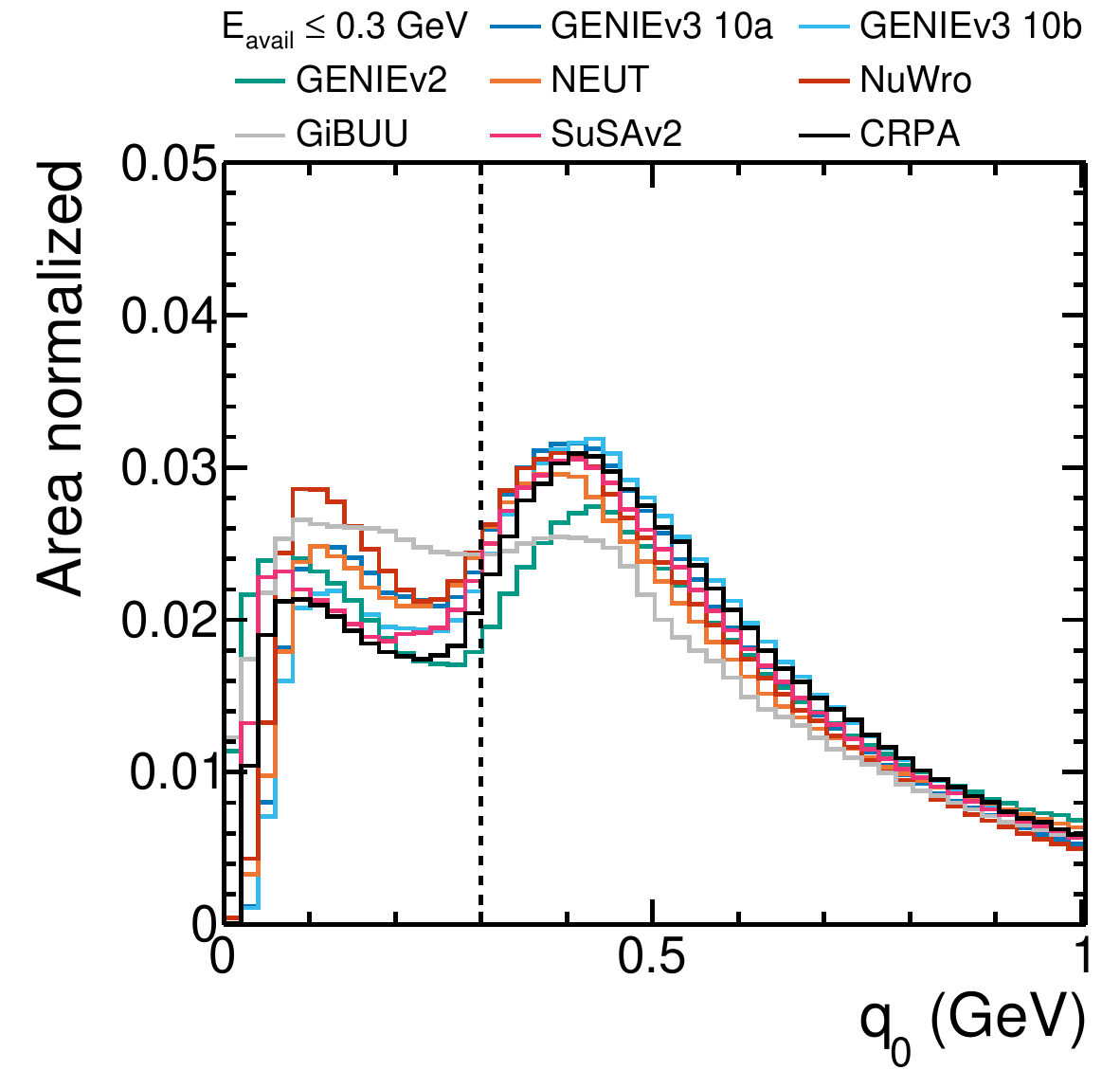}}\\\vspace{-2pt}
  \caption{The \qz distributions for CC-inclusive \numub--\argon events which pass cuts on the three proxy variables which are experimentally observable: \ehadtrue, \ehadreco and \eavail$ \leq$ 0.3 GeV, shown for all the generators models considered, for three different neutrino distributions, which are uniform and centred on 2, 5 and 10 GeV with a $\pm$0.5 GeV width.}
  \label{fig:smear_q0_proxy_numub}
\end{figure*}

\begin{figure*}[htbp]
  \centering
  \captionsetup[subfloat]{captionskip=-1pt}
  \subfloat[\numu--\argon, $\ehadtrue \leq 0.3$ GeV]  {\includegraphics[width=0.3\linewidth]{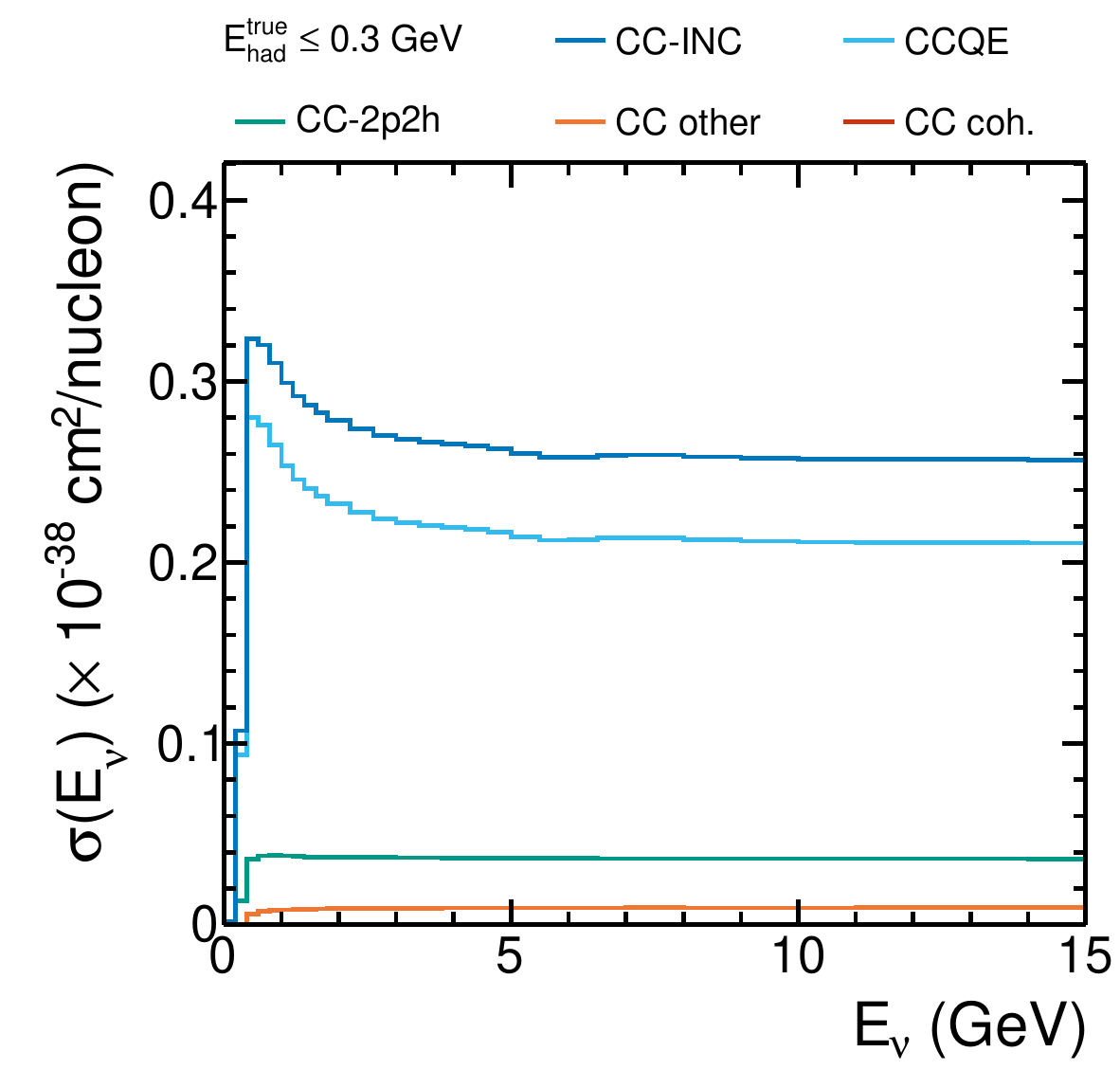}}
  \subfloat[\numu--\argon, $\ehadreco \leq 0.3$ GeV]  {\includegraphics[width=0.3\linewidth]{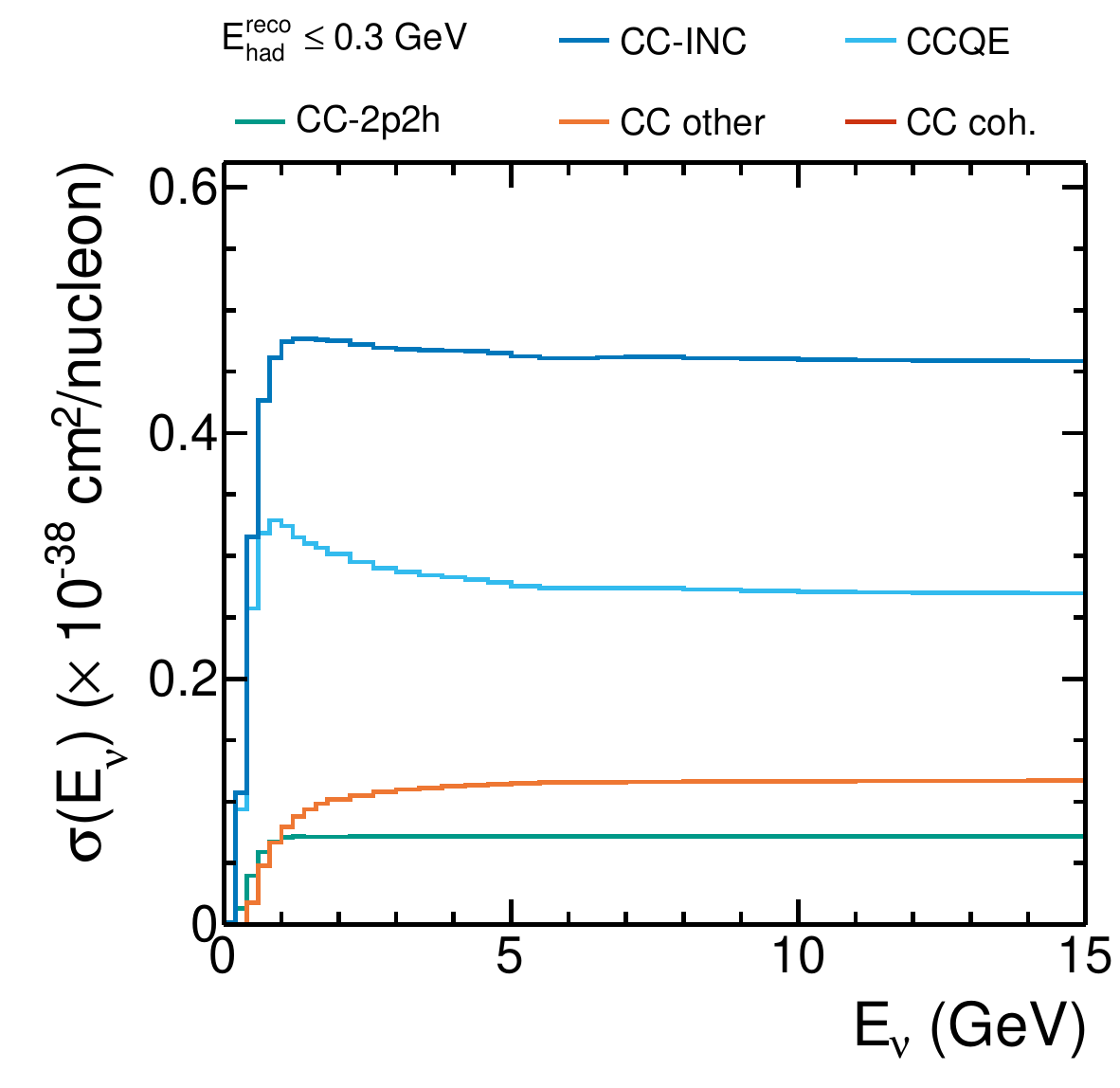}}
  \subfloat[\numu--\argon, $\eavail \leq 0.3$ GeV]    {\includegraphics[width=0.3\linewidth]{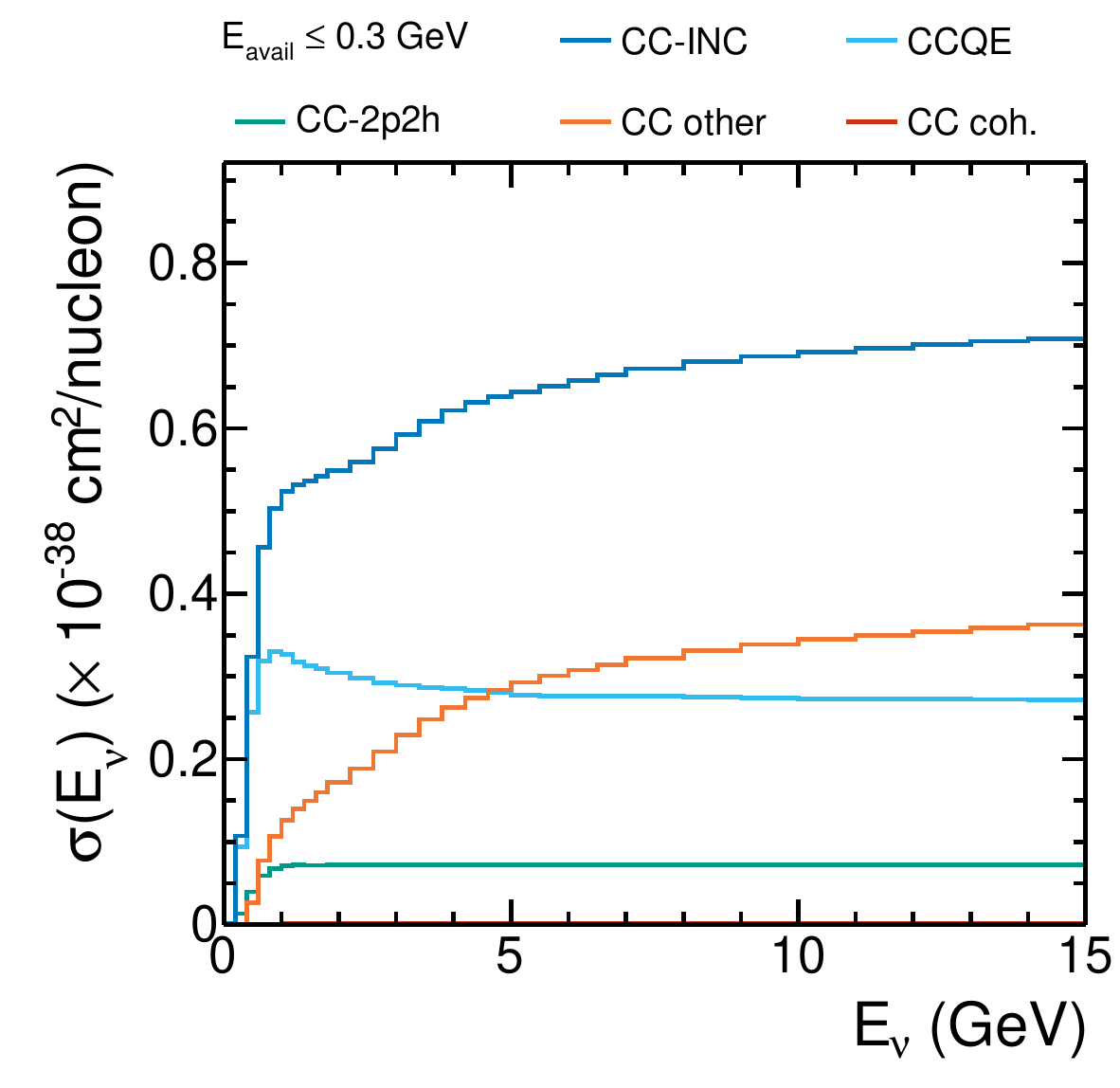}}\\\vspace{-2pt}
  \subfloat[\numub--\argon, $\ehadtrue \leq 0.3$ GeV]  {\includegraphics[width=0.3\linewidth]{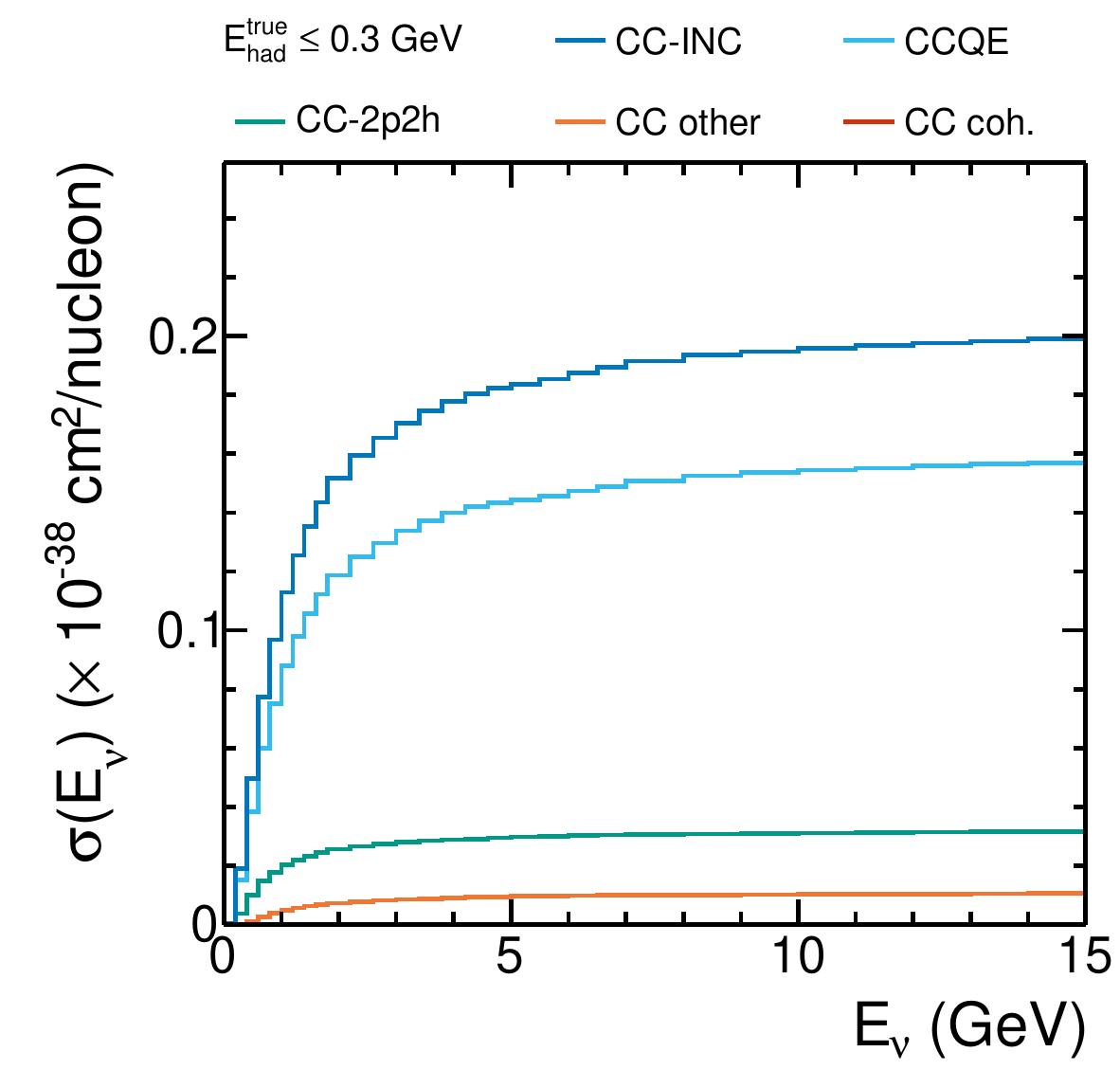}}
  \subfloat[\numub--\argon, $\ehadreco \leq 0.3$ GeV]  {\includegraphics[width=0.3\linewidth]{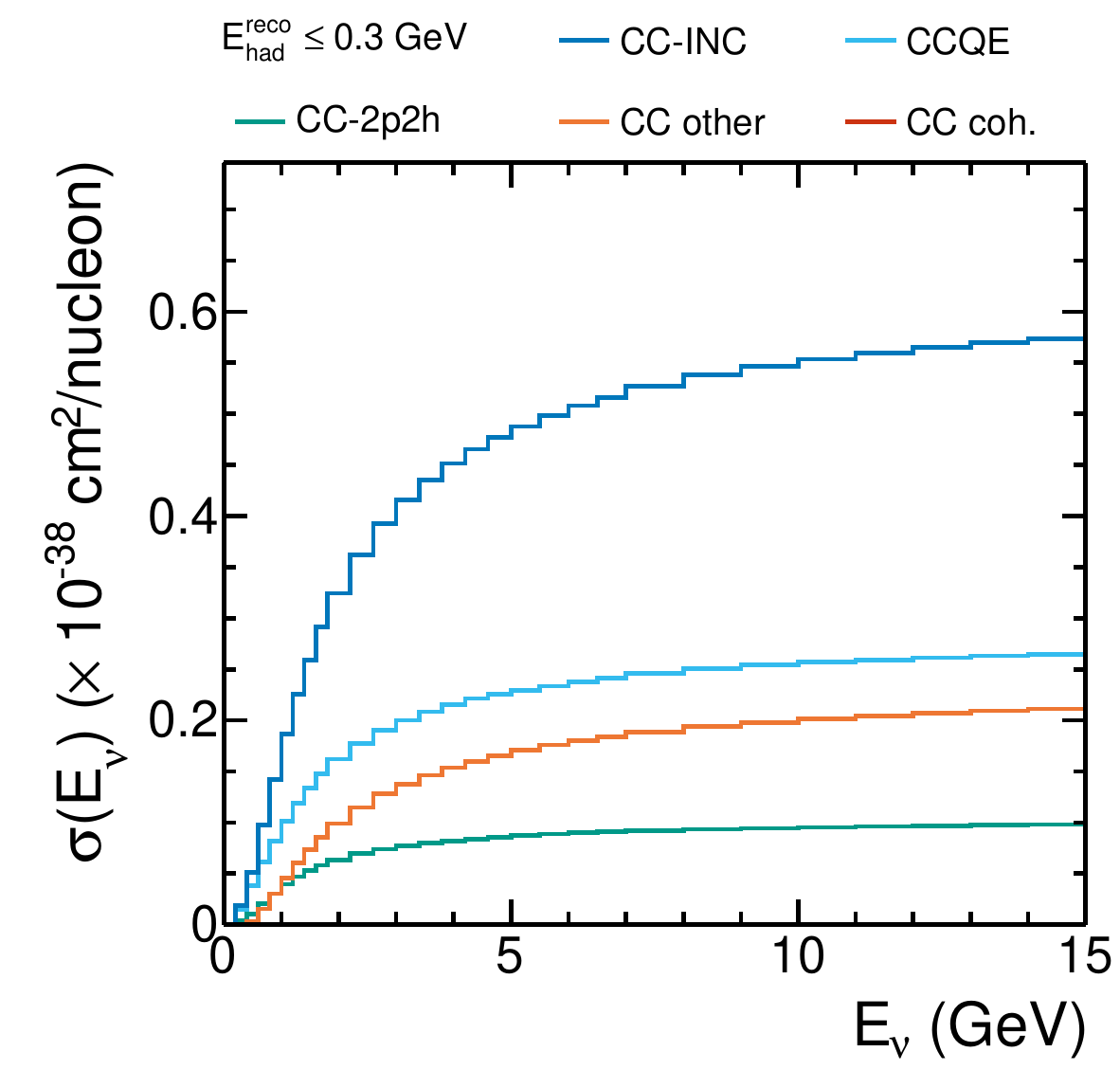}}
  \subfloat[\numub--\argon, $\eavail \leq 0.3$ GeV]    {\includegraphics[width=0.3\linewidth]{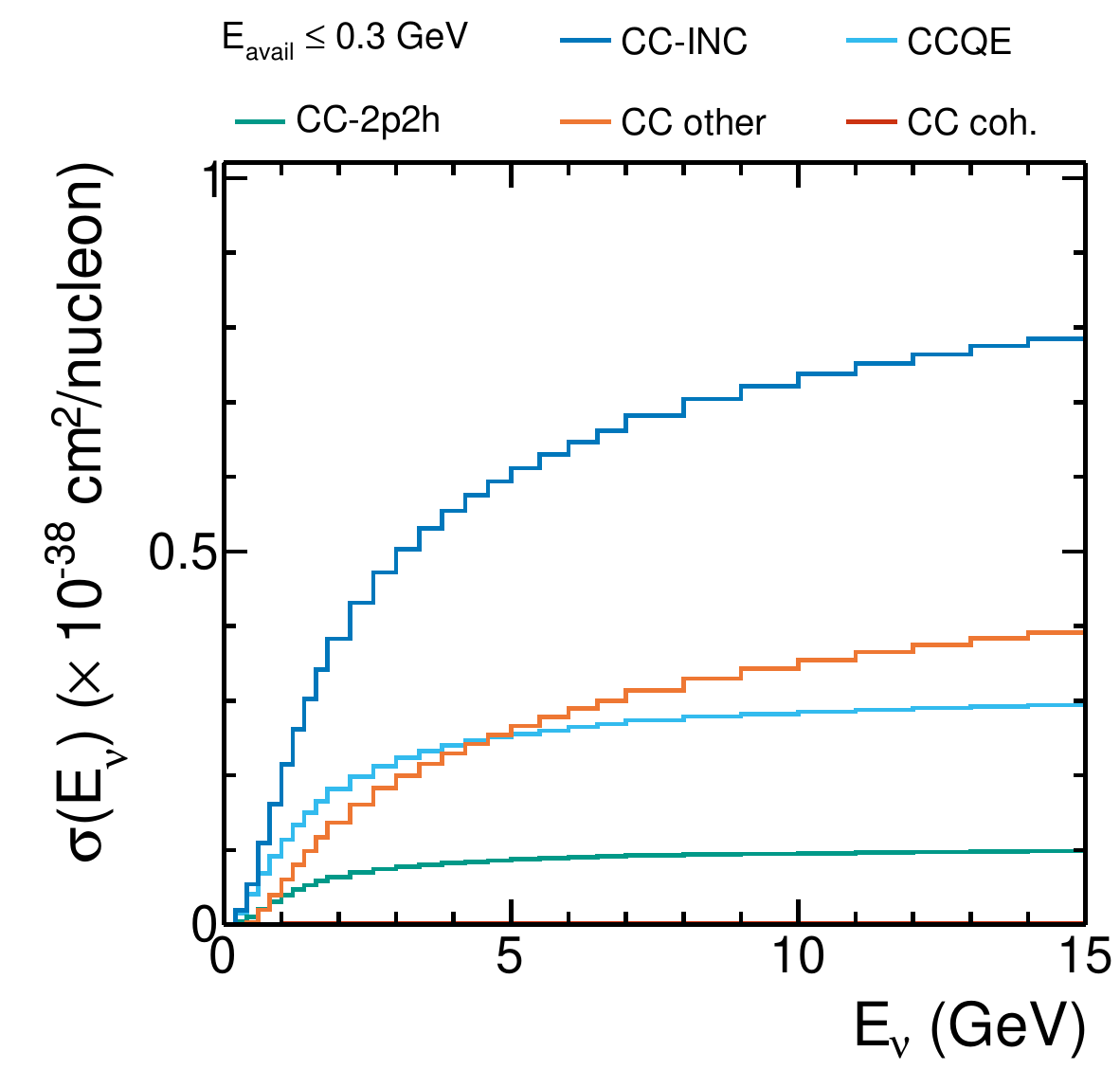}}
  \caption{Contributions to the GENIEv3 10a charged-current \numu--\argon and \numub--\argon cross sections with cuts on the three \qz proxy variables, \ehadtrue, \ehadreco, \eavail $\leq$ 0.3 GeV as a function of \enutrue, and generated with a uniform neutrino distribution from 0--20 GeV. Contributions from separate interaction channels are also shown.}
  \label{fig:contrib_vary_proxy}
\end{figure*}

In this section, we investigate the second and third related requirements for using the \lownu method as set out in \autoref{sec:introduction}: that a low-\qz sample can be isolated experimentally without introducing significant model-dependent corrections; and, the reconstructed neutrino energy for events in the selected sample can be related to the true neutrino energy without introducing further model dependence.

\subsection{Energy Transfer}
\label{sec:en_trans}
Although we refer to a ``\lownu sample'' above, the true energy transfer \qz is not experimentally accessible. An experiment can measure the outgoing charged lepton in a charged-current interaction, but can not know the initial neutrino energy event by event due to the broad spectrum produced by accelerator sources. Additionally, even with a perfect detector, the initial state of the nucleus is inherently dynamic; both the Fermi-motion and removal energies required to liberate nucleons are not trivial at the energy transfers of interest for modern accelerator experiments $\mathcal{O}(0.1$--$5~\text{GeV})$. Finally, detectors are often unable to accurately measure the energy of neutral particles. We define three different \qz proxy variables which are experimentally accessible:

\begin{enumerate}
\item $\ehadtrue = \Big(\sum_{i=n, p} E^i_{\mathrm{kin}}\Big) + \Big( \sum_{i=\pi^{\pm}, \pi^{0}, \gamma}E^i_{\mathrm{total}}\Big)$ --- the true hadronic energy is defined as the sum of the kinetic energies of the protons and neutrons, and the total energy of all pions and photons. Other hadrons can be neglected at the energy transfers of interest. This variable is what could be seen by a ``perfect'' detector capable of measuring all outgoing particles without threshold or uncertainty, including neutrons. This differs from \qz in a number of important ways: the nucleus is a dynamic system and struck nucleons have momentum in the initial state and energy is required to liberate them; and FSI may modify the observable final state. By measuring \ehadtrue, no corrections are made for these processes.

\item $\ehadreco = \Big(\sum_{i=p} E^i_{\mathrm{kin}}\Big) + \Big(\sum_{i=\pi^{\pm}, \pi^{0}, \gamma} E^i_{\mathrm{total}}\Big)$ --- the reconstructed hadronic energy is the same as \ehadtrue, but without including neutrons. Experiments typically recover some fraction of the neutron energy, as demonstrated by MINERvA~\cite{MINERvA:2019wqe}, but reliably measuring the total neutron energy event-by-event is extremely challenging.
  
\item $\eavail = \Big(\sum_{i=\pi^{\pm}, p} E^i_{\mathrm{kin}}\Big) + \Big( \sum_{i=\pi^{0}, \gamma} E^i_{\mathrm{total}} \Big)$ --- the available, or recoil, energy is the calorimetric sum of the outgoing hadronic state. Given the low energy transfers of interest, it is a proxy for the energy seen in a detector with a high tracking threshold, where individual charged-pions are not identified, and no neutron energy is measured. This is equivalent to MINERvA's definition of \eavail~\cite{PhysRevLett.116.071802}.
\end{enumerate}

The experimental observables described above are idealized, and detector-specific reconstruction will incur additional smearing of the observable energies. Fine-grained, calorimetric tracking detectors that can accurately identify charged-pion deposits will be able to observe something between \ehadreco and \eavail. Nevertheless, these observables are illustrative and provide useful proxies without the need for a full detector-specific simulation. It is worth stressing that even with a perfect detector, if the target material is not composed entirely of free nucleons, it would not be possible to cut on \qz, as the incoming neutrino energy cannot be measured exactly.

\autoref{fig:smear_q0_proxy_numu} (\autoref{fig:smear_q0_proxy_numub}) shows the true \qz distributions for CC-inclusive \numu--\argon (\numub--\argon) events produced by cutting on the three different proxy variables considered, all with a cut of $\leq$~0.3~GeV. These distributions are produced using uniform fluxes centred on 2, 5 and 10 GeV, with a $\pm$0.5 GeV width, in order to demonstrate how each proxy's smearing evolves with respect to true \qz changes as a function of neutrino energy. The general trend is to include true \qz contributions significantly above the proxy variable's cut value, for all of the proxy variables and for all energies. This trend is more pronounced for \numub than \numu because of the relative probability for emitting energetic, but unobservable, neutrons. The smearing largely increases with neutrino energy, although this effect is much more pronounced for \numub than \numu. The smearing between \qz and \ehadtrue is due to the initial-state nuclear dynamics and unmeasured energy-loss to the nucleus through final state interactions and nuclear excitations. The broad tail in the \qz distribution when cutting on \ehadreco is due to missed neutron energy. The additional smearing between \qz and \eavail is largely due to the effect of not accounting for the masses of charged pions, which results in the additional high-\qz events migrating into the ``low-\qz'' sample. Although there are pronounced differences between the generators in all cases, they tend to be largest for $\qz<0.3~\text{GeV}$, which is less problematic if the differences in this region are constant as a function of \enutrue, which seems to be the case for \numu--\argon in \autoref{fig:smear_q0_proxy_numu}, although holds less true for \numub--\argon in \autoref{fig:smear_q0_proxy_numub}. A bigger concern is the different migration in from higher \qz, which is clearly not constant with increasing \enutrue for \numub--\argon. Notably, GiBUU predicts significantly different behaviour to the other models, and, in general, model differences above the cut value appear to be larger for \numub than \numu.

The evolution of the proxy smearing as a function of neutrino energy is primarily caused by the changing proportion of non-CCQE interactions (which tend to have a less well reconstructed \qz) within the sample, as is demonstrated in \autoref{fig:contrib_vary_proxy}. This shows the contributions to \numu--\argon and \numub--\argon \lownu samples change for the three different proxy variables cut at $\leq 0.3~\text{GeV}$ as a function of \enutrue and broken down into different interaction channels for the GENIEv3 10a model. \ehadtrue is similar to the true \qz case, shown in \autoref{fig:contrib_vary_q0}, with a small increase in the contribution from CC-other interactions due to nuclear effects. A cut on \ehadreco produces a markedly different sample, with larger CC-2p2h and much larger CC-other contributions. For \numu--\argon, the asymptotic behavior of the CC-inclusive cross section remains broadly similar at high energies and the rise in CC-other contributions masks the turnover in the CCQE cross section at $\approx1~\text{GeV}$. The situation is quite different for \numub--\argon, where ignoring neutrons introduces a much larger CC-other contribution. For \eavail, the cross section is no longer dominated by CCQE across all neutrino energies, with CC-other starting to dominate for $\enutrue\gtrsim5~\text{GeV}$ for both \numu--\argon and \numub--\argon. Notably, this contribution does not become approximately constant by 15~GeV in either case. 

\begin{figure*}[htbp]
  \centering
  \captionsetup[subfloat]{captionskip=-1pt}
  \subfloat[\numu--\argon, $\ehadtrue \leq 0.3$ GeV]  {\includegraphics[width=0.33\linewidth]{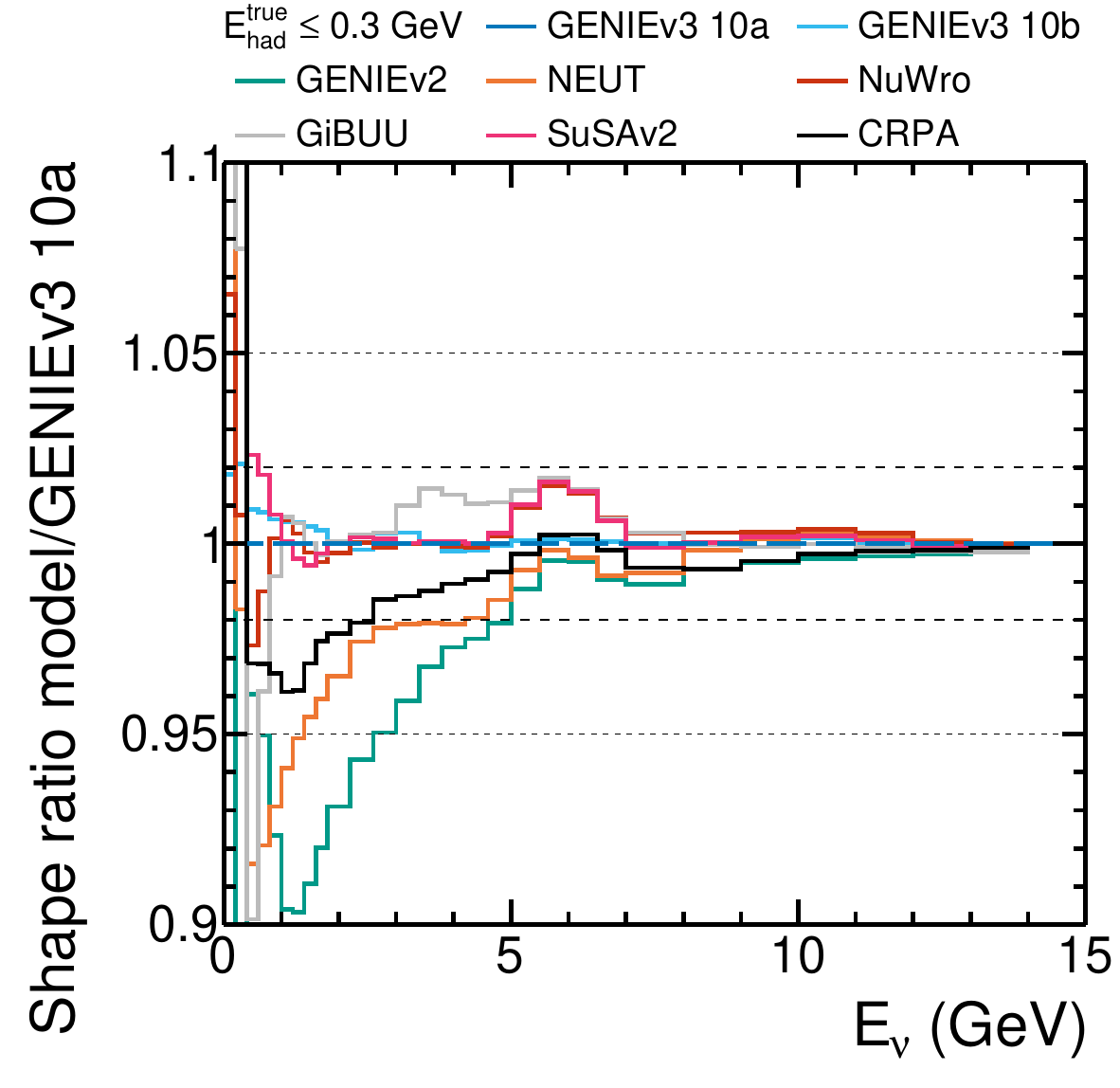}}
  \subfloat[\numu--\argon, $\ehadreco \leq 0.3$ GeV]  {\includegraphics[width=0.33\linewidth]{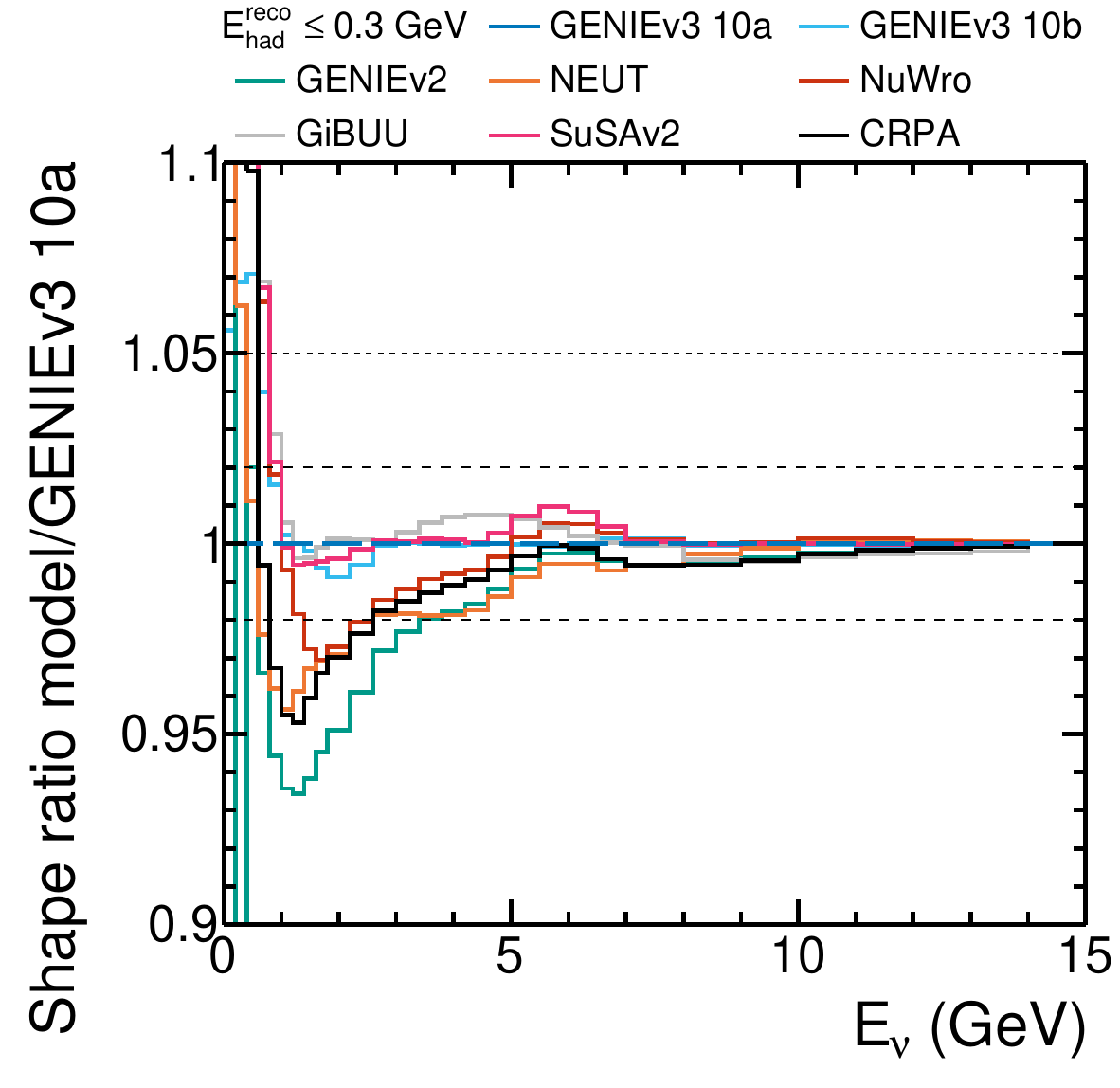}}
  \subfloat[\numu--\argon, $\eavail \leq 0.3$ GeV]    {\includegraphics[width=0.33\linewidth]{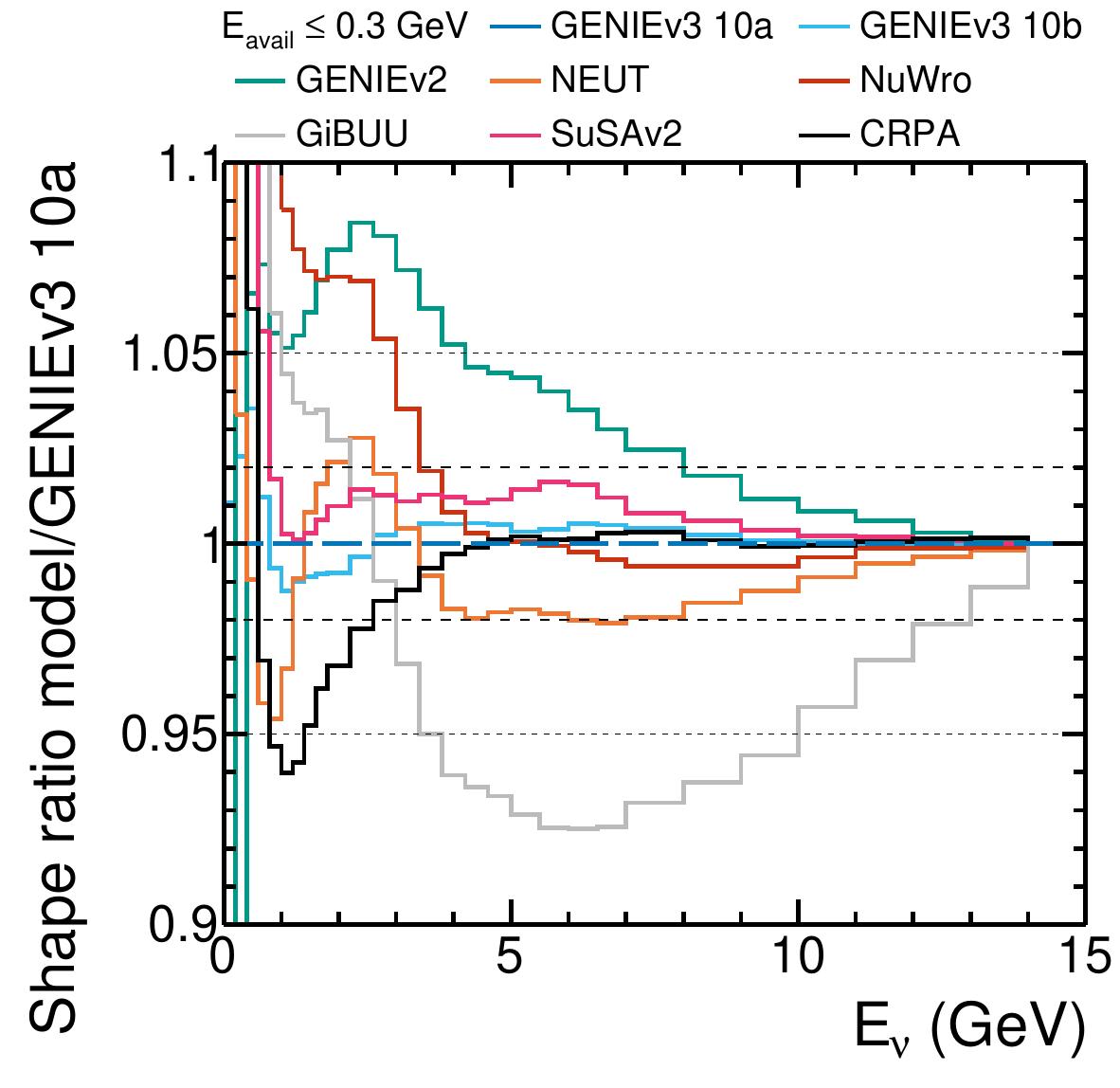}}\\\vspace{-2pt}
  \subfloat[\numub--\argon, $\ehadtrue \leq 0.3$ GeV] {\includegraphics[width=0.33\linewidth]{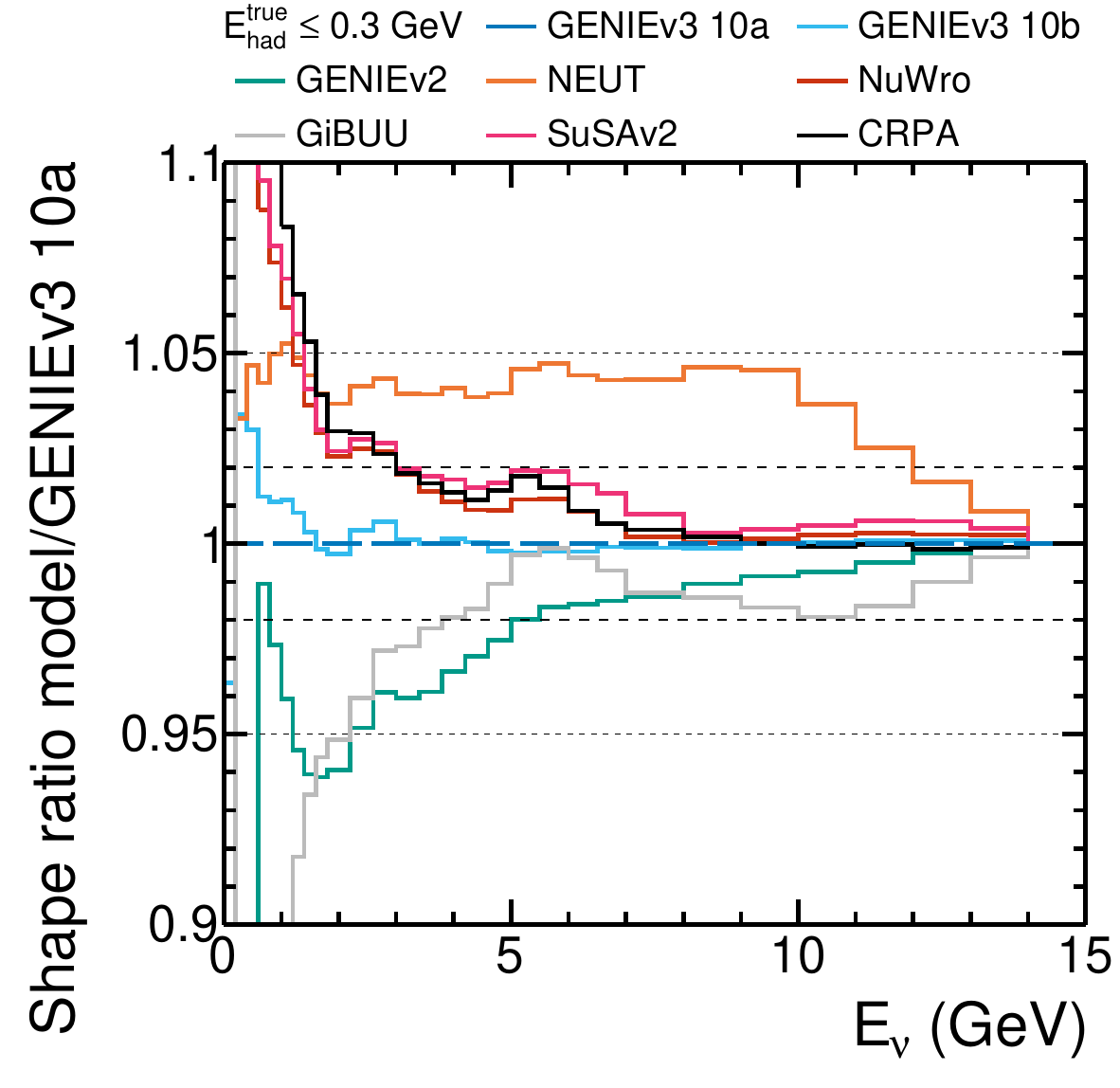}}
  \subfloat[\numub--\argon, $\ehadreco \leq 0.3$ GeV] {\includegraphics[width=0.33\linewidth]{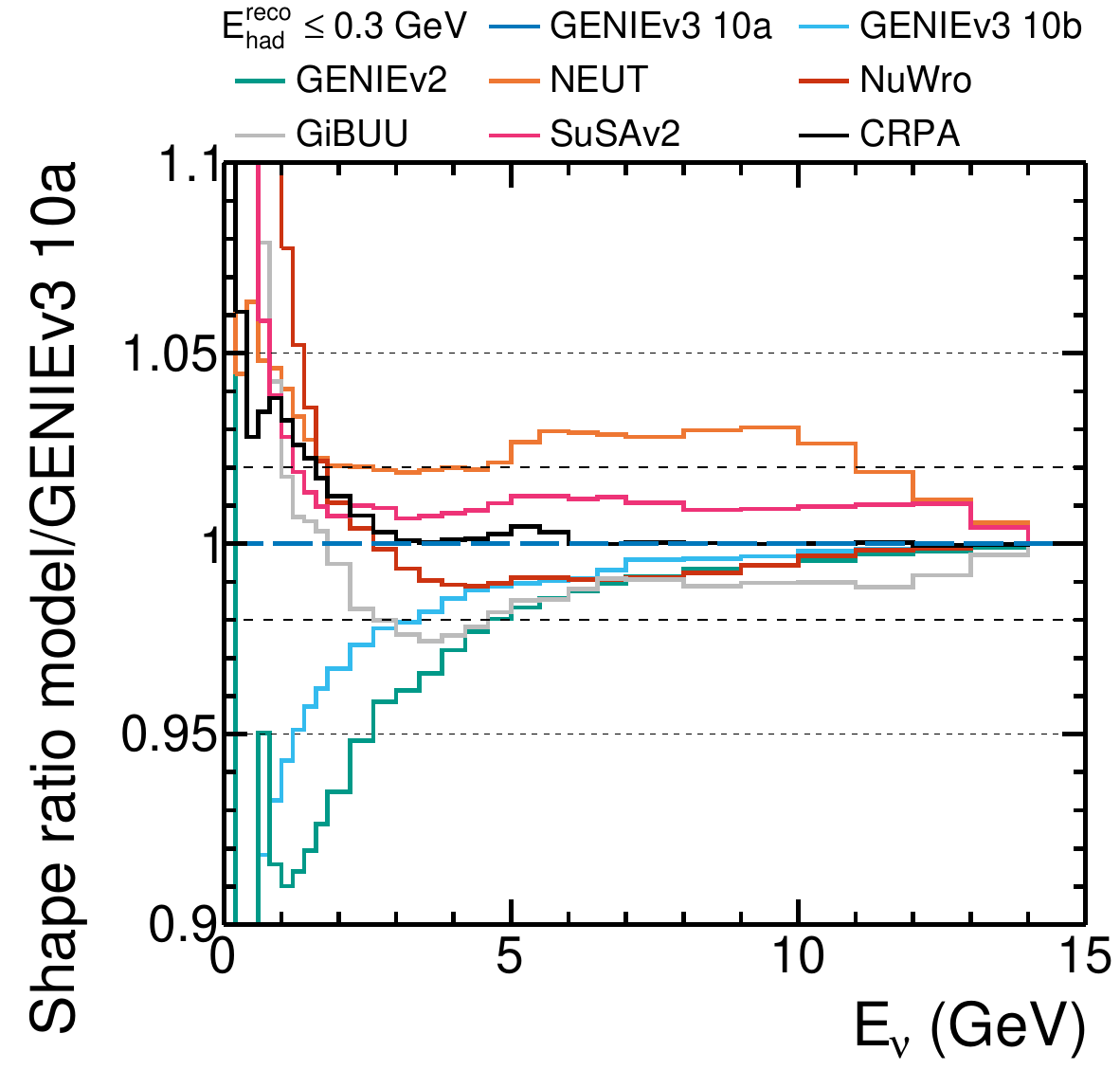}}
  \subfloat[\numub--\argon, $\eavail \leq 0.3$ GeV]   {\includegraphics[width=0.33\linewidth]{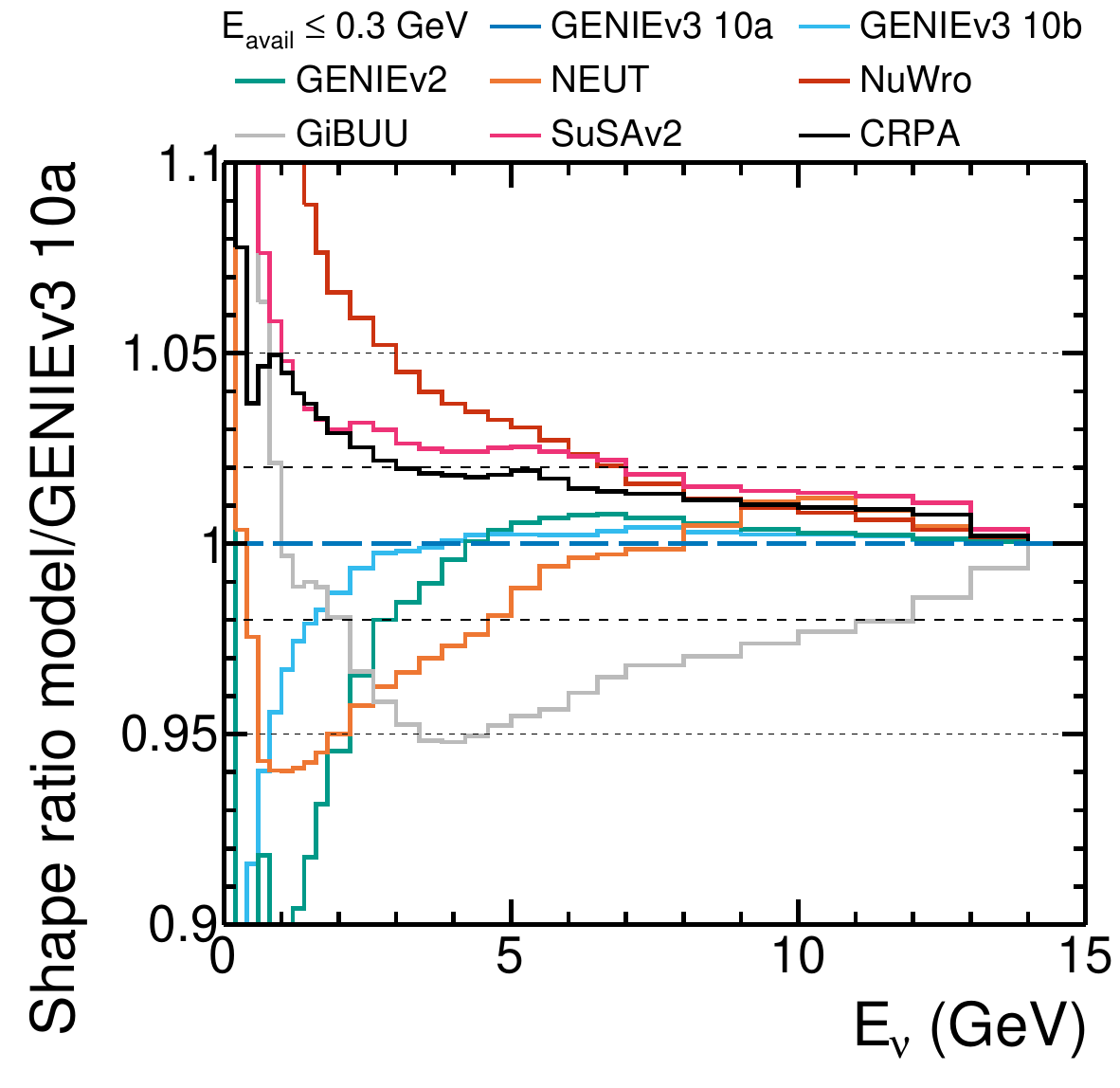}}
  \caption{Comparison of the shape-only ratios (with respect to the GENIE 10a prediction) of the \numu--\argon and \numub--\argon charged-current cross sections with a cut on 0.3 GeV for the three proxy variables, \ehadtrue, \ehadreco and \eavail, as a function of \enutrue. Figures produced using the flat neutrino flux described in \autoref{sec:low-qz-cross-section}. Horizontal long (short) dashed lines have been added at $\pm$2\% ($\pm$5\%), to guide the eye in assessing the scale of the bias.}
  \label{fig:shape_vary_proxy}
\end{figure*}

A real experiment is not able to cut on true \qz, so would perform a \lownu analysis by cutting on an experimentally accessible proxy variable, and then applying model dependent corrections to remove high \qz contributions. It is therefore interesting to ask how the event samples with cuts on these proxy variables differ between models, to assess how problematic that additional model dependence is. \autoref{fig:shape_vary_proxy} shows the shape-only ratios between the various generator models under investigation, with respect to the GENIEv3 10a model, of the charged-current cross sections with \qz, \ehadtrue, \ehadreco and \eavail $\leq 0.3$ GeV, for both \numu--\argon and \numub--\argon, as a function of \enutrue. The model differences for \ehadtrue are similar to \qz. Interestingly, the differences appear to be smaller for \ehadreco, although that may be related to the observation made in \autoref{fig:shape_vary_q0} that the differences between models become {\it smaller} at higher \qz, and there being significant migration from high-\qz into low-\ehadreco as illustrated in \autoref{fig:smear_q0_proxy_numu} and \autoref{fig:smear_q0_proxy_numub}. Unsurprisingly, the model differences become increasingly problematic for a low \eavail sample, where there is more migration from pion-production processes. Although a fixed cut value of 0.3~GeV is used for all proxy variables in \autoref{fig:shape_vary_proxy}, the behaviour of shape-only ratios as a function of cut values is similar for the three proxy variables as for true \qz (see \autoref{fig:shape_vary_q0}), with significantly worse performance for a cut value of 0.1~GeV, and generally improving with an increased cut value.

As further discussed in \autoref{app:chtarget}, the general behavior for $\numu^{\bracketbar}$--\ch and $\numu^{\bracketbar}$--\chtwo interactions is the same as has been presented for the  $\numu^{\bracketbar}$--\argon scattering case. This indicates that the modeling situation is not significantly better understood for nuclei lighter than argon. This is directly relevant to the use of the \lownu method on a \chtwo target at the DUNE near detector~\cite{DUNE:2021tad}, as well as for current and future experiments using lighter targets, such as T2K~\cite{T2K:2011qtm}, NOvA~\cite{NOvA:2007rmc}, and Hyper-Kamiokande~\cite{Hyper-Kamiokande:2018ofw}.

\subsection{Neutrino Energy}

\begin{figure*}[htbp]
  \centering
  \captionsetup[subfloat]{captionskip=-1pt}
  \subfloat[\emu + \ehadtrue = 1 GeV]    {\includegraphics[width=0.3\linewidth]{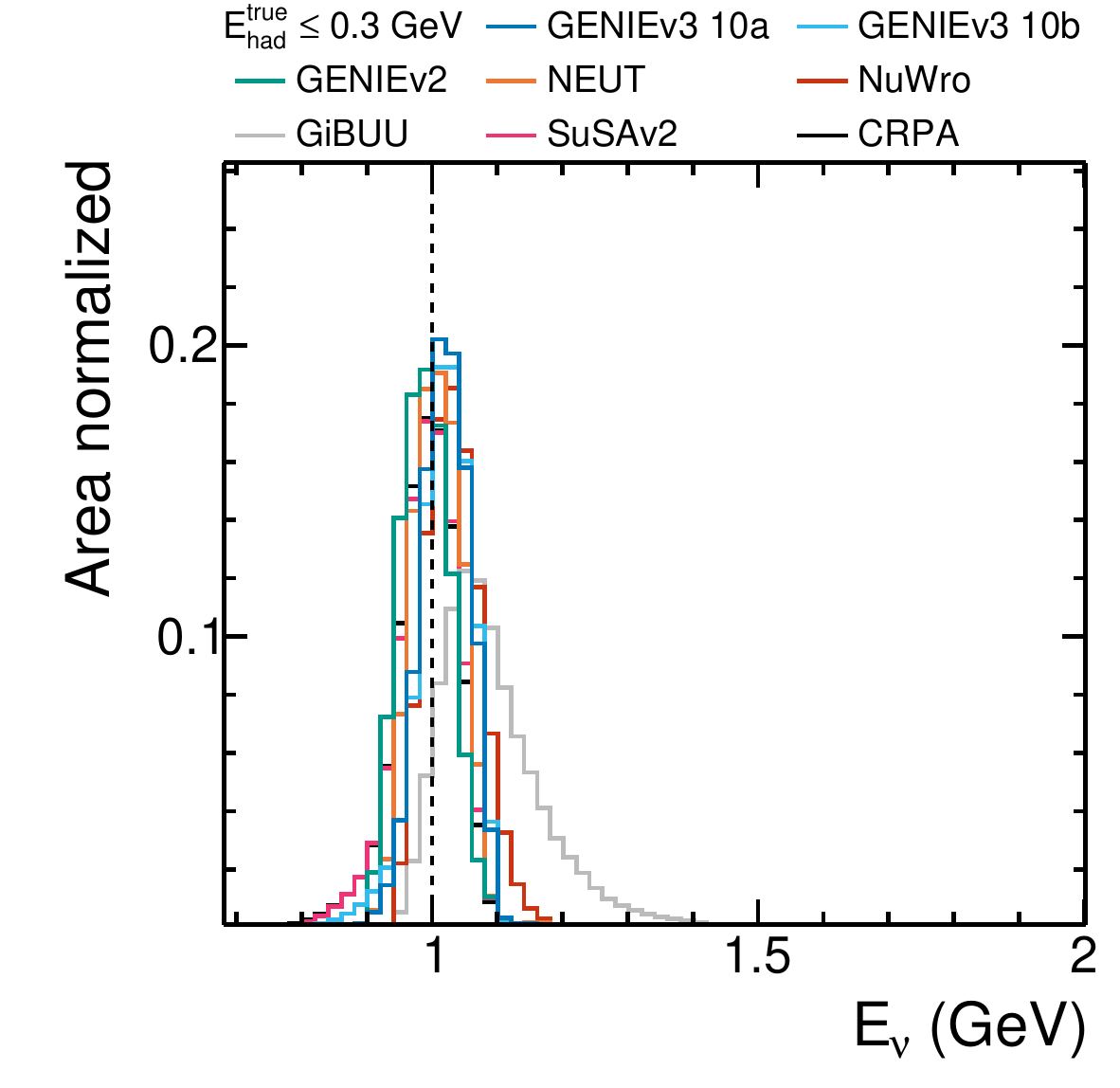}}
  \subfloat[\emu + \ehadtrue = 2.5 GeV]  {\includegraphics[width=0.3\linewidth]{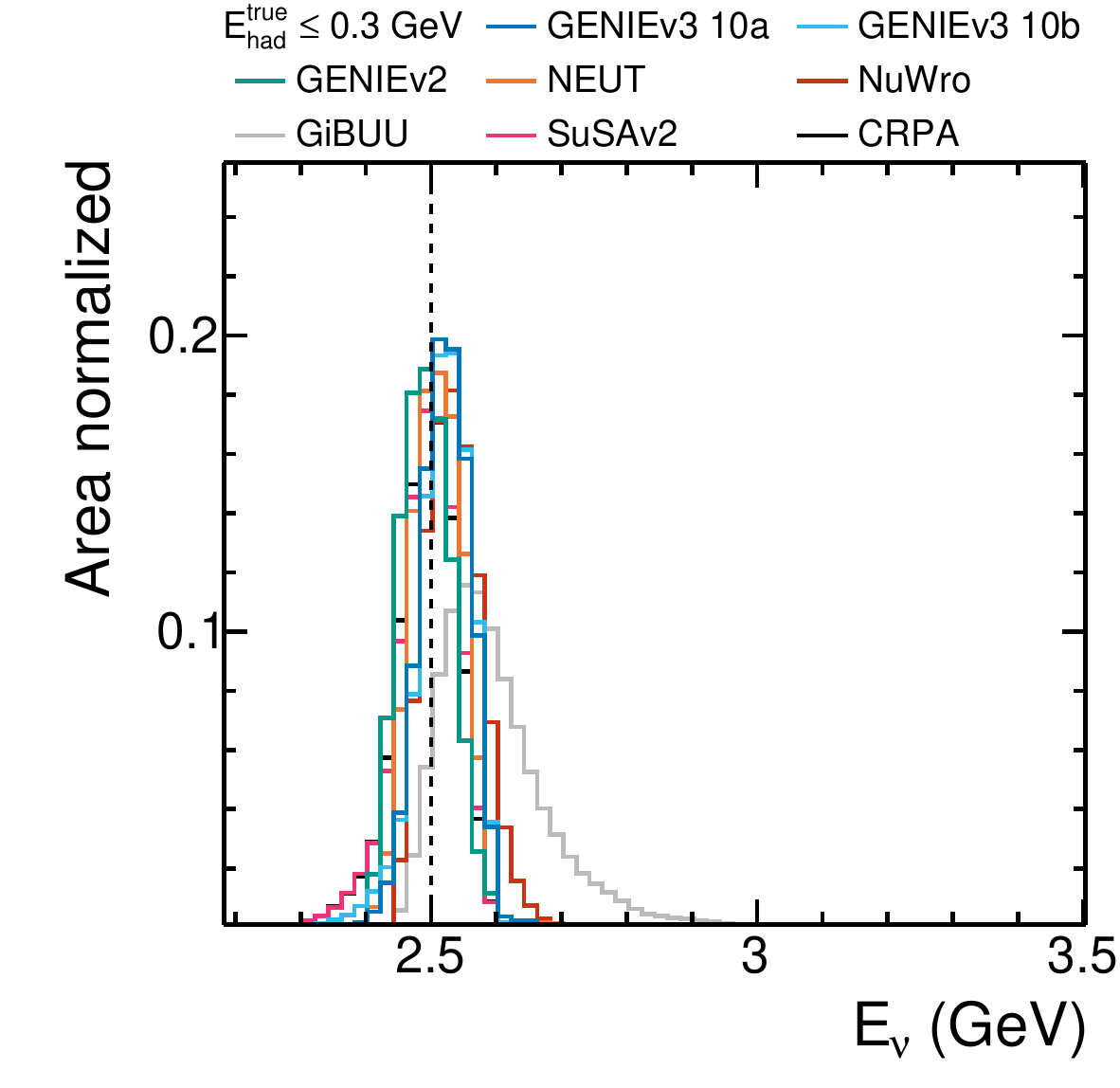}}
  \subfloat[\emu + \ehadtrue = 5 GeV]    {\includegraphics[width=0.3\linewidth]{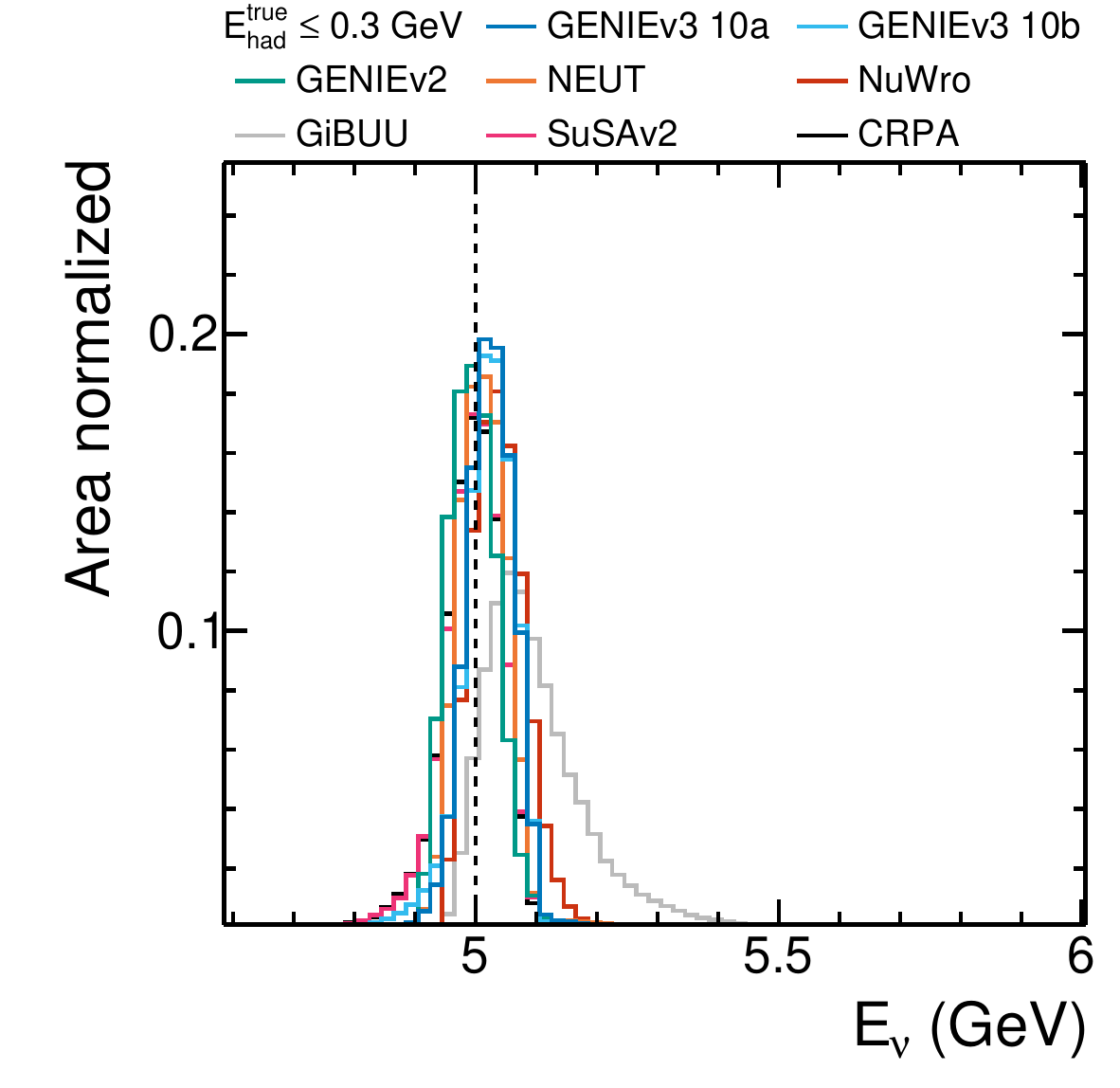}}\\
  \subfloat[\emu + \ehadreco = 1 GeV]    {\includegraphics[width=0.3\linewidth]{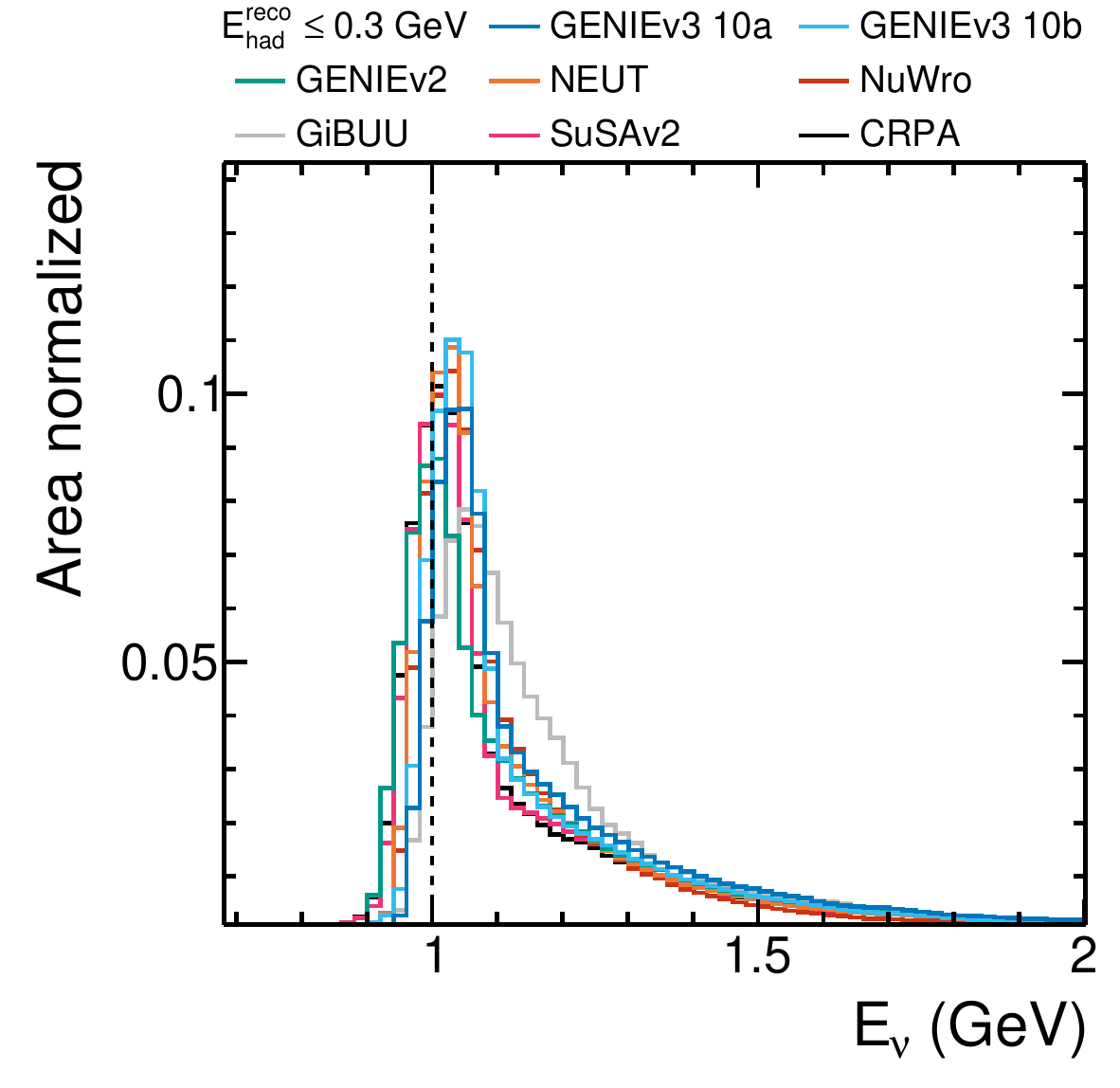}}
  \subfloat[\emu + \ehadreco = 2.5 GeV]  {\includegraphics[width=0.3\linewidth]{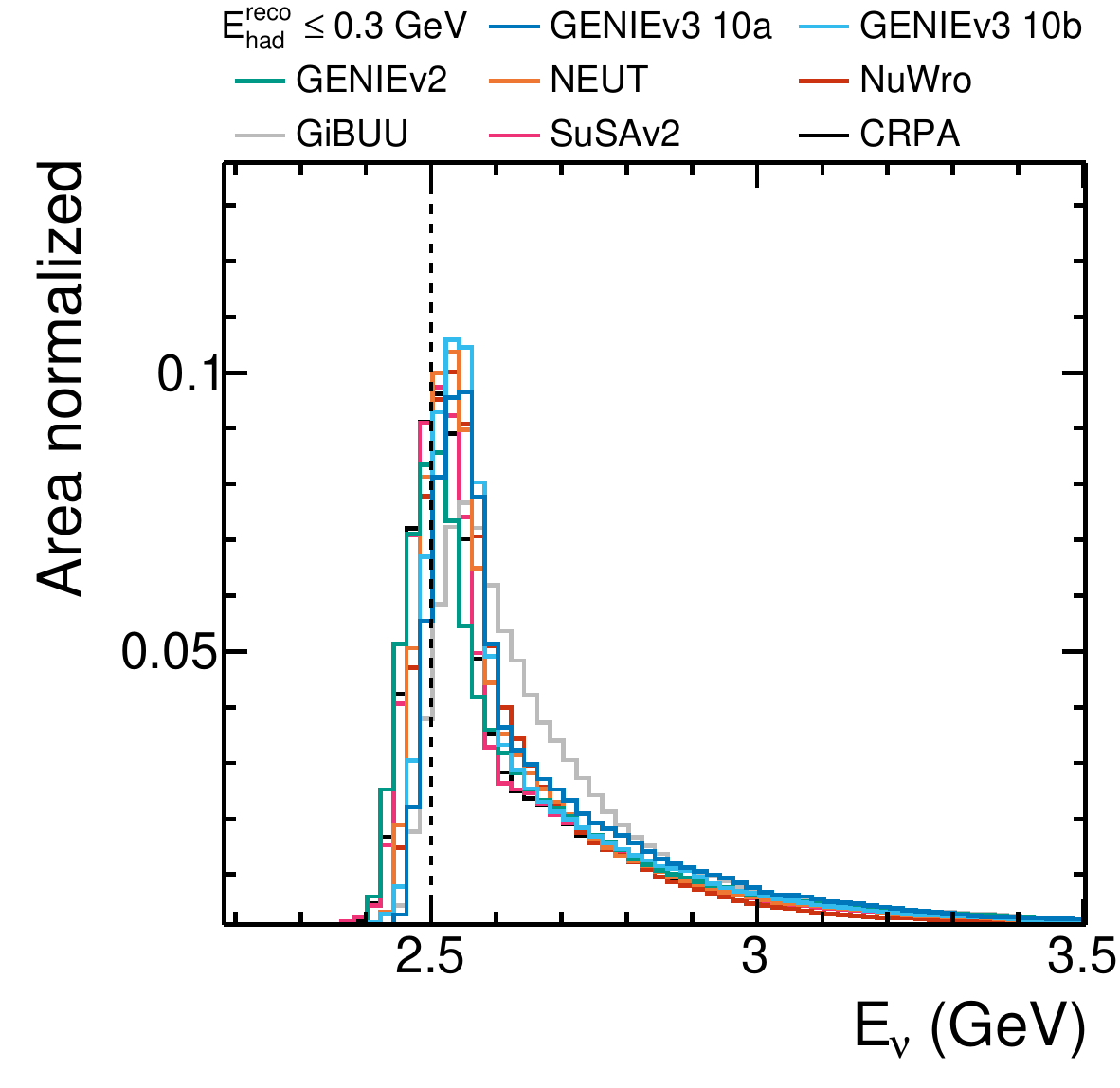}}
  \subfloat[\emu + \ehadreco = 5 GeV]    {\includegraphics[width=0.3\linewidth]{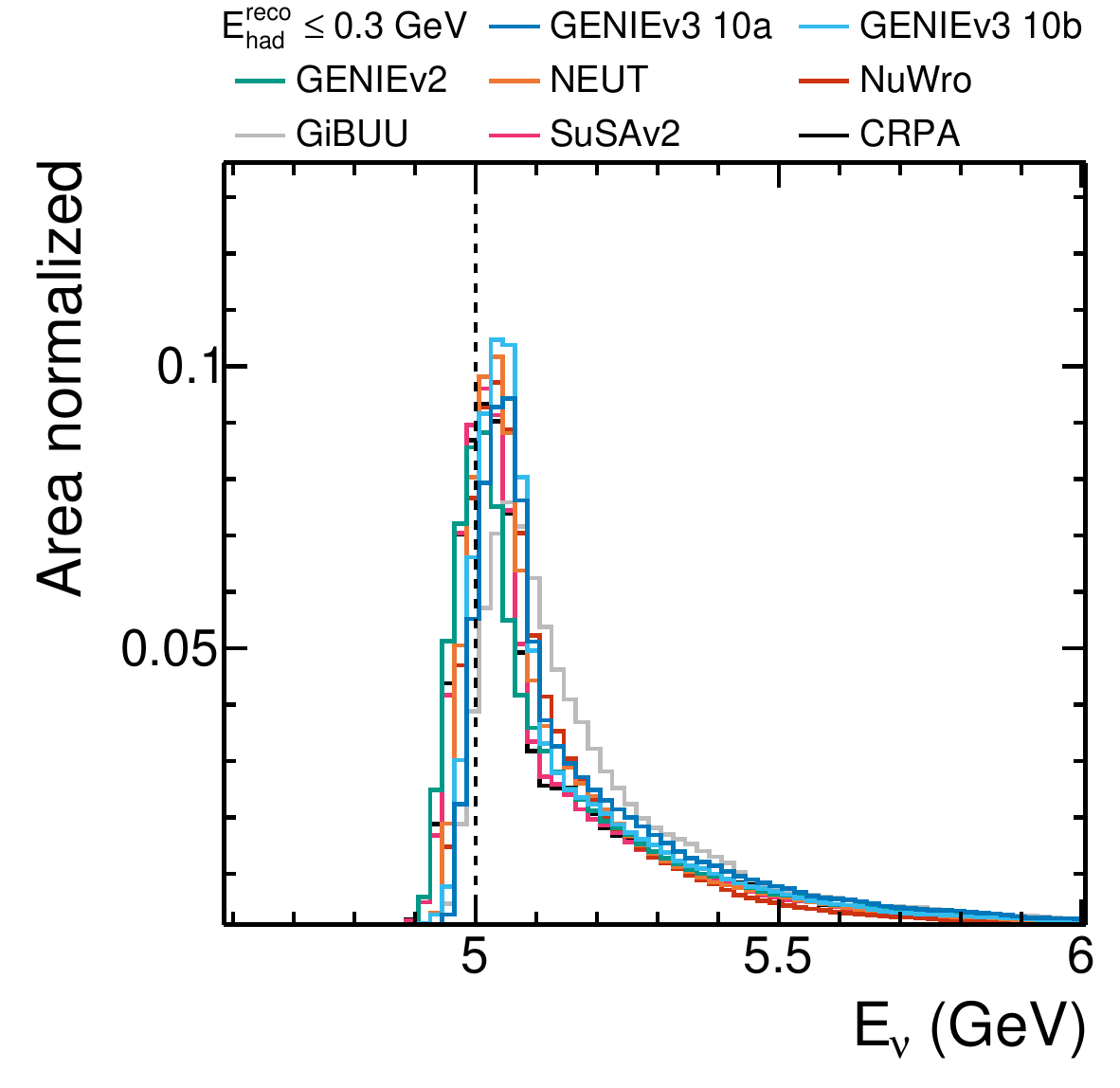}}\\
  \subfloat[\emu + \eavail = 1 GeV]    {\includegraphics[width=0.3\linewidth]{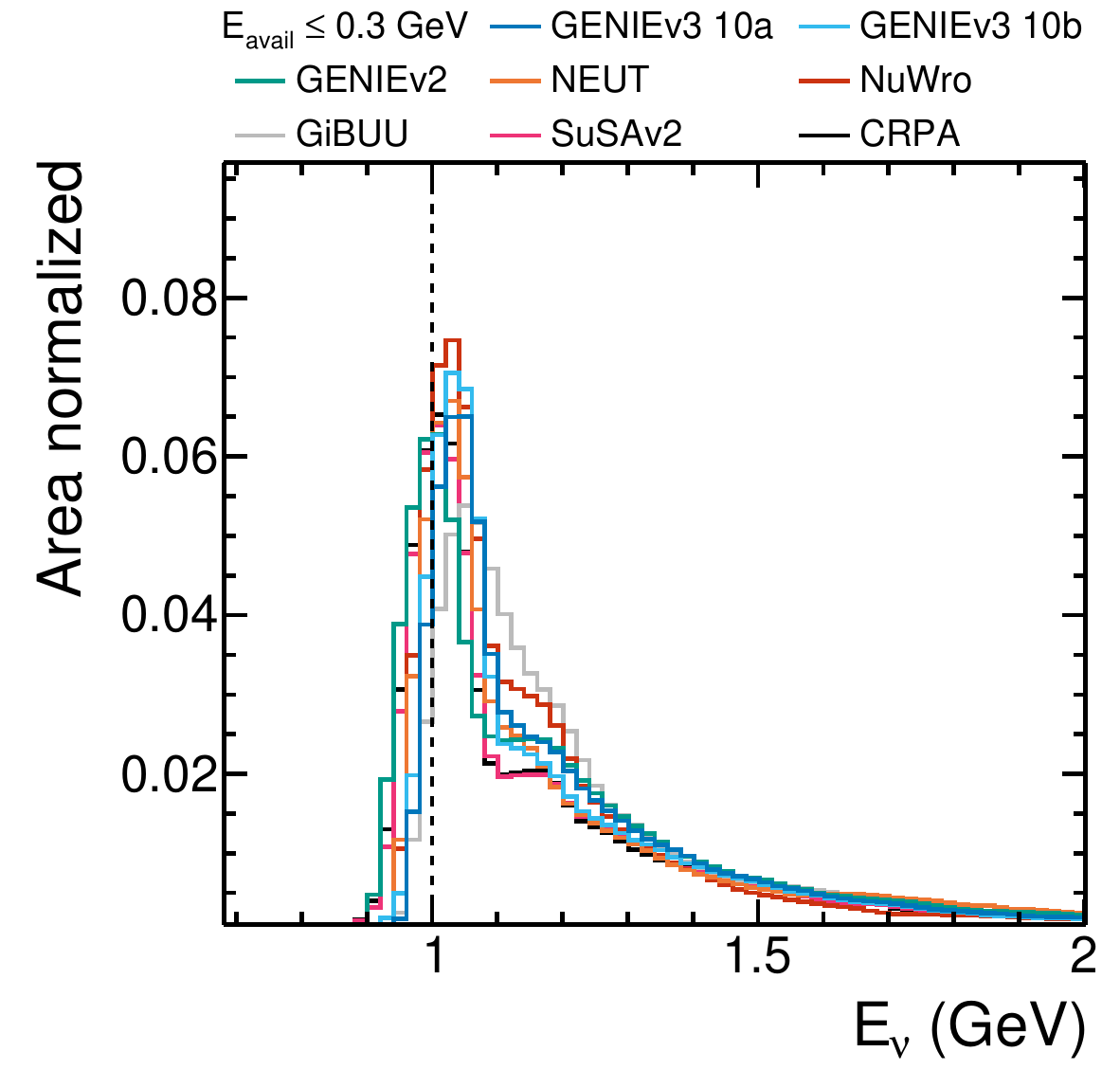}}
  \subfloat[\emu + \eavail = 2.5 GeV]  {\includegraphics[width=0.3\linewidth]{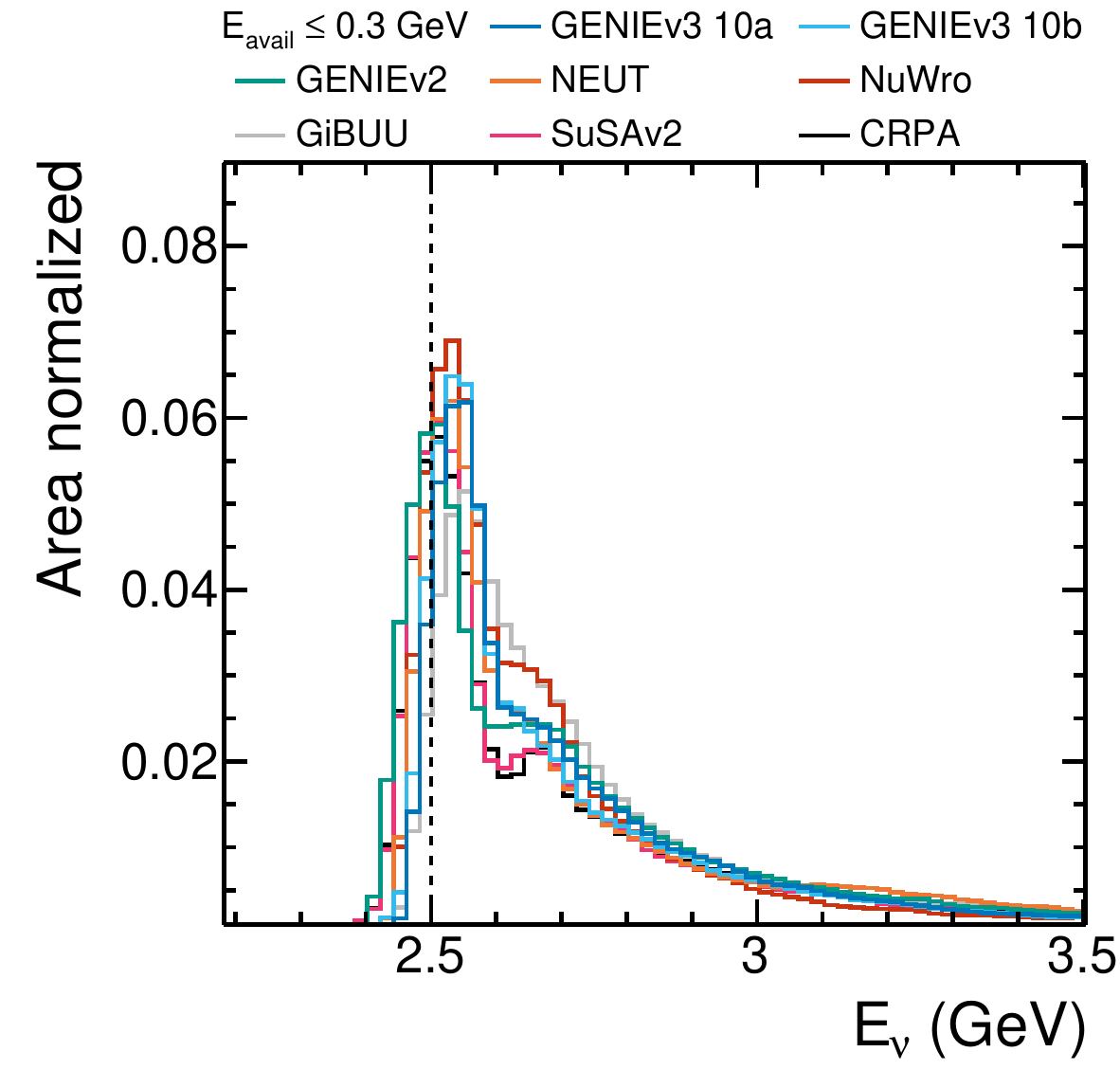}}
  \subfloat[\emu + \eavail = 5 GeV]    {\includegraphics[width=0.3\linewidth]{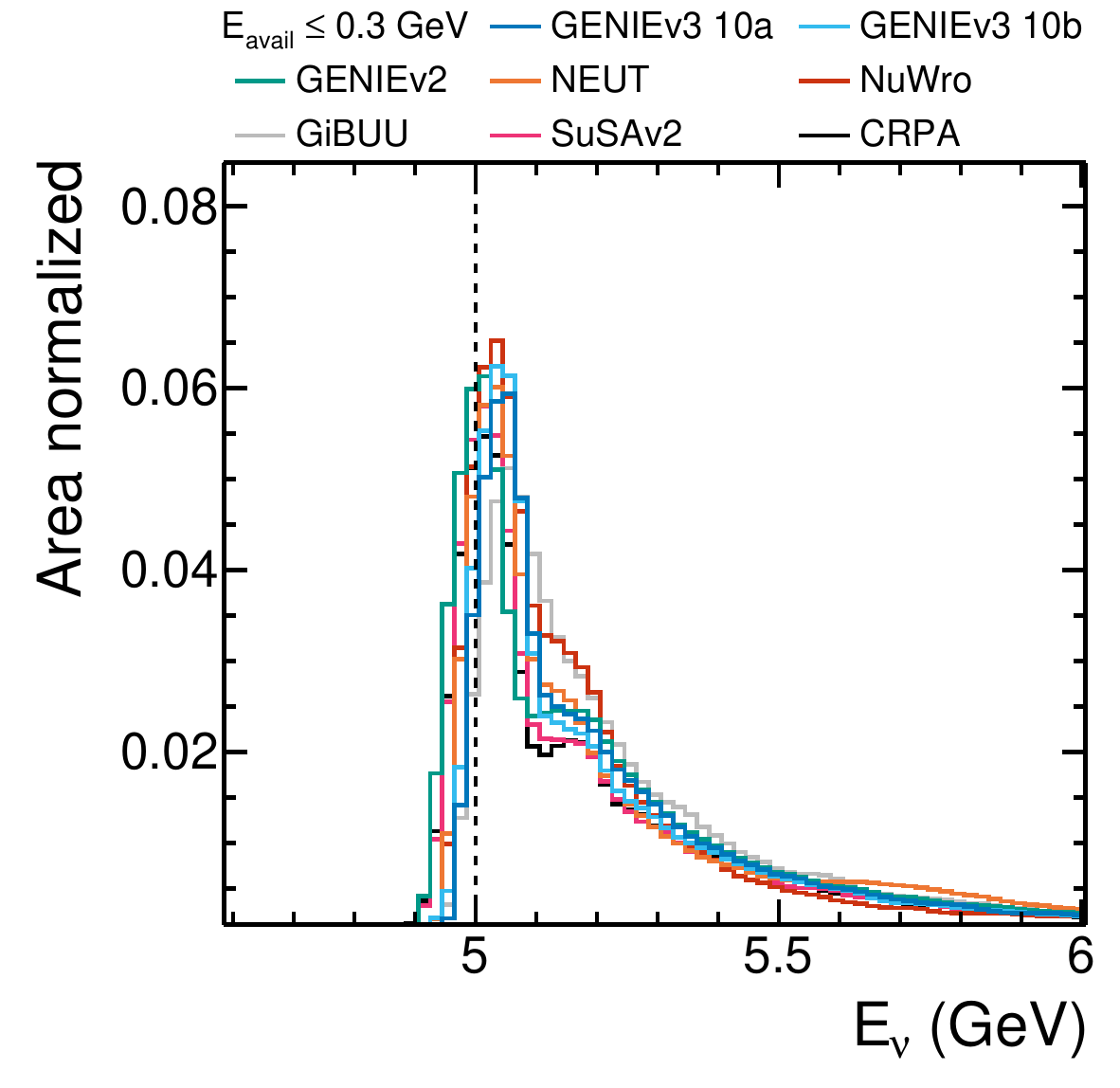}}
  \caption{Area normalized distributions of \enutrue corresponding to fixed values of the three proxy \enureco variables, \emu + \ehadtrue, \emu + \ehadreco and \emu + \eavail, at 1, 2.5 and 5 GeV, shown for \numu--\argon events, for all generator models of interest. A cut on the relevant \qz proxy of $\leq$ 0.3 GeV is made in each case.}
  \label{fig:smear_enu_proxy_numu}
\end{figure*}
\begin{figure*}[htbp]
  \centering
  \captionsetup[subfloat]{captionskip=-1pt}
  \subfloat[\emu + \ehadtrue = 1 GeV]    {\includegraphics[width=0.3\linewidth]{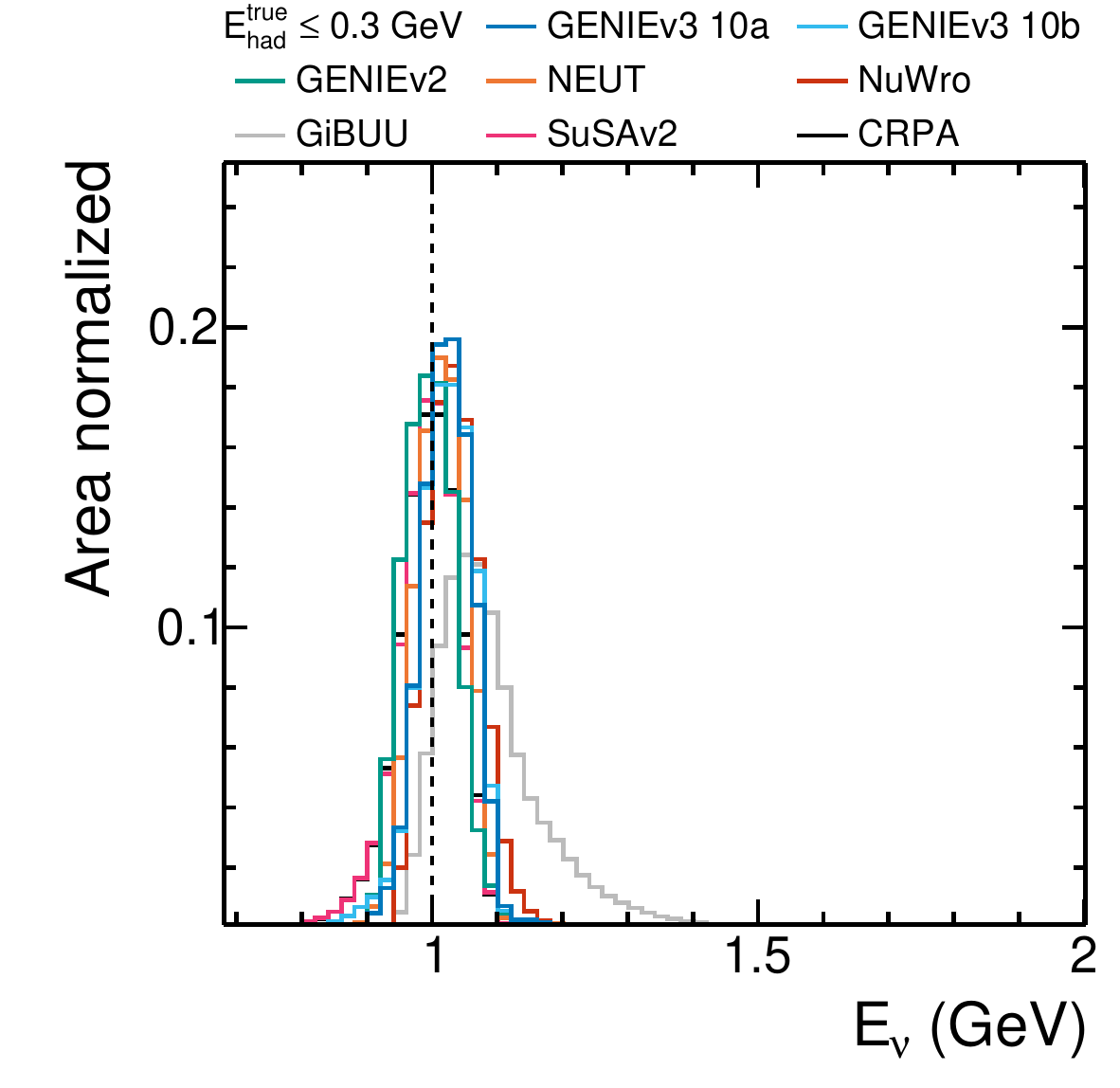}}
  \subfloat[\emu + \ehadtrue = 2.5 GeV]  {\includegraphics[width=0.3\linewidth]{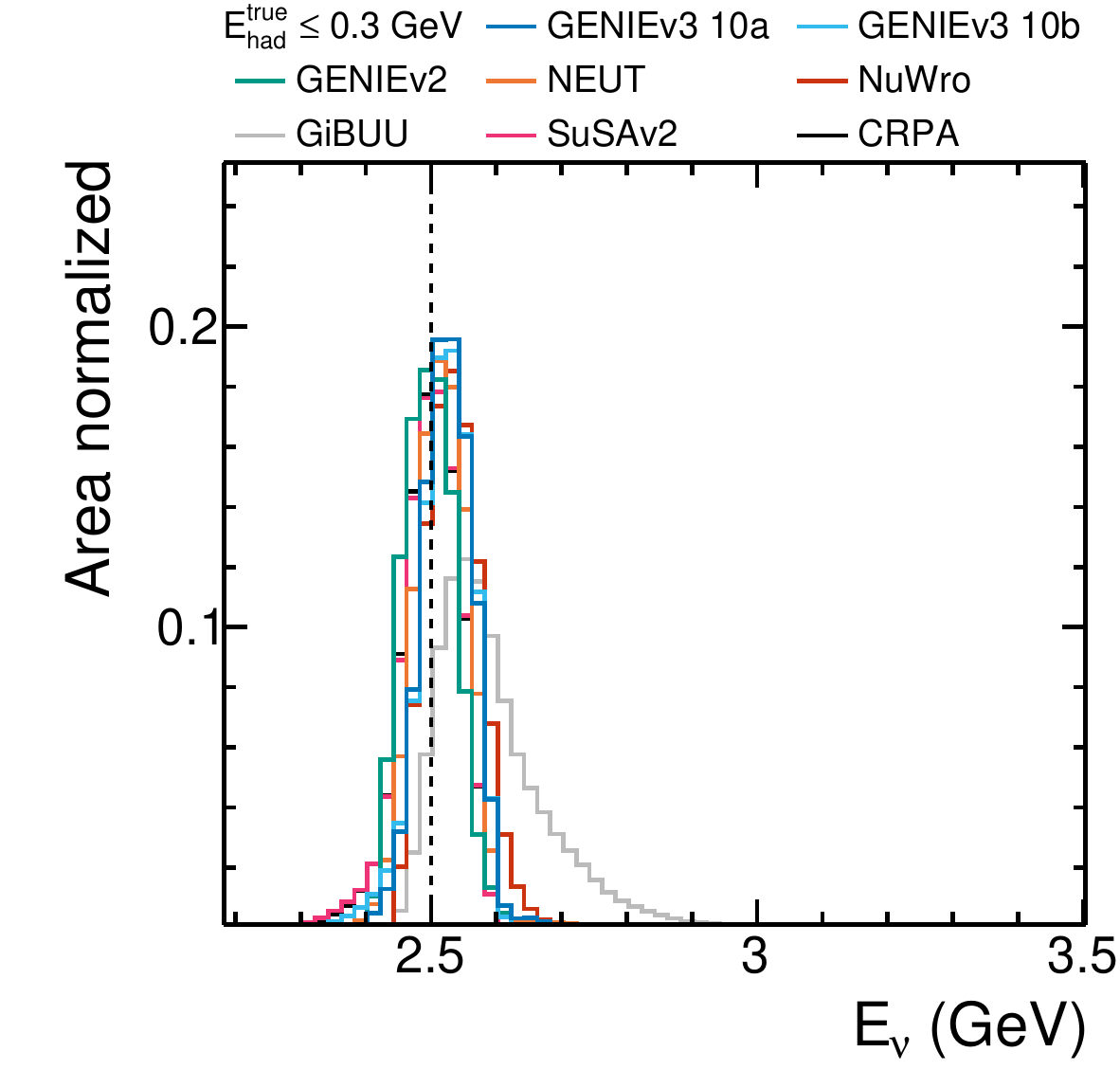}}
  \subfloat[\emu + \ehadtrue = 5 GeV]    {\includegraphics[width=0.3\linewidth]{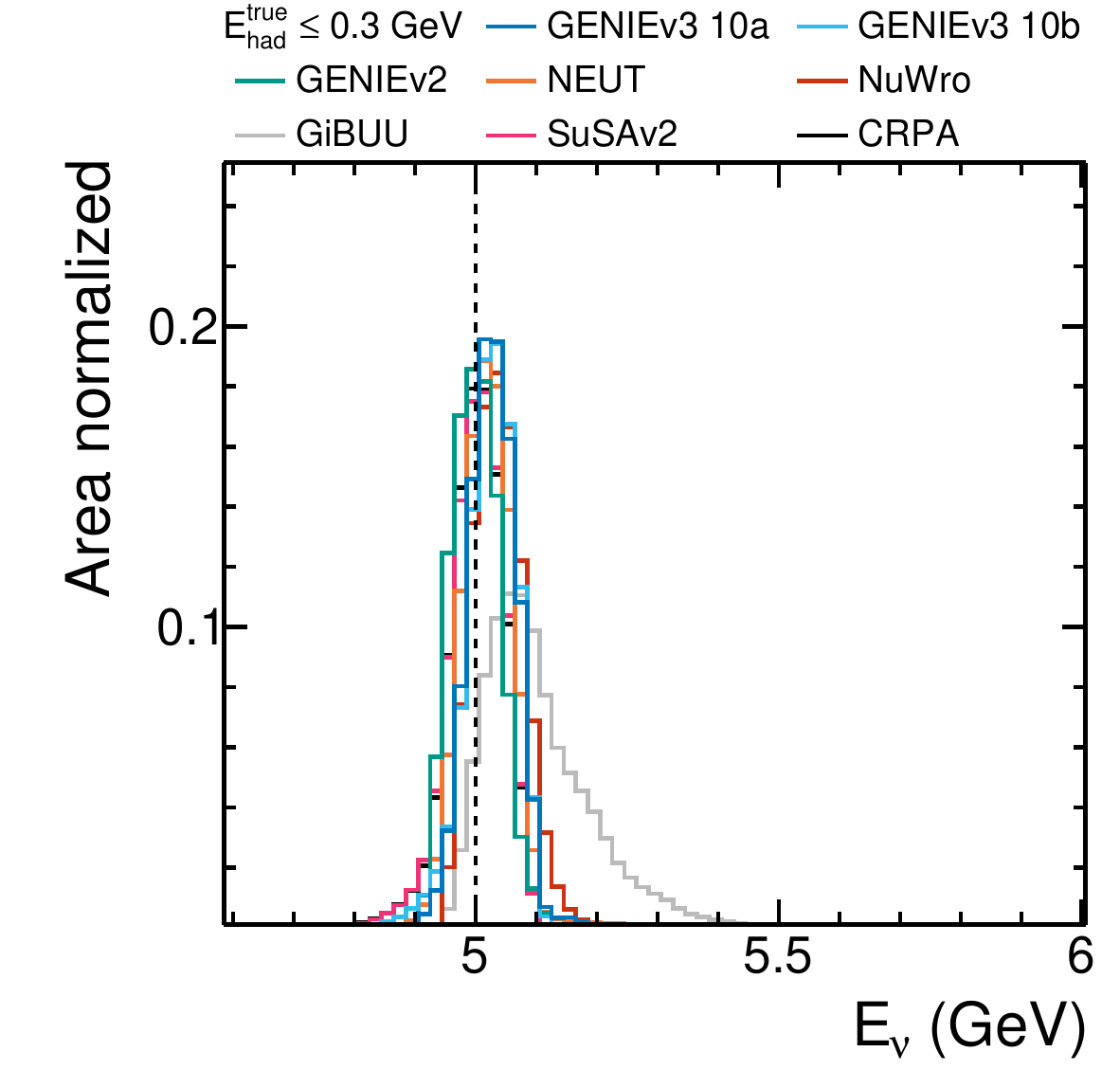}}\\
  \subfloat[\emu + \ehadreco = 1 GeV]    {\includegraphics[width=0.3\linewidth]{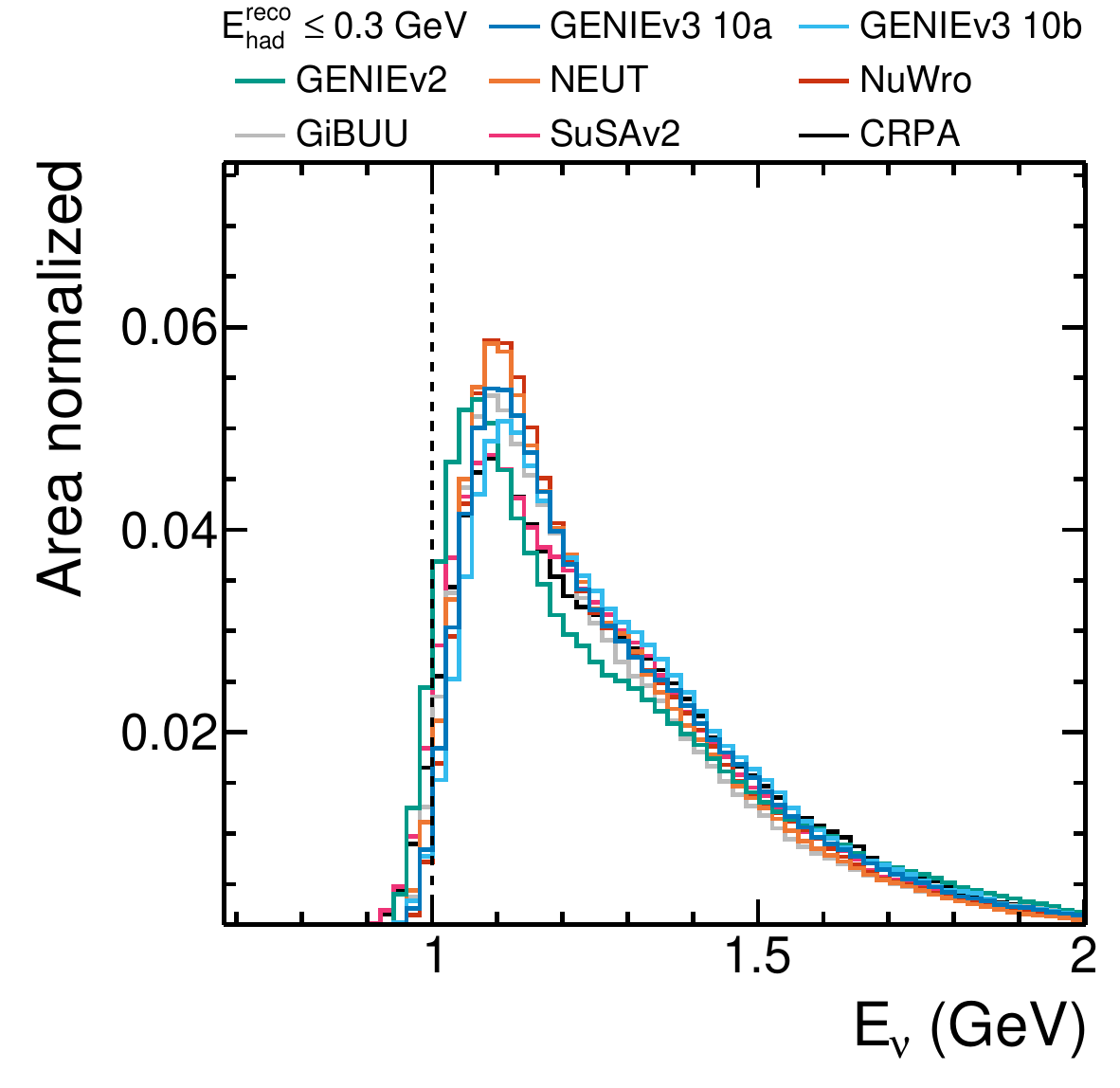}}
  \subfloat[\emu + \ehadreco = 2.5 GeV]  {\includegraphics[width=0.3\linewidth]{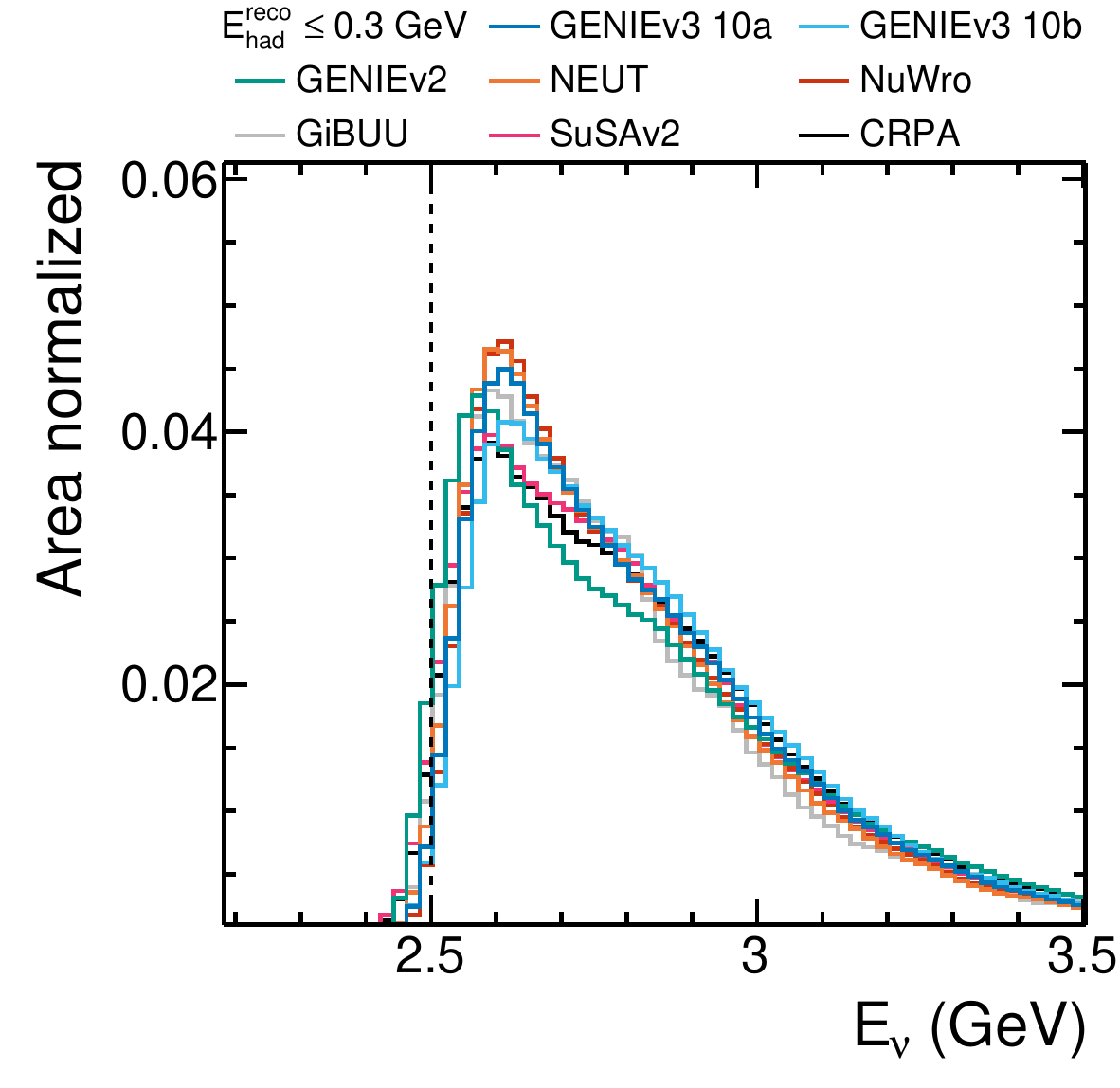}}
  \subfloat[\emu + \ehadreco = 5 GeV]    {\includegraphics[width=0.3\linewidth]{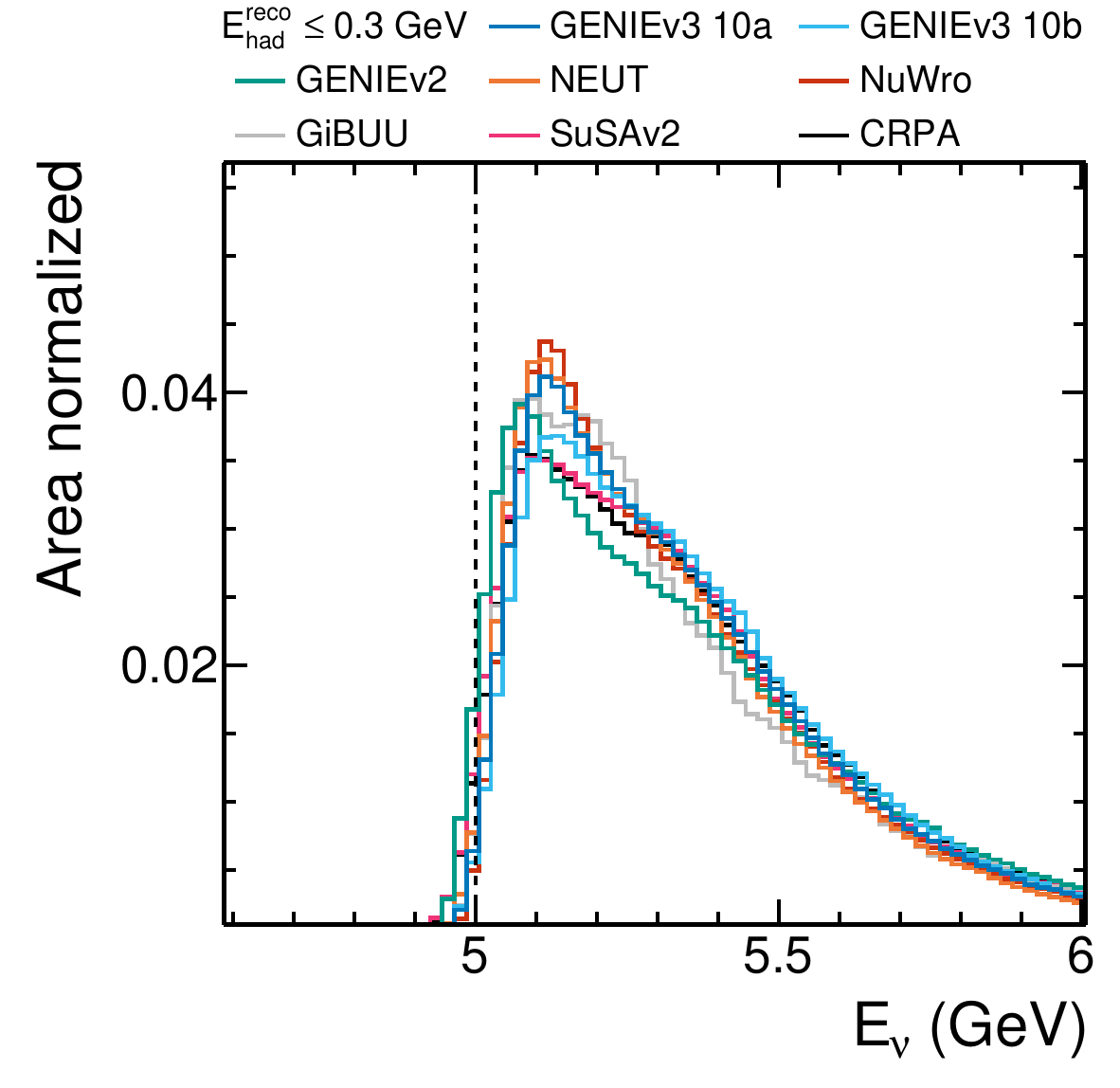}}\\
  \subfloat[\emu + \eavail = 1 GeV]    {\includegraphics[width=0.3\linewidth]{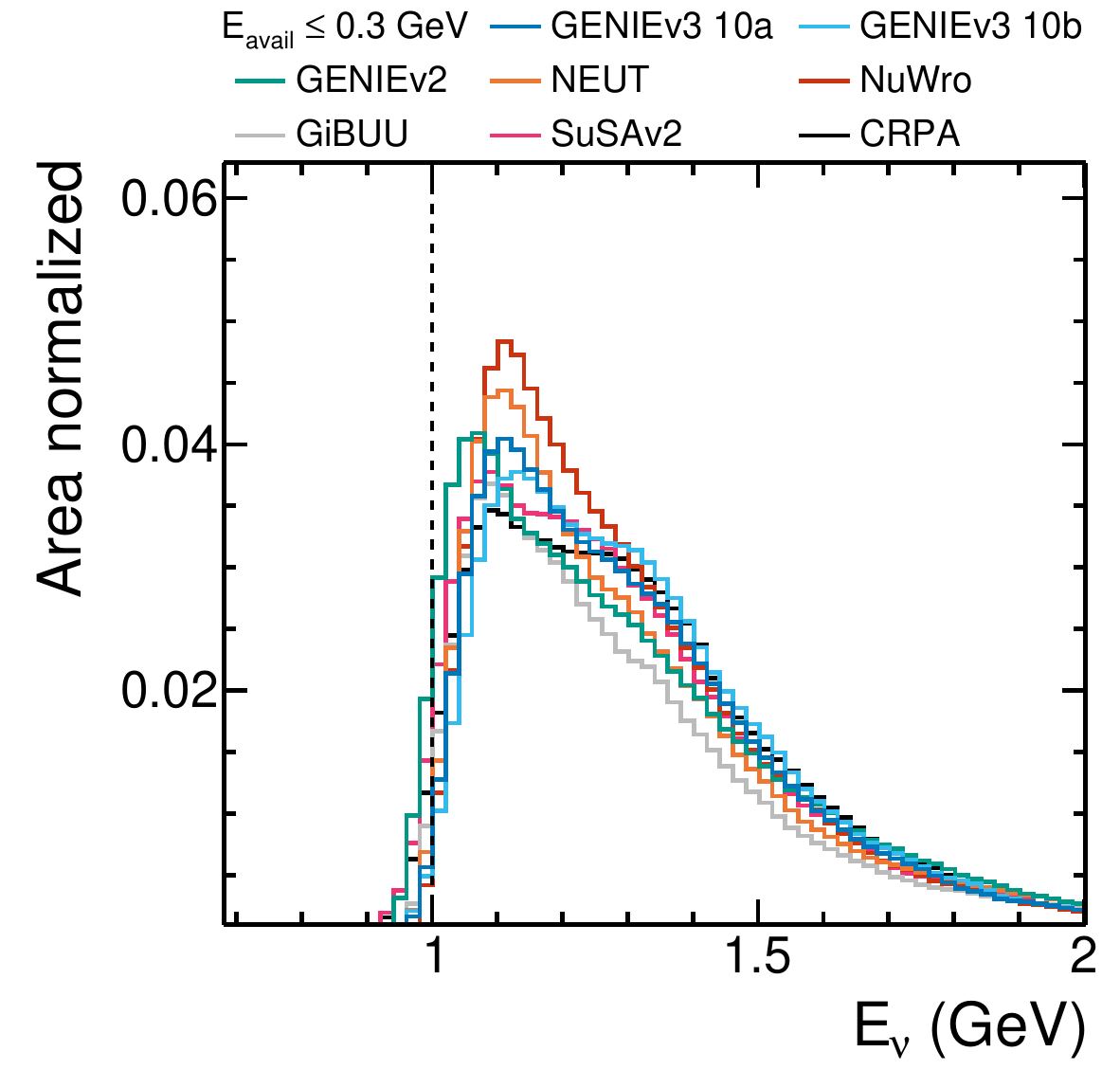}}
  \subfloat[\emu + \eavail = 2.5 GeV]  {\includegraphics[width=0.3\linewidth]{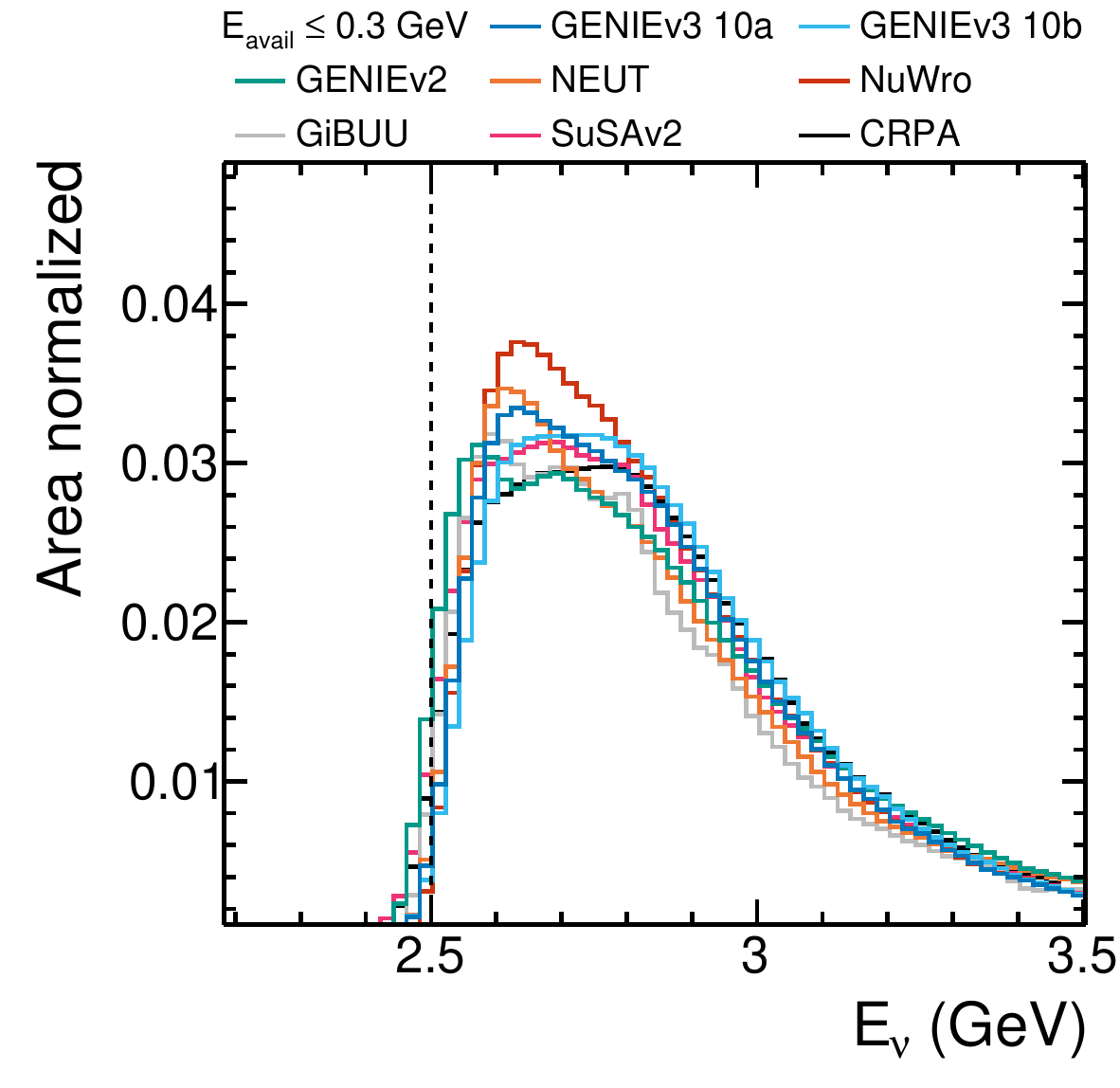}}
  \subfloat[\emu + \eavail = 5 GeV]    {\includegraphics[width=0.3\linewidth]{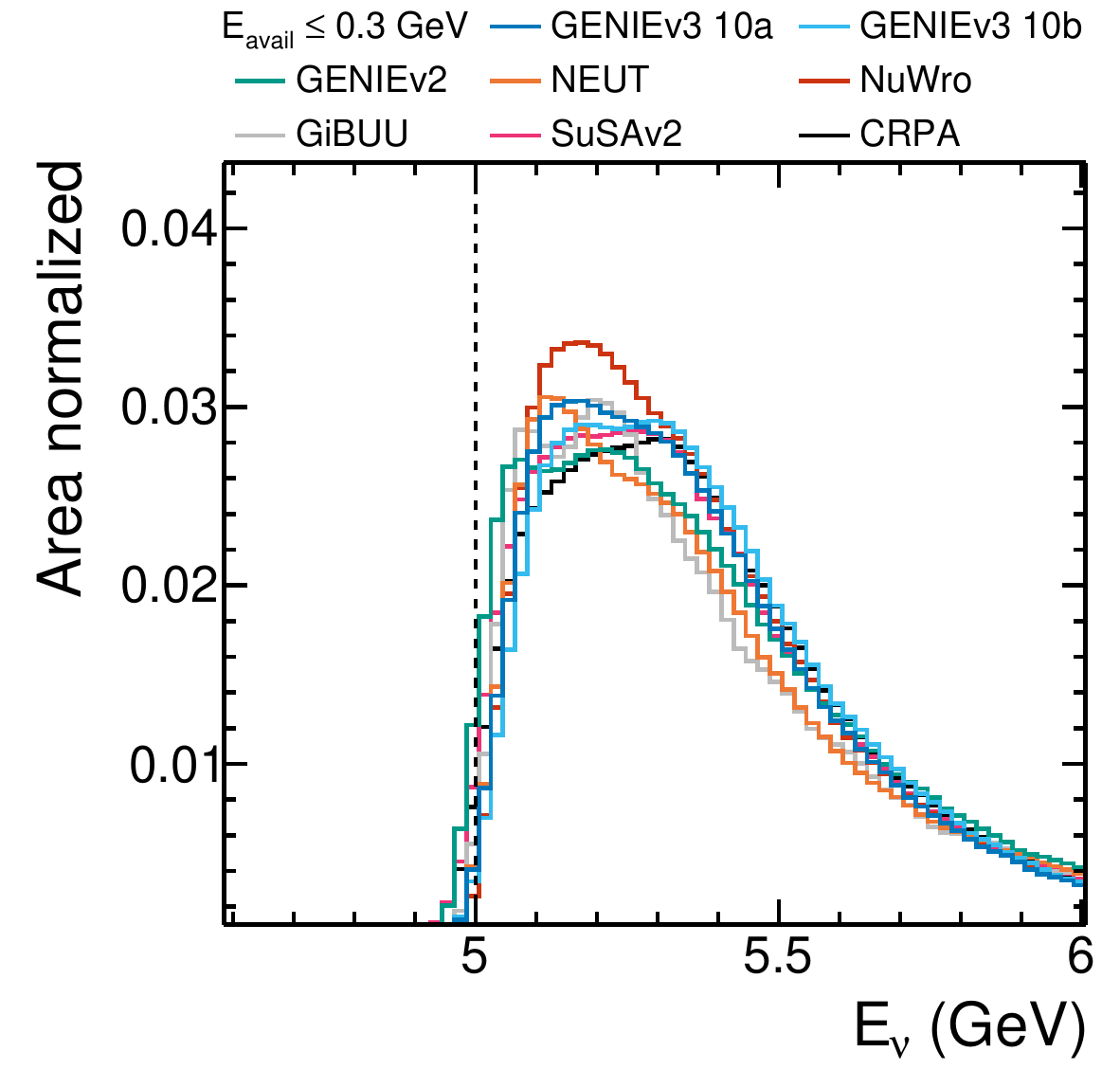}}
  \caption{Area normalized distributions of \enutrue corresponding to fixed values of the three proxy \enureco variables, \emu + \ehadtrue, \emu + \ehadreco and \emu + \eavail, at 1, 2.5 and 5 GeV, shown for \numub--\argon events, for all generator models of interest. A cut on the relevant \qz proxy of $\leq$ 0.3 GeV is made in each case.}

  \label{fig:smear_enu_proxy_numub}
\end{figure*}

The true neutrino energy is at the heart of any neutrino flux measurement; indeed, the goal of the \lownu method is to measure the shape of the incoming neutrino energy spectrum. Therefore, although the results in \autoref{fig:shape_vary_proxy} demonstrate that the cross section in true \qz for a \lownu sample obtained with our proxy variables vary between models, this is not the full story as those comparisons are shown as a function of \enutrue. Here we discuss three proxy variables for the reconstructed neutrino energy \enureco which use the measured muon energy \emu and the energy-transfer proxy variables described in \autoref{sec:en_trans}: $\emu + \ehadtrue$, $\emu + \ehadreco$, $\emu + \eavail$. 
For all three of the \enureco variables defined, there is some non-trivial smearing between \enutrue and \enureco that would need to be corrected for to extract a flux constraint in a \lownu analysis. Therefore, differences between generator models in the mapping between the true and reconstructed neutrino energies will introduce additional model-dependent bias to a \lownu analysis.

\autoref{fig:smear_enu_proxy_numu} and \autoref{fig:smear_enu_proxy_numub} show the relationship between \enutrue and the three different \enureco variables of interest for \numu--\argon and \numub--\argon \lownu candidate samples, respectively. In each case, the \enutrue distribution corresponding to fixed values of the \enureco proxy variables are shown, for all models described in \autoref{sec:models}. For each, a cut of $\leq 0.3~\text{GeV}$ is made on the corresponding \qz proxy to form the \lownu candidate sample. If a higher (lower) cut value had been chosen, more (less) smearing would be introduced because the degree of migration increases (decreases).
For all \enureco variables of interest there is some significant smearing. Even in the ``perfect'' detector case (\enureco = \emu+\ehadtrue) the smearing alone is severe enough to potentially damage the utility of the method for low neutrino energies, especially given the significantly different predictions from GiBUU compared to other models. The degree of smearing for the more realistic proxy variables can be seen to be significantly larger, as expected. In all cases, the bulk of the smearing is such that \enureco is smaller than \enutrue, which is not surprising as the effect is largely to miss energy, although smearing from the initial nuclear state can also increase the reconstructed energy.
As was the case with the smearing between \qz and its proxies in \autoref{fig:smear_q0_proxy_numu} and \autoref{fig:smear_q0_proxy_numub}, the migration effect from higher values of the true variable is more pronounced for \numub--\argon than \numu--\argon for variables without neutron reconstruction.
Although the smearing between both true \qz and its proxy variables, and between \enutrue and the three \enureco definitions is significant, the differences between generator models here is not as striking as the differences between the low-\qz cross sections shown in \autoref{sec:low-qz-cross-section}. However, as discussed in \autoref{sec:models} any apparent agreement between generator predictions of the feed down may be somewhat artificial, since the generators tend to have less model diversity at higher \qz.

\section{Conclusions}

\label{sec:conclusions}

In this work we have introduced the \lownu method and described how it has been used historically, from its first usage by experiments with $\mathcal{O}$(10-100 GeV) neutrino beams, to more recent usage in few-GeV accelerator neutrino experiments. We have investigated its potential use as a tool for constraining the neutrino flux for current and future few-GeV experiments and found that the requirements for a precision application of the \lownu method are not fulfilled. In particular, we find significant differences between the predictions of different neutrino interaction models within the low \qz region that the method requires to be well understood. Furthermore, the differences between the models investigated \textit{increase} for lower \qz regions, counter to what is expected when applying the \lownu method at higher energies. We find that this is largely due to differences in the way each model treats poorly-understood nuclear-medium effects, which are most impactful at low \qz. Far from being a ``standard candle'', we conclude that the low-\qz region actually represents the least consistently predicted region of phase space, undermining the entire motivation for the \lownu method for few-GeV energies. It was further found that model differences for \numub--\argon scattering are often different than those \numu--\argon scattering, which raises particular concerns regarding the bias a \lownu constraint might introduce for experiments hoping to measure CP-violation. We note that the expected uncertainties are of a similar magnitude when repeating the analysis with \ch and \chtwo targets (\autoref{app:chtarget}), and the general conclusions are applicable to all of the current and future accelerator neutrino experiments discussed.

Additionally, we explored the effect of using observable proxies for \qz and \enutrue, as neither are directly accessible experimentally. We found that cutting on any of these proxy variables introduced significant migration from high true-\qz into \lownu candidate samples. Similarly, experimentally observable \enureco proxy variables introduced significant smearing in \enutrue, which would introduce further model-dependence into a \lownu flux constraint. Proxy variables in which neutrons are not reconstructed are more problematic for \numub than \numu. Whilst the use of observable variables imply significant additional challenges in extracting a meaningful flux constraint from a \lownu analysis, we found that model differences in the smearing to observable variables are unlikely to be as damaging as those already observed in the ideal true low-\qz cross-section ratios---largely because the region exhibits the most model dependence.

We caution that the model spread technique used in this work is unlikely to reflect the full uncertainty that should be applied to any discussion of the \lownu method. The apparent lack of model spread at few-GeV neutrino energies observed in Figure~\ref{fig:shape_vary_highq0} with a high (few-GeV) \qz cut is likely due to the lack of diversity in the models. Futhermore, there are additional uncertainties on the nucleon level cross section discussed in Ref.~\cite{Belusevic:1988ab}, particularly in relation to the contribution of heavy resonances to single pion production, which are not reflected in the model spread shown here, but which are as unknown now as they were 35 years ago~\cite{Kabirnezhad:2017jmf}. These should be carefully considered in the context of proposals to apply the \lownu method to hydrogen samples~\cite{Duyang:2019prb}.

To conclude, at the SBN, T2K/Hyper-K, NOvA and DUNE \enupeak, the shape uncertainty from the \lownu method is cross-section model dependent, with model differences similar to or greater than the {\it a priori} flux shape uncertainties for \numu and \numub interactions on \argon, \ch and \chtwo. Given the current landscape of neutrino interaction modeling, and acknowledging the limitations of the model-spread technique used herein, we find that the \lownu method can only provide useful constraints at neutrino energies higher than the region of interest for oscillation measurements, $\enutrue \gtrsim 5~\text{GeV}$ for neutrinos and $\enutrue \gtrsim 12~\text{GeV}$ for antineutrinos.

\section*{Acknowledgements}
The work of C. Wilkinson was supported by the U.S. Department of Energy, Office of Science, Office of High Energy Physics, under contract number DE-AC02-05CH11231
The work of C. Wret was supported by the Department of Energy, Office of Nuclear Physics, under Contract No. DE-SC-0008475. L.~Pickering is supported by a Royal Society University Research Fellowship \\ (URF\textbackslash{}R1\textbackslash{}211661).
This research used resources of the National Energy Research Scientific Computing Center (NERSC), a U.S. Department of Energy Office of Science User Facility located at Lawrence Berkeley National Laboratory, operated under Contract No. DE-AC02-05CH11231 using NERSC award HEP-ERCAP0018205.
S.~Dolan would like to thank A.~Nikolakopoulos for crucial discussions regarding the CRPA model predictions.

\appendix
\section{\lownu with hydrocarbon targets}
\label{app:chtarget}
This appendix repeats some of the key analyses from \autoref{sec:low-qz-cross-section} and \autoref{sec:variables} but using \ch and \chtwo targets rather than an \argon target. These results are relevant for DUNE's plastic scintillator beam monitor (\chtwo), NOvA liquid scintillator (predominantly hydrocarbon), and plastic scintillator components of T2K/Hyper-K's near detector complex (\ch). They can also be generalized to any other few-GeV neutrino experiment with hydrocarbon targets, and can be seen as representative of ``light'' nuclear targets such as water.

\begin{figure*}[htbp]
  \centering
  \captionsetup[subfloat]{captionskip=-3pt}
  \subfloat[\numu--\ch, $\qz \leq 0.1$ GeV]  {\includegraphics[width=0.29\linewidth]{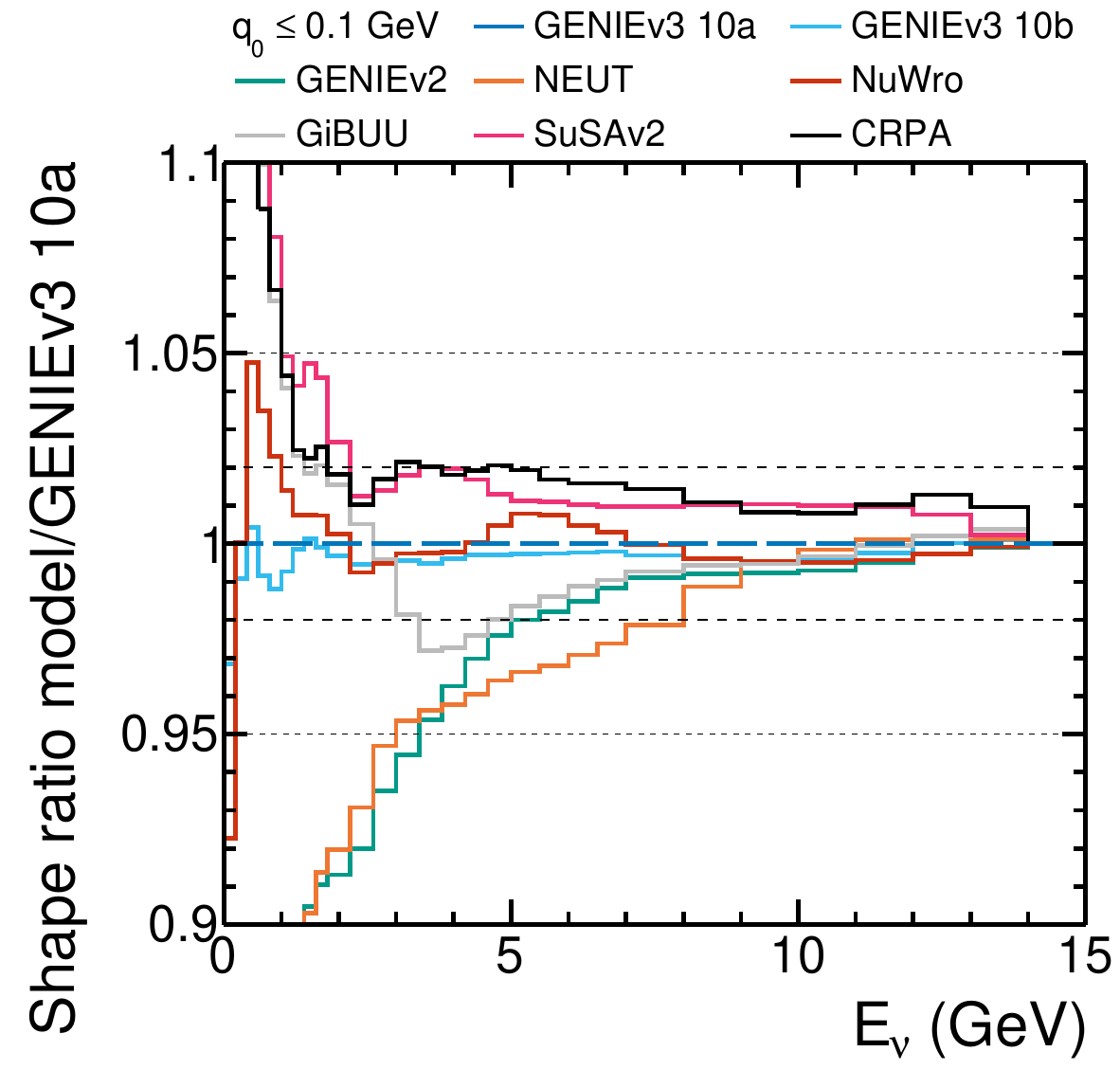}}
  \subfloat[\numu--\ch, $\qz \leq 0.3$ GeV]  {\includegraphics[width=0.29\linewidth]{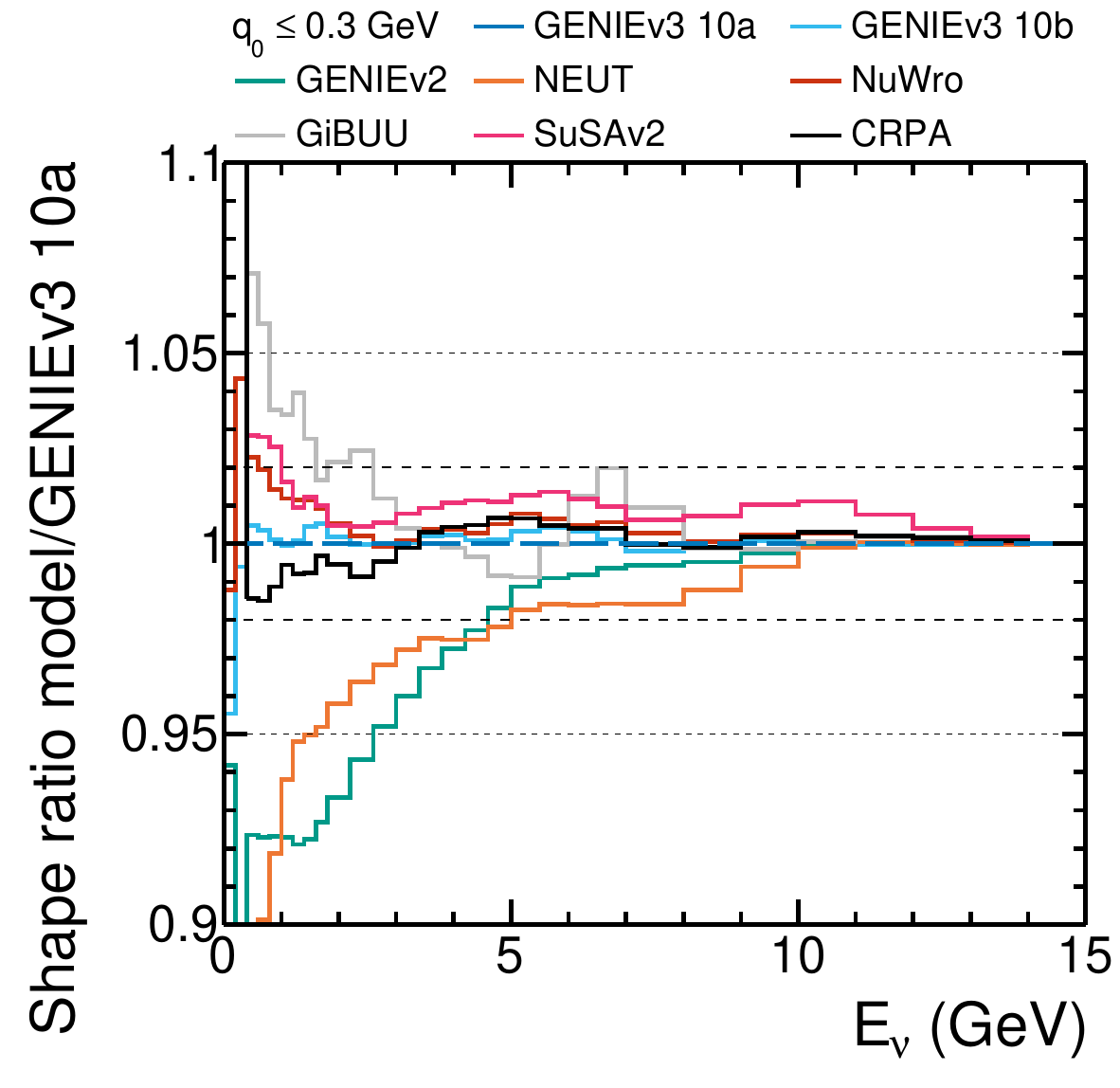}}
  \subfloat[\numu--\ch, $\qz \leq 0.5$ GeV]  {\includegraphics[width=0.29\linewidth]{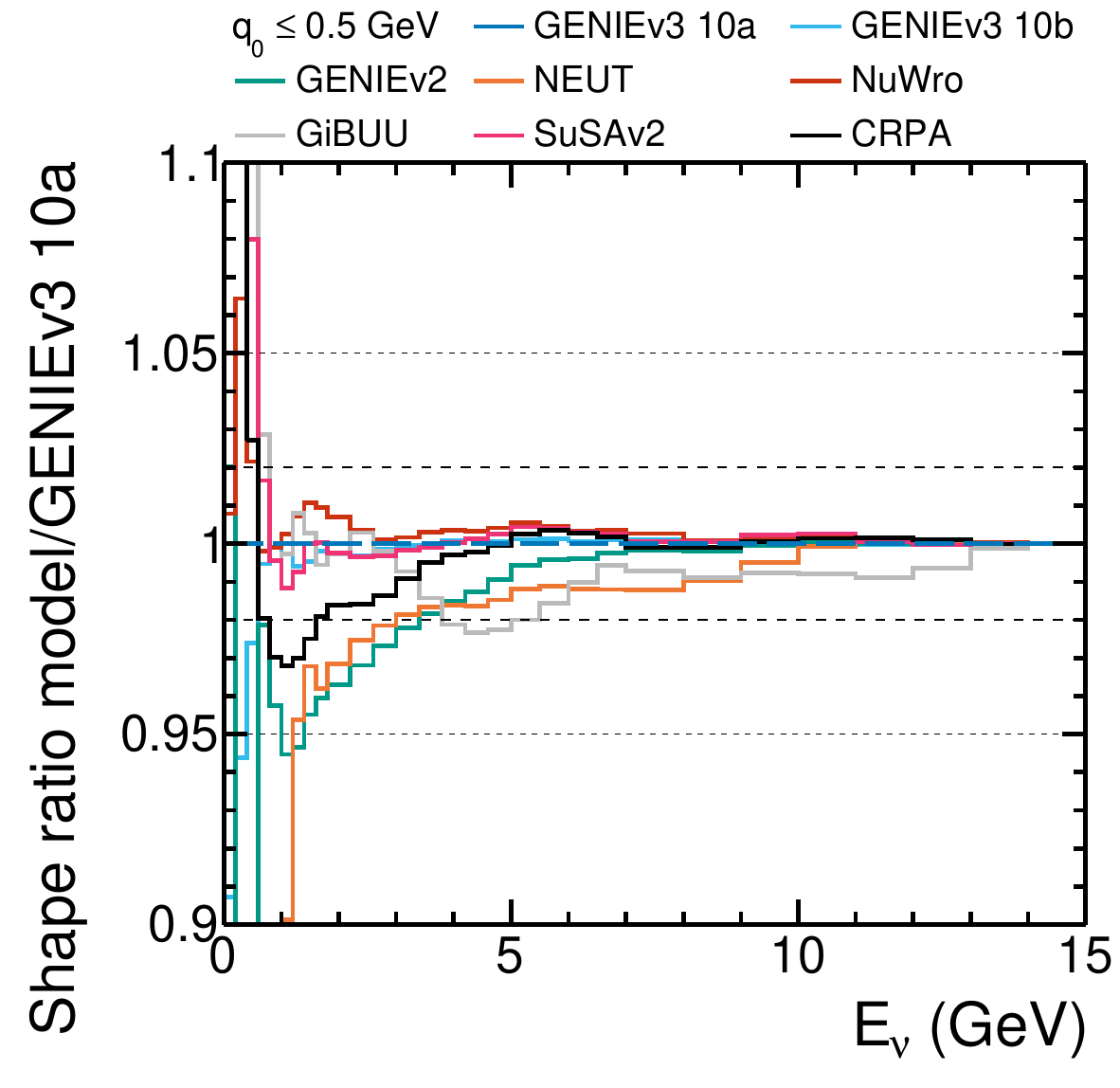}}\\\vspace{-7pt}
  \subfloat[\numub--\ch, $\qz \leq 0.1$ GeV]  {\includegraphics[width=0.29\linewidth]{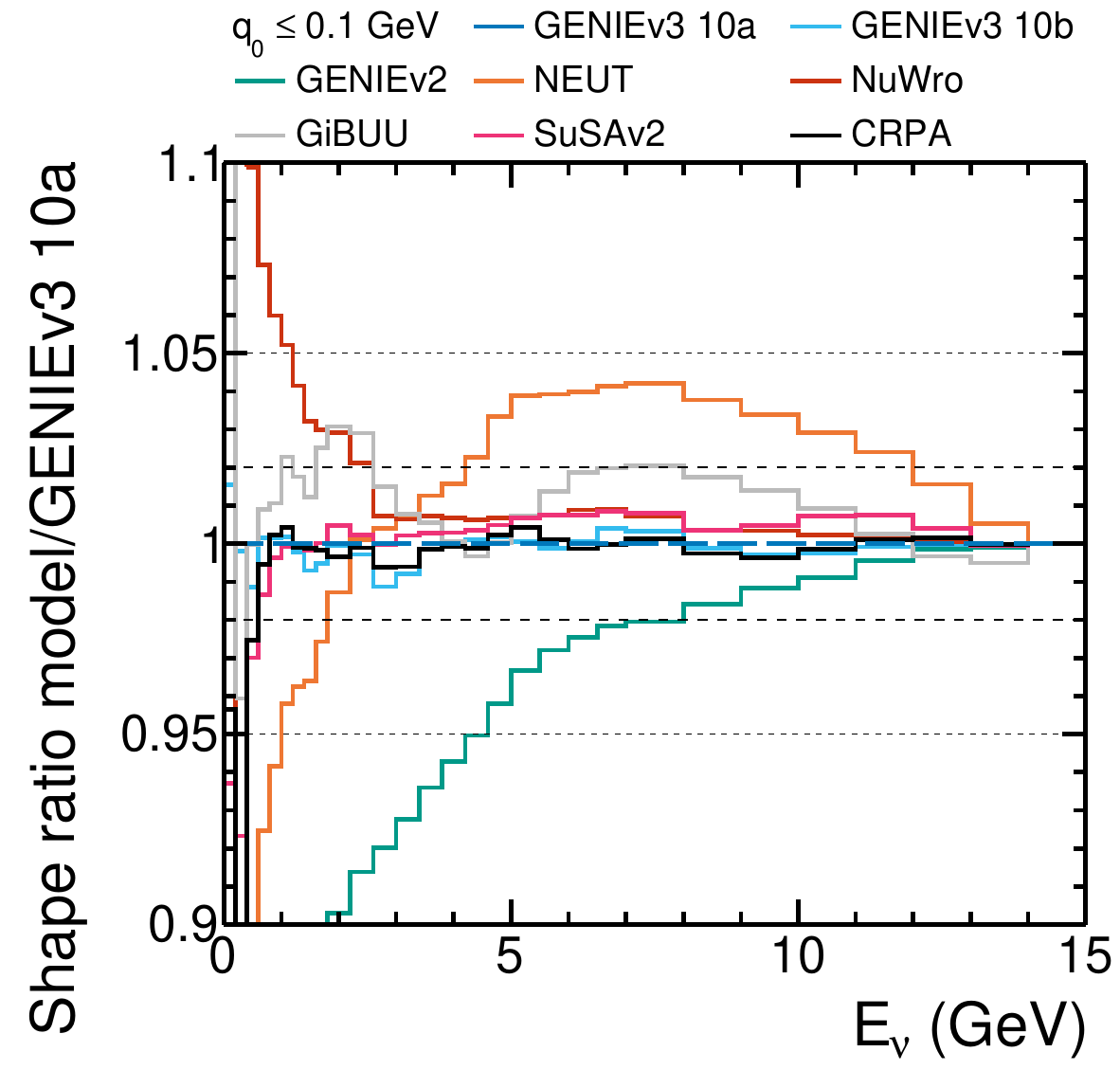}}
  \subfloat[\numub--\ch, $\qz \leq 0.3$ GeV]  {\includegraphics[width=0.29\linewidth]{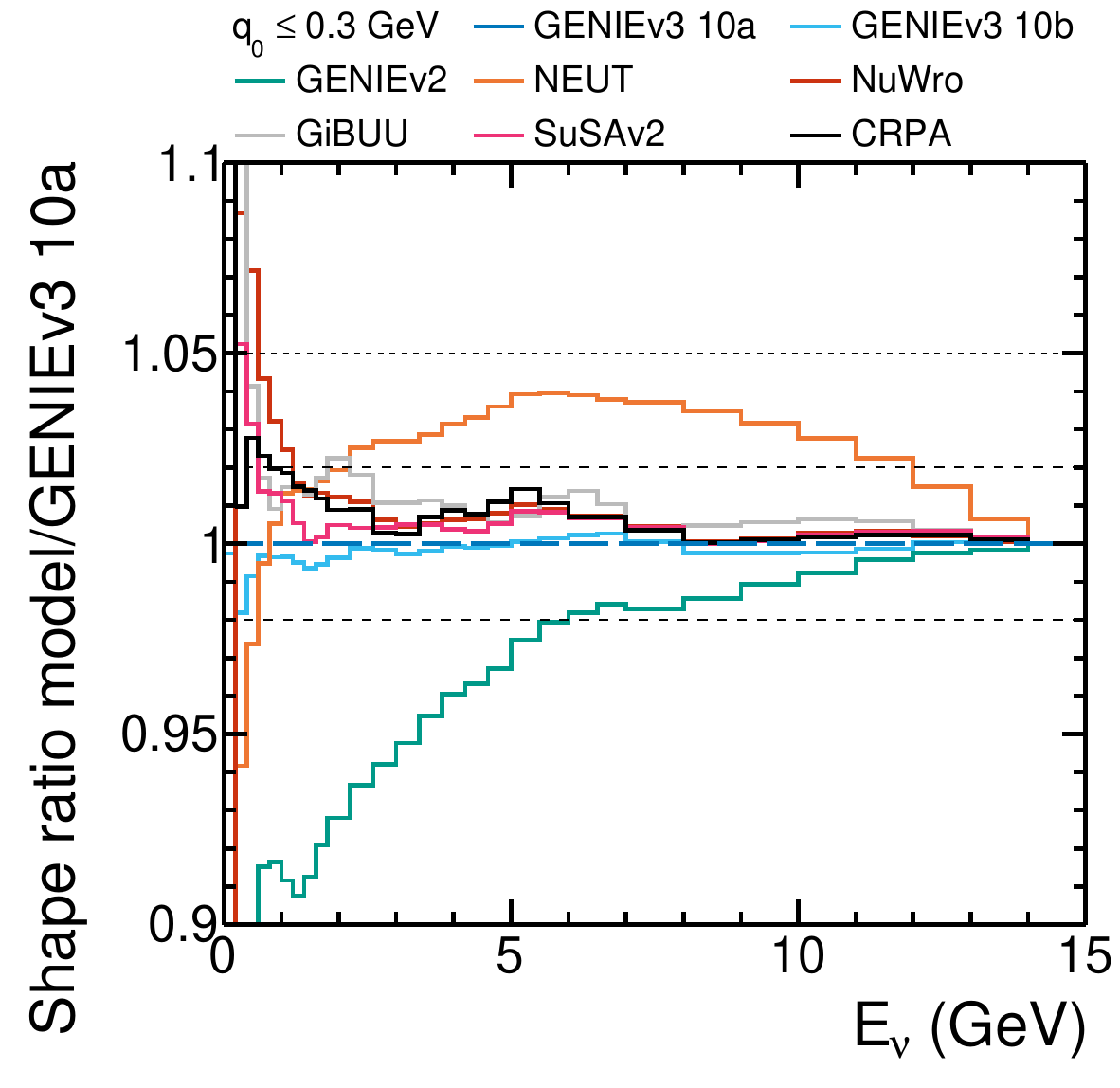}}
  \subfloat[\numub--\ch, $\qz \leq 0.5$ GeV]  {\includegraphics[width=0.29\linewidth]{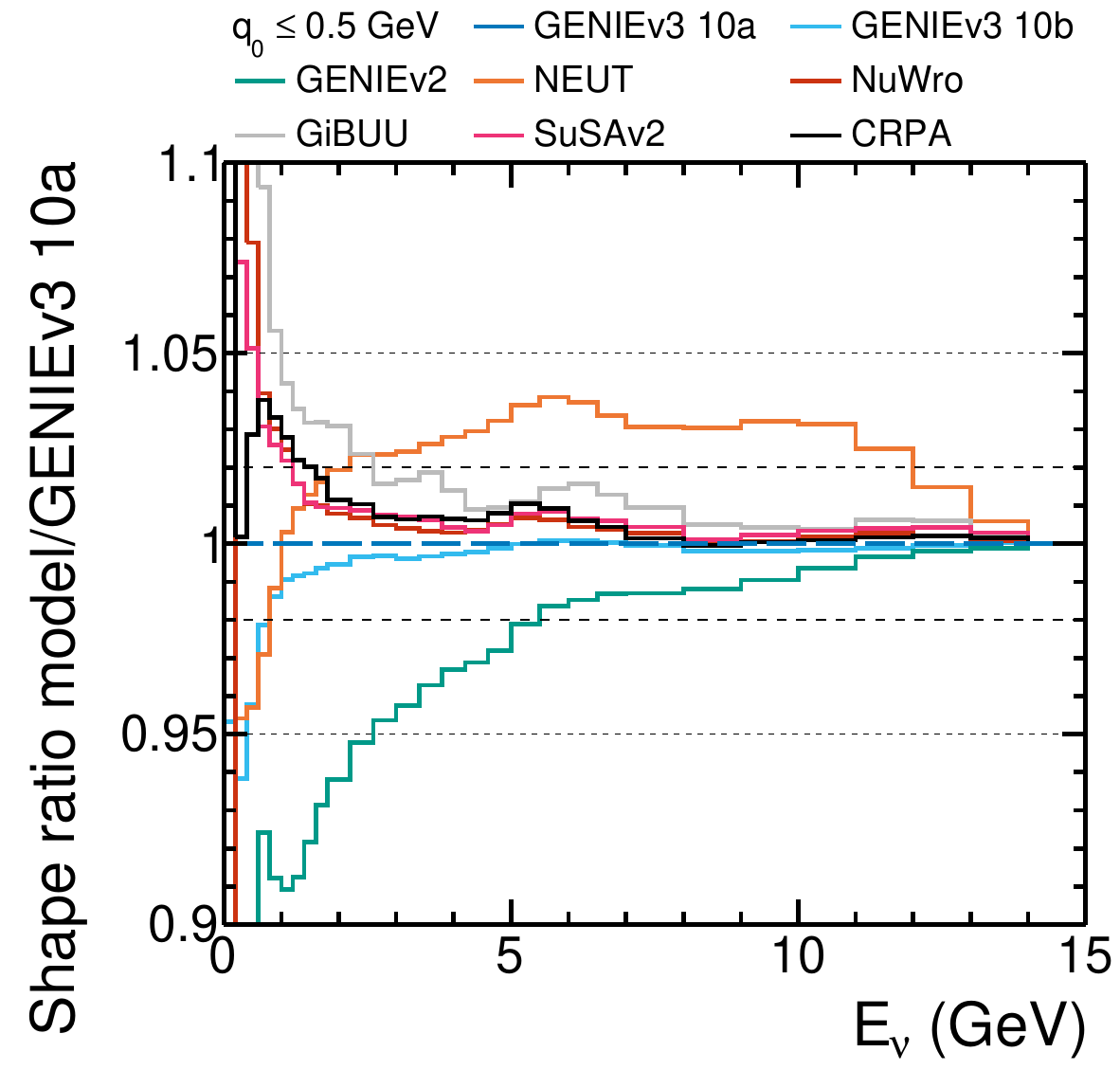}}
  \vspace{-1mm}
  \caption{Comparison of the shape-only ratios of the charged-current cross section with \qz $\leq$ 0.1, 0.3 and 0.5 GeV as a function of \enutrue, with respect to the GENIEv3 10a prediction. Shown for both \numu--\ch and \numub--\ch, and generated with a flat neutrino flux. Horizontal long (short) dashed lines have been added at $\pm$2\% ($\pm$5\%), to guide the eye in assessing the scale of the bias.\vspace{-5mm}}
  \label{fig:shape_vary_q0_CH}
\end{figure*}

\begin{figure*}[htbp]
  \centering
  \captionsetup[subfloat]{captionskip=-3pt}
  \subfloat[\numu--\chtwo, $\qz \leq 0.1$ GeV]  {\includegraphics[width=0.29\linewidth]{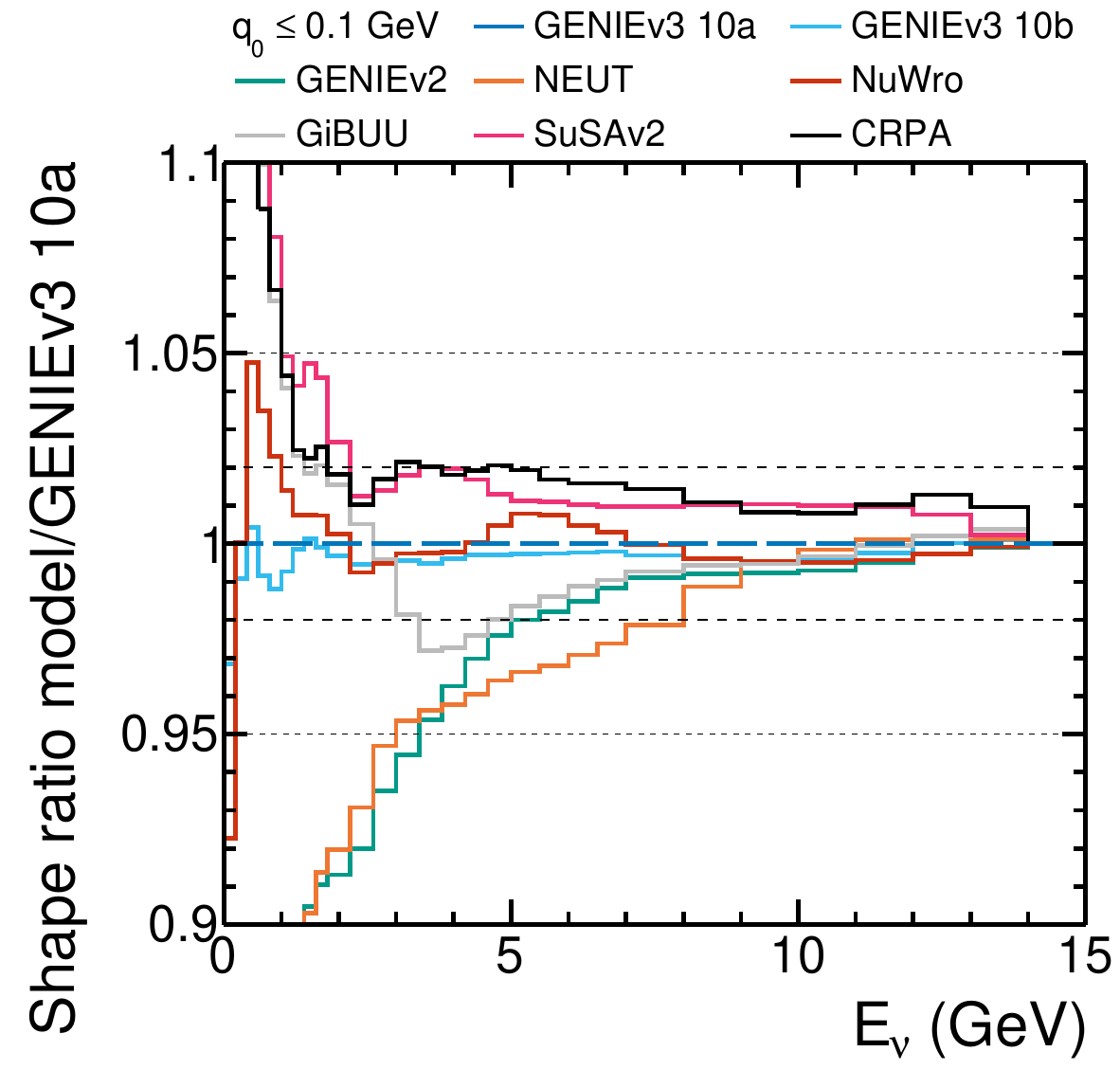}}
  \subfloat[\numu--\chtwo, $\qz \leq 0.3$ GeV]  {\includegraphics[width=0.29\linewidth]{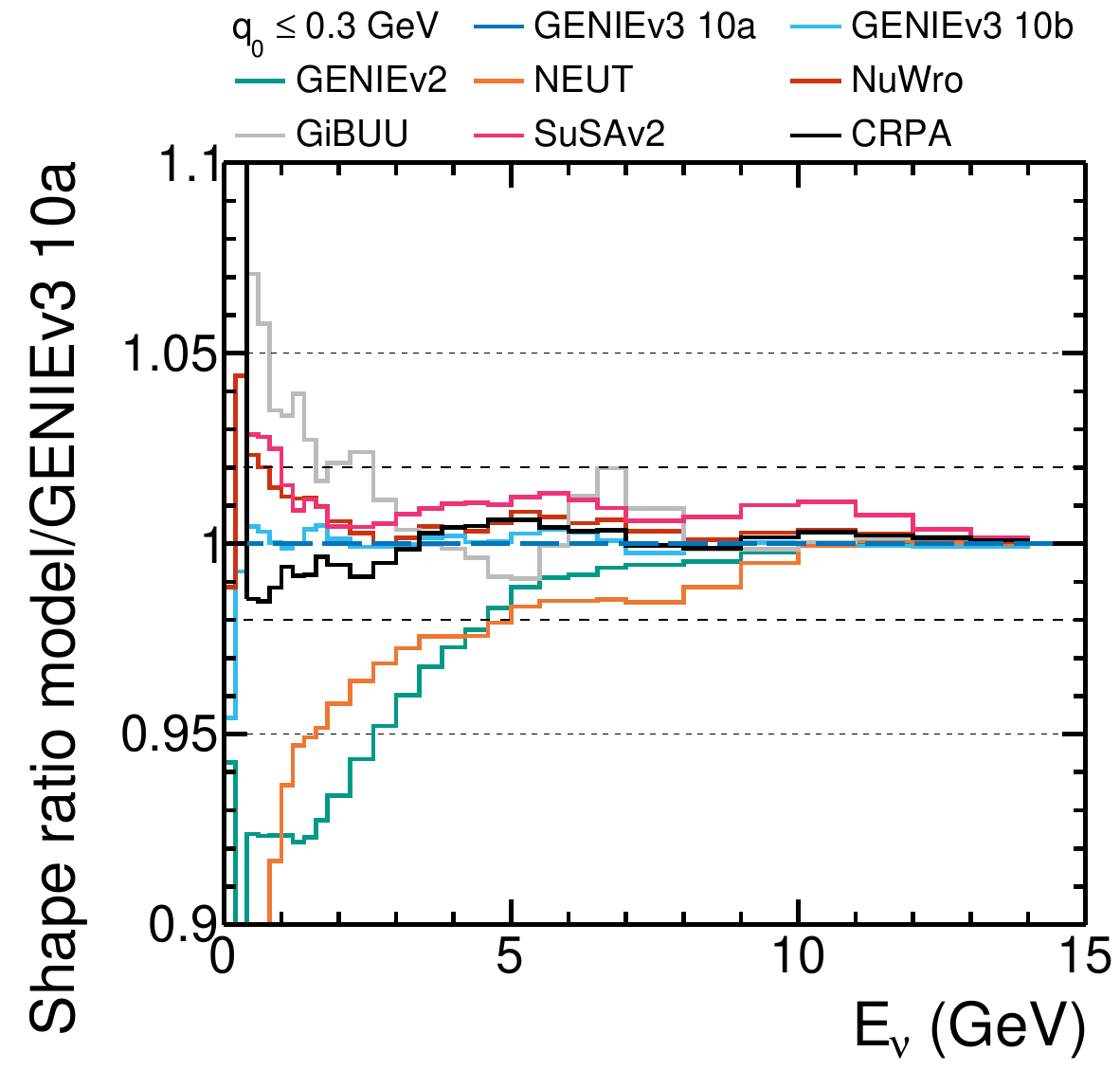}}
  \subfloat[\numu--\chtwo, $\qz \leq 0.5$ GeV]  {\includegraphics[width=0.29\linewidth]{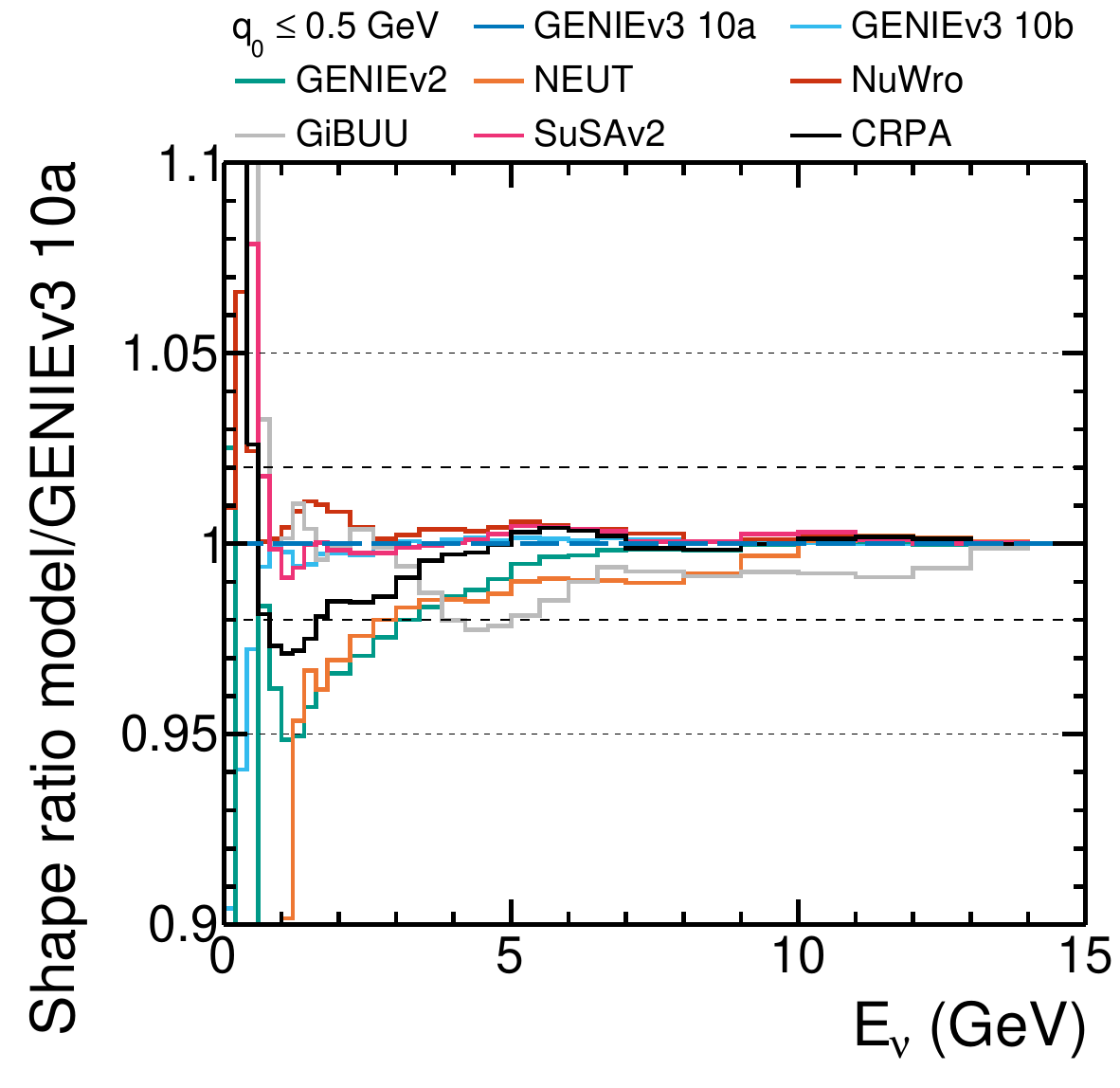}}\\\vspace{-7pt}
  \subfloat[\numub--\chtwo, $\qz \leq 0.1$ GeV]  {\includegraphics[width=0.29\linewidth]{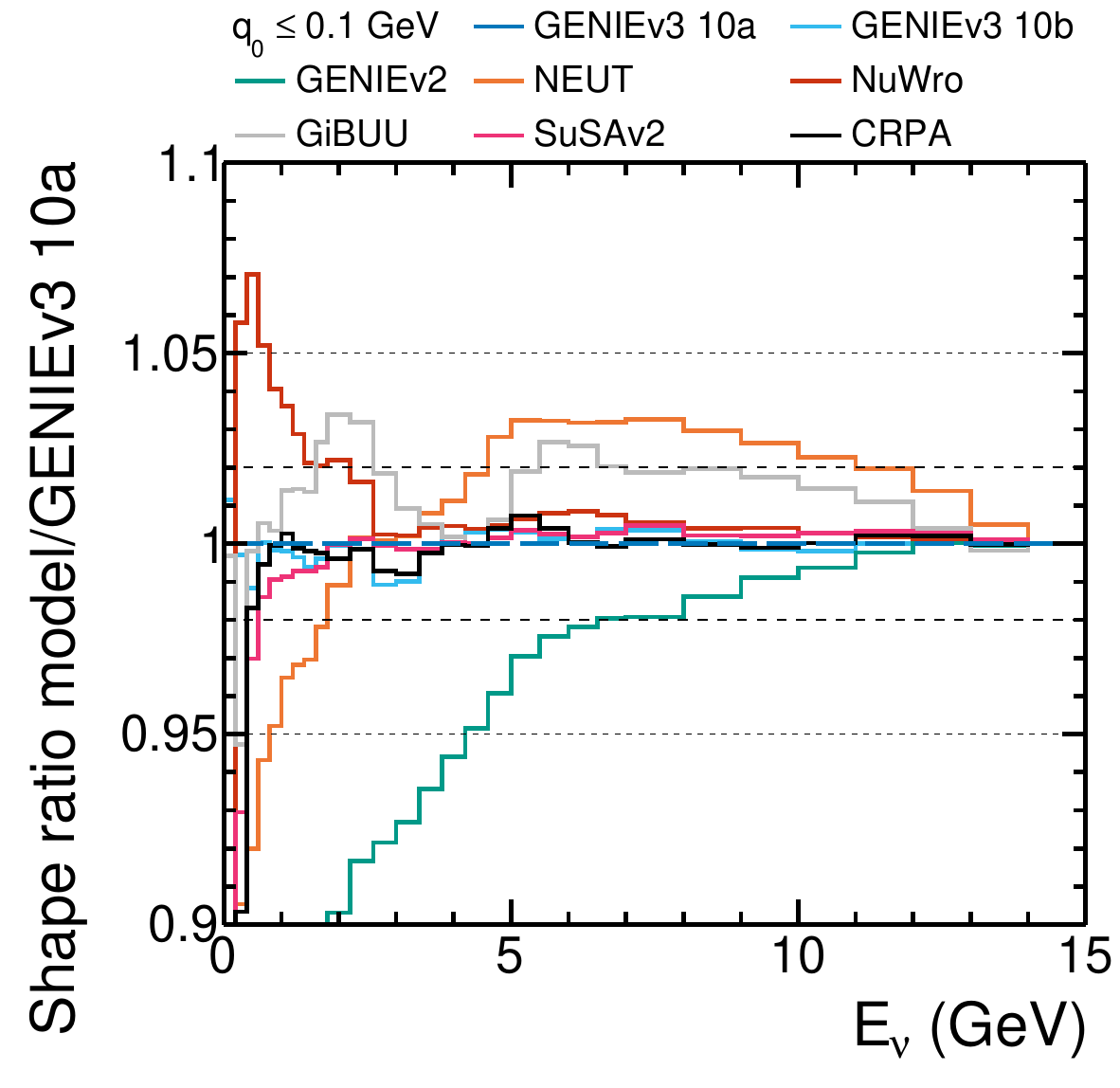}}
  \subfloat[\numub--\chtwo, $\qz \leq 0.3$ GeV]  {\includegraphics[width=0.29\linewidth]{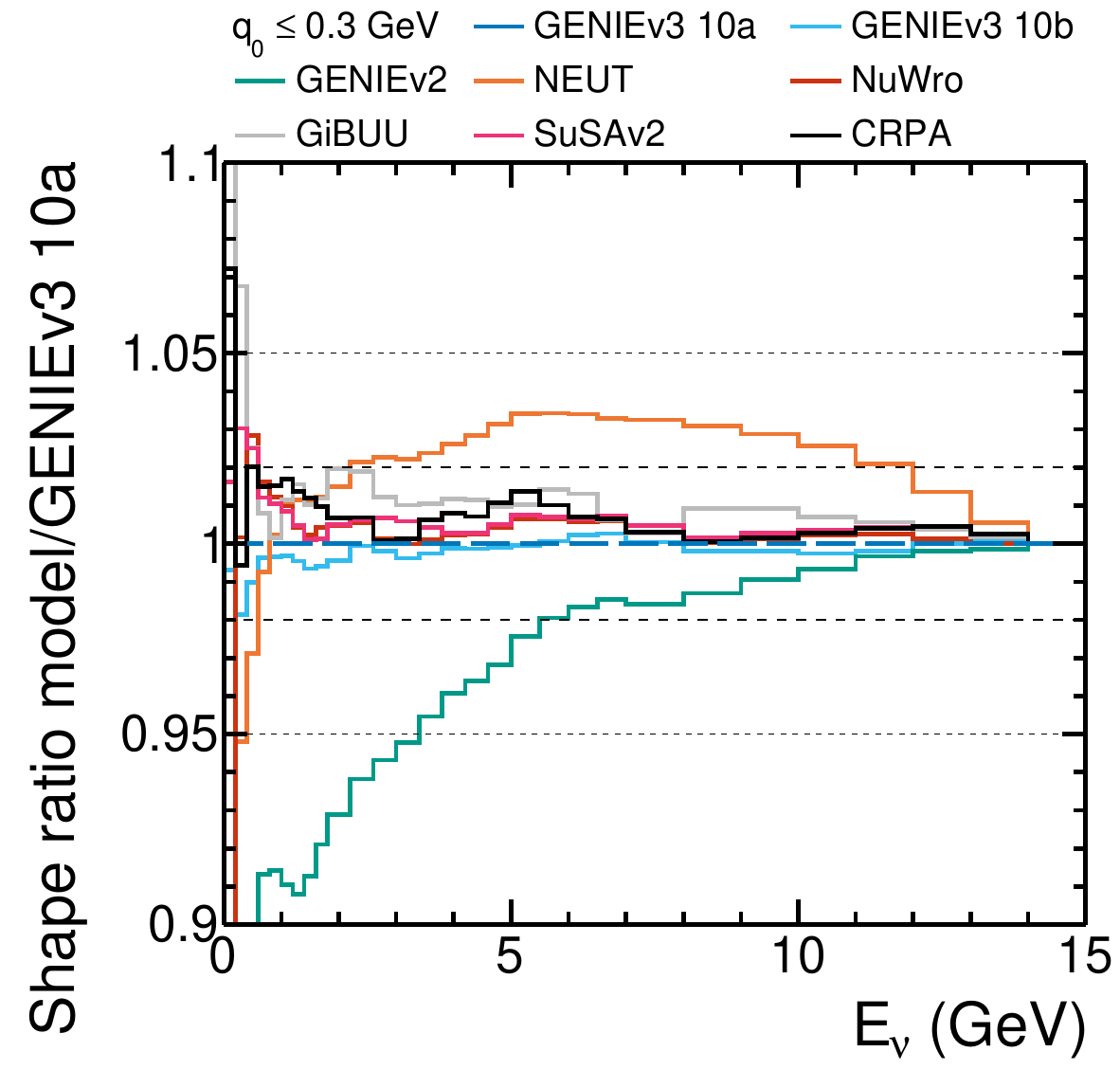}}
  \subfloat[\numub--\chtwo, $\qz \leq 0.5$ GeV]  {\includegraphics[width=0.29\linewidth]{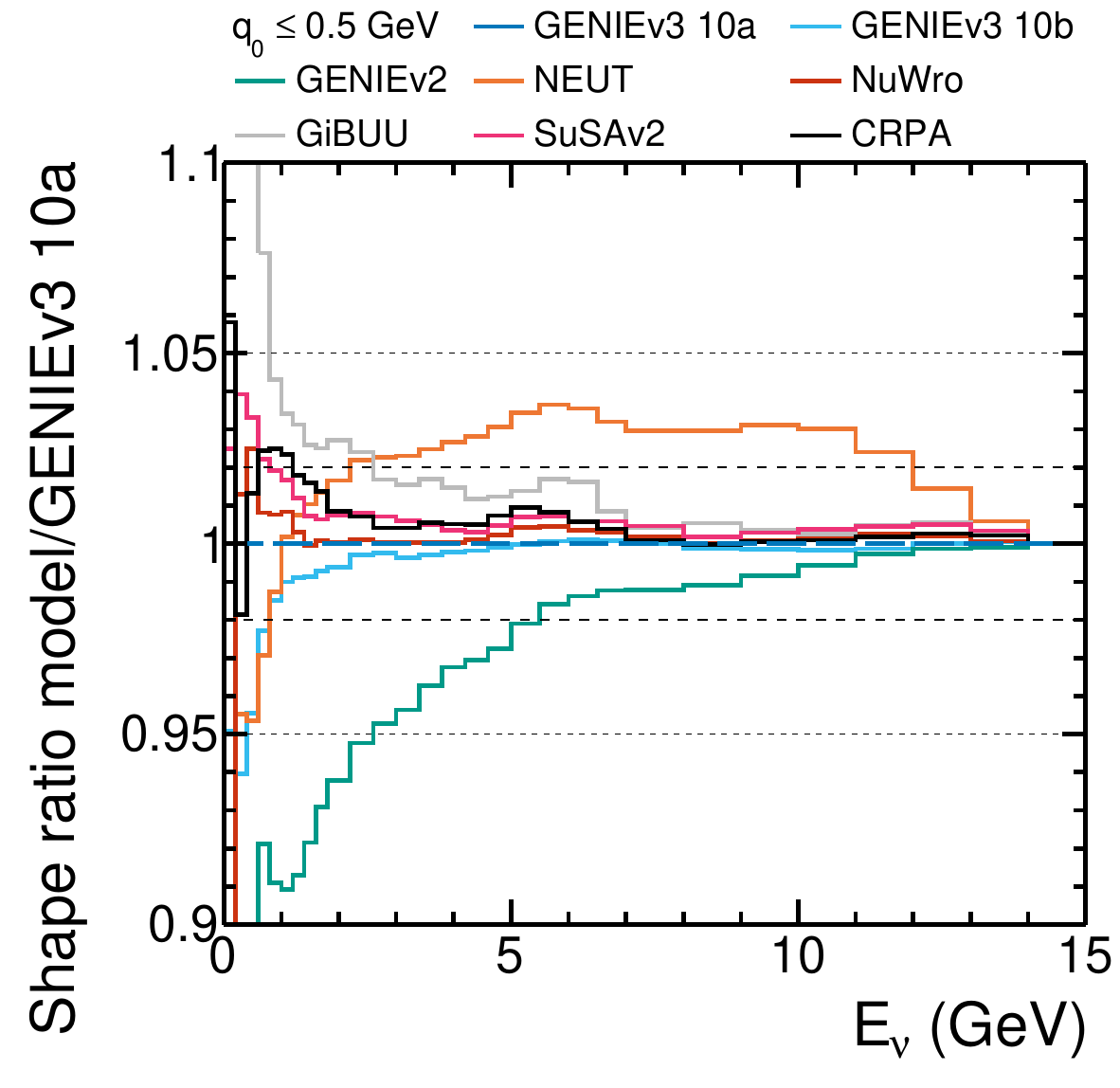}}
  \vspace{-1mm}
  \caption{Comparison of the shape-only ratios of the charged-current cross section with \qz $\leq$ 0.1, 0.3 and 0.5 GeV as a function of \enutrue, with respect to the GENIEv3 10a prediction. Shown for both \numu--\chtwo and \numub--\chtwo, and generated with a flat neutrino flux. Horizontal long (short) dashed lines have been added at $\pm$2\% ($\pm$5\%), to guide the eye in assessing the scale of the bias.\vspace{-5mm}}
  \label{fig:shape_vary_q0_CH2}
\end{figure*}

\autoref{fig:shape_vary_q0_CH} (\autoref{fig:shape_vary_q0_CH2}) shows the \ch (\chtwo) target version of \autoref{fig:shape_vary_q0}, demonstrating that, as for the \argon case, even when analysing the non-constant behavior of model predictions in \enutrue following a cut on \qz the span of model predictions is large relative to the precision required for a \lownu analysis to provide a useful flux constraint. It is similarly evident that lower \qz cuts do not provide better agreement between models, as would naively be expected from \autoref{eq:low-nu}, indicating the premise of the method is likely broken by the corrections induced by relevant nuclear-medium effects, even for a relatively light hydrocarbon targets. The results are qualitatively very similar for the \ch and \chtwo cases, with slightly more variation in the \numub case. This is unsurprising as the low-\qz contributions are CCQE dominated, and there is no \numu--hydrogen CCQE contribution, whereas there is one for \numub--hydrogen.

\begin{figure*}[htbp]
  \centering
  \captionsetup[subfloat]{captionskip=-3pt}
  \subfloat[\numu--\ch, $\ehadtrue \leq 0.3$ GeV]  {\includegraphics[width=0.29\linewidth]{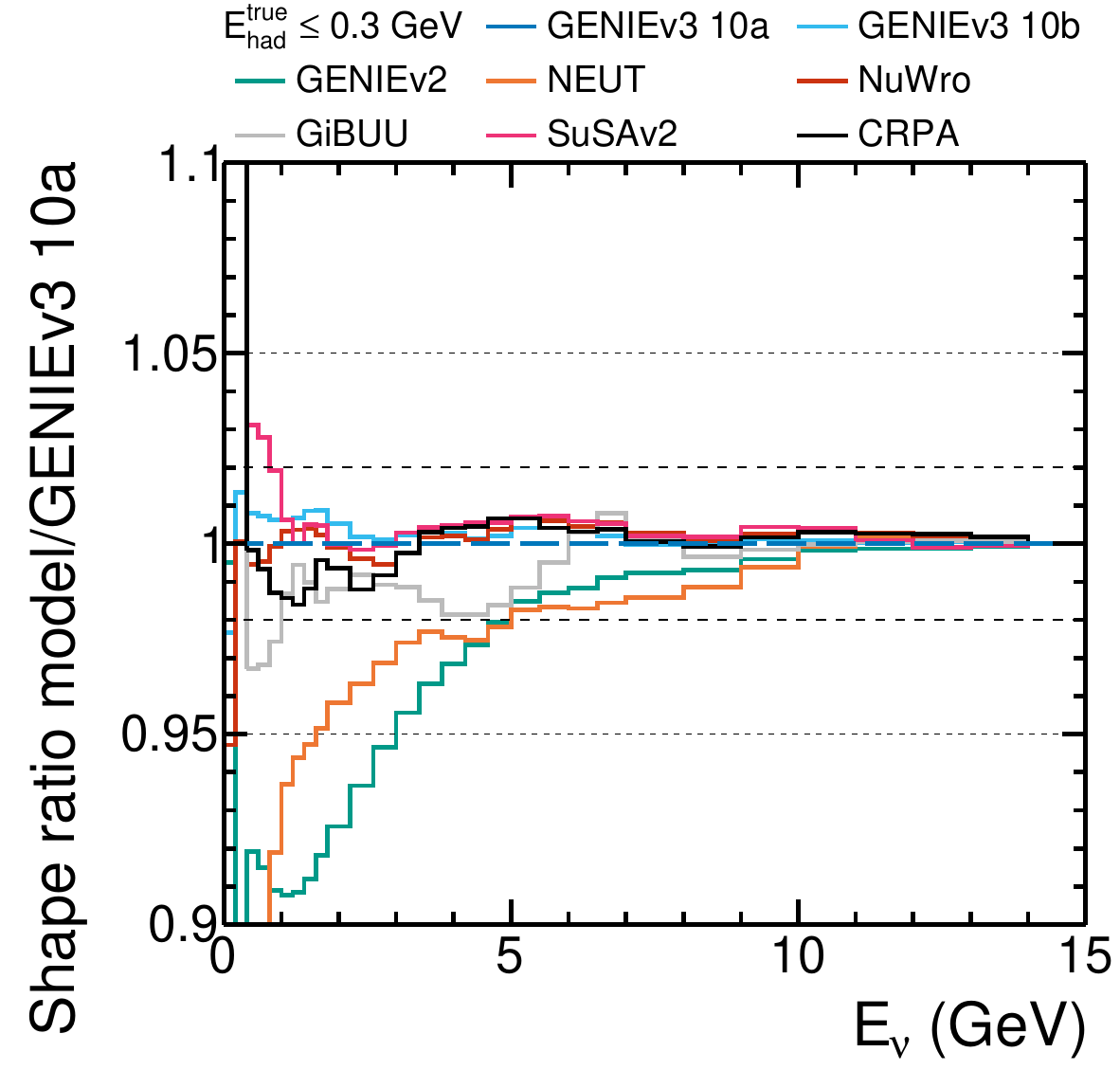}}
  \subfloat[\numu--\ch, $\ehadreco \leq 0.3$ GeV]  {\includegraphics[width=0.29\linewidth]{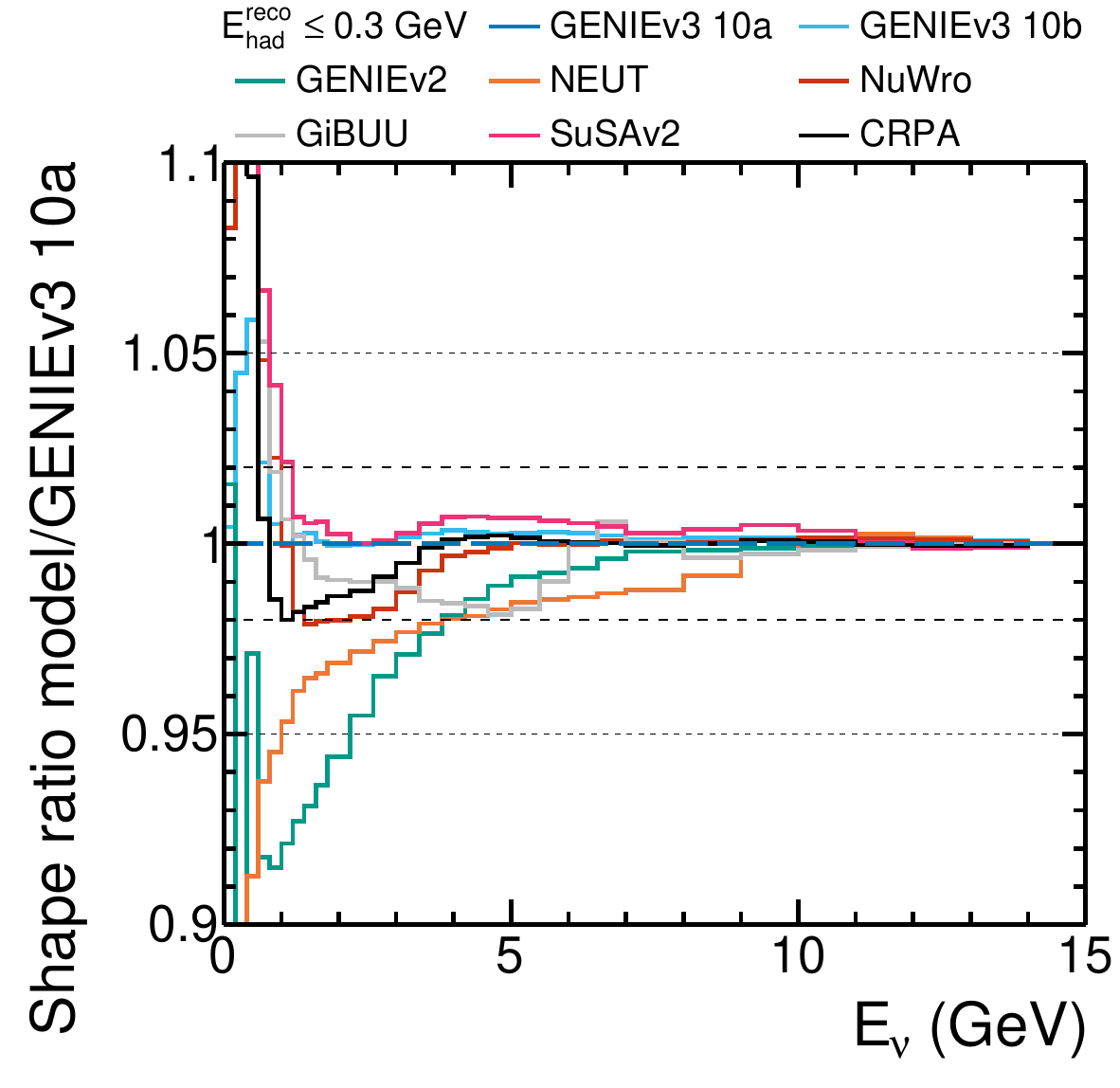}}
  \subfloat[\numu--\ch, $\eavail \leq 0.3$ GeV]    {\includegraphics[width=0.29\linewidth]{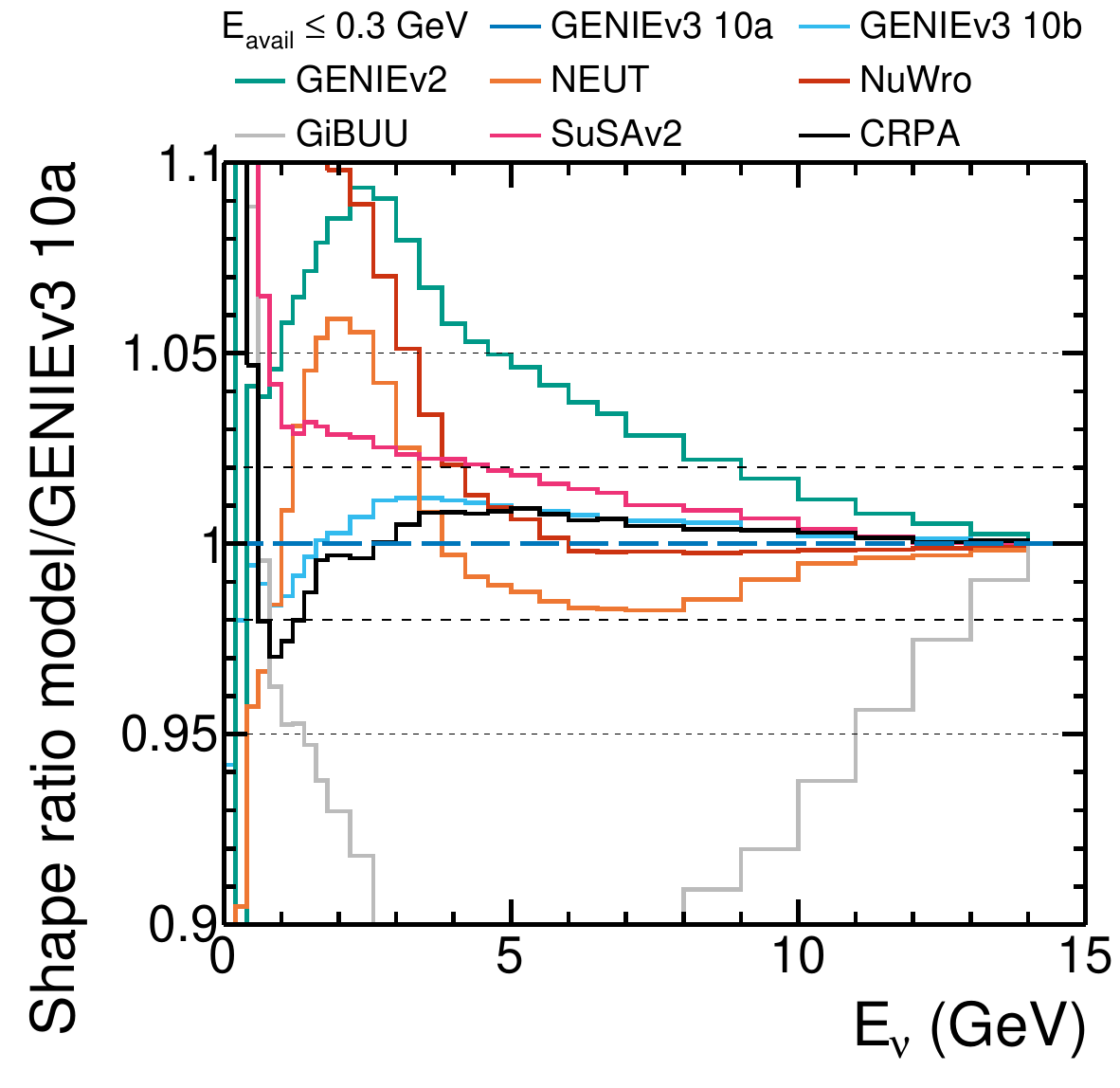}}\\\vspace{-7pt}
  \subfloat[\numub--\ch, $\ehadtrue \leq 0.3$ GeV] {\includegraphics[width=0.29\linewidth]{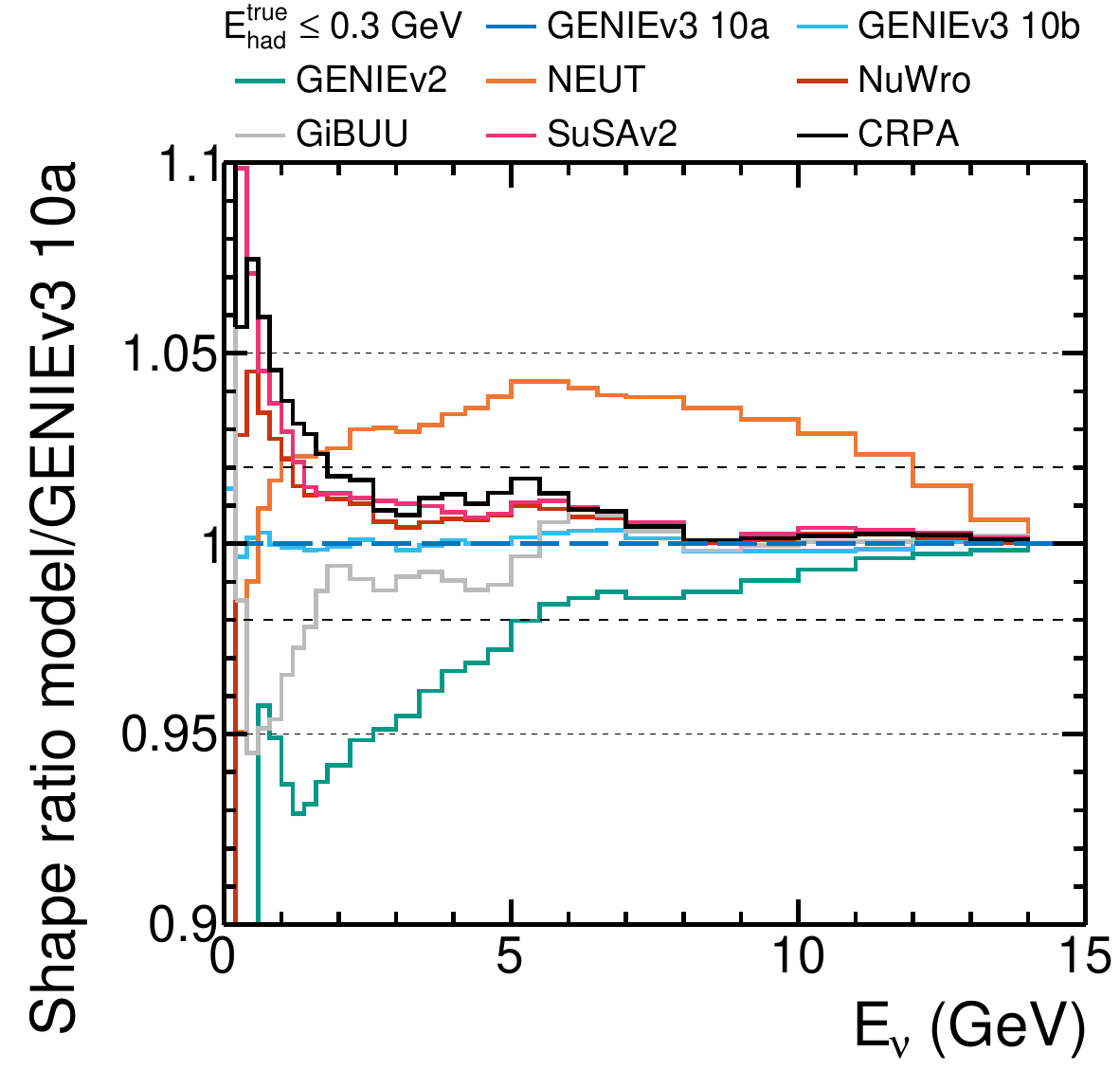}}
  \subfloat[\numub--\ch, $\ehadreco \leq 0.3$ GeV] {\includegraphics[width=0.29\linewidth]{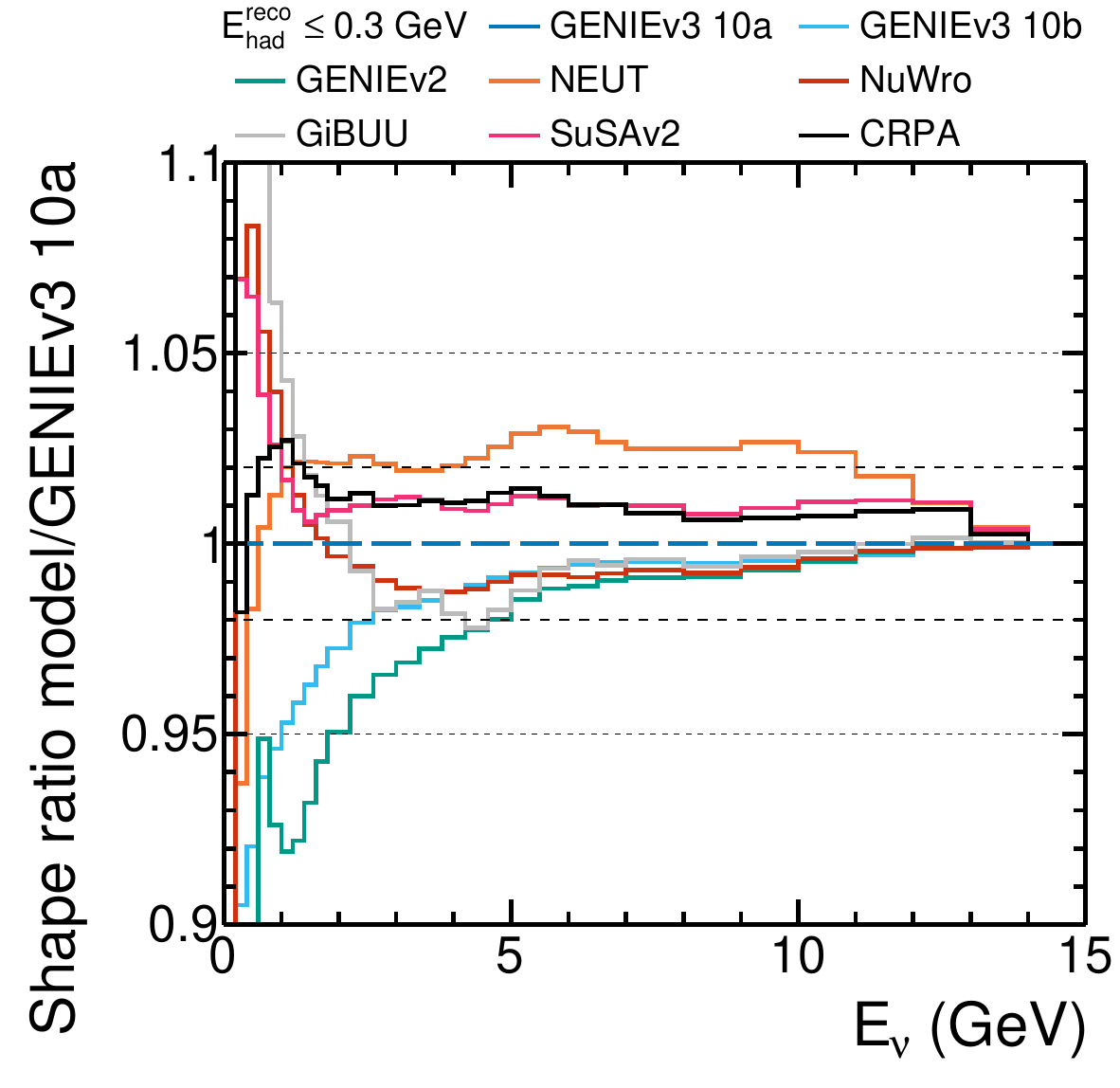}}
  \subfloat[\numub--\ch, $\eavail \leq 0.3$ GeV]   {\includegraphics[width=0.29\linewidth]{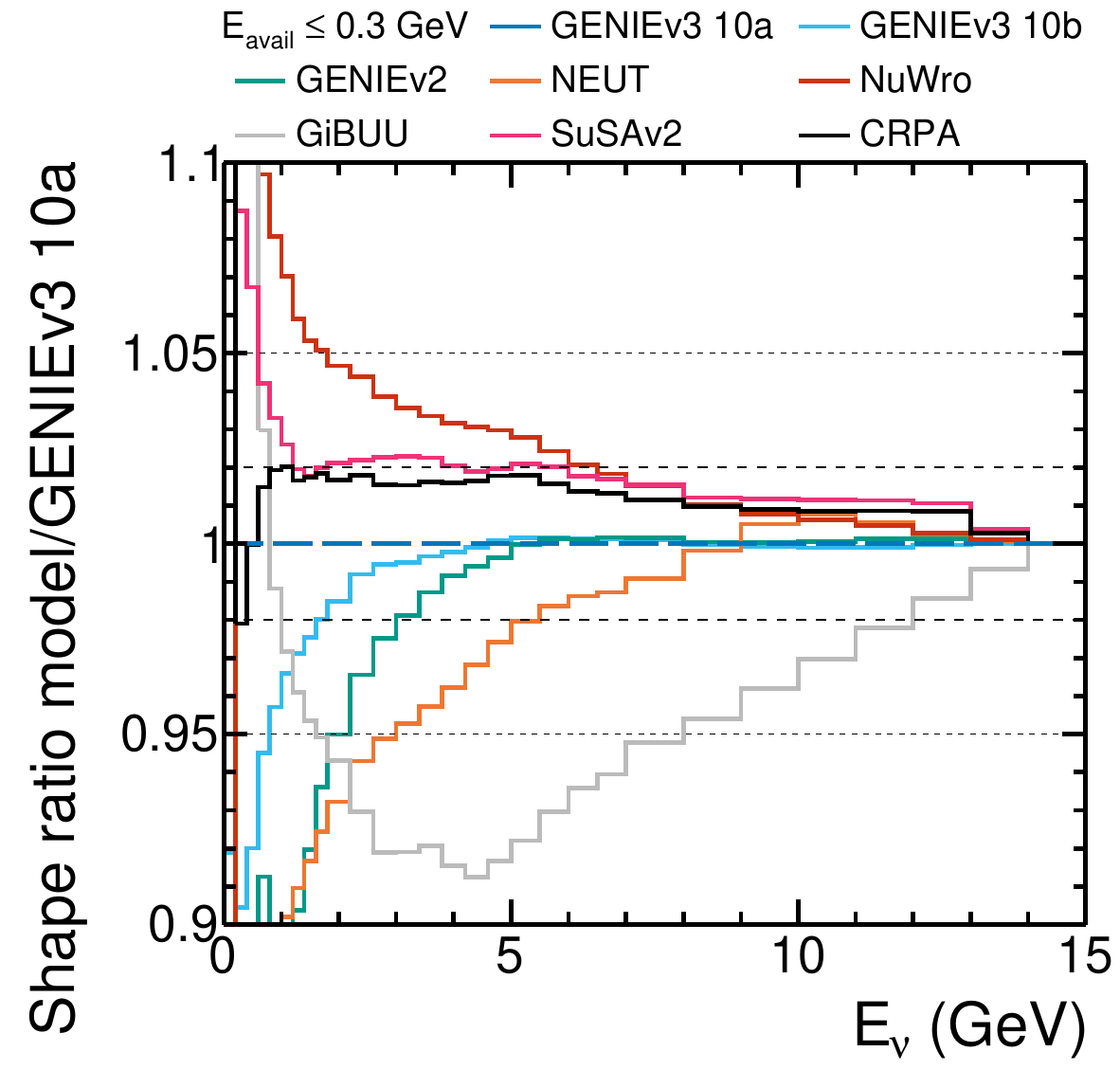}}
  \vspace{-1mm}
  \caption{Comparison of the shape-only ratios (with respect to the GENIEv3 10a prediction) of the \numu--\ch and \numub--\ch charged-current cross section with a cut on 0.3 GeV for the three different proxy variables, \ehadtrue, \ehadreco and \eavail, as a function of \enutrue. Figures produced using the flat neutrino flux described in \autoref{sec:low-qz-cross-section}. Horizontal long (short) dashed lines have been added at $\pm$2\% ($\pm$5\%), to guide the eye in assessing the scale of the bias.\vspace{-5mm}}
  \label{fig:shape_vary_proxy_CH}
\end{figure*}

\begin{figure*}[htbp]
  \centering
  \captionsetup[subfloat]{captionskip=-3pt}
  \subfloat[\numu--\chtwo, $\ehadtrue \leq 0.3$ GeV]  {\includegraphics[width=0.29\linewidth]{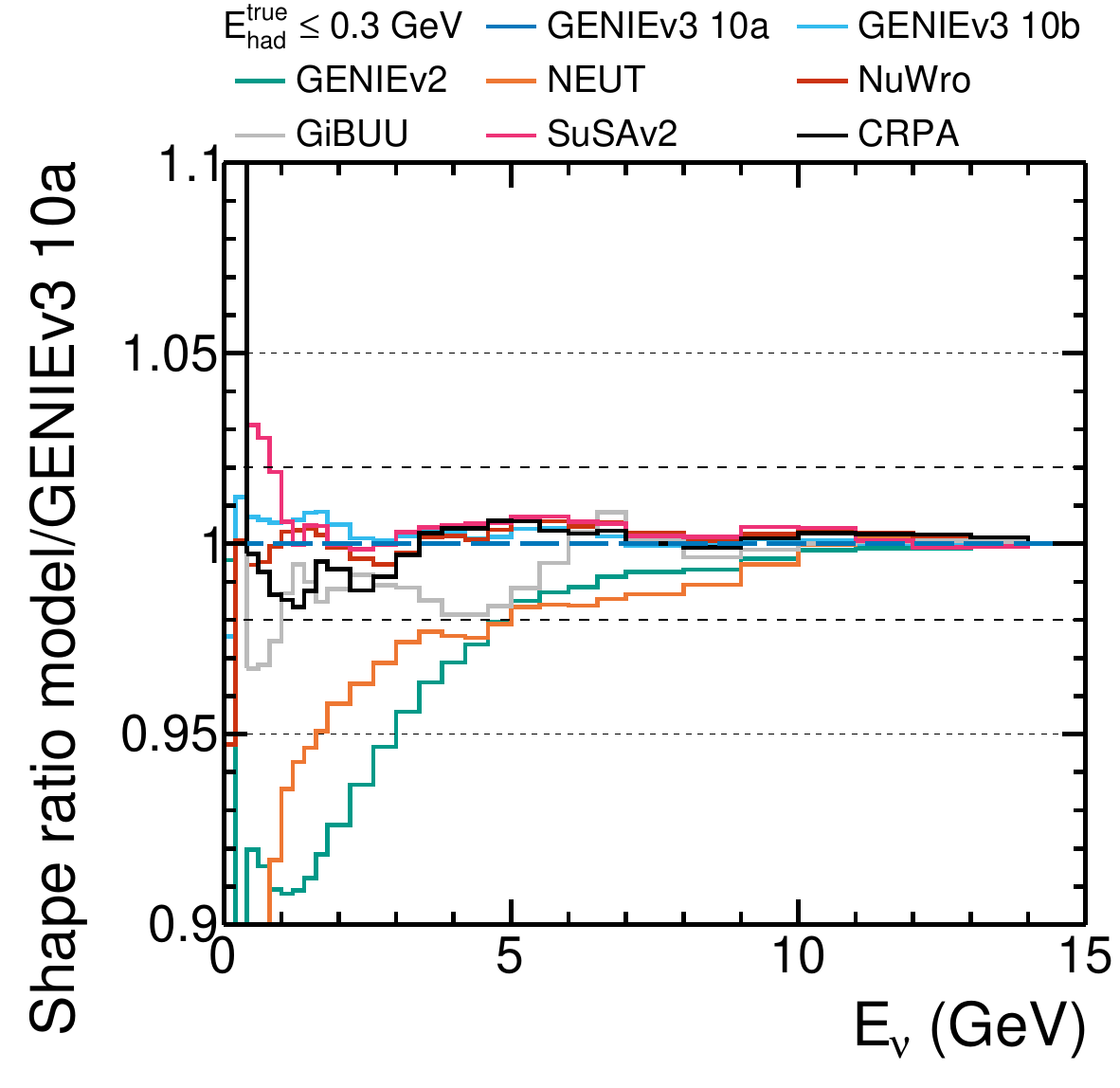}}
  \subfloat[\numu--\chtwo, $\ehadreco \leq 0.3$ GeV]  {\includegraphics[width=0.29\linewidth]{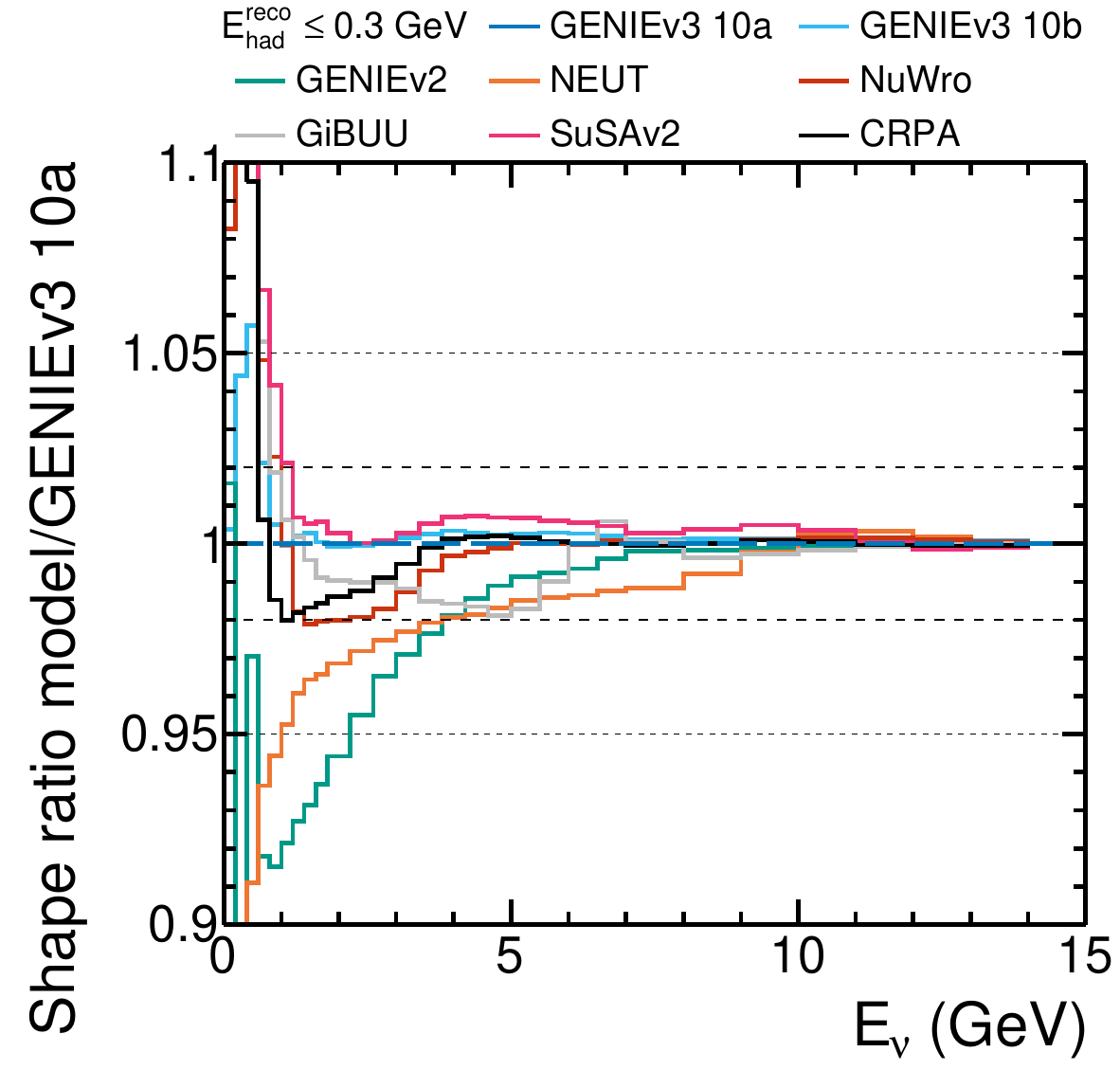}}
  \subfloat[\numu--\chtwo, $\eavail \leq 0.3$ GeV]    {\includegraphics[width=0.29\linewidth]{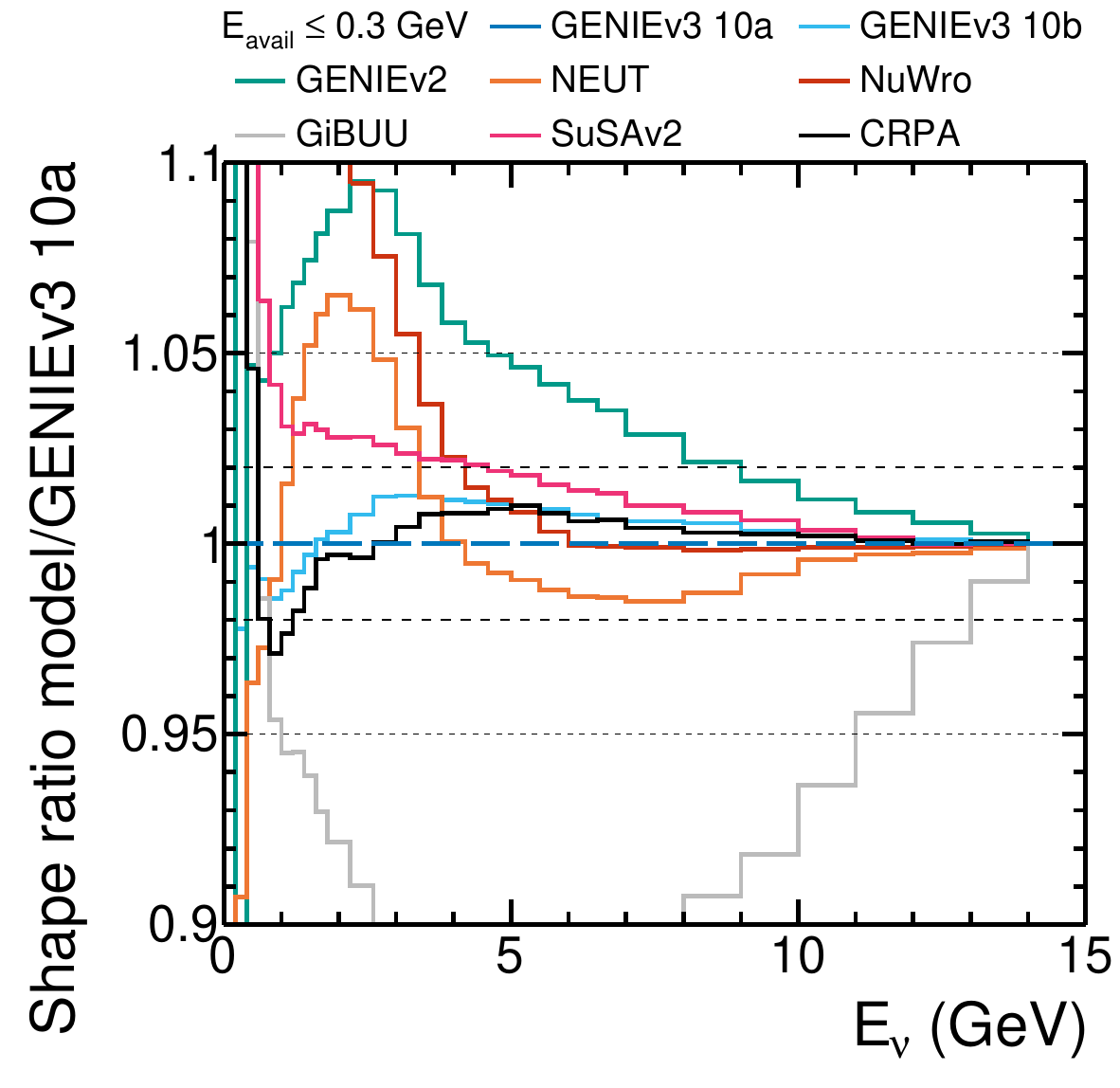}}\\\vspace{-7pt}
  \subfloat[\numub--\chtwo, $\ehadtrue \leq 0.3$ GeV] {\includegraphics[width=0.29\linewidth]{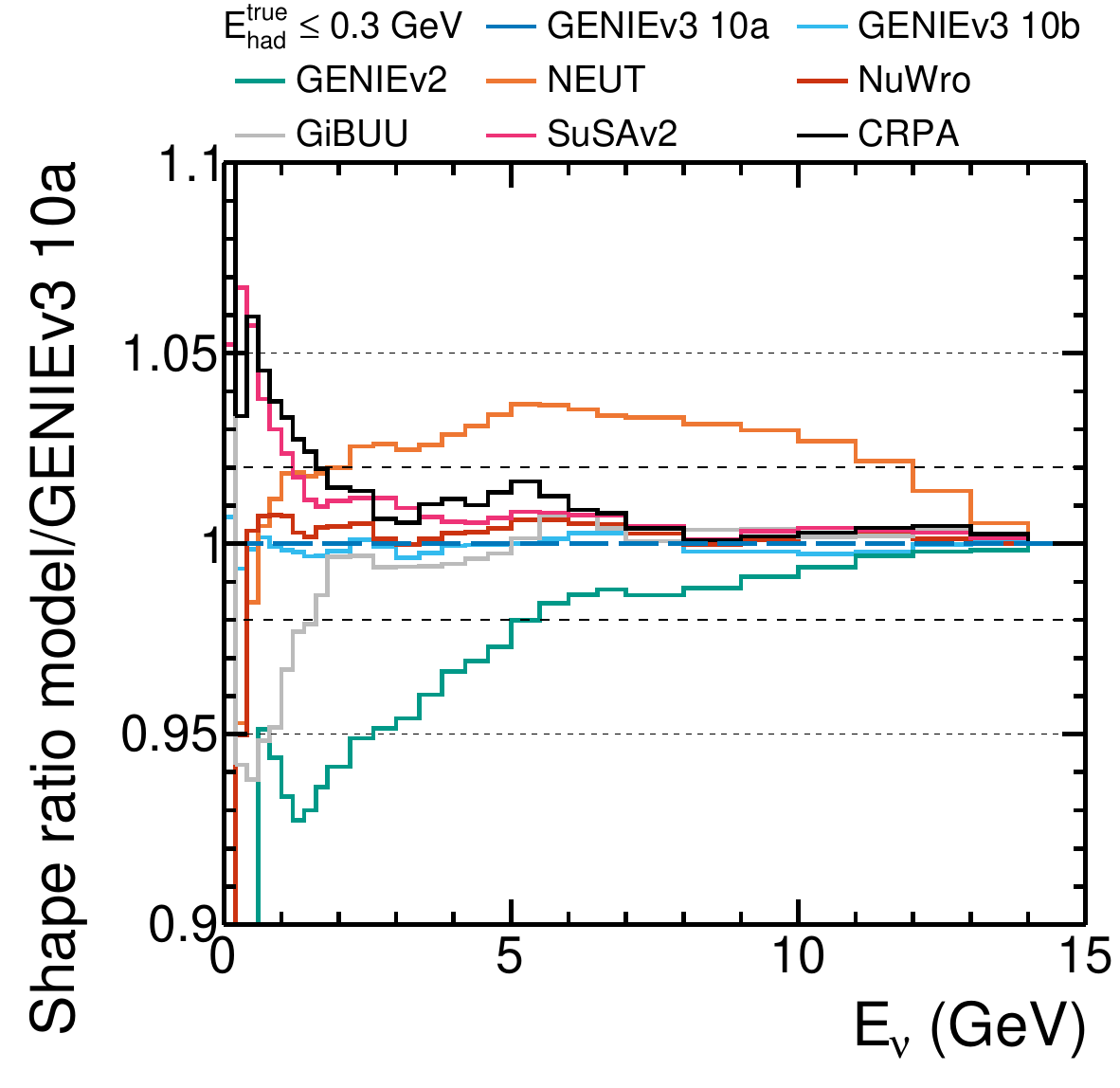}}
  \subfloat[\numub--\chtwo, $\ehadreco \leq 0.3$ GeV] {\includegraphics[width=0.29\linewidth]{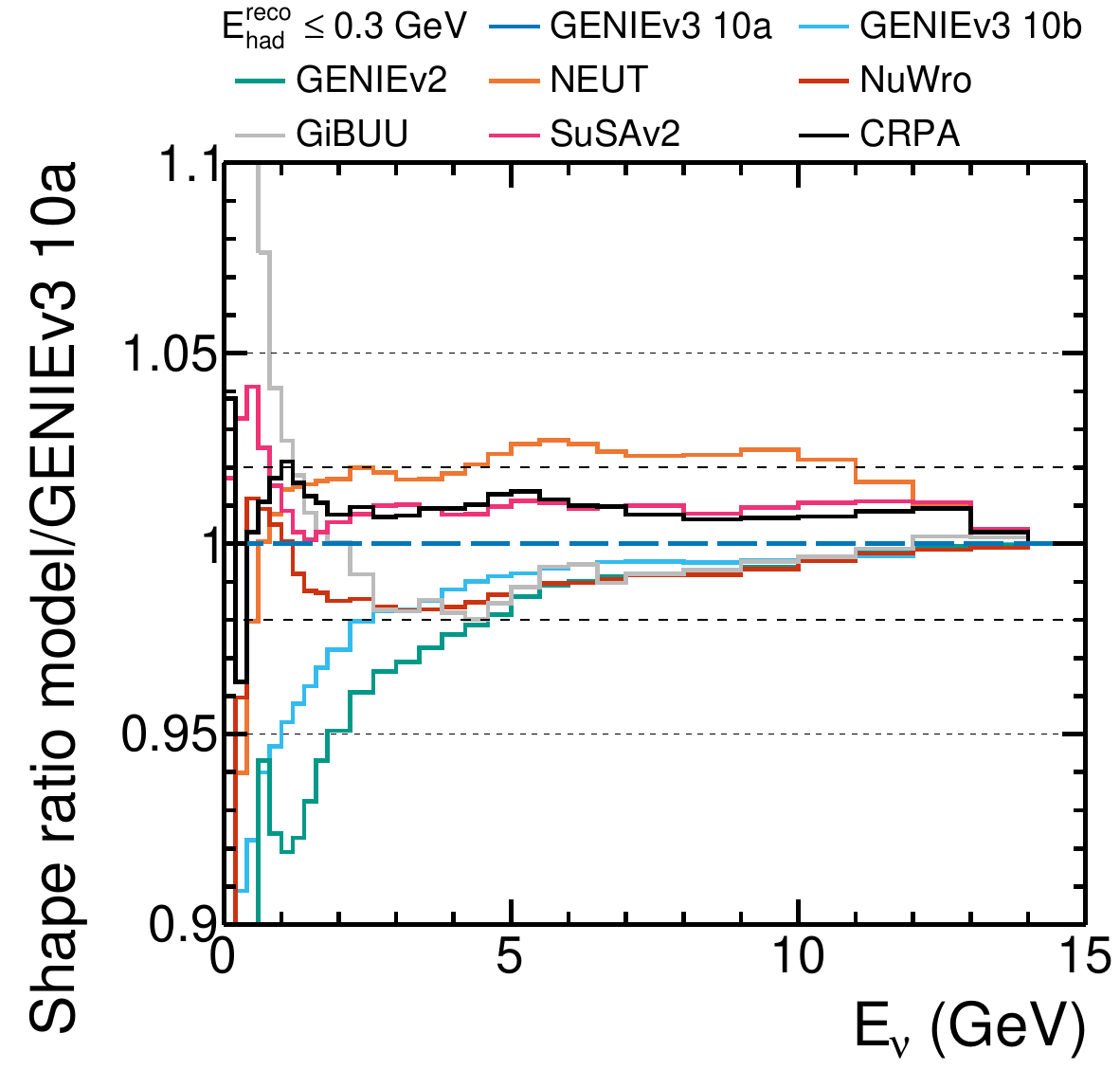}}
  \subfloat[\numub--\chtwo, $\eavail \leq 0.3$ GeV]   {\includegraphics[width=0.29\linewidth]{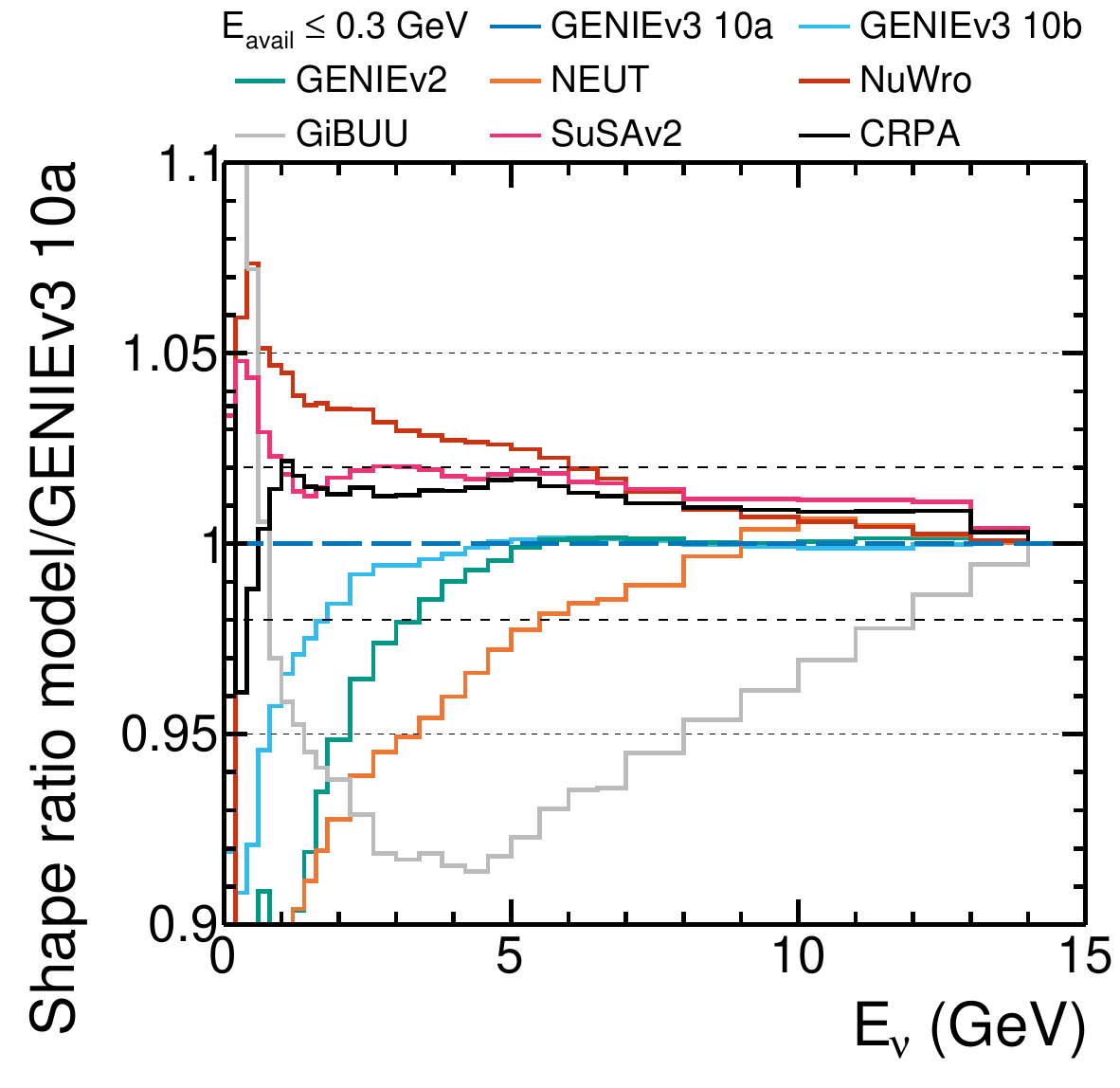}}
  \vspace{-1mm}
  \caption{Comparison of the shape-only ratios (with respect to the GENIEv3 10a prediction) of the \numu--\chtwo and \numub--\chtwo charged-current cross section with a cut on 0.3 GeV for the three different proxy variables, \ehadtrue, \ehadreco and \eavail, as a function of \enutrue. Figures produced using the flat neutrino flux described in \autoref{sec:low-qz-cross-section}. Horizontal long (short) dashed lines have been added at $\pm$2\% ($\pm$5\%), to guide the eye in assessing the scale of the bias.\vspace{-5mm}}
  \label{fig:shape_vary_proxy_CH2}
\end{figure*}

\autoref{fig:shape_vary_proxy_CH} (\autoref{fig:shape_vary_proxy_CH2}) shows the \ch (\chtwo) target version of \autoref{fig:shape_vary_proxy}, analyzing more realistic applications of the \lownu method where a cut is placed on a proxy for \qz rather than \qz itself (as detailed in \autoref{sec:variables}). Whilst significant migration from high \qz is introduced from the use of a proxy variable, particularly at low \enutrue, the model agreement does not become substantially worse for the two more optimistic proxies (\ehadreco and \ehadtrue). In the case of a detector which missed neutron energy and is unable to identify charged pion multiplicity (i.e. where the \eavail proxy is most pertinent) the model spread increases drastically, as it did for the \argon case. The results are qualitatively very similar for the \ch and \chtwo cases, as when cutting on true \qz.

Overall, our assertion that the applicability of the \lownu method for neutrinos with $\enutrue \leq 5~\text{GeV}$ is likely undermined by large uncertainties in nuclear-effect modeling is shown to apply equally to \ch and \chtwo as well as \argon targets. This suggests that issue really lies in the general modeling of relevant nuclear effects before even considering their application to the particularly challenging case of \argon. 

\bibliographystyle{utphys}
\bibliography{low-nu.bib}

\end{document}